\newcommand{\mydriver}{pdflatex} 

\documentclass[12pt,\mydriver ]{thesis}  

\usepackage{titlesec}
   \titleformat{\chapter}
      {\normalfont\large}{Chapter \thechapter:}{1em}{}
\usepackage{tikz}
\usepackage{times}
\usepackage[USenglish]{babel}
\usepackage{subcaption}
\usepackage{graphicx}
\usepackage{lscape}
\usepackage{indentfirst}
\usepackage{latexsym}
\usepackage{multirow}
\usepackage{epstopdf}
\usepackage{tabls}
\usepackage{wrapfig}
\usepackage{slashbox}
\usepackage{float}
\usepackage{caption}
\usepackage{supertabular}
\usepackage[onehalfspacing]{setspace} 

\usepackage[colorlinks=true,urlcolor=magenta,linkcolor=blue,citecolor=red]{hyperref}

\usepackage{tocloft,calc}
\renewcommand{\cftlottitlefont}{\hspace*{\fill}\large}
\renewcommand{\cftafterlottitle}{\hspace*{\fill}}
\renewcommand{\cftloftitlefont}{\hspace*{\fill}\large}
\renewcommand{\cftafterloftitle}{\hspace*{\fill}}
\renewcommand{\cftchapaftersnum}{:\ }
\renewcommand{\cftchappresnum}{\chaptername\space}
\makeatletter
\g@addto@macro\appendix{%
  \addtocontents{toc}{%
    \protect%
  }%
}
\renewcommand{\baselinestretch}{2}

\usepackage{doi}
\usepackage[square,numbers,comma,sort&compress]{natbib}
\usepackage{multirow}
\usepackage{dsfont}
\usepackage{braket}
\usepackage{bm}
\usepackage{dsfont}
\usepackage{dcolumn}
\usepackage{amssymb}
\usepackage{subcaption}
\usepackage{amsmath}
\usepackage{mathtools}
\usepackage{bm}
\usepackage{cancel}
\usepackage{slashed}
\newcommand{\obar}{\underline{1}}
\newcommand{\nui}{\nu_{\underline{1}}}
\newcommand{\num}{\nu_0}
\newcommand{\nuo}{\nu_1}
\newcommand{\hnui}{\hat{\nu}_{\underline{1}}}
\newcommand{\hnum}{\hat{\nu}_0}
\newcommand{\hnuo}{\hat{\nu}_1}
\newcommand{\ladder}{\Gamma}
\newcommand{\ApBp}{ A(\obar)^\dagger \cdot B(1)^\dagger}
\newcommand{\BpAp}{ B(\obar)^\dagger \cdot A(1)^\dagger}

\newcommand{\kket}[1]{\left| \left| #1 \right\rangle \right.}

\newcommand{\goldieotimes}{%
  \mathop{\mathchoice{\textstyle\bigotimes}{\bigotimes}{\bigotimes}{\bigotimes}}%
}
\usepackage{array}
\newcolumntype{C}[1]{>{\centering\let\newline\\\arraybackslash\hspace{2pt}}m{#1}}
\newcolumntype{R}[1]{>{\centering\let\newline\\\arraybackslash\hspace{0pt}}m{#1}}
\usepackage{multirow}
\usepackage{soul}

\begin{document}
\pagestyle{empty}

\hbox{\ } \vspace{.7in}
\renewcommand{\baselinestretch}{1}
\small \normalsize

\begin{center}
\large{{ABSTRACT}}

\vspace{3em}

\end{center}
\hspace{-.15in}
\begin{tabular}{ll}
Title of Dissertation:    & {\large  Theoretical Developments in Lattice Gauge Theory } \\
&                     {\large for Applications in Double-Beta Decay Processes}\\
&                     {\large and Quantum Simulation} \\

\ \\
&                          {\large  Saurabh Vasant Kadam} \\
&                           {\large Doctor of Philosophy, 2023} \\
\ \\
Dissertation Directed by: & {\large  Professor Zohreh Davoudi} \\
&               {\large  Department of Physics } \\
\end{tabular}

\vspace{3em}

\renewcommand{\baselinestretch}{2}
\large \normalsize
Nuclear processes have played, and continue to play, a crucial role in unraveling the fundamental laws of nature.
They are governed by the interactions between hadrons, and in order to draw reliable conclusions from their observations, it is necessary to have accurate theoretical predictions of hadronic systems.
The strong interactions between hadrons are described by quantum chromodynamics (QCD), a non-Abelian gauge theory with symmetry group SU(3).
QCD predictions require non-perturbative methods for calculating observables, and as of now, lattice QCD (LQCD) is the only reliable and systematically improvable first-principles technique for obtaining quantitative results.
LQCD numerically evaluates QCD by formulating it on a Euclidean space-time grid with a finite volume, and requires formal prescriptions to match numerical results with physical observables.

This thesis provides such prescriptions for a class of rare nuclear processes called double beta decays, using the finite volume effects in LQCD framework.
Double beta decay can occur via two different modes: two-neutrino double beta decay or neutrinoless double beta decay.
The former is a rare Standard Model transition that has been observed, while the latter is a hypothetical process whose observation can profoundly impact our understating of Particle Physics.
The significance and challenges associated with accurately predicting decay rates for both modes are emphasized in this thesis, and matching relations are provided to obtain the decay rate in the two-nucleon sector.
These relations map the hadronic decay amplitudes to quantities that are accessible via LQCD calculations, namely the nuclear matrix elements and two-nucleon energy spectra in a finite volume.
Finally, the matching relations are employed to examine the impact of uncertainties in the future LQCD calculations.
In particular, the precision of LQCD results that allow constraining the low energy constants that parameterize the hadronic amplitudes of two-nucleon double beta decays is determined.

Lattice QCD, albeit being a very successful framework, has several limitations when general finite-density and real-time quantities are concerned.
Hamiltonian simulation of QCD is another non-perturbative method of solving QCD that, by its nature, does not suffer from those limitations.
With the advent of novel computational tools, like tensor network methods and quantum simulation, Hamiltonian simulation of lattice gauge theories (LGTs) has become a reality.
However, different Hamiltonian formulations of the same LGT can lead to different computational-resource requirements with their respective system sizes.
Thus, a search for efficient formulations of Hamiltonian LGT is a necessary step towards employing this method to calculate a range of QCD observables.
Toward that goal, a loop-string-hadron (LSH) formulation of an SU(3) LGT coupled to dynamical matter in 1+1 dimensions is developed in this thesis.
Development of this framework is motivated by recent studies of the LSH formulation of an SU(2) LGT that is shown to be advantageous over other formulations, and can be extended to higher-dimensional theories and ultimately QCD.

\thispagestyle{empty} \hbox{\ } \vspace{1.5in}
\renewcommand{\baselinestretch}{1}
\small\normalsize
\begin{center}

\large{{Theoretical Developments in Lattice Gauge Theory

for Applications in Double-beta Decay Processes and Quantum Simulation}}\\
\ \\
\ \\
\large{by} \\
\ \\
\large{Saurabh Vasant Kadam}
\ \\
\ \\
\ \\
\ \\
\normalsize
Dissertation submitted to the Faculty of the Graduate School of the \\
University of Maryland, College Park in partial fulfillment \\
of the requirements for the degree of \\
Doctor of Philosophy \\
2023
\end{center}

\vspace{7.5em}

\noindent Advisory Committee: \\
\hbox{\ }\hspace{.5in}Professor Zohreh Davoudi, Chair/Advisor \\
\hbox{\ }\hspace{.5in}Professor Paulo Bedaque\\
\hbox{\ }\hspace{.5in}Professor Zackaria Chacko\\
\hbox{\ }\hspace{.5in}Professor Manuel Franco Sevilla\ \\
\hbox{\ }\hspace{.5in}Professor Konstantina Trivisa

\thispagestyle{empty}
\hbox{\ }

\vfill
\renewcommand{\baselinestretch}{1}
\small\normalsize

\vspace{.5in}

\begin{center}
\large{\copyright \hbox{ }Copyright by\\
Saurabh Vasant Kadam  
\\
2023}
\end{center}

\vfill

\newpage 

\pagestyle{plain} \pagenumbering{roman} \setcounter{page}{2}

\renewcommand{\baselinestretch}{2}
\small\normalsize
\hbox{\ }
 
\vspace{.5in}

\begin{center}
\large{Dedication}
\end{center} 
\begin{center}
To my parents, \textit{Vasant} and \textit{Lata Kadam}.
\end{center}

\renewcommand{\baselinestretch}{2}
\small\normalsize
\hbox{\ }
 
\vspace{.5in}

\begin{center}
\large{Acknowledgments} 
\end{center} 

\vspace{1ex}

I am profoundly grateful to numerous individuals who played important roles in my path toward becoming a physicist.
This journey would have been unattainable without the support and guidance of my advisor, Prof. Zohreh Davoudi.
Her unwavering commitment to academic excellence and dedication for perfection have been a constant source of inspiration.
Her role as an advisor was not limited to helping me with my research but extended well beyond, making me a better academic.
She guided me in identifying important problems to work on, pushed me when I was slacking, and helped me when I was stuck.
She also made me realize the importance of presenting my work to a wider audience and establishing collaborations.
She encouraged me to take a more active role in teaching when I worked as a teaching assistant for the courses she taught.
This was a valuable opportunity for me to hone my teaching skills - an experience I deeply appreciated due to my passion for teaching.
Finally, Zohreh's expertise and interests in a wide range of physics topics allowed me to explore different branches of physics.
It gave me a broader perspective on how to further scientific advancements across various fronts and work towards achieving shared goals among different physics communities.

I would like to acknowledge the invaluable contributions of my collaborators, Jesse R. Stryker and Indrakshi Raychowdhury.
I thank them for the insights and perspectives they provided during our collaboration whose outcome is now a part of this thesis.

I am indebted to my parents for my achievements.
In a culture that considers education purely as a means of livelihood and does not encourage the pursuit of knowledge for intellectual curiosity and passion, my parents gave me the freedom to make choices for myself and supported me when I decided to pursue my career in physics, a decision questioned and ridiculed by many.
Everything else would not have been possible without their strong faith in me. 

Popular science books punctured the bubble of ignorance surrounding me while growing up in a small town in India.
They inspired me to be a physicist by instilling curiosity in me and introduced me to intriguing concepts like quantum physics and general relativity in fun and exciting ways.
Two in particular that I am thankful for are `\textit{Kimayagar}' by Achyut Godbole and `\textit{Einstein for Everyone}' by Robert L. Piccioni. 

I am grateful to have learned from excellent professors such as: Nabamita Banerjee, Bhas Bapat, Seema Sharma, Arvind Kumar, and many others.
Many thanks to professors Tom Cohen, Paulo Bedaque, Zackaria Chacko and Shmuel Nussinov for their valuable guidance and educating conversations.
I want to express my gratitude towards my undergraduate institute, IISER Pune, for providing me with an enriching and intellectually stimulating environment.
It allowed me to test my abilities to conduct research in physics and gave me the confidence to pursue a PhD.

Finally, I would also like to extend my heartfelt thanks to my friends for making this journey enjoyable.
Shruti Chakravarty and Rajeev Singh Rathour, life would have been much harder without your friendship.
Thank you Mitali Thatte, Kaustubh Deshpande, Kalyani Kataria, Michael Winer, and William Grunow for all the conversations that gave me new perspectives and often changed my views and beliefs.
Thank you Mrunal Korwar, Nandini Hazra, Deepak Sathyan, Sanket Doshi, Yukari Yamauchi, Dan Zhang, Sagar Airen, Edward Broadberry, Emily Jiang, Spandan Pathak, Chung-Chun Hsieh, Batoul Banihashemi, Reza Ebadi, Shahriar Keshvari, Gautam Nambiar, Abu Musa Patoary, and Amit Vikram for your friendship and laughter. 
    \cleardoublepage
    \phantomsection
    \addcontentsline{toc}{chapter}{Table of Contents}
    \renewcommand{\contentsname}{Table of Contents}
\renewcommand{\baselinestretch}{1}
\small\normalsize
\tableofcontents 
\newpage
\addcontentsline{toc}{chapter}{List of Tables}
    \renewcommand{\contentsname}{List of Tables}
\listoftables 
\newpage
\addcontentsline{toc}{chapter}{List of Figures}
    \renewcommand{\contentsname}{List of Figures}
\listoffigures 
\newpage
\addcontentsline{toc}{chapter}{List of Abbreviations}

\renewcommand{\baselinestretch}{1}
\small\normalsize
\hbox{\ }

\vspace{.5in}

\begin{center}
\large{List of Abbreviations}
\end{center} 

\vspace{3pt}

\begin{supertabular}{ll}
QED & Quantum electrodynamics\\
SM & Standard Model\\
QCD & Quantum chromodynamics\\
LQCD & Lattice quantum chromodynamics\\
FV & Finite volume\\
BSM & Beyond standard model\\
$2\nu\beta\beta$ & Two-neutrino double beta\\
$0\nu\beta\beta$ & Neutrinoless double beta\\
EFT & Effective field theory\\
LEC & Low energy coefficients\\
ME & Matrix element\\
$NN$ & Two-nucleon\\
LO & Leading order\\
NLO & Next-to-leading order\\
CM & Center of mass\\
ERE & Effective range expansion\\
${^1S_0}$ & Spin-singlet\\
${^3S_1}$ & Spin-triplet\\
MeV & Mega-electron volt\\
GeV & Giga-electron volt\\
KSW & Kaplan-Savage-Wise\\
MS & Minimal subtraction\\
PDS & Power divergent subtraction\\
CC & Charged-current\\
UV & Ultraviolet\\
IR & Infrared\\
RG & Renormalization group\\
LNV & Lepton-number violating\\
LGT & Lattice gauge theory\\
KS & Kogut-Susskind\\
irrep & irreducible representation\\
ISB & Irreducible Schwinger boson\\
\end{supertabular}

\newpage
\setlength{\parskip}{0em}
\renewcommand{\baselinestretch}{2}
\small\normalsize

\setcounter{page}{1}
\pagenumbering{arabic}

\renewcommand{\thechapter}{1}

\chapter{Introduction 
\label{ch: Intro}
}
\noindent
Quantum field theory is a framework that combines three major ideas of modern physics: the quantum theory, the concept of fields, and the principle of relativity.
Early applications of this framework in the 1920s described the interactions between electrons and photons.
This led to the development of theory of quantum electrodynamics (QED), a theory that describes the electromagnetic force between particles.
To date, QED remains one of the most accurately tested physical theories.
Later developments in particle physics led to the discovery of a plethora of sub-atomic particles along with two more forces, the weak force and the strong force that are responsible for radioactive decays in nuclei, such as the beta-decay, and the binding of atomic nuclei, respectively.
Unlike QED, which is an Abelian gauge theory with symmetry group $U(1)$, these two forces hinted a non-Abelian nature.
The framework of quantum field theory was then successfully extended in the later half of the 20th century to describe these forces and particles, which is the well established standard model (SM) of particle physics.

The SM has enjoyed successful predictions of new particles and their properties with a good precision.
However, not all components of the SM have the ultra-precise predictive power of QED.
The remarkable power of QED predictions relies upon its perturbative nature, that is, the dimensionless fine structure constant which measures the strength of the electromagnetic interactions is small, and observables can be systematically computed order-by-order in powers of this coupling constant.
The smallness of the coupling ensures that the higher order contributions can be ignored, and observables can be calculated up to a finite order in this expansion to achieve the desired precision on its prediction.

Quantum chromodynamics (QCD), the sector of SM that describes the strong force, does not have the precise analytical predictive power at all energy scales.
QCD is a non-Abelian gauge theory with local symmetry group SU(3), and it exhibits \textit{asymptotic freedom} which means the running of QCD coupling through the renormalization group procedure indicates that its value is small at high energies and large at low energies.
Thus, analytical perturbative calculations predict QCD observables only at high energies where it is weakly interacting, and fail to predict low energy observables accurately where QCD becomes strongly interacting.
Furthermore, at energy scales near the confinement scale, $\Lambda_{\rm QCD}\sim \mathcal{O}(10^2) {\rm\;MeV}$, QCD \textit{confines}, meaning the fundamental degrees of freedom, i.e. quarks and gluons,  cannot exist as free particles and always appear as bound states known as \textit{hadrons}, e.g pions, neutrons, protons etc.
The phenomenon of confinement together with running of QCD coupling requires non-perturbative calculations for calculating observables involving interactions between these \textit{effective} hadronic degrees of freedom.
However, as of now, a general non-perturbative analytical way of solving QCD does not exist.
Nonetheless, there exists a numerical method to solve QCD non-perturbatively from first principles known as lattice QCD (LQCD).
Lattice QCD is a lattice gauge theory (LGT) in which the theory of QCD is formulated on a discrete space-time with a finite size, and the observables are calculated using the path integral formulation with an imaginary time.
This allows the path integral to be interpreted as a probability distribution function over various gauge-field configurations, which can be generated via Monte-Carlo sampling methods.
To date, LQCD remains the only non-perturbative method for predicting QCD observables.
This thesis deals with formal topics in LGT with a focus on LQCD, its applications in a few cases, its limitations, and developments in its upcoming and rising alternatives.

Obtaining QCD predictions from numerical LQCD calculation is a non-trivial task due its imaginary time formulation in a finite volume (FV).
Decades worth of progress has led to various methods for obtaining hadron spectrum, hadronic scattering amplitudes, hadron decay amplitudes, etc., more accurately from LQCD.
The first part of this thesis extends one such method, the FV formalism, to provide a prescription for predicting the hadronic transition amplitude of an exotic nuclear decay, known as the double beta decay, from first-principles numerical LQCD calculations.

Lattice QCD, albeit being powerful and successful, has limitations.
Its shortcoming hinders exploring QCD physics in many interesting physical situations, e.g. thermodynamics of QCD at finite baryon density, hydrodynamics of QCD, and highly energetic and highly inelastic scattering processes.
In such cases, the \textit{Hamiltonian formulation} of QCD in conjunction with novel computational tools like tensor networks, quantum simulation, and quantum computation, can pose an alternative to LQCD as another non-perturbative numerical method of solving QCD, as it circumvents some of the limitations of LQCD.
The second half of this thesis contains developments in Hamiltonian formulation of non-Abelian gauge theories with an aim of obtaining computational resource efficient formulation of QCD.

The rest of this chapter contains an overview of the necessary background and puts forward the central problem statements addressed in this thesis.
In Sec.~\ref{sec: QCD overview}, the theory of QCD is reviewed through a historical lens leading up to its current form.
QCD Lagrangian and its degrees of freedom, quarks and gluons, are briefly explained along with its renormalization group analysis demonstrating the feature of asymptotic freedom and the phenomenon of confinement.

Despite being very successful in describing the particle spectrum and their interactions at short distances, the SM is not complete.
One of its limitations is its inability to explain neutrino masses.
Neutrinos are elusive particles that are electrically neutral and only interact with other particles via the weak force.
The existence of such a particle was first postulated by Wolfgang Pauli in 1930 to explain the electron emission spectrum in beta-decays~\cite{Brown:PauliLetter}, and later experimentally confirmed in 1956 by Clyde L. Cowan and Frederick Reines~\cite{Reines:1953pu}.
Further experiments confirmed that neutrinos have masses and they come in three different flavors:  electron neutrinos, muon neutrinos and tau neutrinos associated with the corresponding charged leptons, the electron, muon, and tau, respectively.
Their masses allow them to oscillate between these flavors and measurements of such neutrino flavor oscillations along with other neutrino mass measurements have put constraints on their masses.

The SM predicts that neutrinos have no mass. Thus, the existence of neutrino masses is solid evidence for particle physics beyond the SM (BSM).
There are many BSM scenarios which allow for neutrino masses, and their predictions depend on BSM parameters that model the new physics.
Thus, accurate measurements of (or tighter constraints on) neutrino masses help us understand the allowed possibilities of new physics.
Since neutrinos are charge neutral particles, some BSM models allow neutrinos to have a Majorana mass, which arises from identifying the anti-particle of neutrinos, anti-neutrinos, to be the same as neutrinos.
An unavoidable prediction of allowing neutrinos to have a Majorana mass is the existence of an exotic nuclear transition known as the neutrinoless double beta decay.

Double beta decay is a rare transition between two nuclei with same mass number where two neutrons in the parent nucleus convert to two protons, changing the atomic number by two units.
Chapter~\ref{ch: DBD from LQCD} gives a brief overview of these transitions and their implications.
It elaborates on two modes through which such a transition can occur: the two-neutrino double beta ($2\nu\beta\beta$) decay, and the neutrinoless double-beta ($0\nu\beta\beta$) decay.
The former is allowed within the SM, and it has been observed in a dozen nuclei.
The latter, on the other hand, violates lepton number conservation by two units and is therefore forbidden in the SM.
While several experiments searching for this decay have been performed in the past, such a transition has never been observed, and an extensive experimental program continues to seek evidence for $0\nu\beta\beta$ decays.

The observation of $0\nu\beta\beta$ decay would immediately imply that neutrinos are Majorana particles.
Results from future experiments, either a detection or not, will constrain the BSM parameters for a variety of lepton number violating (LNV) BSM scenarios through bounds on half-life values.
Section~\ref{sec: 0vbb} reviews the light neutrino exchange scenario that is the minimal deviation from the SM.
In order to have precise constraints on BSM parameters, one needs to know the relation between the value of the decay half-life and the corresponding BSM parameters accurately.
Furthermore, since the $2\nu\beta\beta$ decay transition occurs in the same atomic nuclei that allow for a $0\nu\beta\beta$ decay, precise predictions of $2\nu\beta\beta$ decay half-lives within the SM are also important.
This decay forms the dominant background in the $0\nu\beta\beta$ decay search experiments, and its half-life value is also useful in our understanding of nuclear spectroscopy and to test some of the BSM hypotheses regarding neutrino properties like the \textit{bosonic} neutrino and the \textit{self-interacting} neutrino, see Ch.~\ref{ch: DBD from LQCD} for details.

Predictions of double beta decay half-life values involve calculations involving over a wide range of energy scales, from the high energy scale of the new physics to the energy scale of atomic nuclei of experimentally relevant isotopes.
The atomic nuclei are bound states of nucleons, i.e. neutrons and protons, which are again made up of quarks interacting via the strong force.
Thus, the double beta decay half-life predictions involve QCD calculations to predict the decay rate at the nuclei level using the fundamental degrees of freedom like quarks and gluons.
In the case of $2\nu\beta\beta$ decay, the interactions between quarks and gluons are completely determined by the SM, while for the $0\nu\beta\beta$ decay quarks and gluons have additional interactions characterized by the LNV BSM physics and the associated BSM parameters.
As mentioned before, the running of QCD coupling and confinement makes it challenging to predict observables at low energies involving nucleons from the interactions within the high energy degrees of freedom. 

The challenge of describing interactions among hadrons that are the effective degrees of freedom below the confinement scale led to the development of \textit{effective field theories} (EFTs).
EFT is a technique that re-describes a theory at a given energy scale by integrating out higher energy fluctuations and introducing infinitely many unknown parameters known as low energy coefficients (LECs), which capture the effects of the integrated out interactions.
Moreover, it introduces a \textit{counting scheme} that systematically organizes the operators or interactions and suggests a hierarchy for the size of associated LECs.
The predictive power of EFTs lies in the fact that at a given order in the counting scheme, there are only finitely many unknown parameters.
Once the LECs at a given order are fixed by mapping them to some experimental observation, the EFT calculations can then predict other observables involving the same LECs.

Chiral EFT and pionless EFT are two specific examples of EFTs that have been developed for describing nuclear physics.
Chiral EFT is an EFT that is based on the principles of chiral symmetry which is a symmetry that arises in QCD in the limit of zero quark masses.
It is used to describe the low-energy behavior of nuclear systems using the relevant degrees of freedom, like nucleons and pions.
Pionless EFT, on the other hand, is a simplified version of chiral EFT that neglects the explicit inclusion of pions as degrees of freedom.
It is based solely on the properties of nucleons and their interactions, and it is designed to describe the low-energy behavior of nuclear systems in the limit where the momentum transfer between nucleons is much smaller than the pion mass.
An overview of both EFTs is given in Sec.~\ref{sec: Nuclear ETFs} with more emphasis on pionless EFT.
Chiral EFT and pionless EFT have been used to study the properties of few-nucleon systems such as the few nucleon scattering phase shifts, hadron decay rates, binding energies of smaller nuclei, etc., in their respective physical regimes of applicability.
They have provided insights into the nature of the nuclear force at low energies, see Sec.~\ref{sec: Nuclear ETFs} for a review. 

The few-nucleon potentials derived from EFT calculations are used in \textit{ab initio} nuclear many-body methods for solving the nuclear Schr\"odinger equation to calculate observables in larger nuclei.
Chapter~\ref{ch: DBD from LQCD} reviews methods for obtaining double beta decay transitions in experimentally relevant isotopes from few-nucleon potentials.
Furthermore, the precision of predictions in heavier nuclei through the \textit{ab initio} methods significantly depends on the accuracy of few-nucleon physics obtained from EFTs.
However, the mapping of a high energy theory into an EFT is not unique.
The choice of different EFTs for the same underlying theory can lead to varying degrees of accuracy in predicting experimental outcomes depending on factors such as the type of observable being considered and the energy scale of the experiment.
Moreover, achieving a given accuracy on EFT predictions for a process involves knowing the strength of all interactions that appear up to the order in the counting scheme associated with that accuracy.
The corresponding LECs are constrained using the existing measurements of other processes where the coefficients of interest contribute.
Thus, a lack of experimental observation can leave some LECs undetermined, amounting to a failure of EFTs predictive power.
It will be shown in Sec.~\ref{subsec: 0vbb amplitude in pionless EFT} that this indeed is the case for evaluating two-nucleon $0\nu\beta\beta$ decay amplitude within non-relativistic pionless EFT for the light neutrino exchange scenario.
This leaves even the most dominant (leading order) contribution to the hadronic amplitude undetermined.
Similarly, the two-nucleon $2\nu\beta\beta$ decay amplitude in pinonless EFT contains an LEC that appears at the sub-dominant (next to leading order) contribution which is a dominant source of error in its predictions, as discussed in Sec.~\ref{subsec: 2vbb amplitude in pionless EFT}.

These limitations again point to a lack of non-perturbative analytical way of solving QCD.
There is, however, a numerical method of solving QCD non-perturbatively which is LQCD.
Its origin traces back to Kenneth Wilson's paradigm shifting approach of studying gauge theories using a discrete lattice, known as LGTs
In his approach, Euclidean-time gauge theory is formulated on a discrete hyperrectangular lattice with a finite extent.
The observables are calculated using path integral formalism, where the Euclidean-time action acts as a probability distribution function over various gauge field configurations which can be generated via Monte-Carlo sampling methods.
Averaging the observable over this generated gauge ensemble approximates calculating them by summing over quantum fluctuations in gauge fields.
The continuum limit of evaluated observables can be obtained by performing calculations with different lattice sizes and spacings and using the renormalization group procedure resulting in a non-perturbative QCD calculation.
The method of LQCD and its successes will be discussed in more details in Sec.~\ref{sec: LQCD}.

Lattice QCD has become an essential tool for understanding the hadron physics.
However, it is not a straightforward process to map physical correlation functions (i.e. observables) to the ones calculated using LQCD. 
This is because the former are defined with a Minkowski time while the latter are calculated with a Euclidean time.
The mapping becomes even more non-trivial for observables like scattering amplitudes and decay rates and was believed to be impossible for general kinematics in the infinite-volume limit.
But seminal work by Martin L\"uscher in the late 1980s showed that the two-particle physical scattering amplitude can be obtained from analyzing their FV energy spectrum.
Such a mapping, now known as the FV or L\"uscher's formalism, is reviewed in Sec.~\ref{sec: FV Formalism} with an instructive application to two-hadron scattering amplitude in Sec.~\ref{subsec: NN from FV}.
Furthermore, the FV formalism for obtaining transition amplitudes of processes involving external currents was developed by Lellouch and L\"uscher in 2001.
In the past couple of decades, the FV formalism has been extended to hadronic electroweak decays, three-particle scattering amplitudes and many other interesting processes, see Sec.~\ref{sec: FV Formalism} for a review.

Chapter~\ref{ch: DBD from LQCD} extends the FV formalism further for obtaining the two-nucleon $2\nu\beta\beta$ decay amplitude within the SM and the two-nucleon $0\nu\beta\beta$ decay amplitude in the light neutrino exchange scenario from the corresponding nuclear matrix elements (MEs) and FV energy eigenvalues calculated using LQCD.
The amplitudes are expressed in pionless EFT at next-to-leading order (NLO) and at leading order (LO), respectively, and by matching them to first-principles LQCD calculations, the corresponding LECs can be constrained.
The complications of matching the Euclidean-time correlation functions containing two time-separated electroweak current insertions to the corresponding Minkowski-time scattering amplitude has been addressed in both cases.
Furthermore, the issues arising from the presence of a light neutrino mode in the $0\nu\beta\beta$ case are carefully resolved.

For the numerical LQCD implementation of the prescription for $0\nu\beta\beta$ decay amplitude in near future, it is important to ask if anticipated uncertainties in calculating the LQCD ingredients required to perform such matching are sufficiently small to achieve the desired precision on the LEC to be constrained .
Given the complexity of the matching relation, it is not straight-forward to obtain the sensitivity of LEC constraints to uncertainties in LQCD calculations.
An analysis is performed to find the accuracy requirements of the upcoming LQCD studies at the physical quark masses to reach the precision goal of the LEC involved in $0\nu\beta\beta$ decay amplitude.
Synthetic data is generated mimicking the uncertainties on LQCD MEs and the FV energy eigenvalues, and a statistical analysis is performed on it using the matching relations for $0\nu\beta\beta$ to obtain the associated uncertainties on the corresponding LEC.

Chapter~\ref{ch: DBD from LQCD} is based on the following publications:
\begin{itemize}
    \item Z. Davoudi and S. V. Kadam, \textit{Two-neutrino double-$\beta$ decay in pionless effective field theory from a Euclidean finite-volume correlation function}, Phys. Rev. D 102 11, 114521, arXiv: 2007.15542 (2020)
    \item
    Z. Davoudi and S. V. Kadam, \textit{Path from Lattice QCD to the Short-Distance Contribution to $0\nu\beta\beta$ Decay with a Light Majorana Neutrino}, Phys.Rev.Lett. 126 15, 152003 arXiv: 2012.02083 (2021)
    
    \item Z. Davoudi and S. V. Kadam, \textit{Extraction of low-energy constants of single- and double-$\beta$ decays from lattice QCD: A sensitivity analysis}, Phys. Rev. D 105 9, 094502, arXiv: 2111.11599 (2022)
\end{itemize}

Lattice QCD has come a long way since its conception both in the theoretical developments and numerical calculations.
Further progress and more precise calculations will continue to improve our understanding of QCD physics.
This method is well-suited for studying equilibrium properties and static observables, but because of its Euclidean-time formulation it becomes challenging to put it to use for studying real-time observables, like scattering amplitudes, decay rates, transport coefficients etc., and out-of-equilibrium phenomena.
Matching relations like those in Ch.~\ref{ch: DBD from LQCD} map LQCD simulations to physical processes with Minkowski time.
But similar matching relations get increasingly complicated for multi-particle and/or more involved processes.
Studying QCD at finite temperature and chemical potential to understand the quark-gluon plasma and the phase diagram of QCD with LQCD also has an issue.
At nonzero chemical potential, the fermionic path integral becomes oscillatory and non-positive-definite which hampers the application of standard Monte Carlo sampling techniques.
This notorious \textit{sign problem} remains a significant challenge in non-perturbative studies of QCD phase diagram using LQCD.
Section~\ref{sec: Hamiltonian for LGT} will discuss these issues in more details and also review its alternative: the Hamiltonian simulation of QCD.

The Hamiltonian formulation of QCD was discovered and developed around the same time as LQCD.
However, it deals directly with the Hilbert space of QCD which scales exponentially in system sizes, and thus, the computational implementation of Hamiltonian simulation of QCD with the traditional computation methods was impractical.
Recent developments in tensor network methods and the advent of quantum simulation and quantum computing have revived interest in the Hamiltonian simulation of QCD, since these tools allow for handling such systems with polynomial scaling in their respective computational resources.
Even though a predictive non-perturbative Hamiltonian simulation of QCD is a distant future, theoretical developments in that direction are gaining interest.
One of the challenges on its theoretical front is to find a computationally resource efficient formulation of the QCD Hamiltonian and its Hilbert space.
Towards that aim, Ch~\ref{ch: LSH} provides a loop-string-hadron (LSH) formulation of an SU(3) gauge theory formulated in one spatial dimension with real time.
Its derivation, advantages, and numerical validation from comparing it with other Hamiltonian formulations is given in the sections therein.
These results are based on the following publication
\begin{itemize}
    \item  S. V. Kadam, I. Raychowdhury and J. R. Stryker \textit{Loop-string-hadron formulation of an SU(3) gauge theory with dynamical quarks}, Phys. Rev. D 107 9, 094513, arXiv:  2212.04490 (2022)
\end{itemize}

The following sections of this chapter provide the required background and give a brief review of the field prior to these publications for a complete discussion on the results presented here.
\section{Quantum Chromodynamics
\label{sec: QCD overview}
}
This section gives a tentative historical account that led to the now known theory of strong force, \textit{qauntum chromodyamics}, which is largely based on a recent review in Ref.~\cite{Gross:2022hyw}. 
The basics of QCD in its current form are summarized later in this section. 

The history of QCD starts with hints of a strong force after the discovery of neutron in 1932~\cite{Chadwick:1932wcf}.
It indicated that atomic nuclei are made up of neutrons and protons, but it left behind a puzzle: how are these particles bound together in the nucleus, where the electromagnetic force does not interact with charge-neutral neutrons and would normally repel the positively charged protons
This pointed towards the presence of a force strong enough to overcome electromagnetic repulsion.
Next important developments in explaining this force came from Heisenberg, who introduced isospin as the symmetry of strong force~\cite{Heisenberg:1932dw} and Yukawa, who proposed a spinless particle exchanges, a \textit{meson}, that would generate the nuclear force~\cite{Yukawa:1935xg}.
The existence of this hypothetical particle was confirmed experimentally in 1947 by studying the particle interactions in cosmic-rays~\cite{Lattes:1947mx,Lattes:1947my}.

The extension of isospin symmetry to a larger Lie group with the introduction of strangeness quantum number~\cite{Gell-Mann:1953hzm} paved the road for understanding the structure underneath strong interaction.
It was known that the energy gap at low energies is small: the lightest state, pion, has mass $m_\pi\simeq 135$ MeV which is small compared to the mass of proton, $M_p\simeq 938$ MeV.
Nambu explained this using the spontaneous isospin symmetry breaking where pions are the supposedly massless modes~\cite{Nambu:1960xd}.
However, since the isospin symmetry is not exact, pions are not exactly but only approximately massless.
Combining this with strangeness, Gell-Mann and Ne'eman extended the $SU(2)$ isospin group to $SU(3)$ and proposed the \textit{Eightfold Way}~\cite{Neeman:1961jhl,Gell-Mann:1962yej} which led to the understanding of baryon and meson masses.
Its modern understating through flavor symmetries between quarks will be discussed later in \ref{sec: Nuclear ETFs}.

Gell-Mann~\cite{Gell-Mann:1964ewy} and Zweig~\cite{Zweig:1964ruk}, independently hypothesized that baryons are bound states of three constituent particles, now known as the \textit{up} $(u)$, \textit{down} $(d)$, and \textit{strange} $(s)$ quarks, to explain their classification in the Eightfold Way.
Further developments proposed a new quantum number for the constituents to accommodate the spin-statistics theorem~\cite{Bogolubov:1965xla,Han:1965pf,10.1143/PTPS.E65.187}, which was later coined as the \textit{color} quantum number.
But these hypothesized constituents never showed up in experiments.
Nonetheless, Bjorken predicted scaling laws in the electron-proton deep inelastic scattering cross-sections assuming nucleons contain point-like constituents~\cite{Bjorken:1968dy} which were later confirmed in scattering experiments by MIT-SLAC collaboration~\cite{Bloom:1969kc}.
These experiments also suggested that there exist three distinct colors, now labeled as \textit{red} $(r)$, \textit{green} $(g)$, and \textit{blue} $(b)$ .
To reconcile the successes of the quark model but a lack of experimental observation of individual quarks, a new interaction by a gauge field, now known as gluon, analogous to the electromagnetic force was considered that would tightly hold these quarks together.
And if the gluons carry the color charge, then this \textit{confined} nature of quarks can be applied to them as well, leading to only the color-neutral states in the theory.
These requirements pointed towards a gauge theory with a non-Abelian local gauge symmetry~\cite{Yang:1954ek}, and such features for quantized non-Abelian gauge theories were demonstrated in the 1970s by Wilson, Gross, Wilczeck, Politzer, and 't Hooft~\cite{Wilson:1974sk, Gross:1973id, Gross:1973ju, Gross:1974cs, Politzer:1974fr,tHooft:1971qjg,tHooft:1971akt}.

\subsection{QCD Lagrangian
\label{subsec: QCD Lagrangian}
}
The following section reviews the QCD Lagrangian with quarks and gluons degrees of freedom, how it leads to the phenomenon of confinement, and its confined spectrum based on the symmetry breaking analysis. This section is based on Refs.~\cite{Peskin:1995ev, Halzen:1984mc, Davoudi:2014uxa}.

Gauge theories obey a local gauge invariance, that is the Lagrangian density remains unchanged under space-time dependent rotations in the internal space of matter fields.
In the case of QCD, the local rotations belong to an $SU(3)$ group and matter fields are the fermionic quarks, $q_{f, \alpha}^{a}(x)$, where $f$ denotes the flavor index, $f=u,\, d,\, s$, and $a$ indicates the color index, $a=1,\,2,\,3$.
\footnote{Colors will be indexed by 1, 2, and 3 instead of $r$, $g$ and $b$ from here onwards.}
The space-time coordinate is labeled by $x$, and $\alpha$ is its Dirac index.
Internal rotations in the color space are given by
\footnote{Repeated indices are summed over, unless shown by explicit sums.}
\begin{equation}
	q_{f, \alpha}^{a} (x)\to q_{f, \alpha}^{'a} (x) = \Omega^{a}_{b}(x)\; q_{f, \alpha}^{b}(x) \equiv (e^{-igT^{\ell} \theta^{\ell}(x)})^{a}_{b} \; q_{f, \alpha}^{b} (x).
	\label{eq: guage rotation of quarks}
\end{equation}
Here, $\Omega^{a}_{b}(x)$ is a $3\times 3$ unitary matrix with unit determinant and thus belongs to the $SU(3)$ Lie group.
$T^{\ell}$s, with $\ell=1,\,2,\cdots\,8$, are the generators of this group normalized as ${\rm Tr}[T^iT^j] = \frac{1}{2}\delta^{ij}$, and $g$ is the strong coupling constant.
The generators form a Lie algebra under commutation as
\begin{equation}
	[T^i,T^j] = i f_{ijk}T^k,
	\label{eq: generator Lie algebra}
\end{equation}
where $ f_{ijk}$ are the $SU(3)$ structure constants, and they are related to the Gell-Mann matrices, $\lambda^i$ as $T^i=\frac{\lambda^i}{2}$.

It is clear that the free Lagrangian density\footnote{All instances of Lagrangian in this thesis are Lagrangian densities, which for brevity, will be called Lagrangian throughout.} of quark multiplets do not obey the gauge invariance because the derivative term in
\begin{equation}
	\mathcal{L}_{\rm free}(x) = \sum_{f=1}^{N_f} \bar{q}^{a}_{f,\alpha}(x)\, (i\gamma^{\mu}\partial_{\mu}-m_{f})_{\alpha,\beta}\,  q_{f, \beta}^{a}(x),
	\label{eq: free quark Lagrangian density}
\end{equation}
transforms as
\begin{equation}
	\partial_{\mu} q_{f, \alpha}^{a} \to \partial_{\mu}(q_{f, \alpha}^{'a}) = \partial_{\mu}(\Omega^{a}_{b}(x))\; q_{f, \alpha}^{b} + \Omega^{a}_{b}(x)\; (\partial_{\mu}q_{f, \alpha}^{b})
	\label{eq: derivative term under Gauge}.
\end{equation}
Here, $N_f$ is the total number of quark flavors each with quark mass $m_f$.
To impose the invariance under Eq.~\eqref{eq: guage rotation of quarks}, there must exist another field that would absorb the first term in Eq.~\eqref{eq: derivative term under Gauge} so that the form in Eq.~\eqref{eq: free quark Lagrangian density} is restored.
The independence of generators in determining $\Omega$ implies that there should a field associated with each generator $T^i$.
A four-vector gauge field, $A^{i}_\mu(x)$, for each generator $T^i$ can be introduced such that it transforms as 
\begin{equation}
	A_\mu(x) \equiv A^{a}_\mu(x) T^a \to A'_\mu(x) = \Omega(x)A_{\mu}(x)\Omega^\dagger(x)+\frac{i}{g}(\partial_{\mu}\Omega(x))\Omega^{\dagger}(x),
	\label{eq: transformation of gauge field}
\end{equation}
where the matrix indices are omitted.
A covariant derivative can then be constructed as 
\begin{equation}
	D_\mu q^a = \partial_{\mu} q^a + i g A^i (T^i)^a_b q^b,
	\label{eq: covariant derivative definition}
\end{equation}
(unchanged indices are suppressed to avoid clutter) such that Eq.~\eqref{eq: derivative term under Gauge} together with Eq.~\eqref{eq: transformation of gauge field} implies that
\begin{equation}
	D_\mu q^a \to \Omega(x)D_\mu q^a.
	\label{eq: covariant derivative transformation}
\end{equation}
Hence, replacing ordinary derivative, $\partial_{\mu}$, with the covariant derivative, $D_\mu$, ensures gauge invariance in Eq.~\eqref{eq: free quark Lagrangian density}:
\begin{equation}
	\mathcal{L}_{q}(x) = \sum_{f=1}^{N_f}\bar{q}^{a}_{f,\alpha}(x)\, (i\gamma^{\mu}D_{\mu}-m_{f})_{\alpha,\beta}\,  q_{f, \beta}^{a}(x).
	\label{eq: quark covariant derivative interaction}
\end{equation}

To obtain the dynamics of gauge fields $A^a_\mu(x)$, a Lagrangian containing only the gauge fields needs to be constructed such that it respects Lorentz and gauge invariance while keeping the theory renormalizable.
Quantization of $A^a_\mu$ field leads to particles that mediate the strong force called gluons.
Since a mass term for $A^a_\mu$, which is given by $A^a_\mu A^{\mu,a}$, violates the gauge symmetry, as seen from Eq.~\eqref{eq: transformation of gauge field}, the gluons remain massless.

Motivated by the gauge theory of electromagnetic interactions, QED, one can form a field strength tensor to construct interaction terms for $A^a_\mu(x)$, 
\begin{equation}
	F_{\mu\nu} = F_{\mu\nu}^a T^a \equiv \frac{-i}{g}[D_\mu,D_\nu] = (\partial_{\mu}A_\nu^a -\partial_{\nu}A_\mu^a - g f_{abc} A^b_\mu A^c_\nu)T^a,
	\label{eq: gluon field strength tensor definition}
\end{equation}
which can be used to construct two types of interaction terms for gluons.
One violates the $CP$-symmetry, while the other preserves it.
Violation of $CP$-symmetry is observed in the weak-force sector of the SM~\cite{Christenson:1964fg}, however, direct $CP$-violation in the strong-force sector has not been observed yet.
We thus briefly mention the $CP$-violating term without going into much details:
\begin{equation}
	\mathcal{L}^{\text{ \cancel{CP}}}_{g} \propto \theta\, \epsilon_{\alpha\beta\mu\nu}{\rm Tr}\left[ F^{\alpha\beta}F^{\mu\nu}\right],
	\label{eq: gauge interaction CP odd}
\end{equation}
where $\theta$ measures its strength and $\epsilon_{\alpha\beta\mu\nu}$ is the fully anti-symmetric Levi-Civita tensor.

The $CP$-even interaction is given by
\begin{equation}
	\mathcal{L}_{g} = -\frac{1}{2}{\rm Tr}\left[ F_{\mu\nu}F^{\mu\nu}\right]=-\frac{1}{4}F^a_{\mu\nu}F^{a\,\mu\nu},
	\label{eq: gauge interaction CP even }
\end{equation}
where the choice of prefactor ensures a normalized kinetic energy term.

Putting everything together, the QCD Lagrangian is given by
\begin{equation}
	\mathcal{L}_{\rm QCD}(x) = \sum_{f=1}^{N_f} \bar{q}^{a}_{f,\alpha}(x)\, (i\gamma_{\mu}D^{\mu}-m_{f})_{\alpha,\beta}\,  q_{f, \beta}^{a}(x)
	-\frac{1}{2}{\rm Tr}\left[ F_{\mu\nu}F^{\mu\nu}\right].
	\label{eq: full QCD Lagrangian}
\end{equation}

Two key features of Eq~\eqref{eq: full QCD Lagrangian} due to its non-Abelian gauge symmetry are: $1)$ Equation~\eqref{eq: gluon field strength tensor definition} implies that the last term in Eq.~\eqref{eq: full QCD Lagrangian} contains self interacting vector boson particles which is different than QED.
The self-interacting gluons are responsible for color confinement in QCD since long-range interactions mediated by self interacting gluons become energetically unfavorable.
$2)$ The gauge symmetry completely determines the gluon-quark interaction strength just from the strength of the self-interacting gluons.
It prohibits different interaction strengths for different quark flavors, since they all couple to the same gauge fields which also transform under the local gauge symmetry.

More non-trivial features of the Lagrangian in Eq.~\eqref{eq: full QCD Lagrangian} can be understood by applying the renormalization group procedure to study the running of the coupling $g$ with the renormalization scale $\mu$.
This running is described by the QCD $\beta$ function for $\alpha_s(\mu) = g^2(\mu)/4\pi$, which is the QCD analogue of the fine-structure constant.
The $\beta$ function has a perturbative expansion in the limit $g\to 0$ as
\begin{equation}
    \beta(\alpha_s) = \mu \frac{d}{d\mu} \alpha_s(\mu) \xrightarrow{g\to 0} -\sum_{k=0}^{\infty} b_k \alpha_s^{k+2}.
    \label{eq: QCD beta function}
\end{equation}
The first two coefficients, $b_0$ and $b_1$, are given by $b_0 = \frac{1}{12\pi}(33-2 N_f)$ and $b_1 = \frac{1}{24\pi2}(153-19 N_f)$ when calculated in perturbation theory up to two-loops using the Feynman rules for the Lagrangian in Eq.~\eqref{eq: full QCD Lagrangian}~\cite{ParticleDataGroup:2022pth}.
Similarly $b_2$, $b_3$ and $b_4$ corresponding to 3-, 4- and 5-loop diagrams have been computed~\cite{vanRitbergen:1997va, Czakon:2004bu, Baikov:2016tgj, Luthe:2016ima,Herzog:2017ohr, Luthe:2017ttg, Chetyrkin:2017bjc}.
The negative sign in Eq.~\eqref{eq: QCD beta function} suggests that for $N_f\leq16$, the overall sign of the $\beta$ function is negative, meaning the strong coupling becomes weak for large momentum transfer processes characterized by large $\mu$.
This is the property of \textit{asymptotic freedom}~\cite{Gross:1973id,Gross:1973ju,Gross:1974cs,Politzer:1974fr}.
Solving Eq.~\eqref{eq: QCD beta function} for obtaining a relation between $\alpha$ at two different scales, $\mu$ and $\mu'$, by keeping only the $b_0$ term and ignoring higher order contributions gives
\begin{equation}
    \alpha_s(\mu')  = \frac{\alpha_s(\mu)}{1+b_0 \alpha_s(\mu)\ln{\frac{\mu'}{\mu}}}.
    \label{eq: confinement}
\end{equation}
This implies that for $b_0>0$, $\alpha(\mu')<\alpha(\mu)$.
Furthermore, if there exists a scale $\mu = \Lambda_{\rm QCD}$, such that $b_0 \alpha_s(\Lambda_{\rm QCD})\ln{\frac{\mu'}{\Lambda_{\rm QCD}}} =1$, then Eq.~\eqref{eq: QCD beta function} implies that $\alpha_s({\mu'})\to0$ assumption breaks near $\mu'\simeq\Lambda_{\rm QCD}$.
Experimental data and LQCD calculations point to $\Lambda_{\rm QCD}\sim 300$ MeV~\cite{ParticleDataGroup:2022pth}, which is referred as the QCD scale.

Further investigation of low energy QCD suggested that at low energies, the potential energy between two quarks increases linearly with distance between them~\cite{Wilson:1974sk}, which leads to formation of bound states of quarks that are color neutral.
This is known as \textit{color confinement} in QCD.

\subsection{Chiral symmetry in QCD
	\label{subsec: chiral symmetry in QCD}
}
\begin{figure}[t]
	\centering
	\includegraphics[scale=1]{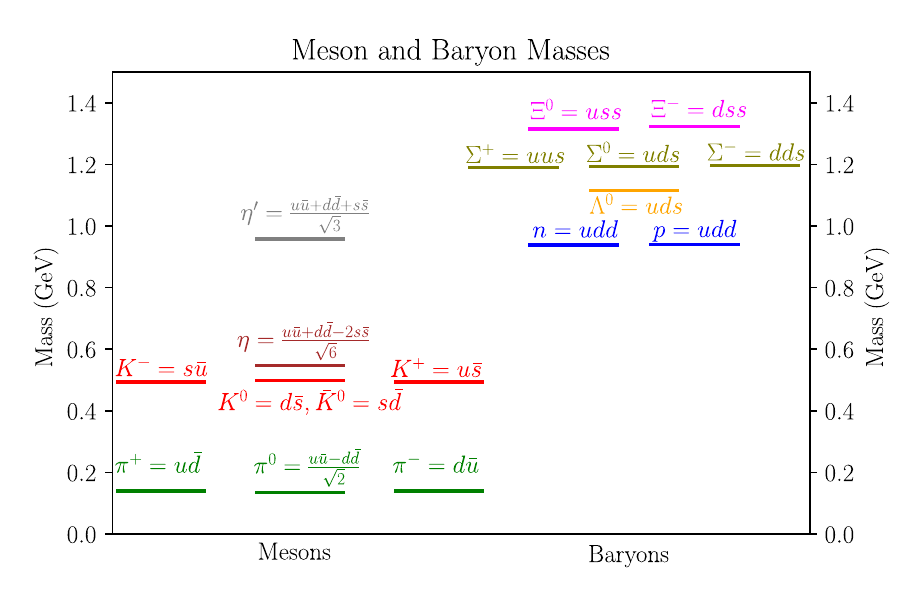}
	\caption{
    Central values of masses of a few light mesons and baryons with $J^P=0^-$ and $J^P=\frac{1}{2}^+$, respectively, are shown in this figure along with their quark content. The values are taken from Ref.~\cite{ParticleDataGroup:2022pth}.
		\label{fig: hadron_spectrum}}
\end{figure}
Confinement leads to color neutral bounds states called hadrons that are made up of gluons or quarks and gluons.
The energy spectrum of low energy QCD is dominated by two-quark bound states called mesons, and three-quark bound states called baryons.
They can be characterized and organized by their mass, total angular momentum $(J)$ and parity $(P)$.
Figure~\ref{fig: hadron_spectrum} shows the masses of a few of the lightest mesons and baryons with $J^P=0^-$ and $J^P=\frac{1}{2}^+$, respectively.
It shows interesting organization patterns.
First, notice that mesons are significantly lighter compared to baryons with meson masses $\sim \mathcal{O}(100)$ MeV and baryon masses $\gtrsim1$ GeV.
Second, the lightest mesons, pions, are not exactly but almost massless.
Third, there is a degeneracy pattern in meson masses and baryon masses, and these two patterns are slightly different.
Fourth, there is a big mass gap between $\eta$ and $\eta'$ mesons, making the $\eta'$ almost as heavy as baryons.
All these features can be understood via the fundamental theory of QCD.

\begingroup
\renewcommand{\arraystretch}{2}
\begin{table}[h!]
	\centering
	\begin{tabular}{c||c|c|c|c|c|c}
		Flavor & up & down & strange & charm & bottom & top\\
		\hline
		Mass (MeV) & $2.16^{+0.49}_{-0.26}$ & $4.67^{+0.48}_{-0.17}$ & $93.4^{+8.6}_{-3.4}$ & $1270\pm 20$ & $4180^{+30}_{-20}$ & $172,690\pm 300$ 
	\end{tabular}
	\caption{Masses of different quark flavors. Values are taken from Ref. ~\cite{ParticleDataGroup:2022pth}. The $u$-, $d$- and $s$-quark masses are given at the renormalization scale $\mu$ = 2 GeV in the $\overline{\rm MS}$ scheme, while for the rest the scale is set to the their respective $\overline{\rm MS}$ masses.}
	\label{tab: quarkmasses}
\end{table}
\endgroup

In the SM, the QCD Lagrangian in Eq.~\eqref{eq: full QCD Lagrangian} has six flavors, i.e. $N_f=6$, and the experimental results strongly exclude the possibility of additional quark flavors~\cite{Kuflik:2012ai}.
Apart from the previously mentioned $u$, $d$, and $s$ flavors, three more flavors \textit{charm} $(c)$, \textit{bottom} $(b)$ and \textit{top} $(t)$ quarks were discovered experimentally~\cite{SLAC-SP-017:1974ind,E598:1974sol,E288:1977xhf,CDF:1995wbb,D0:1995jca}.
Masses of different quark flavors are summarized in Tab.~\ref{tab: quarkmasses}.
The $c$-, $b$- and $t$-quarks are much heavier than $\Lambda_{\rm QCD}$, and do not play a role in the low energy physics.
Restricting the discussion to three light quark flavors: $u$, $d$ and $s$, the low energy spectrum can be analyzed by understanding the symmetries of the quark interaction term in Eq.~\eqref{eq: full QCD Lagrangian}.
Noticing that the quark masses are much below the lightest mesons, see Fig.~\ref{fig: hadron_spectrum} and Tab.~\ref{tab: quarkmasses}, the quark mass term can be neglected to a good approximation.
The massless quark interaction can then be written in terms of left-handed, $q_L=\frac{1-\gamma^5}{2} q$, and right-handed, $q_R=\frac{1+\gamma^5}{2} q$, quark components where $q$ is the flavor multiplet, $q=(u,d,s)^T$, as 
\begin{equation}
	\mathcal{L}_{\rm chiral} = \bar{q}\, i\gamma_{\mu}D^{\mu}\,  q = \bar{q}_L \, i\gamma_{\mu}D^{\mu}\,  q_L + \bar{q}_R \, i\gamma_{\mu}D^{\mu}\,  q_R,
	\label{eq: chiral quark Lagrangian}
\end{equation}
where the subscript `chiral' stands for the chiral limit, i.e. massless quarks limit.
Equation~\eqref{eq: chiral quark Lagrangian} is invariant under the independent \textit{global} rotations of left-handed and right-handed quarks as $q_L\to L q_L $ and $q_R \to R q_R$, where $L$ and $R$ are $3\times 3$ unitary matrices.
Thus, the overall symmetry of Eq.~\eqref{eq: chiral quark Lagrangian} is $U(3)_L \times U(3)_R$ which can be decomposed by separating out the phases of $U(3)$ as $SU(3)_L \times SU(3)_R \times U(1)_L \times U(1)_R$.
This is the full chiral symmetry of the QCD Lagrangian in the chiral limit.
The corresponding $18$ expected conserved currents by Noether's theorem are given by
\begin{equation}
	L_\mu^i = \bar{q}_L \gamma_\mu T^i q_L, \quad R_\mu^i = \bar{q}_R \gamma_\mu T^i q_R, \quad L_\mu = \bar{q}_L \gamma_\mu q_L, \quad R_\mu = \bar{q}_L \gamma_\mu q_L.
	\label{eq: left right currents}
\end{equation}
However, as we will see, not all of them are conserved.

To understand this, the currents can be written in a more useful way by looking at linear combinations of left- and right-handed quark rotations as
\begin{eqnarray}
	V_\mu^i = R_\mu^i + L_\mu^i = \bar{q} \gamma_\mu T^i q, 
	\label{eq: isovector vector current}\\
	A_\mu^i = R_\mu^i - L_\mu^i = \bar{q} \gamma_\mu \gamma_5 T^i q,
	\label{eq: isovector axial-vector current}\\
	V_\mu = R_\mu + L_\mu = \bar{q} \gamma_\mu q,
	\label{eq: isosinglet vector current}\\
	A_\mu = R_\mu - L_\mu = \bar{q} \gamma_\mu \gamma_5 q.
\end{eqnarray}
Here $V_\mu^i$ ($A_\mu^i$) are isovector (axial-)vector currents and  $V_\mu$ ($A_\mu$) is an isosinglet (axial-)vector current.
The vector currents correspond to the same left- and right-handed quark rotations in their respective symmetry groups, while the axial-vector currents correspond to opposite rotations.
Furthermore, quarks fields transform as $q(t,{\bm x}) \to \gamma^0 q(t,-{\bm x})$ under parity, resulting in $V_\mu^i (t,{\bm x})\to V_\mu^i (t,-{\bm x})$ and $A_\mu^i (t,{\bm x}) \to -A_\mu^i (t,-{\bm x})$, and similarly for their isosinglet counterparts\footnote{Boldface letters denote a three-vector.}.

The $U(1)_A$ symmetry associated with $A_\mu$ current is broken anomalously.
That is, even though the classical Lagrangian has this symmetry, the quantum fluctuations in the quantized theory break this symmetry~\cite{Adler:1969gk,Adler:1969er}.
This is an explanation of why the isosinglet pseudo-scalar meson $\eta'$ is heavier than that of  isovector pseudo-scalars~\cite{tHooft:1976rip,tHooft:1986ooh}.
This reduces the symmetry group of Eq.~\eqref{eq: chiral quark Lagrangian} to $SU(3)_V \times SU(3)_A \times U(1)_V$.
However, the QCD vacuum does not respect the $SU(3)_A$ symmetry indicating a spontaneous symmetry breaking of $SU(3)_A$ symmetry.
There are two reasons to justify this: $1)$ If the vacuum obeyed the $SU(3)_A$ symmetry, there would have been a parity doubling in the hadron spectrum which is not the case~\cite{ParticleDataGroup:2022pth}.
$2)$ The spontaneous symmetry breaking mechanism leads to massless scalar modes, one mode per broken generator.
In this case, it would lead to eight massless modes corresponding to eight generators of $SU(3)_A$.
Furthermore, the eight pseudo-scalar mesons are very light compared to their opposite parity meson partners, vector mesons.
Thus, pseudo-scalar mesons are candidates for Nambu-Goldstone modes of broken symmetry and their observed light masses are attributed to explicit chiral symmetry breaking due to non-zero quark masses, making them pseudo Nambu-Goldstone modes.

The inclusion of quark masses in Eq.~\eqref{eq: chiral quark Lagrangian} leads to
\begin{equation}
	\mathcal{L}_{\rm quark} = \bar{q}_L \, i\gamma_{\mu}D^{\mu}\,  q_L + \bar{q}_R \, i\gamma_{\mu}D^{\mu}\,  q_R, - \bar{q}_R \mathds{M} q_L - \bar{q}_L \mathds{M}^\dagger q_R,
	\label{eq: mass quark Lagrangian left right}
\end{equation}
where, $\mathds{M} = {\rm diag}(m_u, m_d, m_s)$.
This term explicitly breaks $SU(3)_A$.
However, the $U(1)_V$ symmetry is still preserved for any quark masses.
This symmetry is the result of flavor independence of QCD interactions and it conserves the total baryon number.
On the other hand, $SU(3)_V$ is preserved only when all quark masses are equal.
Such a scenario was originally proposed by Gell-Mann and Ne'eman~\cite{critchfield1965theoretical}.
The next level of symmetry comes by observing that $m_u\simeq m_d<<m_s$.
Thus, assuming $m_u = m_d \neq m_s$ reduces the $SU(3)$ flavor symmetry to $SU(2)$ isospin symmetry.
Finally, taking $m_u \neq m_d$ leads to isospin symmetry breaking.
These symmetry breaking patterns are discussed in great detail in Ref.~\cite{Pagels:1974se}.

Explanation of hadron spectrum using the symmetries of the QCD Lagrangian is encouraging.
Next comes the question of calculating these masses and other low energy observables using the fundamental degrees of freedom given in Eq.~\eqref{eq: full QCD Lagrangian}, especially regarding atomic nuclei and nuclear matter.
When dealing with the non-perturbative regime of QCD, calculating nuclear observables from QCD becomes a challenging task. 
One approach to solve QCD in this regime is by using the numerical method of LQCD, see Sec.~\ref{sec: LQCD} for a review.
However, LQCD calculations are computationally expensive, especially when dealing with large systems.
Additionally, there are several technical challenges associated with the numerical simulations, such as the need to extrapolate results to the continuum limit and to control systematic errors.
Because of this, the use of LQCD for studying nuclear physics problems becomes costly with increasing atomic number. 
There exists another approach to perform nuclear physics calculations that utilizes the principle of EFT and the symmetries of the QCD Lagrangian to calculate nuclear observables perturbatively.
The use of EFT is justified by noticing that the potential energies of nuclear systems are small compared to the typical scale of QCD interactions.
This separation of scales allows a systematic field theory description consisting of the low energy degrees of freedom that uses the ratio of these two scales as a small parameter for expansion.
The next section reviews the method of EFT, its use in nuclear physics, and a few of its applications to calculate observables.

\section{Nuclear Effective Field Theories
\label{sec: Nuclear ETFs}
}

\textit{Effective field theory} is a powerful framework that provides a systematic way to study physical phenomena at a given energy scale by considering only the relevant degrees of freedom at that scale.
This section reviews the this framework for nuclear physics and is largely based on Refs.~\cite{Scherer:2002tk, Epelbaum:2010nr, vanKolck:1998bw, Davoudi:2014uxa}.

The central idea behind the working of EFTs is the concept of decoupling where the description of a system at low energies is insensitive to the details of its behavior at higher energies.
Thus, for the development of an EFT it is important to identify separation of scales.
In the case of QCD hadron spectrum, a large gap between the lightest pseudo-scalar mass of pions and the masses of the vector mesons, like $\rho$ meson with $m_\rho\simeq770$ MeV, sets the scale separation.
The previous section argued associating light pseudo-scalar mesons with Nambu-Goldstone modes of spontaneous chiral symmetry breaking in the chiral limit.
Thus, for the construction of a nuclear EFT it is natural to identify the high energy scale, $\Lambda_{\chi}$, with $m_\rho$ and low energy scale, $Q$, with $m_\pi$.

Next, one needs to find the relevant degrees of freedom.
As already noticed, the degrees of freedom at low energies are nucleons and pions to build the low energy atomic nuclei and study nuclear reactions.
Thus, the appropriate low energy degrees of freedom are the hadrons, however, they need to be connected to the fundamental theory of QCD which is expressed in terms of quarks and gluons.
This connection is established through the symmetries of the underlying theory as stated in a `folk theorem' by Weinberg~\cite{Weinberg:1978kz}:

\begingroup
\setlength{\leftskip}{1.5cm}
\setlength{\rightskip}{1.5cm}
\begin{quote}
If one writes down the most general possible Lagrangian, including all terms consistent with assumed symmetry principles, and then calculates matrix elements with this Lagrangian to any given order of perturbation theory, the result will simply be the most general possible S-matrix consistent with analytical, perturbative unitarity, cluster decomposition, and the assumed symmetry principles.
\end{quote}
\endgroup

With this, and the QCD symmetry breaking patterns discussed in the previous section, the construction of a nuclear EFT then involves writing the most general Lagrangian with hadronic degrees of freedom such that it respects the $SU(3)_V \times SU(3)_A \times U(1)_V$ symmetry and a ground state of spontaneously broken $SU(3)_A$ symmetry that respects the remnant $SU(3)_V \times U(1)_V$ symmetry.

\subsection{EFT for meson-meson and meson-nucleon interactions
\label{subsec: meson meson and meson nucleon EFT}
}
To achieve this for an EFT of interacting mesons, one can construct a field $\Sigma(x)$ that governs the excitation of mesons, and transforms under the remnant subgroup of the full symmetry group.
This is done by defining~\cite{Weinberg:1968de,Coleman:1969sm,Callan:1969sn}
\begin{equation}
	\Sigma(x) = \exp\left(\frac{i\phi(x)}{f_{\pi}}\right),
	\label{eq: U field definition ChPT}
\end{equation}
with
\begin{equation}
	\phi = \sum_{a=1}^{8}\phi_a \lambda_a =
	\begin{pmatrix}
		\frac{\pi^0}{\sqrt{2}} + \frac{\eta}{\sqrt{6}} & \pi^+ & K^+ \\
		\pi^- & -\frac{\pi^0}{\sqrt{2}} + \frac{\eta}{\sqrt{6}} & K^0 \\
		K^- & \bar{K}^0 & -\frac{2\eta}{\sqrt{6}}.
	\end{pmatrix}.
\end{equation}
Here, $f_{\pi}$ is an unknown constant that is determined by matching it with an observable, e.g., the decay rate of leptonic pion decays.
The field $\Sigma(x)$ transforms under the $L$ and $R$ rotations in the quark flavor space as $\Sigma(x) \to R \Sigma(x) L^\dagger$.
The ground state corresponds to no meson excitation, that is $\phi(x) = 0 \implies \Sigma(x)=\mathds{1}\equiv \Sigma_0$ $\forall x$.
$\Sigma_0$ is clearly not invariant under axial rotations, $R = L^\dagger$, as $\Sigma_0\to L^\dagger \Sigma_0 L^\dagger =L^\dagger \mathds{1} L^\dagger \neq \Sigma_0$, but is invariant under vector rotations $R = L$, as $\Sigma_0\to L \Sigma_0 L^\dagger =L \mathds{1} L^\dagger = \Sigma_0$.

The most general Lagrangian with this field contains infinitely many terms.
Thus, calculating any observables requires a sense of importance of these terms, which is given by an organization scheme that can distinguish between more and less important contributions.
This is called a power-counting scheme.
Guided by the expansion according to a power-counting scheme, one can calculate Feynman diagrams for the problem under consideration to the desired accuracy.

The power-counting scheme proposed by Weinberg uses the naturally available scales, $Q$ and $\Lambda_{\chi}$ to expand observables in powers of $Q/\Lambda_{\chi}$.
In the chiral limit, when pions are massless, the soft scale $Q$ is identified with the momentum involved in the process of interest, $p$.
Since the Nambu-Goldstone pseudo-scalar mesons have vanishing scattering amplitude in the zero momentum limit, the lowest order Lagrangian interaction term must consist of only momentum dependent interactions of pseudo-scalar mesons.
This is achieved by considering $\partial_\mu \Sigma(x)$ which when combined with an expansion of the exponential in Eq.~\eqref{eq: U field definition ChPT} implies that the leading contribution is the spatial derivatively coupled $\phi(x)$ field.
Thus, the most general Lagrangian containing minimal derivatives of $\Sigma(x)$ that is consistent with all symmetries is
\begin{equation}
	\mathcal{L}_{\pi\pi} = \frac{f_{\pi}^2}{8}{\rm Tr}\left[\partial_\mu \Sigma \partial^\mu \Sigma^\dagger\right],
	\label{eq: LO massless mesonic chiral Lagrangian}
\end{equation}
where the prefactor ensures correct normalization for the kinetic term of $\phi$ upon expansion. 
To include the explicit chiral breaking effects due to quark masses, one can think of $\mathds{M}$ in Eq.~\eqref{eq: mass quark Lagrangian left right} as an object that transforms as $\mathds{M}\to R\mathds{M}L^\dagger $ such that Eq.~\eqref{eq: mass quark Lagrangian left right} remains invariant under $L$ and $R$ rotations.
Using this, $\mathds{M}$ can be coupled with $\Sigma(x)$ based on their transformations to get the leading order (LO) effective Lagrangian for meson interactions as
\begin{equation}
	\mathcal{L}_{\pi\pi} = \frac{f_{\pi}^8}{4}{\rm Tr}\left[\partial_\mu \Sigma \partial^\mu \Sigma^\dagger + 2B (\mathds{M} \Sigma^\dagger + \Sigma\mathds{M}^\dagger)\right]. 
	\label{eq: LO massive mesonic chiral Lagrangian}
\end{equation}
The sign between the last two terms is fixed by the parity symmetry.
Here $B$ and $f_\pi$ are the low energy constants (LECs).
EFTs have infinitely many LECs but their predictive power is assured by having only finitely many LECs at a given order in the power-counting scheme.
LECs need to be fixed by matching them to one or several observables, but once fixed they can be used to predict other observables.
In the case of $f_\pi$ and $B$, they are related to weak decay rate of pion and mass of pions, respectively.
One way to make such connections is by calculating observables via a Feynman diagrammatic expansion using the EFT Lagrangian.
In the chiral power-counting scheme, the power of the expansion parameter $Q/\Lambda_{\chi}$ is given by $\nu$ which can be calculated from the Feynman diagram.
This can be achieve by counting the powers of small momenta associated with derivatives at the interaction vertices coming from the expansion  of $\Sigma(x)$ in Eq.~\eqref{eq: LO massive mesonic chiral Lagrangian}, pion propagators, integrations over loop momenta and the $\delta$-functions.
This gives $\nu = 2 +2L + \sum_i V_i\Delta_i$ where $L$ denotes the number of loops, $\Delta_i=(d_i-2)$ with $d_i$ being the dimension of $i$ type vertex interaction and $V_i$ refers to number of such vertices.
As an example of the success of chiral perturbation theory, the isoscalar $\pi\pi$ scattering length, $a_0^0$, in the $s$-channel evaluated up to two-loop gave $a_0^0 = 0.220\pm0.005$ ~\cite{Colangelo:2000jc} which is in a good agreement with combine experimental data $a_0^0=0.217\pm0.008 (\rm exp) \pm 0.006 (\rm th)$~\cite{Colangelo:2008sm}.

\textit{Single-nucleon EFT}

To extend the EFT framework to include nucleons, it is more convenient to restrict the discussion to the isospin subgroup of the baryon octet, $\Psi=(n\;p)^T$.
A new matrix $u$ can be defined as $\Sigma=u^2$, which is more useful in defining the transformation properties of $\Psi$.
It can be shown that the multiplet $(\Sigma\;\Psi)^T$ defines a non-linear realization of the chiral group if $\Psi$ transforms as $\Psi\to h\Psi$ where $h$ is defined through the transformation of $u$: $u\to u' = Luh^{-1} = huR^\dagger$, see Refs.~\cite{Coleman:1969sm, Callan:1969sn} for more details.
With this, $\Psi$ transforms linearly as an isospin doublet under the $SU(2)_V$ subgroup, however, the transformation of $\Psi$ depends on $\Sigma$ which is space dependent.
This implies that $\partial_\mu \Psi$ does not transform covariantly, and it is resolved by defining a `connection' term $\Gamma^\mu\equiv\frac{1}{2}\left(u^\dagger\partial_\mu u+u\partial_\mu u^\dagger\right)$.
The most general Lagrangian in the chiral limit to first order in the derivatives is then given by
\begin{equation}
	\mathcal{L}_{\pi N} = \bar{\Psi}\left(i\gamma^\mu D_\mu - M + \frac{g_A}{2}\gamma^\mu\gamma_5 u_\mu\right)\Psi,
	\label{eq: nucleon pion Lagrangian relativistic}
\end{equation}
where $D_\mu \equiv \partial_\mu + \Gamma_\mu$, $M$ is the nucleon mass, and $g_A$ is a new LEC called the axial-vector coupling constant, or simply the axial charge of the nucleon, which can be matched to the neutron semi-leptonic weak decay.
Its value is given by $g_A=1.2670\pm0.0035$~\cite{ParticleDataGroup:2022pth}

Unlike pions, the nucleon mass does not vanish in the chiral limit, and it introduces an additional scale in the problem.
The time-derivative of a relativistic baryon field generates an energy factor which is of the order of nucleon mass.
When used in an expansion scheme, this gives rise to $M/\Lambda_{\chi}$ factor which is close to one.
Thus, the notion of power-counting like the pionic EFT seems to have been lost.

The solution to this problem is known as the heavy baryon chiral perturbation theory~\cite{Jenkins:1990jv,Bernard:1992qa} which treats baryons as heavy static sources, such that the momentum transfer between baryons by pion exchange is small compared to the baryon mass which serves as the soft scale $Q$.
This is achieved by parameterizing the four momentum of the heavy baryon $p^\mu$ as $p^\mu = M v^\mu + l^\mu$, where $v^\mu$ is the four-velocity with $v^2=1$, and $l^\mu$ is the small residual momentum such that $v\cdot l<< M$.
One can then define projection operators $P^\pm_v =\frac{1\pm\gamma_\mu v^\mu}{2}$ that obey $P^+ + P^- = 1$ to introduce velocity dependent fields $N$ and $h$ as
\begin{equation}
	N = e^{iM v\cdot x} P^+_v\Psi, \quad h = e^{iM v\cdot x} P^-_v\Psi,
\end{equation}
such that
\begin{equation}
	\Psi = e^{-iMv\cdot x}(N+h).
\end{equation}
In the nucleon rest-frame with $v_\mu = (1,0,0,0)$, $N$ and $h$ correspond to the large and small components of the free positive-energy Dirac field as
\begin{eqnarray}
	N = \sqrt{\frac{E+M}{2M}}
	\begin{pmatrix}
	\chi_s\\
	\vec{0}
	\end{pmatrix}
	e^{-i(E-M)t +i{\bm p}\cdot {\bm x}},\label{eq: heavy baryon N definition} \\
	h = \sqrt{\frac{E+M}{2M}}
	\begin{pmatrix}
		\vec{0}\\
		\frac{{\bm \sigma}\cdot{\bm p}}{E+M}\chi_s
	\end{pmatrix}
	e^{-i(E-M)t +i{\bm p}\cdot {\bm x}},
\end{eqnarray}
where $E$ is the baryon energy, $\vec{0} = (0 , 0)^T$, and $\chi_s$ is the ordinary two-component Pauli spinor that represents the nucleon spin.
It indicates that the $h$ component is suppressed by $1/M$.
Finally, Eq.~\eqref{eq: nucleon pion Lagrangian relativistic} can be simplified further since the $e^{-iMt}$ cancels out the nucleon mass term and Eq.~\eqref{eq: nucleon pion Lagrangian relativistic} reduces to 
\begin{equation}
	\mathcal{L}_{\pi N} = \bar{N}\left(i\gamma^\mu D_\mu + \frac{g_A}{2}\gamma^\mu\gamma_5 u_\mu\right)N + \cdots,
\end{equation}
where the dots indicate additional terms involving the suppressed $h$ field. 
Keeping only the non-zero upper components in $N$, the above equation simplifies further to
\begin{equation}
 	\mathcal{L}_{\pi N} = N^\dagger\left(i D_0 - \frac{g_A}{2}{\bm \sigma} \cdot{\bm u}\right)N + \mathcal{O}(1/M),
 	\label{eq: NR EFT nucleon pion}
\end{equation}
where ${\bm \sigma}$ are the three Pauli matrices in the nucleon spin space.
Further expansion of ${\bm u}$ in terms of pion fields up to the LO in pion-nucleon interactions gives
\begin{equation}
	\mathcal{L}_{\pi N} = N^\dagger\left(i \partial_0 - \frac{g_A}{2f_\pi}{\bm \tau}\cdot({\bm \sigma} \cdot{\bm \nabla}){\bm \pi}- \frac{1}{4f^2_\pi}{\bm \tau}\cdot({\bm \pi}\times\partial_0{\bm \pi})\right)N + \cdots,
	\label{eq: NR EFT nucleon pion expanded in pion fields}
\end{equation}
This Lagrangian can be used to calculate many single nucleon observables like the pion-nucleon form factor, corrections to nucleon mass, axial and induced pseudo-scalar form factors, etc., and are reviewed in Ref.~\cite{Scherer:2002tk}.
It also captures the long and intermediate range nucleon interactions, however, it does not completely capture the nucleon interactions as nucleons also have a short-range component to their force that is described by the two-nucleon (NN) and higher-nucleon Lagrangians.

\subsection{EFT for nucleon-nucleon interactions 
\label{subsec: NN EFT}
}

To complete the nuclear EFT description of NN interactions, one needs to add the NN contact interactions made up of four nucleon fields without any mesons.
These terms are needed to renormalize loop integrals and model the unresolved short distance dynamics of the nucleon-nucleon force.
The terms are written in terms of the heavy baryon field in Eq.~\eqref{eq: heavy baryon N definition} and the parity symmetry restricts the number of derivatives in each term to even numbers.
Furthermore, the lowest order NN Lagrangian has no derivative since the nucleon interactions are not restricted by the chiral symmetry breaking like in the case of Nambu-Goldstone mesons. 
In fact, the low energy interactions among the nucleons are strong enough to bind them together and lead to shallow bound states like deuteron and triton which represent a non-perturbative phenomena.

\begin{figure}
	\centering
	\includegraphics[scale=1]{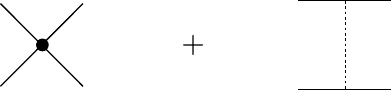}
	\caption{
    The LO Feynman diagrams contributing to the NN potential in the Weinberg power-counting. Solid line denotes a nucleon while dotted line indicates a pion.
    The first term is a four-nucleon contact term from interactions in Eq.~\eqref{eq:  Cs Ct definitions contact interaction}, where the solid circle denotes $C_S$ or $C_T$, while the second diagram is the one pion exchange coming from the interaction in Eq.~\eqref{eq: NR EFT nucleon pion expanded in pion fields} 
		\label{fig: chpt nucleon potential}}
\end{figure}

Since nucleons are fermions, the Pauli exclusion principle implies that the NN wavefunction must be anti-symmetric.  
Thus, in the isospin symmetric case, the NN wavefunction for the $S$-wave channel can either be spin-triplet and isospin-singlet or spin-singlet and isospin-triplet.
With this, the NN contact interaction at the LO is given by
\begin{equation}
	\mathcal{L}_{NN} = -\frac{1}{2} C_S (N^\dagger N)(N^\dagger N) -\frac{1}{2} C_T (N^\dagger {\bm \sigma}N) \cdot (N^\dagger {\bm \sigma}N),
	\label{eq:  Cs Ct definitions contact interaction}
\end{equation}
where $C_S$ and $C_T$ are the contact LECs that model the nucleon short-range force.
By extending the Weinberg power-counting described earlier for mesons to Feynman diagrams involving nucleons, the NN potential can be constructed order by order using the one pion exchange Lagrangian in Eq.~\eqref{eq: NR EFT nucleon pion expanded in pion fields} and the NN contact interaction in Eq.~\eqref{eq:  Cs Ct definitions contact interaction}.
At the LO, the NN amplitude is given by diagrams shown in Fig.~\ref{fig: chpt nucleon potential}, where the contact interaction is denoted by a four-nucleon-legged operator and the second contribution shows a one pion exchange diagram.
The one pion exchange provides the tensor force that is required to describe the deuteron, and explains NN scattering amplitude for high orbital angular momentum $l$ partial waves.
However, when this LO potential is used in the Lippmann-Schwinger equation to calculate the scattering amplitude, the iterative loops arising from two consecutive one pion exchanges are not suppressed and have singularities~\cite{Weinberg:1990rz,Weinberg:1991um}.
These singularities cannot be absorbed by introducing new contact interactions at this order since they do not appear in the LO order NN interaction used in the Lippmann-Schwinger equations.
This issue with the Weinberg power-counting can be avoided by introducing a power-counting directly at the amplitude level.

To see this, let us look at the analytic properties of the non-relativistic NN scattering amplitude.
For two nucleons in the isospin symmetric limit that are interacting via a potential $V$, the $S$-matrix for unmixed partial wave channel with orbital angular momentum $l$ is parameterized with a single phase shift $\delta_l$ and can be written in terms of the scattering amplitude, $\mathcal{M}$, as
\begin{equation}
	S_l = e^{2i\delta_l(p)} = 1+\left(\frac{Mp}{2\pi}\right) \mathcal{M}_l(p),
\end{equation} 
where $p$ is the NN relative momentum in the center of mass (CM) frame that is related to the NN CM energy $E$ as $p=\sqrt{M E}$.
$\mathcal{M}_l(p)$ is related to $\delta_l$ through 
\begin{equation}
	\mathcal{M}_l(p) =\frac{4\pi}{M}\frac{1}{p\cot{\delta_l}-ip}.
	\label{eq: M in pcot del deifinition}
\end{equation}
The function $p\cot{\delta_l}$ can be shown to be a real meromorphic function of $p^2$ near the origin for non-singular potentials of a finite range~\cite{Bethe:1949yr,Blatt:1949zz}.
Thus, it has a Taylor-expansion below the $t$-channel cut and around the origin that leads to the well-known effective range expansion (ERE):
\begin{equation}
	p\cot{\delta_l} = -\frac{1}{a_l} + \frac{1}{2} r_l p^2 + v_{2,l} p^4 + v_{3,l} p^6 + \cdots,
	\label{eq: ERE definition}
\end{equation}
with $a_l$, $r_l$ and $v_{i,l}$s being the scattering length, effective range and the shape parameters, respectively.
Then, the scattering amplitude can be re-written as
\begin{equation}
	\mathcal{M}_l(p) =\frac{4\pi}{M}\frac{1}{\left(-\frac{1}{a_l} + \frac{1}{2} r_l p^2 + v_{2,l} p^4 + v_{3,l} p^6+ \cdots\right)-ip}.
\end{equation}
One might think that in low energy scattering of nucleons, the  scattering parameters being length scales of low energy QCD would be comparable to $m_\pi^{-1}$ and thus $pa_l$, $pr_l$ or $pv_{i,l}$ would be good expansion parameters for expanding the scattering amplitude.
This turns out to be true in many scattering channels, however, the $S$-wave NN scattering, i.e. $l=0$, has unnatural features. The value of the $S$-wave scattering length in the spin-singlet $({^1}S_0)$ channel is $a_0^{({^1}S_0)} =-23.714\pm0.013$ fm~\cite{Noyes:1972xkg}, which is much greater than $m_\pi^{-1}\simeq 1.5$ fm.
Similarly, for the spin-triplet channel which couples S and D partial waves as $^3S_1-{^3}D_1$, the scattering length is $a_0^{(^3S_1-{^3}D_1)}  =5.423\pm0.005$ fm~\cite{Noyes:1972xkg} which gives rise to a near threshold bound state of deuteron with binding energy $\sim 2$ MeV that is much smaller than any QCD scale discussed before.
Thus, the correct momentum expansion for $\mathcal{M}_0$ in both channels is by expanding in momentum $p$ but keeping $a_0p\sim1$ as
\begin{equation}
	\mathcal{M}_0(p) =-\frac{4\pi}{M}\frac{1}	{\left(1/a_0+ip\right)}\left[1+\frac{r_0/2}{\left(1/a_0+ip\right)}p^2 + \frac{v_{2,0}}{(1/a_0+ip)} p^4
	+ 
	+\cdots
	\right],
	\label{eq: scattering amplitude expansion in KSW}
\end{equation}
which implies that to get this amplitude through a power-counting scheme, the LO must behave as $p^{-1}$.
In the Weinberg power-counting, the LO contribution is momentum independent contact interaction that scales as $p^0$.
Thus, a different power-counting scheme is needed to capture the LO $p^{-1}$ behavior.

A power-counting scheme that achieves the required LO scaling was developed by Kaplan-Savage-Wise (KSW)~\cite{Kaplan:1998tg,Kaplan:1998we}, which is described below. Instead of proceeding with pionful chiral EFT of two-nucleons, the discussion in this section is restricted to a much simpler case in which the pions are not explicit degrees of freedom.
This is a good approximation for the processes with energies $Q<<m_\pi$, where the only relevant degrees of freedom are the nucleons.
This is known as the pionless EFT with the hard scale, $\Lambda_{\cancel{\pi}}\sim m_\pi$.
The processes considered in this thesis fall in this energy region, and thus, the EFT employed for describing nucleons is the pionless EFT.

\subsection{Two-nucleon scattering in pionless EFT with KSW power-counting
\label{subsec: NN in pionless}
}

We now describe the KSW power-counting in pionless EFT by employing it to calculate the NN $S$-wave scattering for energies well below $m_\pi$ up to the LO $(\mathcal{O}(p^{-1}))$ and NLO $(\mathcal{O}(p^0))$.
This will set up and introduce notations that will be used in Ch.~\ref{ch: DBD from LQCD}.
The NN scattering is considered in the spin-singlet and the spin-triplet channels, and the partial-wave mixing will be neglected in the latter channel.
This is justified~\cite{Kaplan:1998tg,Kaplan:1998we} since the scattering amplitudes that mix $^3S_1$ to $^3D_1$ and for transitions $^3D_1$ to $^3D_1$ scattering result from one pion exchange potentials.
Moreover, the contact operators for $^3S_1-$$^3D_1$ mixing appear at $\mathcal{O}(p^1)$, and operators for $^3D_1$ to $^3D_1$ transition appear at $\mathcal{O}(p^4)$, which are at higher order than considered here.
Thus, the spin-triplet channel will be denoted by $^3S_1$ instead of $^3S_1-{^3}D_1$.

For the small energies considered, Galilean invariance is required for the non-relativistic systems.
In the absence of any external sources, the most general effective Lagrangian that is consistent with Galilean invariance, baryon number conservation and isospin symmetry is given by
\begin{equation}
	\mathcal{L} = \mathcal{L}^{(1)} + \mathcal{L}^{(2)} +\cdots,
	\label{eq: EFT Lagrangian expansion}
\end{equation}
where $\mathcal{L}^{(n)}$ is comprised of $n-$nucleon operators, and ellipsis denotes higher-nucleon operators.
Terms up to the next-to-leading (NLO) in this Lagrangian are given below.
The single-nucleon Lagrangian for the field $N=(n\; p)^T$ is
\begin{equation}
	{\cal L}^{(1)}=N^{\dagger }\bigg(i\partial_{0}+\frac{{\nabla}^2}{2M}\bigg)N+\cdots
	\label{eq: EFT 1 nucleon Lagrangian}
\end{equation}
which is the non-relativistic kinetic-energy operator of the nucleon at the LO, and the ellipsis denotes relativistic corrections.
The Lagrangian with two-nucleon operators is given by
\begin{figure}[t]
	\centering
	\includegraphics[scale=1]{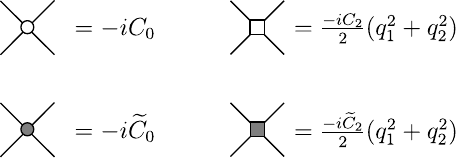}
	\caption{
    Interaction vertices for NN Lagrangian in Eq.~\eqref{eq: EFT 2 nucleon Lagrangian} are shown here.
    The (un)shaded vertices indicate the $(^1S_0)$ $^3S_1$ channel.
    Circles indicate the momentum independent LO interactions, while squares denote the momentum dependent NLO interactions.
    Here, ${\bm q}_1$ (${\bm q}_2$) is the relative momentum in the CM frame for the incoming (outgoing) NN system.
	\label{fig: C0 C2 Feynman diagram notation}}
\end{figure}
\begin{align}
	{\cal L}^{(2)}=
	&-C_{0}(N^{T}P_{i}N)^{\dagger }(N^{T}P_{i}N)
	-\widetilde{C}_{0}(N^{T}\widetilde{P}_{i}N)^{\dagger }(N^{T}\widetilde{P}_{i}N)
	\nonumber \\
	&+{\frac{C_{2}}{8}}\left[ (N^{T}P_{i}
	N)^{\dagger }[N^{T}(\overleftarrow{{\bf\nabla}}^{2}P_{i}-2
	\overleftarrow{{\bf\nabla}}\cdot P_{i}\overrightarrow{{\bf\nabla}}+
	P_{i}\overrightarrow{{\bf\nabla}}^{2})N]+{\rm h.c.}\right]
	\nonumber\\
	&+{\frac{\widetilde{C}_{2}}{8}}\left[ (N^{T}\widetilde{P}_{i}N)^{\dagger }(N^{T}[
	\overleftarrow{{\bf\nabla}}^{2}\widetilde{P}_{i}-2\overleftarrow{{\bf\nabla}}\cdot \widetilde{P}_{i}
	\overrightarrow{{\bf\nabla}}+\widetilde{P}_{i}\overrightarrow{{\bf\nabla}}^{2}]N)+{\rm h.c.}\right]+\cdots,
	\label{eq: EFT 2 nucleon Lagrangian}
\end{align}
where the overhead arrow indicates which nucleon fields are being acted by the derivative operator, and ellipsis denotes higher-derivative operators. 
Here $P_{i}$ and $\widetilde{P}_{i}$ are the spin-isospin projection operators for $^1S_0$ and $^3S_1$ channels, respectively
\footnote{Overhead tilde are used throughout the thesis to denote two-nucleon quantities in the $^3S_1$ channel, unless stated otherwise.}%
, with definition and normalization
\begin{eqnarray}
	P_{i} &\equiv &\frac{1}{\sqrt{8}}\sigma _{2}\tau _{2}\tau_{i},
	\quad \text{Tr}\left[P_{i}^{\dagger }P_{j}\right]=\frac{1}{2}\delta _{ij},
	\nonumber\\
	\widetilde{P}_{i} &\equiv &\frac{1}{\sqrt{8}}\sigma _{2}\sigma _{i}\tau _{2},\quad 
	\text{Tr}\left[\widetilde{P}_{i}^{\dagger }\widetilde{P}_{j}\right]=\frac{1}{2}\delta _{ij},
	\label{eq: spin-isospin projection operators}
\end{eqnarray}
where $\sigma_i$ $(\tau_i)$ are the Pauli matrices acting in spin (isospin) space.
The term proportional to $C_0\;(\widetilde{C}_0)$ corresponds to the LO contact interaction in the $^1S_0$ ($^3S_1$) channel while the terms proportional to $C_2\;(\widetilde{C}_2)$ describe the NLO momentum-dependent interactions.
Feynman rules for these interactions are shown in Fig.~\ref{fig: C0 C2 Feynman diagram notation}.
Furthermore, $C_0$ and $\widetilde{C}_0$ are realted to $C_S$ and $C_T$ in Eq.~\eqref{eq:  Cs Ct definitions contact interaction} via the relations $C_0 = C_S-3C_T$ and $\widetilde{C}_0 = C_S+C_T$.

The $S$-wave scattering amplitudes in $^1S_0$ and $^3S_1$ channels are denoted by $\mathcal{M}(p)$ and $\widetilde{\mathcal{M}}(p)$, with the corresponding phase shifts $\delta$ and $\widetilde{\delta}$, respectively
\footnote{Angular momentum subscript, $l$, is dropped from the notation for brevity.}
.
Following the expansion in Eq.~\eqref{eq: scattering amplitude expansion in KSW}, $\mathcal{M}(p)$ and $\widetilde{\mathcal{M}}(p)$ can be expanded up to NLO as
\begin{align}
    \mathcal{M}(p) &=\frac{4\pi}{M}\frac{-1}	{\left(1/a+ip\right)}+\frac{4\pi}{M}\frac{-r/2}{\left(1/a+ip\right)^2}p^2 +\cdots,
    \nonumber \\
    & = \mathcal{M}^{(\rm LO)}(p) + \mathcal{M}^{(\rm NLO)}(p) + \cdots,
	\label{eq: M in 1S0 expansion KSW}
\end{align}
and
\begin{align}
    \widetilde{\mathcal{M}}(p) &=\frac{4\pi}{M}\frac{-1}	{\left(1/\widetilde{a}+ip\right)}+\frac{4\pi}{M}\frac{-\widetilde{r}/2}{\left(1/\widetilde{a}+ip\right)^2}p^2 +\cdots,
    \nonumber \\
    & = \widetilde{\mathcal{M}}^{(\rm LO)}(p) + \widetilde{\mathcal{M}}^{(\rm NLO)}(p) + \cdots,
	\label{eq: M in 3S1 expansion KSW}
\end{align}
respectively.
Here, $a\;(\widetilde{a})$ is the scattering length and $r\;(\widetilde{r})$ is the effective range in $^1S_0\;(^3S_1)$ channel. 
Denoted below are their central values obtained using NN phase shifts for $S$-wave scattering generated by the Nijmegen phenomenological NN potential~\cite{Stoks:1994wp}, that are the result of fits to NN scattering data in Ref.~\cite{NNonline}:
\begin{align}
	a  &=-23.5\;\text{fm}, \hspace{2cm} r = 2.75\;\text{fm},
    \label{eq: 1S0 ERE values}\\
    \widetilde{a}  &=5.42\;\text{fm}, \hspace{2.35cm} \widetilde{r} = 1.75\;\text{fm}.
    \label{eq: 3S1 ERE values}
\end{align}

\begin{figure}
	\centering
	\includegraphics[scale=1]{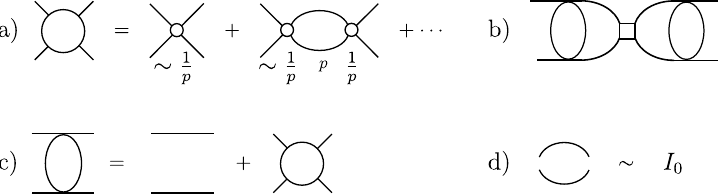}
	\caption{
    a) and b) denote the Feynman diagrammatic expansion of $\mathcal{M}^{\rm (LO)}$ and $\mathcal{M}^{\rm (NLO)}$, respectively.
    The momentum $p$ labels in a) denote the scaling of quantities according to the KSW power-counting, see the text.
    The fully dressed NN propagator appearing on the either side of $C_2$ interaction in b) is defined in c), and the NN $s$-channel loop , $I_0$, is defined in d).
    The vertex rules are according to the Fig.~\ref{fig: C0 C2 Feynman diagram notation}, and the diagrams corresponding to the $^3S_1$ channel can be obtained by replacing the unshaded parts with their shaded analogue. 
    \label{fig: LO and NLO NN in KSW}}
\end{figure}

To match $a$ $(\widetilde{a})$ with the pionless EFT LECs in Eq.~\eqref{eq: EFT 2 nucleon Lagrangian}, the Feynman diagrammatic expansion at the LO will involve $C_0$ $(\widetilde{C}_0)$, which is momentum independent.
Moreover, the only loop involved in the elastic scattering regime are the $s$-channel two nucleon loops, $I_0$, as shown in Fig.~\ref{fig: LO and NLO NN in KSW}-d.
It has the following expression for the CM $E$ (each nucleon has the energy $E/2$) based on the nucleon propagator given by Eq.~\eqref{eq: EFT 1 nucleon Lagrangian}~\cite{Kaplan:1998we}:
\begin{align}
    I_0(E) & \equiv -i \left(\frac{\mu}{2} \right)^{4-d} \int \frac{d^dq}{(2\pi)^d}\frac{i}{E/2+q^0-\frac{{\bm q}^2}{2M}+i\epsilon}\frac{i}{E/2-q^0-\frac{{\bm q}^2}{2M}+i\epsilon} \nonumber\\
    & = -i \left(\frac{\mu}{2} \right)^{4-d} \int \frac{d^{(d-1)}q}{(2\pi)^{(d-1)}}\frac{i}{E-\frac{{\bm q}^2}{M}+i\epsilon} \nonumber\\
    & = -M (-ME-i\epsilon)^{(d-3)/2} \Gamma\left(\frac{3-d}{2}\right)\frac{(\mu/2)^{4-d}}{(4\pi)^{(d-1)/2}},
    \label{eq: I0 definition}
\end{align}
where $\mu$ is the renomarlization scale, and $d$ is the dimension.
Note, $q^0$ integral is performed to get from the second step from the first.
For defining a theory, a subtraction scheme is needed to regulate integrals in the theory which amounts to dividing between contributions from the vertices and contributions from the ultraviolet (UV) part of the loop integration.
The widely used minimal subtraction scheme $(MS)$ amounts to subtracting any divergence of the sort $1/(d-4)$ before taking the $d\to4$ limit.
Since Eq.~\eqref{eq: I0 definition} does not have any such pole, the result in $MS$ scheme is given by
\begin{equation}
    I^{MS}_0(E) = \left(\frac{M}{4\pi}\right)\sqrt{-ME-i\epsilon}=-i\left(\frac{M}{4\pi}\right)ME \propto p.
\end{equation}
This suggests that in order to achieve $p^{-1}$ scaling of $\mathcal{M}^{\rm(LO)}$, one needs to perform a geometric sum of $I_0(E)$ loops, as shown by the sum of bubble Feynman diagrams in Fig~\ref{fig: LO and NLO NN in KSW}-a.
But this has a natural problem, the coefficient $C_0$ does not scale as $p^{-1}$ to keep the first diagram in Fig~\ref{fig: LO and NLO NN in KSW}-a at the same order as the rest of the diagrams in the sum.
To fix this, a power divergent scheme was proposed in Refs.~\cite{Kaplan:1998tg,Kaplan:1998we}, which amounts to subtracting from the dimensionally regulated loop all divergences corresponding to poles in lower dimensions than $d=4$ in addition to the $MS$ scheme poles.
This means removing even the $d=3$ pole in Eq.~\eqref{eq: I0 definition} by adding $\delta I_0 = \frac{-M\mu}{4\pi(d-3)}$ resulting in 
\begin{equation}
    I^{PDS}_0(E) = I_0(E) + \delta I_0 =-\left(\frac{M}{4\pi}\right)(\mu +ip).
    \label{eq: I0 expression in PDS}
\end{equation}
Using this to evaluate diagrams for the in Fig~\ref{fig: LO and NLO NN in KSW}-a, gives the expression
\begin{align}
    i\mathcal{M}^{\rm(LO)}(p) &= (-iC_0) + (-iC_0) I_0(E)C_0 + (-iC_0) (I_0(E)C_0)^2 + \cdots
    \nonumber\\
    &=\frac{-iC_0}{1-C_0I_0(E)} \xrightarrow{PDS}\frac{-iC_0}{1+M(\mu+ip)C_0/4\pi} = \frac{4\pi}{M}\frac{-i}{\frac{4\pi}{MC_0}+\mu+ip}.
    \label{eq: MLO expression}
\end{align}
Matching this to $\mathcal{M}^{\rm (LO)}$ in Eq.~\eqref{eq: M in 1S0 expansion KSW} gives
\begin{equation}
    a = \frac{1}{\mu + \frac{4\pi}{MC_0}}.
\end{equation}
But $a$ is an observable, and thus, it is renormalization scale independent.
This implies $C_0$ depends on the renormalization scale $\mu$ and must satisfy the running equation~\cite{Kaplan:1998tg,Kaplan:1998we}
\begin{equation}
    \mu\frac{d}{d\mu} C_0 = \frac{M}{4\pi}C^2_0 \mu
    \label{eq: C0 scale dependence}
\end{equation}
which leads to 
\begin{equation}
    C_0(\mu) = \frac{4\pi}{M}\left(\frac{1}{-\mu+1/a}\right).
\end{equation}
By setting $\mu=p$ in the above equation, one achieves the desired $p^{-1}$ scaling for $C_0$, such that the EFT expansion is consistent.
Same results are also true for the $^3S_1$ channel upon substituting $a\to\widetilde{a}$, $C_0\to\widetilde{C_0}$, and $\mathcal{M}\to\widetilde{\mathcal{M}}$.

This is the KSW power-counting scheme in which the LO amplitude is an infinite sum of diagrams that appear at the same order.
The NLO order is given by one NLO contact operator $C_2$ $(\widetilde{C}_2)$ with any number of $C_0$ $(\widetilde{C}_0)$ insertions on its either side that connected by $I_0$ loops, as shown in Fig~\ref{fig: LO and NLO NN in KSW}-b and~\ref{fig: LO and NLO NN in KSW}-c.

Below, the NLO amplitude is derived using the Lippmann-Schwinger equation which is equivalent to the Feynman diagram approach~\cite{Kaplan:1996xu}.
The purpose of this, is to review this method which will be used in Ch.~\ref{ch: DBD from LQCD}.

\textit{Lippmann-Schwinger approach}

To study the strongly interacting NN system, a relation between eigenstates of the non-interacting Hamiltonian and eigenstates of the strongly interacting Hamiltonian is required.
The strong Hamiltonian, $\hat{H}_S$, for the NN system is given by 
\begin{equation}
	\hat{H}_{S} \, = \, \hat{H}_0 + \hat{V}_{S} ,
	\label{eq: Strong Hamiltonian}
\end{equation}
with $\hat{H}_0$ and $\hat{V}_S$ being the free two-nucleon Hamiltonian and the strong interaction potential, respectively. 
These are derived straightforwardly from the strong-interaction Lagrangian in pionless EFT presented in Eq.~\eqref{eq: EFT Lagrangian expansion}. 
In order to treat strong interactions perturbatively, one needs to find the Feynman amplitude between two-nucleon eigenstates of $\hat{H}_S$, which are constructed from the eigenstates of $\hat{H}_0$.
In the CM frame of the two-nucleon system, the $\hat{H}_0$ eigenstate in channel $NN$ ($^1S_0$ or $^3S_1$) with relative momentum $\bm{p}$ and energy $E$ is given by
\begin{equation}
	\hat{H}_0| \bm{p},NN \rangle  = \frac{\hat{\bm p}^2}{M}\, | \bm{p},NN \rangle=
	\frac{p^2}{M} | \bm{p},NN \rangle  = E\, | \bm{p},NN \rangle ,
	\label{eq: two-nucleon eigenstates of free Hamiltonian}
\end{equation}
where $\hat{\bm p}$ is the relative-momentum operator in the CM frame, and $p^2={\bm p} \cdot {\bm p}$.
These states can be constructed from the non-interacting Hamiltonian vacuum, $|0\rangle$, using nucleon field $N(x)$ and the NN channel projection operator in a spatial volume $V$, as shown for the $^1S_0$ state
\begin{equation}
    | \bm{p}, {^1}S_0 \rangle= \frac{1}{\sqrt{V}} \int d^3{\bm x}\;d^3 {\bm y}\; e^{i{\bm p}\cdot({\bm x}-{\bm y})}\left[N^{T}({\bm x})P_{i}N({\bm y})\right]^{\dagger }|0\rangle,
\end{equation}
and similarly for the $^3S_1$ with $\widetilde{P}_i$ operator.
This state is normalized according to
\begin{equation}
	\langle \bm{p}',NN |\bm{p},NN \rangle = (2\pi)^3 \delta^3({\bm p}-{\bm p}'),
	\label{eq:free-state-norm}
\end{equation}
consistent with the normalization convention for non-relativistic two-body states with zero total momentum.
The free retarded Green's function
\begin{equation}
	G_0 (E^+) = \frac{1}{E-H_0 + i\epsilon},
	\label{eq: Green's function: Free}
\end{equation}
plays the role of NN $s$-channel loops, and it is related to $I_0(E)$ through
\begin{align}
    I_0(E) &= \int \frac{d^3{\bm q}}{(2\pi)^3}\frac{d^3{\bm q}'}{(2\pi)^3} \langle{\bm q},NN| G_0 (E^+) |{\bm q}',NN\rangle 
    \label{eq: I0 LS equivalence}\\
    &= \int \frac{d^3{\bm q}}{(2\pi)^3}\frac{d^3{\bm q}'}{(2\pi)^3} \frac{(2\pi)^3 \delta^3({\bm q}-{\bm q}')}{E-\frac{{\bm q}^2}{M} + i\epsilon},
    \nonumber
\end{align}
which agrees with Eq.~\eqref{eq: I0 definition} after performing the ${\bm q}'$ integral.

Assuming $\epsilon > 0$ and recalling the T-matrix for strong interaction
\begin{equation}
	T_S(E^+) = \hat{V}_{S} + \hat{V}_{S}\, G_0 (E^+)\,T_S(E^+),
	\label{eq: T-matrix for strong interaction}
\end{equation}
the $\hat{H}_S$ eigenstate in two-nucleon channel $NN$ with energy $E$ and relative momentum $\bm{p}$ is given by
\begin{equation}
	|\phi(E^+),NN \rangle = | {\bm p},NN \rangle + G_0 (E^+)T_S(E^+) \, | {\bm p},NN \rangle .
	\label{eq: Strong Hamiltonian Eigenstate}
\end{equation}

The Feynman amplitude for NN scattering is the non-trivial part of the full scattering amplitude between the interacting states, and it is related to the MEs of the T-matrix between the eigenstate of $\hat{H}_0$ as
\begin{equation}
	i\mathcal{M}_{NN}(E) \;=\; -i\langle{\bm q}_1,{{^1}S_0}|T_S(E^+) | {\bm q}_2,{{^1}S_0}\rangle ,
	\label{eq: NN amplitude in terms of T-matrix}
\end{equation}
where the elastic amplitude for CM energy $E$ is given by setting $q^2_1 = q^2_2 = p^2$.
From here onward, the argument for $\mathcal{M}$ and $\widetilde{\mathcal{M}}$ has been changed to the CM energy $E$ instead of relative momentum as defined earlier in Eq.~\eqref{eq: M in 1S0 expansion KSW} and~\eqref{eq: M in 3S1 expansion KSW}, respectively.

Finally, the strong retarded Green's function, $G_S (E^+)$, is defined as
\begin{equation}
	G_S (E^+) = \frac{1}{E-H_S + i\epsilon}
	= G_0 (E^+) + G_0 (E^+)\, T_S(E^+) \, G_0 (E^+).
	\label{eq: Green's function: Strong}
\end{equation}

Equations \eqref{eq: T-matrix for strong interaction}, and \eqref{eq: NN amplitude in terms of T-matrix} imply that the NN Feynman amplitude can be expressed in terms of MEs of $\hat{V}_{S}$ between free eigenstates, defined in Eq.~(\ref{eq: two-nucleon eigenstates of free Hamiltonian}).
From the Lagrangian in Eq.~\eqref{eq: EFT 2 nucleon Lagrangian}, the MEs of $\hat{V}_S$ between such momentum eigenstates are given by
\begin{align}
	&\langle {\bm q}_1,{^1S_0}| \; \hat{V}_S^{\rm (LO)} \; | {\bm q}_2, {^1S_0} \rangle = C_0,\quad
	\label{eq: Strong potential MEs I}
	\\
	&\langle {\bm q}_1,{^1S_0}| \; \hat{V}_S^{\rm (NLO)} \; | {\bm q}_2, {^1S_0} \rangle = 
	\frac{C_{2}}{2}(q^2_1+q^2_2),
	\label{eq: Strong potential MEs II}
\end{align}
for the LO and the NLO interactions, respectively.
Similar relations can be written for the spin-triplet channel with the corresponding couplings.

Using Eqs.~\eqref{eq: Strong potential MEs I} and (\ref{eq: Strong potential MEs II}) and inserting a complete set of free single- and multi-particle states in the iterative sum in Eq.~\eqref{eq: T-matrix for strong interaction}, one can obtain the $T_S$ MEs up to NLO:
\begin{align}
	\langle{\bm q}_1,{{^1}S_0}| T_S^{\rm (LO)}(E^+) | {\bm q}_2,{{^1}S_0}\rangle  &=\langle{\bm q}_1,{{^1}S_0}| \left(\hat{V}^{\rm (LO)}_{S} + \hat{V}^{\rm (LO)}_{S}\, G_0 (E^+)\,\hat{V}^{\rm (LO)}_{S} +\cdots \right)| {\bm q}_2,{{^1}S_0}\rangle \nonumber \\
    &=\frac{C_0}{1-I_0(E)\,C_0},
	\label{eq: ME of strong T-matrix I}
	\\
	\langle{\bm q}_1,{{^1}S_0}| T_S^{\rm (NLO)}(E^+) | {\bm q}_2,{{^1}S_0}\rangle &=\langle{\bm q}_1,{{^1}S_0}|\left( \hat{V}^{\rm (NLO)}_{S} + \hat{V}^{\rm (LO)}_{S}\, G_0 (E^+)\,\hat{V}^{\rm (NLO)}_{S} +\cdots\right) | {\bm q}_2,{{^1}S_0}\rangle \nonumber \\
	&=\frac{C_2}{2}\frac{(q^2_1+q^2_2-2p^2)}{1-I_0(E)\,C_0}+ \frac{C_2\,p^2}{[1-I_0(E)\,C_0]^2},
	\label{eq: ME of strong T-matrix II}
\end{align}
where in the third line, only the terms up to one $\hat{V}^{\rm (NLO)}$ are kept.
Similar expressions can be formed for the spin-triplet channel upon replacements ${{^1}S_0} \to {{^3}S_1}$, $T_S \to \widetilde{T}_S$, $C_0 \to \widetilde{C}_0$, and $C_2 \to \widetilde{C}_2$.
The on-shell NN elastic scattering amplitude is defined as
\begin{align}
	i\mathcal{M}^{\rm (LO+NLO)}(E) &\equiv -i \; \langle {^1S_0},{\bm p}| \left( T_S^{\rm (LO)}(E^+)+T_S^{\rm (NLO)}(E^+)\right) | {\bm p},{^1S_0}\rangle ,
	\label{eq: LO+NLO 2to2 Amplitude}\\
    & = \frac{-iC_0}{1-I_0(E)\,C_0}+ \frac{-iC_2\,p^2}{[1-I_0(E)\,C_0]^2},\nonumber
\end{align}
with a similar expression for the amplitude in the spin-triplet channel upon replacements ${^1S_0} \to {^3S_1}$, $\mathcal{M} \to \widetilde{\mathcal{M}}$, and $T_S \to \widetilde{T}_S$.
The LO part agrees with Eq.~\eqref{eq: MLO expression}.
From the requirement of renormalization-scale invariance of this amplitude, the scale dependence of the strong interaction couplings can be deduced. The scale dependence of $C_0$ is same as derived earlier in Eq.~\eqref{eq: C0 scale dependence} using the Feynman diagram method, and the scale dependence of $C_2$ is given by
\begin{align}
	\mu\frac{dC_2}{d\mu} = \frac{M \mu}{2 \pi}C_0 C_2,
	\label{eq: C2 scale dependence}
\end{align}
with similar relations for the spin-triplet channel upon replacements $C_0 \to \widetilde{C}_0$, and $C_2 \to \widetilde{C}_2$.

The pionless EFT with KSW power-counting has been generalized to higher order and other operators.
For example, it has been applied at higher order, ${\rm N}^2$LO and ${\rm N}^3$LO, to obtain shape parameters in NN scattering channels~\cite{Kaplan:1998tg,Kaplan:1998we,vanKolck:1998bw}.
It has also been extended to include pions systematically~\cite{Kaplan:1998we}.
Although being successful in the ${^1}S_0$ channel, it was pointed out in Ref.\cite{Fleming:1999ee} that the KSW power-counting does not converge at all in the ${^3}S_1$ channel.
On the other hand, the Weinberg power-counting that treats pions non-perturbatively has better convergence in the ${^3}S_1$ channel.
These observations led to the proposal of another systematic power-counting scheme for multi-nucleon systems in Ref.~\cite{Beane:2001bc}, which requires an expansion about the chiral limit.
For a recent review on pionless EFT and other nuclear EFTs, see Ref.~\cite{Hammer:2019poc}.

The objective of achieving a renormalizable quantum field theory for nuclear physics has enjoyed success from pionless EFT. Although its range of applicability is limited, it can be used for bound states with smaller atomic mass numbers.
But there are some unresolved questions regarding extension to larger nuclei and other power-counting schemes that could potentially yield improved calculations of observables~\cite{Vanasse:2014kxa}.
Moreover, limited experimental data and un-constrained or poorly constrained LECs are fundamental limitations of any EFT.
In Sec.~\ref{sec: FV Formalism}, we discuss the method of constraining certain LECs using the first-principles calculations in LQCD.

\subsection{External currents in pionless EFT
\label{subsec: external currents in EFT}
}
Not only the hadrons interact among themselves via the strong force but they also interact via the electroweak force.
These interactions are governed by the external currents constructed in relation to the electroweak theory.
The external currents can be systematically included in the EFT construction, and they play a crucial role in understanding the dynamics of hadronic systems at various energy scales.
External currents act as probes that help in unveiling the hidden structure and symmetries of the underlying physical theory making them essential in matching the EFT predictions to measurable quantities such as decay rates and response functions.
This section briefly reviews inclusion of external current in chiral and pionless EFT, with more emphasis on isovector electroweak currents that will be used in Sec.~\ref{subsec: NN to NN  with 1 J} and Ch.~\ref{ch: DBD from LQCD}.

Developments in applications of external currents in EFT happened at the same time of building nucleon potential from EFT~\cite{PhysRevC.42.830,RISKA1996251,Hockert:1973fot, Chemtob:1971pu,CHEMTOB1969540,D_O_Riska_1985,PhysRevC.36.1928,riska1972meson,Towner:1981hz}.
The formalism developed in Refs.~\cite{Park:1993jf,Park:1995pn,Park:2002yp} included the external gauge currents in the heavy baryon chiral EFT by extending the $R$ and $L$ transformations from global to local.
Then the definition of connection $\Gamma_\mu$ in the covariant derivative in Eq.~\eqref{eq: nucleon pion Lagrangian relativistic} needs to be changed to have $D_\mu\to h D_\mu$ as:
\begin{equation}
    \Gamma_\mu = \frac{1}{2}\left(u^\dagger\partial_\mu u + u \partial_\mu u^\dagger\right) -\frac{i}{2}\left(u^\dagger \,r_\mu u + u \,l_\mu u^\dagger\right)
    \label{eq: nucleon covariant derivative with currents}
\end{equation}
where $l_\mu$ and $r_\mu$ denote the external left-handed and right-handed gauge fields, respectively.
Their transformation $r_\mu(x)\to R(x) \, r(x) R^\dagger(x) -i (\partial_\mu R(x)) R^\dagger(x)$ and $l_\mu(x)\to L(x) \, l(x) L^\dagger(x) -i (\partial_\mu L(x))L^\dagger(x)$ ensures the additional terms from $(\partial_\mu h(x))\Psi$ with local $R(x)$ and $L(x)$ are canceled.
This definition of covariant derivative can then be used to build higher-body nucleon-current interaction operators.

External currents can also be included in pionless EFT by resitricting to the identity term in the expansion of $u(x)$ field.
Applications of pionless EFT for studying Coulomb effects, isospin breaking effects, three and four nucleon systems have been performed~\cite{Kong:2000px,Kong:1999mp, Kong:1999tw, Kong:1998sx, Gegelia:2003ta, Ando:2007fh, Ando:2010wq, Ando:2013yya, Konig:2014ufa, Konig:2015aka, Konig:2016iny, Bedaque:1997qi,Bedaque:1999vb, Bedaque:1999ve,Vanasse:2014kxa,Kirscher:2011zn,Platter:2004zs}.
Applications of electroweak external currents in pionless EFT have also been studied~\cite{Chen:1999tn,Kaplan:1998sz,Kong:1998sx,Chen:1999bg,Chen:1999vd,Kong:1999tw,Kong:1999mp,Kong:2000px,Butler:1999sv,Butler:2000zp,Butler:2002cw,Butler:2001jj,Savage:1998ae,Chen:2012hm,PhysRevC.67.025801,Chen:2005ak,Balantekin:2003ep,Chen:2004wwa}.

The weak interactions are included in pionless EFT via the current-current interactions, where the currents have the well-known$(V-A)$ form.
In this thesis, only the charged-current weak interactions are considered.
The effective Lagrangian for the charged-current (CC) weak interaction is given by
\begin{equation}
	{\cal L}_{\rm CC}\ =-G_{F} \; l_{-}^{\mu
	}J_{\mu }^{+}+{\rm h.c.} ,
	\label{eq: Charged current Lagrangian}
\end{equation}
where $G_{F}$ is the Fermi's constant. The leptonic current
\begin{equation}
    l_{-}^{\mu }=\overline{e}\gamma ^{\mu }(1-\gamma _{5})\nu
    \label{eq: Leptonic current}
\end{equation}
contains electron, $e$, and neutrino, $\nu$, fields, and the hadronic current can be written in terms of vector and axial contributions as
\begin{eqnarray}
	J_{\mu }^{+} =V_{\mu }^{+}-A_{\mu }^{+}
	=\frac{V_{\mu }^{1}+iV_{\mu }^{2}}{\sqrt{2}}-\frac{A_{\mu }^{1}+iA_{\mu }^{2}}{\sqrt{2}}.
	\label{eq: Full Hadronic current}
\end{eqnarray}
The superscript (subscript) in the hadronic (leptonic) current denotes isovector components while the subscript (superscript) denotes the space-time vector components. The vector current mediates Fermi transitions, while the axial current governs Gamow-Teller transitions, which correspond to different isospin $(\Delta I)$ and spin $(\Delta S)$ selection rules.

The non-relativistic one-body vector $(V^a_{\mu(1)})$ and axial-vector $(A^a_{\mu(1)})$ isovector current operators as shown below:~\cite{Butler:1999sv}:
\begin{align}
    V^a_{0(1)} &= \frac{1}{2} N^\dagger \tau^a N,
    \label{eq: scalar vector isovector 1 body}\\
    V^a_{k(1)} &= \frac{-\kappa^{(1)}}{2M} N^\dagger \tau^a \, \epsilon_{kij} \, \sigma_i (\overleftarrow{\nabla}_j+\overrightarrow{\nabla}_j) N,
    \label{eq: vector vector isovector 1 body}\\
    A^a_{0(1)} &= \frac{ig_A}{4M} N^\dagger \tau^a \, {\bm\sigma} \cdot (\overleftarrow{\nabla}+\overrightarrow{\nabla}) N,
    \label{eq: scalar axial-vector isovector 1 body}\\
    A^a_{k(1)} &= \frac{g_A}{2} N^\dagger \tau^a \sigma_k N,
    \label{eq: vector axial-vector isovector 1 body}
\end{align}
where $\kappa^{(1)} = \frac{1}{2}(\kappa_p-\kappa_n)$ being the isovector nucleon magentic moment in nuclear magnetons, with $\kappa_p=2.79285$ and $\kappa_n=-1.91304$~\cite{Butler:1999sv}, and
$g_A$ is the nucleon axial charge defined in Sec.~\ref{subsec: meson meson and meson nucleon EFT}.

The processes in this thesis are considered in the non-relativistic pionless EFT up to NLO.
At this order, only operators in Eq.~\eqref{eq: scalar vector isovector 1 body} and Eq.~\eqref{eq: vector axial-vector isovector 1 body} contribute since Eq.~\eqref{eq: vector vector isovector 1 body} and Eq.~\eqref{eq: scalar axial-vector isovector 1 body} are suppressed by $1/M$.
Furthermore, only the vector part of axial-vector isovector operator $(A^a_{k(2)})$ in needed at the NLO that is given by:~\cite{Butler:1999sv}:
\begin{equation}
    A^a_{k(1)} = L_{1,A}(N^{T}\widetilde{P}_{k}N)^{\dagger }(N^{T}\widetilde{P}_{a}N)
    \label{eq: vector axial-vector isovector 2 body}
\end{equation}
where LEC $L_{1,A}$ depends on the renormalization scale $\mu$.

The LEC $L_{1,A}$ will be discussed in detail in Sec.~\ref{subsubsec: singleIV}.
Furthermore, this LEC contributes to the $2\nu\beta\beta$ decay amplitude which is the central focus of Ch.~\ref{ch: DBD from LQCD}.
Constraining the value of $L_{1,A}$ using the \textit{first-principles} LQCD calculations will be a focus of sensitivity analysis performed in Sec.~\ref{sec: DBD sensitivity analysis}.
The discussion on EFT with electroweak currents and its matching relation from LQCD will be revisited in Secs.~\ref{subsubsec: singleV} and~\ref{subsubsec: singleVIV}, while the next section reviews the method LQCD.

\section{Lattice Quantum Chromodynamics
\label{sec: LQCD}
}
Quantum chromodynamics becomes strongly interacting near the QCD scale, as discussed in Sec.~\ref{sec: QCD overview}, leading to confinement where quarks and gluons cannot be observed as free particles, but are always confined within hadrons.
Perturbation theory fails in calculating observables in this energy regime, and the method of EFT, although being very useful, has severe limitations.
Lattice quantum chromodynamics (LQCD) is a non-perturbative method of numerically solving QCD that does not make any assumptions about the strength of the interaction and can be used to calculate properties of hadrons directly from the fundamental equations of QCD.
A brief overview of LQCD is provided here that is largely based on Ref.~\cite{Gattringer:2010zz,Davoudi:2014uxa}.

The key idea in performing LQCD calculations is to use Monte Carlo methods to approximately estimate an observable in QCD.
An outline of this method for a general quantum field theory is given here which starts by discretizing space-time into a four-dimensional Euclidean lattice within a FV and using the definition of expectation value of an operator $\hat{O}$ in the path integral formalism:
\begin{equation*}
    \langle \hat{O} \rangle = \frac{1}{Z} \int \mathcal{D}[\{\Phi\}] O[\{\Phi\}] e^{-S[\{\Phi\}]},
\end{equation*}
where $\{\Phi\}$ denotes the set of all quantum fields present in the underlying theory with action $S[\{\Phi\}]$.
$\int\mathcal{D}[\{\Phi\}]$ indicates the measure of the path integral, that is, integral over all possible field configurations on all of space-time, and
\begin{equation*}
     Z = \int \mathcal{D}[\{\Phi\}] e^{-S[\{\Phi\}]}
\end{equation*}
ensures proper normalization of this measure.
This can be interpreted as the expectation value of $\hat{O}$ is a weighted average over all possible field configurations, where the weight of each configuration is given by the probability distribution function $e^{-S[\{\Phi\}}]/Z$.

The discretization and FV ensures that the infinite degrees of freedom of the quantum field theory being calculated are truncated, and thus, the number of possible field configurations is finite.
This makes it possible to use Monte Carlo sampling methods to generate a large enough set of field configurations, $\mathcal{S}_{\rm MC}=\{\{\Phi_1\},\{\Phi_2\},\cdots\}$, that are sampled from a distribution with the probability distribution function given by $e^{-S[\{\Phi\}]}/Z$.
Then, $\langle \hat{O} \rangle$ can be approximated as
\begin{equation*}
    \langle \hat{O} \rangle \approx \langle \hat{O} \rangle_{\rm MC} \equiv \frac{1}{N_{\rm MC}}\sum_{\{\Phi\}\in \mathcal{S}_{\rm MC}} O[\{\Phi\}],
\end{equation*}
where $N_{\rm MC}$ is the total number of configurations generated.
The accuracy of this approximation depends on the number of sampled configurations, how well $\mathcal{S}_{\rm MC}$ mimics the desired probability distribution, etc.

Lattice QCD applies this method for QCD observables where the quantum fields are the gluon gauge fields, $A^i_\mu$, and fermionic quark and anti-quark fields, $q^a$, discussed in Sec.~\ref{sec: QCD overview}. 
Let us begin by discretizing space-time into a four-dimensional lattice, $\Lambda$, with sites labeled by $n=(n_1, n_2,n_3,n_4)$, and spacings $a$.
Here $n_4=0,1,\cdots,N_T-1$ is the Euclidean time coordinate, and $n_1,\,n_2,\,n_3=0,1,\cdots,N$ are three space coordinates, such that the hyperrectangle has volume $V=N_TN^3\,a^4$.

The next task is to construct a gauge-invariant action of quarks and gluons on this space-time such that it maps to the QCD action corresponding to the Lagrangian density in Eq.~\eqref{eq: full QCD Lagrangian} when the continuum limit, $a\to0$, is taken.
Observing that quark fields transform locally according to Eq.~\eqref{eq: guage rotation of quarks} under the gauge transformation, while gluon gauge fields transform according to Eq.~\eqref{eq: transformation of gauge field}, it makes sense to put quarks on lattice sites since the latter has derivative term in it that connects neighboring lattice sites.
Thus, matter fields reside on lattice sites as $q(n)$, where the color, flavor, and Dirac indices are suppressed for brevity.
The quarks transform under the gauge transformations as
\begin{equation}
    q(n) \to q'(n) = \Omega(n)\,q(n) \hspace{2 cm} q^\dagger(n) \to q^{\dagger'}(n) = q^\dagger(n)\Omega^\dagger(n),
    \label{eq: gauge rotation of quarks on the lattice}
\end{equation}
where $\Omega(n)$ is the discretized version of $\Omega(x)$ in Eq.~\eqref{eq: guage rotation of quarks}.
The lattice version of the covariant derivative in Eq.~\eqref{eq: covariant derivative definition} needs to be introduced to construct the discretized version of Eq.~\eqref{eq: quark covariant derivative interaction}.
Equation~\eqref{eq: quark covariant derivative interaction} provides a gauge invariant interaction between the quark at $x$ and the neighboring quark in $\mu$ direction.
Thus, to couple $q(n)$ with $q(n+\hat{\mu})$, where $\hat{\mu}$ is the directional index, one needs an object $U_\mu(n)$ that transforms as
\begin{align}
    U_\mu(n) &\rightarrow U'_\mu(n) = \Omega(n) \;U_\mu(n) \;\Omega^\dagger(n+\hat{\mu}),\nonumber\\
    U^\dagger_\mu(n) &\rightarrow U^{\dagger'}_{\mu}(n) = \Omega(n+\hat{\mu}) \; U^\dagger_{\mu}(n) \; \Omega^\dagger(n),
    \label{eq: gauge transformation of link variable lattice}
\end{align}
such that, the gauge-invariant action for quarks can be constructed as
\begin{equation}
    S_F[q,\bar{q},U] = a^4 \sum_{f=1}^{N_f} \sum_{n\in \Lambda} \bar{q}_f(n) \left(\sum_{\mu=1}^4 \gamma^{(E)}_\mu \frac{U_{\mu}(n)\,q_f(n)-U_{-\mu}(n)\,q_f(n-\hat{\mu}) }{2a} + m_f \;q_f(n)\right),
    \label{eq: lattice fermion action}
\end{equation}
where $U_{-\mu}(n)\equiv = U^\dagger(n-\hat{\mu})$ and $\gamma^{(E)}_\mu$ are the Euclidean versions of Dirac matrices.
%
\begin{figure}[t]
	\centering
	\includegraphics[scale=1]{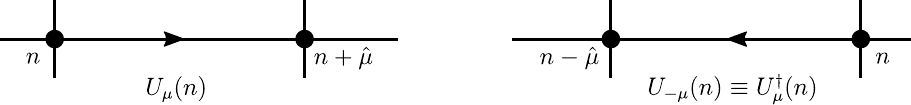}
	\caption{
    Pictorial representation of link variables $U_\mu(n)$ and $U_{-\mu}(n)$.
	\label{fig: link variables pictorially}}
\end{figure}
%

To get the action corresponding to Eq.~\eqref{eq: quark covariant derivative interaction} in the continuum limit from Eq.~\eqref{eq: gauge transformation of link variable lattice}, $U_\mu(n)$ needs to be related to the gauge field.
As shown by Wilson~\cite{Wilson:1974sk}, there indeed exists a quantity in the continuum limit that depends on the gauge field and has the same transformation properties as Eq.~\eqref{eq: gauge transformation of link variable lattice} under the gauge transformation.
This is known as the Wilson line operator or a gauge transporter that is given by
\begin{equation}
    U(x,y)\equiv\mathcal{P} \exp \left[ ig \int_x^y dz^\mu A_\mu(z) \right],
    \label{eq: Wilson line definition}
\end{equation}
where $A_\mu(x)$ is defined in Eq.~\eqref{eq: transformation of gauge field}, and $\mathcal{P}$ denotes the path ordering operation that ensures
\begin{equation}
    U(x,y) \to U'(x,y) = \Omega(x) \;U(x,y) \;\Omega^\dagger(y),
    \label{eq: Wilson line gauge transformation}
\end{equation}
under a gauge transformation.
Comparing it with Eq.~\eqref{eq: gauge transformation of link variable lattice}, it is obvious that $U_{\mu}(n)$ is a discretized Wilson line with a path that joins lattice site $n$ with $n+\hat{\mu}$
\begin{equation}
    U_{\mu}(n) = \exp\left({i\,a\,g\,A^a_\mu(n)T^a}\right),
    \label{eq: link variable in terms of gauge field}
\end{equation}
where the gauge field is taken to be constant over the path.
Thus, $U_{\mu}(n)$ is (appropriately) called a link variable which resides on the link that connects lattice sites at $n$ and $n+\hat{\mu}$, see Fig.~\ref{fig: link variables pictorially}.
Furthermore, since $T_a$s are the generators of the SU(3) group, Eq.~\eqref{eq: link variable in terms of gauge field} implies that the link variable is a general element of the SU(3) group parameterized by the value of gauge field at $n$.
Thus, $U(n)$ is a unitary matrix, $U^\dagger(n)U(n) = U(n)U^\dagger(n)=\mathds{1}_{3\times 3}$ with a unit determinant, $\det{U(n)}=1$.

It can be verified that upon expanding $U_{\mu}(n)$ and $U_{-\mu}(n)$ in Eq.~\eqref{eq: lattice fermion action} for small lattice spacing using Eq.~\eqref{eq: link variable in terms of gauge field}, one gets the discretized Euclidean action corresponding to Eq.~\eqref{eq: quark covariant derivative interaction} up to $\mathcal{O}(a)$ which approaches the continuum action for quarks in the limit $a\to0$.
%
\begin{figure}[t]
	\centering
	\includegraphics[scale=1]{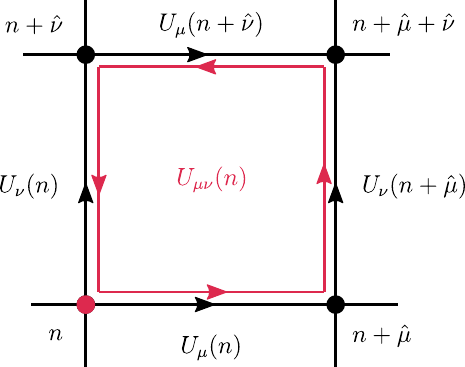}
	\caption{
    Plaquette operator $U_{\mu\nu}(n)$ is depicted here in red square along with the red lattice point that labels the plaquette, see Eq.~\eqref{eq: plaquette definition}.
    The links involved in its definition are also shown.
	\label{fig: plaquette picture}}
\end{figure}
%

To obtain the full discretized QCD action corresponding to the Lagrangian in Eq.~\eqref{eq: full QCD Lagrangian}, one still needs to find the lattice action corresponding to the gauge Lagrangian in Eq.~\eqref{eq: gauge interaction CP even }.
Such an action is expected to arise from just the link variables, which implies that one needs to find gauge-invariant terms involving only the link operators.
This can be achieved by noticing that a Wilson line in Eq.~\eqref{eq: Wilson line definition}, with its gauge transformation given in Eq.~\eqref{eq: Wilson line gauge transformation}, becomes gauge-invariant if the path is closed, i.e. $x=y$, and a trace is performed over the color indices.
This is known as a Wilson loop operator~\cite{Wilson:1974sk}.
Similarly, it can be seen from Eq.~\eqref{eq: gauge transformation of link variable lattice} that the trace of any closed loop of link variables with a path over lattice sites is gauge-invariant.
It was shown by Wilson that the discretized Euclidean gluon action corresponding to Eq.~\eqref{eq: gauge interaction CP even } can be approximated using the shortest non-trivial closed loop, also known as the plaquette, $U_{\mu\nu}(n)$, which is a product of four link variables defined as
\begin{align}
    U_{\mu\nu}(n) &=  U_\mu(n) U_\nu(n+\hat{\mu}) U_{-\mu}(n+\hat{\mu}+\hat{\nu}) U_{-\nu}(n+\hat{\nu}) \nonumber\\
    &=U_\mu(n) U_\nu(n+\hat{\mu}) U^\dagger_\mu(n+\hat{\nu}) U^\dagger_\nu(n).
    \label{eq: plaquette definition}
\end{align}
It is pictorially depicted in Fig.~\ref{fig: plaquette picture}, and the Wilson gauge action, $S_G[U]$, is then given by sum over all plaquettes, with each plaquette counted with only one orientation~\cite{Wilson:1974sk}.
This sum can be realized by a sum over all lattiec points $n$ where the plaquettes are located, combined with a sum over the Lorentz indices $1\leq\mu<\nu\leq4$,
\begin{align}
    S_G[U] &= 2\sum_{n\in\Lambda} \sum_{\mu<\nu} \mathrm{Re}\mathrm{Tr}\left[ \mathds{1} -   U_{\mu\nu}(n) \right]
    \label{eq: gluon action definition}\\
    & = \frac{a^4}{2}\sum_{n\in\Lambda} \sum_{\mu,\nu} \mathrm{Tr}\left[ F_{\mu\nu}(n)F^{\mu\nu} (n)\right] + \mathcal{O}(a^2),
    \label{eq: gluon action expanded}
\end{align}
where the second line is obtained by expanding link variables for small lattice spacings using Eq.~\eqref{eq: link variable in terms of gauge field}.
It can be seen that Eq.~\eqref{eq: gluon action expanded} agrees with the discretized Euclidean action corresponding to gluon Lagrangian in Eq.~\eqref{eq: gauge interaction CP even }, where $F_{\mu\nu}(n)$ is the discretized field stress tensor in Eq.~\eqref{eq: gluon field strength tensor definition} located at lattice point $n$.
Putting everything together, LQCD action is given by
\begin{equation}
    S_{QCD}[q,\bar{q},U] =  S_F[q,\bar{q},U] + S_G[U].
    \label{eq: LQCD action}
\end{equation}

The expectation value of an observable $\mathcal{O}[q,\bar{q},U]$ is then given by
\begin{equation}
    \langle \hat{O} \rangle = \frac{1}{Z_{QCD}} \int \mathcal{D}[q,\bar{q}] \mathcal{D}[U]O[q,\bar{q},U] e^{-S_{QCD}[q,\bar{q},U]},
    \label{eq: LQCD observable expectation value definition}
\end{equation}
where 
\begin{equation}
     Z_{QCD} = \int \mathcal{D}[q,\bar{q}] \mathcal{D}[U] e^{-S_{QCD}[q,\bar{q},U]}.
    \label{eq: LQCD Z definition}
\end{equation}

There are two issues in Eq.~\eqref{eq: LQCD observable expectation value definition} that prohibits its direct use for calculating QCD observables.
The first one concerns lattice formulation of fermions presented in Eq.~\eqref{eq: lattice fermion action}.
It suffers from certain lattice artifacts, so-called \textit{doublers}, which lead to a fermion doubling problem.
The fermion doubling problem can be better understood by examining the lattice propagator for fermions, which is the inverse of the lattice Dirac operator.
The Dirac operator between lattice points at $n$ and $m$ can be obtained from Eq.~\eqref{eq: lattice fermion action} as
\begin{equation}
    D(n|m)_{\substack{\alpha\beta\\ab}} = \sum_{\mu=1}^4 (\gamma^{(E)}_\mu)_{\alpha\beta}\frac{U_\mu(x)\delta_{n+\hat{\mu},m} - U_{-\mu}(n)\delta_{n-\hat{\mu},m}}{2a} + m \delta_{\alpha\beta}\delta_{ab}\delta_{n,m}.
    \label{eq: Dirac operator}
\end{equation}
Computing the Fourier transform of the lattice Dirac operator $D(n|m)_{\substack{\alpha\beta\\ab}}$ for trivial gauge fields $U_{\mu}(n)=\mathds{1}$ gives the Dirac propagator for free fermions:
\begin{equation}
    \widetilde{D}(p) = m\mathds{1}+\frac{i}{a}\sum_{\mu=1}^4\gamma_\mu \sin{(a p_\mu)}
    \label{eq: Dirac propagator definition}
\end{equation}
where $p_\mu$ are the momentum components. 
The inverse of the Dirac propagator governs the behavior of $n$-point functions and therefore plays an important role in calculating observables.
In a simpler case of massless fermions where $m=0$, the inverse of the naive Dirac propagator is given by
\begin{equation}
    \left.\widetilde{D}^{-1}(p)\right|_{m=0} = \frac{-ia\sum_\mu \gamma_\mu \sin{(a p_\mu)}}{\sum_\mu \sin^2{(a p_\mu)}}
    \label{eq: Dirac propagator inverse}
\end{equation}
This has the correct naive continuum limit for a fixed $p$, and it has a pole at $p=(0,0,0,0)$ which corresponds to the single fermion state.
However, the lattice inverse Dirac propagator in Eq.~\eqref{eq: Dirac propagator inverse} has additional poles that are given by $p=(p_1,p_2,p_3,p_4)$ where $p_i$ are the momenta in the first Brillouin zone, and can be 0 or $\pi/a$.
This gives rise to 15 unwanted poles that are called the doublers, and they survive in the continuum limit.
One needs to remove these doublers before performing any calculations in the discretized theory to ensure the right continuum limit for fermions.

Several approaches have been developed to address the fermion doubling problem.
The first solution was suggested by Wilson~\cite{Wilson:1974sk} which is now known as the Wilson fermions.
This method adds an operator, called the Wilson term, to the lattice Dirac operator.
This term gives the doubler fermions a mass of order of the inverse lattice spacing which leads to very heavy doublers in the continuum limit that effectively decouple from the theory.
The other approach is from Kogut and Susskind~\cite{Kogut:1974ag} that reduces the number of doublers by staggering the fermion fields which distributes the fermionic degrees of freedom across the lattice sites leading to a modified dispersion relation for lattice fermions.
This modified dispersion relation has 4 poles instead of 16 and can be manipulated to obtain the correct continuum limit, see a review in Ref.~\cite{Kronfeld:2007ek}.
A two-dimensional case of the staggering method will be revisited for the Hamiltonian formulation of lattice gauge theories in Sec.~\ref{sec: Hamiltonian for LGT}.
There are also other approaches of dealing with the fermion doubling problem that are known and are employed in calculations, like domain wall fermions~\cite{Kaplan:1992bt,Shamir:1993zy,Shamir:1993bi,Furman:1994ky}, overlap fermions~\cite{Neuberger:1997fp}, Ginsparg-Wilson fermions~\cite{Ginsparg:1981bj,Hernandez:2001yd}, twisted-mass fermions~\cite{Frezzotti:2000nk}, however, they are not the focus of this thesis and will not be discussed any further. 
While these solutions address the fermion doubling problem, they do not necessarily respect chiral symmetry in the continuum limit, which protects the fermion mass from being corrected by other mass scales in the theory.
Thus, the lattice chiral symmetry plays an important role in determining which lattice fermion formulation to use for calculations.
For example, the Wilson fermions break the lattice chiral symmetry while several others, like the domain wall fermions, respect the lattice chiral symmetry.
For more details, see a review in Ref.~\cite{Chandrasekharan:2004dph}.

Coming back to Eq.~\eqref{eq: LQCD observable expectation value definition}, the modified lattice fermion action that takes care of the fermion doubling problem resolves one of the issues of using Eq.~\eqref{eq: LQCD observable expectation value definition} for calculating QCD observables.
The second issue arises from the Fermi statistics of fermionic quark fields.
To accommodate the Pauli's principle in the path integral definition, the quark fields are needed to turn into anti-commuting variables, known as the Grassmann numbers.
The path integrals over Grassmann variables, $\int \mathcal{D}[q,\bar{q}]$, can be performed analytically due to their properties arising from anti-commuting behavior, see Ref.~\cite{Gattringer:2010zz} for details.
Equation~\eqref{eq: LQCD observable expectation value definition} then becomes
\begin{equation}
    \langle \hat{O} \rangle = \frac{1}{Z_{QCD}} \int \mathcal{D}[U] e^{-S_G[U]} Z_F[U]\langle \hat{O} \rangle_F,
    \label{eq: LQCD observable expectation OF and G}
\end{equation}
with
\begin{equation}
    \langle \hat{O} \rangle_F \equiv \frac{1}{Z_{F}} \int \mathcal{D}[q,\bar{q}] e^{-S_F[q,\bar{q},U]}O[q,\bar{q},U] ,
    \label{eq: LQCD observable expectation OF definition}
\end{equation}
being the form of the operator after performing path integrals over Grassmann variables.
$\langle \hat{O} \rangle_F$ is normalized with the measure of the fermionic path integrals $Z_{F}$ given by
\begin{equation}
    {Z_{F}} \equiv \int \mathcal{D}[q,\bar{q}] e^{-S_F[q,\bar{q},U]} = \prod_f\det{D_f}.
    \label{eq: ZF definition}
\end{equation}
Since the fermionic action is bilinear in quark fields, it can be evaluated analytically using properties of Grassman variables and Gaussian integration to obtain the last equality in Eq.~\eqref{eq: ZF definition}, where $\det{D_f}$ is the determinant of the Dirac operator defined in Eq.~\eqref{eq: Dirac operator} for flavor $f$ with appropriate modifications that remove the doublers.
Thus, by looking at the Eq.~\eqref{eq: LQCD observable expectation OF and G}, it is clear that by treating $\langle \hat{O} \rangle_F$ as a modified observable, this equation can be calculated using Monte Carlo sampling integration with the probability measure given by $\frac{1}{Z}e^{-S_G[U]}\prod_f\det{D_f}$.
This is achieved by generating a set of gauge field configurations, $\mathcal{S}_{MC} = \{\{U\}_1,\{U\}_2,\cdots\}$, where each configuration $\{U\}_i$ is a collection of $3\times3$ special unitary matrices at each lattice site, using Monte Carlo sampling method to sample from the above-mentioned distribution function, and then performing an average over the values of $\langle \hat{O} \rangle_F$ evaluated for each field configuration:
\begin{equation}
    \langle \hat{O} \rangle \approx \frac{1}{N_{\rm Config}}\sum_{\{U\}_i\in \mathcal{S}_{MC}} \langle \hat{O} \rangle_F[\{U\}_i],
    \label{eq: operator average LQCD}
\end{equation}
where $N_{\rm Config}$ is the total number of gauge configurations generated. 

Since the distribution depends on the fermion determinant, each Monte Carlo update involves its evaluation which is a determinant of a large matrix and is computationally intensive to calculate.
Note that, for $\frac{1}{Z}e^{-S_G[U]}\prod_f\det{D_f}$ to be a well defined probability measure, the factor $\prod_f\det{D_f}$ needs to be real.
This is ensured by a property of the Dirac operator known as the $\gamma_5-$hermiticity~\cite{Gattringer:2010zz}.

Calculating $\langle \hat{O} \rangle_F$ is also a challenging task.
It involves analytical evaluation of the fermionic path integrals for quark fields to obtain a form that depends on gauge fields which is then numerically calculated and averaged over the gauge field configurations.
The analytical calculation involves the application of Wick's theorem to quark fields that results in expressions involving the inverse of the Dirac operator.
This step is referred to as a fermion or quark contraction.
For example, consider obtaining the mass of the lightest meson with quantum numbers given by the operator $O^{\Gamma}(n)=\bar{d}(n)\Gamma u(n)$, where $u$ and $d$ are the up and down quark fields, and $\Gamma$ is a monomial of gamma matrices.
$O^{\Gamma \dagger}$ is then called an interpolating operator that creates a (multi-)hadron state from the QCD vacuum, $|0\rangle$, while $O^{\Gamma}$ is the interpolator that annihilates it.
The two-point function involving these interpolators is then given by
\begin{align}
    \langle O^{\Gamma}(n_t) O^{\Gamma\dagger}(0)\rangle &= \sum_k \langle0|O^{\Gamma}|k\rangle \langle k| O^{\Gamma\dagger}|0\rangle e^{-n_t a E_k}\nonumber\\
    & = A e^{-n_t a E_H} (1+\mathcal{O}(e^{-n_ta\Delta E}),
    \label{eq: meson mass from LQCD example}
\end{align}
where only the time component in operator labels are shown for brevity.
The first equality is obtained by inserting a complete set of states between the interpolators that are labeled by $k$ and energy $E_k$ and expressing operators in the Heisenberg picture.
Large $n_t$ limit is considered in arriving at the second line, where $A=|\langle 0|O^\Gamma|H\rangle|^2$ is a constant, $E_H$ is the energy of the state $|H\rangle$ that is the lightest state with a non-zero overlap with the interpolating operators, and $\Delta E$ is the energy difference to the first excited state.
Eq.~\eqref{eq: meson mass from LQCD example} implies that the mass of the lightest meson corresponding to the interpolating operator $O$ can be obtained by evaluating the left hand side for large $n_t$ and studying its dependence on $n_t$.

LQCD can be employed to evaluate the left hand side of Eq.~\eqref{eq: meson mass from LQCD example} using Eq.~\eqref{eq: operator average LQCD}.
However, it requires performing quark contractions in Eq.~\eqref{eq: meson mass from LQCD example} which turns out to be:
\begin{equation}
    \langle O^{\Gamma}(n_t) O^{\Gamma\dagger}(0)\rangle_F = -{\rm Tr}[\Gamma D^{-1}_u(n_t|0)\Gamma D^{-1}_d(0|n_t)],
    \label{eq: contraction example}
\end{equation}
where $D_u$ and $D_d$ are the Dirac operators for the $u$ and $d$ quarks, respectively.

This summarizes the method of LQCD.
As shown in the example, the mass of the lightest meson corresponding to the interpolating operator $O$ can be obtained from a first-principles calculation, and this procedure is used for obtaining the QCD hadron spectrum using LQCD.
Similarly, nuclear matrix elements of operators between hadrons can also be obtained by studying the correlation functions of operators surrounded by appropriate creation and annihilation interpolating operators.
However, there are many steps involved in actually performing such calculation using a computer.
Examples are, finding efficient algorithms for Monte Carlo sampling~\cite{Luscher:2010ae,Knechtli:2017sna,10.1007/3-540-28504-0_4}, alternative discretization known as the Symanzik improvement that can achieve smaller discretization errors~\cite{Symanzik:1983dc,Symanzik:1983gh,Luscher:1985zq}, algorithmic~\cite{Collins:2007mh,Alexandrou:2013wca,Bali:2009hu,Gambhir:2016jul,Bali:2015qya,Brannick:2007ue,Frommer:2013fsa} and hardware improvements~\cite{Clark:2009wm,Babich:2011np} for efficient evaluation of the quark propagators.
Furthermore, LQCD calculations yield dimensionless values, such as ratios of masses, that need to be converted to physical units like MeV for comparison with experimental data.
Similarly, matching lattice results that are regularized by the lattice spacing to the physical observables require renormalization analysis of operators and the QCD coupling.
The former process is known as the scale setting while the latter is called scale/renormalization analysis which are reviewed in detail in Ref.~\cite{FlavourLatticeAveragingGroupFLAG:2021npn}.

Numerical implementations of LQCD have enjoyed many successes over the years, like calculating the hadron spectrum~\cite{Bulava:2022ovd}, nuclear matrix elements~\cite{Davoudi:2020ngi}, exploring QCD phase diagram~\cite{Guenther:2020jwe}, and a historical overview of LQCD can be found in Ref.~\cite{Gross:2022hyw}.
However, often these studies are performed at larger quark masses to reduce the computational cost.
Calculations at the physical quark masses, albeit computaional-resource intensive, are already available, see the reviews mentioned above.
For the processes considered in this thesis, inputs of baryon physics from LQCD are concerned, particularly in the single-nucleon and two-nucleon sectors.
Thus, status and challenges in obtaining those observables are briefly reviewed here.

Recently, there have been developments in LQCD research where properties of individual nucleons can be accurately extrapolated to the continuum, infinite volume, and physical pion mass limits.
The first successful reproduction of the ground state spectrum of octet and decuplet baryons was obtained in Ref.~\cite{BMW:2008jgk}, and later neutron-proton mass splitting was achieved including QED effects with an accuracy of $\sim300$ keV  in Ref.~\cite{BMW:2014pzb}.
In the recent years, static charges of nucleon have been determined with full
continuum, infinite volume and physical pion mass extrapolations~\cite{Chang:2018uxx,Gupta:2018qil, Walker-Loud:2019cif, Lin:2018obj, Freeman:2012ry,Gupta:2018lvp, Alexandrou:2014sha} .
More importantly, the nucleon axial coupling, $g_A$, that is the most commonly occurring nucleon quantity in weak interactions, and plays a role in nuclear processes considered in Ch.~\ref{ch: DBD from LQCD}, has been obtained from LQCD~\cite{Chang:2018uxx,Gupta:2018qil, Walker-Loud:2019cif}.

The situation is a lot different in the two-nucleon sector.
Calculations involving two-nucleons~\cite{Beane:2006mx,Yamazaki:2015asa,NPLQCD:2012mex,Yamazaki:2012hi,NPLQCD:2013bqy,Orginos:2015aya,Berkowitz:2015eaa,Wagman:2017tmp}, and light (hyper-)nuclei up to atomic mass number four have been performed~\cite{NPLQCD:2020lxg,Green:2021qol,Green:2022rjj,Yamazaki:2015asa,NPLQCD:2012mex,Yamazaki:2012hi,HALQCD:2019wsz}, but at unphysical quarks masses such that $m_\pi\geq 300$ MeV.
Moving toward the physical point makes the resolution of the nature of two-nucleon spectrum even more challenging.
This is due to the infamous signal-to-noise problem in which statistical fluctuations grow exponentially with both the lighter pion mass, larger atomic number, and larger Euclidean time~\cite{Parisi:1983ae,Lepage:1989hd,Beane:2009gs,Wagman:2016bam}.
The increase in statistical fluctuations with Euclidean time can be especially problematic if a sub-optimal set of two-nucleon interpolating operators is used. 
In this case, it can take a very long time to reach the ground state, by which point the noise overwhelms the signal.
To obtain reliable results, it will be necessary to improve the overlap with the low-lying states by using a better interpolating operator basis.
However, this problem could be partially mitigated by carefully examining excited-state contributions at earlier Euclidean times where the data is more accurate~\cite{Drischler:2019xuo,Davoudi:2020ngi}.
Furthermore, the application of variational methods, in which a mutually orthogonal set of approximate energy eigenstates is obtained from generalized eigenvalue problem solutions for positive-definite Hermitian two-point correlation matrices constructed using a set of NN interpolating operators, removes excited-state contamination from determinations of NN ground state energy resulting from the lowest-energy set of states that have a significant overlap with the set of the interpolating operators under consideration~\cite{Amarasinghe:2021lqa,Green:2021qol,Francis:2018qch,Horz:2020zvv}.
Until now, most lattice-QCD computations of two-nucleon scattering have been carried out at only one lattice spacing due to the significant challenge posed by statistical noise in such calculations.
A recent study has conducted the first examination of the continuum limit for the binding energies of two-baryon systems with large quark masses~\cite{Green:2021qol}, where it was revealed that discretization effects could cause the binding energy of the H-dibaryon to deviate significantly from the continuum value. Therefore, additional studies of discretization effects are necessary.

The two-nucleon energy spectrum and nuclear matrix elements obtained from LQCD act as inputs to analytical formalisms that match them to the two-nucleon physical observables which are not directly accessible from LQCD, such as scattering phase shifts and nuclear decay rates.
This formalism is reviewed in the next section.

\section{Finite Volume Formalism
\label{sec: FV Formalism}
}
LQCD is a powerful numerical approach for investigating the properties and interactions of quarks and gluons.
However, it is performed with truncated degrees of freedom to make the quantum field theory finite for its numerical implementation.
This is achieved by discretizing space-time and confining QCD to a finite volume (FV), as seen in Sec.~\ref{sec: LQCD}.
The discretization effects are related to the renormalization group procedure of the QCD coupling and require systematic matching of results with the continuum theory, see Sec.~\ref{sec: LQCD}.
They characterize the behavior of the short distance physics, and thus, they are sensitive to the UV physics.
The FV effects, on the other hand, deal with the size of the system and hence sensitive to the long-distance or infra-red physics.
Furthermore, instead of being just a systematic error source, the FV effects provide a bridge to the physics in infinite volume that otherwise would have been inaccessible.
Understanding this connection will be the focus of discussion in this section, and thus, the continuum limit will be assumed throughout, focusing only on the FV effects.

The analysis of FV physics has been a subject of understanding phase transitions in the statistical physics community since a long time~\cite{doi:10.1142/4146}.
However, understanding FV effects for LQCD studies started from works by Martin L\"uscher in the 1980s~\cite{Luscher:1985dn,Luscher:1986pf,Luscher:1990ux,Luscher:1983rk,Luscher:1982uv}.
This started a framework of obtaining infinite volume physics from the FV, which was developed in the past couple of decades and is still being actively pursued to this day.
For the LQCD calculations with a periodic boundary condition imposed on the lattice, the FV framework (or formalism) utilizes the fact that particles which characterize the long-distance physics can sample the boundary and interact with their ``reflections''.
In the case of QCD, even though the gluons are massless, they are not the true degrees of freedom that sample the boundary due to confinement.
The lightest degrees of freedom are the light pseudo Nambu-Goldstone mesons, particularly pions being the lightest among them all.
If the time direction of the space-time volume is taken to be sufficiently large such that the FV effects are restricted to the spatial confinement in a cubic volume of size $L$, the nature of the FV effects can be classified into two types, as shown in Refs.~\cite{Luscher:1985dn,Luscher:1986pf}:
1) boundary sampling by virtual off-shell particles, like the self energy effects, and 2) by stable on-shell particles, like two interacting particles.
The former were shown to be exponentially suppressed in the volume size $L$ with rate determined by the lightest virtual particle, that is $e^{-m_\pi L}$, while the latter were shown to fall off power law suppressed in $L$, that is $L^{-n}$ for $n>0$.
This section gives a brief sketch of the FV effects from virtual particles by looking at the self-energy corrections to nucleon mass from the virtual pion cloud in a finite volume.
Later in Sec.~\ref{subsec: NN from FV}, the FV effects in NN scattering and its use for obtaining infinite volume scattering amplitude will be discussed in more detail.

Consider the nucleon-pion interaction in Eq.~\eqref{eq: NR EFT nucleon pion expanded in pion fields} for calculating the self-energy from pion loops in 1PI self energy diagrams, $\Sigma^{\rm(1PI)}$.
The diagrams that contribute to these corrections are the quark mass contact interaction for the nucleon at the LO and a pion loop with a nucleon or the $\Delta$-resonance at NLO, see Refs.~\cite{Beane:2011pc, Beane:2004tw,QCDSF-UKQCD:2003hmh} for details.
\begin{figure}[t]
	\centering
	\includegraphics[scale=1]{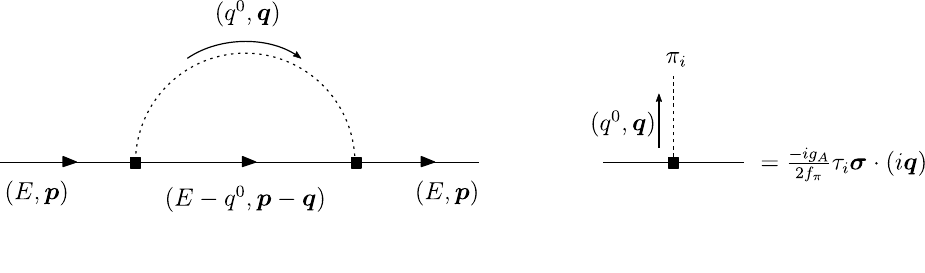}
	\caption{
    Contribution to nucleon self energy from a pion loop is shown in the figure on the left.
    Solid line indicates a nucleon, and a dotted line denotes a pion.
    The Feynman rule for the interaction vertex is shown on the right, which can be obtained from the $1/f_\pi$ contribution in Eq.~\eqref{eq: NR EFT nucleon pion expanded in pion fields}.
	\label{fig: N pi loop}}
\end{figure}
To see the $e^{-m_\pi L}$ effect derived in Ref.~\cite{Luscher:1985dn}, we consider only the diagram with a nucleon and a pion in the loop, as shown in Fig.~\ref{fig: N pi loop}, while similar results can be obtained for the $\Delta$-resonance loop as well, see Ref.~\cite{Beane:2004tw} and also Ref.~\cite{Davoudi:2014uxa} for details.

In Fig.~\ref{fig: N pi loop}, the interaction vertices can be read off from the $1/f_\pi$ term in Eq.~\eqref{eq: NR EFT nucleon pion expanded in pion fields}, and the diagram can be evaluated using the nucleon propagator given by Eq.~\eqref{eq: EFT 1 nucleon Lagrangian} and the pion propagator obtained from expanding $\Sigma(x)$ in Eq.~\eqref{eq: LO massive mesonic chiral Lagrangian}.
This gives
\begin{equation}
    -i\Sigma^{{\rm(1PI)}}_{(NLO)} = \frac{-g_A^2}{4 f_\pi^2}\int\frac{d^4q}{(2\pi)^4} \left(({\bm\sigma} \cdot {\bm q})^2 {\rm Tr}[\tau_i\tau_i]\right)
    \frac{i}{E-q_0-\frac{({\bm p}-{\bm q})^2}{2M}+i\epsilon}
    \frac{i}{q_0^2-q^2-m_\pi^2+i\epsilon} + \cdots,
    \label{eq: single nucleon self energy corrections from pion loop 1}
\end{equation}
where dots denote a similar contribution from the $\Delta$-resonance.
Dropping the $\mathcal{O}(1/M)$ contributions in the denominator and performing the $q_0$ contour integral results in:
\begin{equation}
    -i\Sigma^{{\rm(1PI)}}_{(NLO)} = i\frac{3g_A^2}{2 f_\pi^2}\int\frac{d^3{\bm q}}{(2\pi)^3}
    \frac{q^2}{-2\omega_q(E+\omega_q)} + \cdots,
    \label{eq: single nucleon self energy corrections from pion loop 2}
\end{equation}
with $\omega_q=\sqrt{q^2+m_\pi^2}$.
This integral is UV divergent and needs to be regularized.
However, since we are interested in the FV effects to the nucleon mass, which is an IR effect, this UV divergence will be canceled from its FV counterpart.

If this system is now confined into a finite cubical spatial volume of size $L^3$ with periodic boundary condition, the values of the three-momenta taken by the constituent particles cannot be continuous due to the quantization of momentum in each spatial component as
\begin{equation}
    {\bm q} = \frac{2\pi}{L} {\bm n}, \quad {\bm n}\in \mathbb{Z}^3.
    \label{eq: quantization of momentum}
\end{equation}
Thus, the self energy correction still has the form as in Eq.~\eqref{eq: single nucleon self energy corrections from pion loop 2}, except the three-momentum integral is now replaced with a sum over the quantized momenta as
\begin{equation}
    -i\Sigma^{{\rm(1PI)}V}_{(NLO)} = i\frac{3g_A^2}{2 f_\pi^2}\sum_{\bm{q}  \in \frac{2\pi}{L}\mathbb{Z}^3}
    \frac{q^2}{-2\omega_q(E+\omega_q)} + \cdots,
    \label{eq: single nucleon self energy corrections from pion loop in FV}
\end{equation}
where the superscript $V$ indicates quantities in a FV.
The FV effects, $\delta\Sigma_{(NLO)}$, are then characterized by its deviation from its infinite volume counterpart as
\begin{equation}
    \delta\Sigma_{(NLO)} =  -i(\Sigma^{{\rm(1PI)}V}_{(NLO)}-\Sigma^{{\rm(1PI)}}_{(NLO)}) = i\frac{3g_A^2}{2 f_\pi^2} \frac{1}{L^3}\sum_{\bm{q}}\hspace{-.5cm}\int \frac{-q^2}{2\omega_q(E+\omega_q)} + \cdots,
    \label{eq: single nucleon self energy corrections from pion loop in sum-integral}
\end{equation}
where the sum-integral difference notation is defined as
\begin{align}
	\frac{1}{L^3}\sum_{\bm{q}}\hspace{-.5cm}\int
	&\equiv \frac{1}{L^3}\sum_{\bm{q}  \in \frac{2\pi}{L}\mathbb{Z}^3}-\int\frac{d^3\bm{q}}{(2\pi)^3}.
	\label{eq: sumintegral definition}
\end{align}
Equation~\eqref{eq: single nucleon self energy corrections from pion loop in sum-integral} can be further simplified by replacing the sum-integral difference using the Poisson summation formula:
\begin{equation}
    \frac{1}{L^3}\sum_{\bm{q}  \in \frac{2\pi}{L}\mathbb{Z}^3} f({\bm q})= \int\frac{d^3\bm{q}}{(2\pi)^3} f({\bm q}) + \sum_{{\bm m}\neq {\bm 0}}\int \frac{d^3\bm{q}}{(2\pi)^3} f({\bm q}) e^{i {\bm q}\cdot{\bm m} L},
    \label{eq: Poisson re-summation definition}
\end{equation}
where ${\bm m}\in \mathbb{Z}^3$, and ${\bm m}\neq {\bm 0}$ indicates that ${\bm m}= (0,0,0)$ is excluded from the sum on the right hand side.
With this and assuming nucleon at rest by taking $E=0$ for simplicity,
Eq.~\eqref{eq: single nucleon self energy corrections from pion loop in sum-integral} becomes
\begin{align}
    \delta\Sigma_{(NLO)} &= -i\frac{3g_A^2}{4 f_\pi^2}  \sum_{\bm{m}\neq 0}\int \frac{d^3\bm{q}}{(2\pi)^3} \frac{q^2 e^{i {\bm q}\cdot{\bm m} L}}{q^2 + m^2_\pi} + \cdots
    \label{eq: example of exponential converging FV inf V integral}\\
    & =  -i\frac{3g_A^2}{4 f_\pi^2}  \sum_{\bm{m}\neq 0}\int_{0}^{\infty} \frac{dq}{(2\pi)^2}\frac{q^3}{i |{\bm m}| L(q^2 + m^2_\pi)} (e^{iq|{\bm m}|l}-e^{-iq|{\bm m}|l}) +\cdots \nonumber\\
    & =  \frac{3g_A^2 m^2_\pi}{16\pi f_\pi^2 L}  \sum_{\bm{m}\neq 0}\frac{e^{-m_\pi |{\bm m}| L}}{|{\bm m}|} +\cdots\nonumber
\end{align}
where $q = |{\bm q}|$, and the $q$ integral is performed for getting from the second to the third step.
Thus, the difference between the infinite volume and the FV self-energy falls off exponentially in $L$ with the slowest rate factor from terms with $|{\bm m}|=1$, and the rate in those terms is determined by $m_\pi$ which is the mass of the lightest particle in the theory.
Similar self-energy analysis for other particles can be found in Refs.~\cite{Colangelo:2003hf,Colangelo:2005cg,Colangelo:2010ba}.

The exponential suppression in $L$ of the FV effects in self-energy is a universal feature for theories with a mass gap, as shown in Ref.~\cite{Luscher:1985dn}.
For theories with a massless particle that can sample the boundary, the FV effects in self-energy do not enjoy this universal exponential suppression in $L$.
Such cases have been studied for QCD+QED systems, where photon is the lightest particle which is massless~\cite{Blum:2007cy,Blum:2010ym,Ishikawa:2012ix, Aoki:2012st,Budapest-Marseille-Wuppertal:2013rtp,Borsanyi:2014jba,Horsley:2015eaa,Horsley:2015vla,Fodor:2016bgu,deDivitiis:2013xla,Blum:2017cer,RBC:2018dos,Carrasco:2015xwa,Lubicz:2016xro,Giusti:2017dwk,Christ:2017pze,Giusti:2017jof, Boyle:2017gzv,Hermansson-Truedsson:2022tgw,Boyle:2022lsi,CSSM:2019jmq,CSSMQCDSFUKQCD:2019sxl,Blum:2023vlm,RCstar:2022yjz,Christ:2022rho,MILC:2018ddw}.
However, the IR behavior of the photon needs to be regulated to be consistent with the periodic boundary conditions.
The IR-regulated theories then exhibit power law suppression in $L$, instead of exponential, and the form of FV effects depends on the type of regularization, see Refs.~\cite{Borsanyi:2014jba, Lubicz:2016xro,Davoudi:2014qua,Endres:2015gda,Lucini:2015hfa,Davoudi:2018qpl,Feng:2018qpx}.

The next section looks at the second type of FV effect arising from two interacting on-shell particles sampling the boundary.

\subsection{Finite volume effects in two-nucleon scattering
\label{subsec: NN from FV}
}

For multi-particle physics, the FV effects are due to interacting on-shell particles sampling the boundary.
Thus, unlike the single hadron results where an off-shell particle contributes to the FV effect, the FV effects depend on the interactions between these particles.
If the range of interactions between these particles, $R$, is less than $L/2$, such that they do not interact with each other and their reflections simultaneously, the FV effects are related to their infinite volume scattering amplitude.

The first realization of this effect was by Huang and Yang~\cite{PhysRev.105.767}, where they established a connection between two-body scattering length and the energy eigenvalues in a FV for a quantum mechanical two-particle system.
L\"uscher generalized this result to quantum field theory for its application to LQCD, and derived a non-perturbative relation between the FV energy eigenvalues and two-particle scattering amplitude for a system of scalar particles with zero CM momentum~\cite{Luscher:1986pf,Luscher:1990ux}.
This formalism was then extended over the years to systems in boosted frame~\cite{Rummukainen:1995vs,Christ:2005gi, Kim:2005gf,Gockeler:2012yj}, asymmetric lattices~\cite{Feng:2004ua,Li:2003jn,Lee:2017igf}, two-particle coupled channels~\cite{Bernard:2010fp,Briceno:2014oea, Briceno:2012yi,Hansen:2012tf,He:2005ey,Lage:2009zv,Li:2014wga,Li:2012bi,Liu:2005kr}, systems of non-equal masses~\cite{Davoudi:2011md,Fu:2011xz,Leskovec:2012gb}, twisted boundary conditions~\cite{Bedaque:2004kc,Briceno:2013hya}, with different total spins~\cite{Briceno:2014oea,Beane:2003da}, and different partial wave scattering phase shifts~\cite{Luu:2011ep,Briceno:2013lba,Briceno:2013bda}, see recent reviews in Refs.~\cite{Briceno:2017max,Davoudi:2020ngi}.

The key result that relates the FV energy eigenvalues of a two-hadron system in a finite cubic volume with periodic boundary conditions, to its physical scattering amplitude, $\mathcal{M}$, is called the L\"uscher's quantization condition, which is given by~\cite{Davoudi:2020ngi}
\begin{equation}
    {\rm Det}[\mathcal{M} + F^{-1}] = 0,
    \label{eq: Luscher's quantization condition}
\end{equation}
where the determinant is over all kinematically-allowed two-hadron channels, as well as over the total angular momentum $J$ and its azimuthal component $m_J$, the total parital wave $l$, and the spin $S$ of the system.
$F$ is a kinematic function defined by
\begin{align}
    [F]&_{Jm_J,lS;J'm_{J'},l'S';\rho,\rho'} = \frac{i n_{\rho}p_{\rho}}{8\pi E} \;\delta_{S,S'}\;\delta_{\rho,\rho'}
    \Big[
    \delta_{J,J'} \delta_{m_J,m_{J'}} + \frac{2i}{\pi\gamma} \sum_{l'',m''}(\tilde{p}_\rho)^{-l''-1}\mathcal{Z}^{\bm d}_{l''m''}[1;(\tilde{p}_\rho)^2] \nonumber\\
    \times &\sum_{m_l,m_{l'},m_S}\langle lS,Jm_J | lm_l,Sm_S \rangle
    \langle l'm_{l'},Sm_S | l'S,J'm_{J'}\rangle\int d\Omega Y^*_{l,m_l} Y^*_{l'',m_{l''}} Y_{l',m_{l'}}\Big].
    \label{eq: general F definition}
\end{align}
Here, $\tilde{p}_\rho=p_\rho L/2\pi$, where $p_\rho$ is the magnitude of the relative momentum of two hadrons in channel $\rho$ in the CM frame, $E$ and $E_{\rm Lab}$ are the CM and laboratory-frame energies, respectively, $\gamma=E_{\rm Lab}/E$ is the relativistic $\gamma-$factor, and $n_\rho = 1/2$ $(1)$ if the particles in channel $\rho$ are identical (distinguishable).
L\"uscher's $\mathcal{Z}-$function is defined as
\begin{equation}
    \mathcal{Z}^{\bm d}_{lm}[s;x^2] = \sum_{\vec{r}}\frac{r^lY_{l,m}({\bm r})}{(r^2-x^2)^s},
    \label{eq: Luschers Z function full definition}
\end{equation}
where $r=|{\bm r}|$.
The sum is performed is performed over ${\bm r}=\hat{\gamma}^{-1}({\bm n}-\alpha_{\rho}{\bm d})$, where ${\bm n}\in \mathbb{Z}^3$, ${\bm d}$ is the normalized boost vector
${\bm d}={\bm P}L/2\pi$, $\alpha_{\rho}=\frac{1}{2}\left[1+\frac{m^2_{\rho,1}-m^2_{\rho,2}}{E^{2}}\right]$, and $\hat{\gamma}^{-1}{\bm x}\equiv\gamma^{-1}{\bm x}_{||} + {\bm x}_{\perp}$, with ${\bm x}_{||}$ $({\bm x}_{\perp})$ denoting the component of ${\bm x}$ that is parallel (perpendicular) to the total momentum, ${\bm P}$, $m_{\rho,1}$ and $m_{\rho,2}$ denote the masses of each hadron in channel $\rho$.
When twisted boundary conditions are used, or the volume has asymmetric extents, a modified $\mathcal{Z}$ function is required~\cite{Feng:2004ua,Detmold:2004qn,Bedaque:2004kc,Li:2003jn,Lee:2017igf}.

The result in Eq.~\eqref{eq: Luscher's quantization condition} for LQCD is derived under the following assumptions:
1) The CM energy of the two-hadron system is below the three-hadron threshold.
2) The exponentially suppressed effects from virtual particle exchanges, like in Eq.~\eqref{eq: example of exponential converging FV inf V integral}, have been ignored. Thus the fully dressed FV single hadron propagators involved are exponentially close to their infinite volume counterparts.
3) Lattice discretization effects are ignored. 4) thermal effects due to a finite length in the time direction in LQCD calculations are ignored.

L\"uscher's formalism has enabled LQCD determination of phase shifts in variety of two-hadron channels, see Refs.~\cite{Beane:2009py,Wetzorke:2002mx,NPLQCD:2010ocs,NPLQCD:2011naw,NPLQCD:2013bqy,Francis:2018qch,Wagman:2017tmp,Berkowitz:2015eaa,Beane:2012ey,Orginos:2015aya,Junnarkar:2019equ,Wilson:2015dqa,Lang:2016hnn,Briceno:2016mjc,Wu:2017qve, Brett:2018jqw, Guo:2018zss,Skerbis:2018lew,Andersen:2018mau,Dudek:2016cru,Woss:2019hse,Beane:2006mx,Horz:2020zvv,Amarasinghe:2021lqa}, and has been extended for electroweak nuclear matrix elements, as will be briefly discussed later in Sec.~\ref{subsec: NN to NN  with 1 J}. The rest of this section derives a simpler limit of Eq.~\eqref{eq: Luscher's quantization condition} to introduce the notation and set the stage for the derivation of the matching relation between FV three- and four-point functions in the two-nucleon systems and the corresponding hadronic transition amplitudes in Ch.~\ref{ch: DBD from LQCD}.

The derivation presented here is drawn primarily from the approach of Refs.~\cite{Kim:2005gf,Beane:2003da,Briceno:2015csa,Briceno:2015tza,Briceno:2019opb}. 
In all FV quantities considered, the temporal extent of space-time is taken to be infinite while the spatial extents are taken to be finite.
The spatial geometry is taken to be cubic with extents $L$, and periodic boundary conditions are assumed on the fields resulting in quantization of momemtum as shown in Eq.~\eqref{eq: quantization of momentum}.
In comparing assumptions of this work with the conditions on LQCD correlation functions, several differences can be highlighted.
First, the space-time is discretized in LQCD computations. In the following, therefore, it is assumed that the continuum limit of LQCD quantities are taken before matching to physical amplitudes through the matching relations provided.
Second, unless specified, the matching relations do not make any reference to the Euclidean space-time and correlation functions/MEs are represented in Minkowski space-time in both finite and infinite volume.
Nonetheless, quantities computed with LQCD correspond to Euclidean space-time. The distinction between Minkowski and Euclidean correlation functions becomes important only for the four-point functions in Ch.~\ref{ch: DBD from LQCD}, where the currents are separated in time, and the analytic structure of the MEs may differ significantly between different time signatures~\cite{Briceno:2019opb}.
As a result, despite the two- and three-point functions, a straightforward analytic continuation does not transform quantities from one time signature to the other.

Consider the two-nucleon systems in the spin-singlet channel.
Similar relations can be obtained identically for the spin-triplet, or any two-hadron channels for that matter, below the lowest three-hadron inelastic threshold.
Non-relativistic kinematic is assumed throughout although generalization to relativistic kinematic is straightforward and known~\cite{Kaplan:1998we}.
The relations presented are general for all partial-wave amplitudes that, besides possible mixing of the physical amplitudes in the spin-singlet channel, get further mixed in a finite cubic volume.
However, the s-wave limit will be made explicit when necessary, which is relevant for two-nucleon processes of this work that occur at low energies.
The key idea of relating the finite-volume MEs to infinite-volume amplitudes is to equate two equivalent definitions of the $n$-point correlation function in a finite volume~\cite{Briceno:2015csa,Briceno:2015tza}, as will become evident shortly.

Consider the momentum-space finite-volume two-point correlation function of two nucleons projected to zero spatial momentum, 
\begin{align}
	C_L(P) &=
	\int_{L}d^3x\int dx_0\,e^{iP\cdot x}
	\left[\langle 0|\, T[B(x)B^\dagger(0)]\, |0\rangle \right]_L,
	\label{eq: two-point correlation (P)}
\end{align}
which can be expressed, alternatively, in the mixed momentum-time representation:
\begin{align}
	\label{eq: two point correlation(x-y) in C(P)}
	C_L(y_0-x_0,{\bm P}=0)
	&\equiv L^3 \int \frac{dE}{2\pi}\;e^{-iE(y_0-x_0)} \;C_L (P)
	\\
	\label{eq: two point correlation(x-y) in <BB>}
	&= \int_Ld^3x\;d^3y 
	\left[\langle 0|\, T[B(y)B^\dagger(x)]\, |0\rangle\right]_L
	\\
	&=L^6\sum_{E_n}\;e^{-iE_n(y_0-x_0)}\; \left[\langle 0|\,B(0) \vphantom{B^\dagger} \,| E_n;L\rangle\right]_L \left[\langle E_n;L|\,B^\dagger(0)\, |0\rangle\right]_L.
	\label{eq: two point correlation: dispersion}
\end{align}
In these equations, $B^{\dagger}$ and $B$ are two-nucleon source and sink interpolating operators for the spin-singlet channel, respectively.
The operators in the second line have space-time dependences and $x\equiv(x_0,{\bm x})$ and $y\equiv(y_0,{\bm y})$.
$P$ denotes the total four-momentum of the two-nucleon system and $P=(E,\bm{P}=\bm{0})$.
The Minkowski metric $g_{\mu\nu}={\rm diag}(1,-1,-1,-1)$ is assumed throughout. $T$ denotes the Minkowski time-ordering operator, and without loss of generality, it is assumed that that $y_0>x_0$.
The subscript $L$ on the spatial integral denotes the integral is performed over a finite cubic volume, and the $[\cdots]_L$ is used to denote the finite-volume nature of the ME enclosed.
In going from the second to the third line, the relation $B(x) = e^{i\hat{P}_0x_0-i\hat{\bm P}\cdot  \bm{x}} B(0) e^{-i\hat{P}_0x_0+i\hat{\bm P}\cdot  \bm{x}}$ is used, where $\hat{P}_0$ and $\hat{\bm{P}}$ are, respectively, energy and momentum operators acting on the adjacent states.
Furthermore, a complete set of discretized finite-volume two-nucleon states are inserted, with $n$ denoting the state index.
These finite-volume states in the CM frame are characterized by the total energy of two nucleons, $E_n$, and an $L$ argument to remind the finite-volume nature of these states.
They are chosen to satisfy the normalization condition
\begin{equation}
	\langle E_m;L|E_n;L\rangle  = \delta_{n,m},
	\label{eq: Finite V states normalization}
\end{equation}
to be compared with the normalization of the infinite-volume states in Eq.~\eqref{eq:free-state-norm}.
Note that for on-shell states, $E_n=p_n^2/M$, where $p_n \equiv |\bm{p}_n|$ is the corresponding  interacting momentum of each nucleon in the CM frame.
\begin{figure}
\centering
\includegraphics[scale=1]{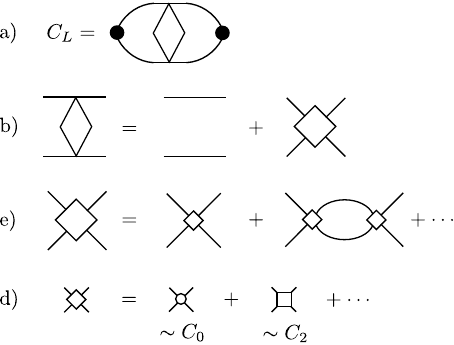}
    \caption{a) Diagrammatic representation of the finite-volume two-point correlation function corresponding to the expansion in Eq.~\eqref{eq: two point correlation analytic 1}.
    The black dots are the interpolating operators for the two-nucleon state in the spin-singlet channel. 
    The diamond represents fully dressed two-nucleon propagator defined in b) and c) where the two-nucleon Bethe-Salpeter kernel, $\mathcal{K}$, in pionless EFT in the $^1S_0$ channel is defined in d).
    The s-channel loops in c) are evaluated as a sum over discretized momenta, as discussed in the texts.
    Ellipsis in d) denotes higher-order interactions in the EFT appearing in Eq.~\eqref{eq: EFT Lagrangian expansion}.
    In all figures, the line represents the nucleon.
    The counterparts of these diagrams for the $^3S_1$ channel are exactly the same, and they are distinguished from the $^1S_0$ diagrams by the gray filled symbols in the upcoming figures.
 \label{fig: two point correlation diagram}}
\end{figure}

The goal now is to express $C_L(P)$ in terms of a diagrammatic representation illustrated in Fig.~\ref{fig: two point correlation diagram}, with the building blocks provided therein.
One can then perform the inverse Fourier integral in Eq.~\eqref{eq: two point correlation(x-y) in C(P)} to arrive at an equivalent form that can be compared against Eq.~\eqref{eq: two point correlation: dispersion}.
The correlation function $C_L(P)$ contains an infinite coupled sum with s-channel two-nucleon loops, $iI_0^V$, as depicted in Fig.~\ref{fig: two point correlation diagram}-b and -c, connected via the two-nucleon Bethe-Salpeter kernels, $\mathcal{K}$, as shown in Fig.~\ref{fig: two point correlation diagram}-d for pionless EFT.
Here $I_0^V$ is the finite volume counterpart of $I_0$ in Eq.~\eqref{eq: I0 definition} obtained after the replacement:$ \int \frac{d^4q}{(2\pi)^4} \to \frac{1}{L^3}\sum_{\bm q}\int\,\frac{dq_0}{2\pi}$

This expansion can be written in the compact notation of Refs.~\cite{Hansen:2019nir,Briceno:2019opb} as
\begin{equation}
	C_L(P) = \bar{B} \otimes iI^V_0 \sum_{n=0}^{\infty} \left[\otimes \; \mathcal{K} \otimes I^V_0 \right]^n \otimes \bar{B}^{\dagger}.
	\label{eq: two point correlation analytic 1}
\end{equation}
Here, all kinematic dependence of the functions are suppressed, but these will be made explicit shortly.
$\bar{B}^\dagger$ and $\bar{B}$ are the zero-spatial-momentum projected source and sink interpolating operators for the two-nucleon system in the $^1S_0$ channel, i.e., $\bar{B}(P)=\int d^3x \, B(x)$ and $\bar{B}^\dagger(P)=\int d^3x \, B^\dagger(x)$.
The symbol $\otimes$ should be interpreted as an sum/integral convolution of the adjacent functions.
Explicitly, for two arbitrary functions $\chi$ and $\xi$ that are smooth in $q=(q^0,\bm{q})$:
\begin{eqnarray}
	\label{eq:I-convolution-1}
	\chi \otimes I^V_0 \otimes \xi &\equiv& 
	\frac{i}{L^3}\sum_{\bm{q}  \in \frac{2\pi}{L}\mathbb{Z}^3}\int\,\frac{dq_0}{2\pi}\; \chi(q) \frac{1}{q_0-\frac{{\bm q}^2}{2M}+i\epsilon}\;\frac{1}{E-q_0-\frac{{\bm q}^2}{2M}+i\epsilon}\xi(q) 
	\\
	\label{eq:I-convolution-2}
	&=& \frac{1}{L^3} \sum_{\bm q} \; \chi(E,\bm{q})\frac{1}{E-\frac{{\bm q}^2}{M}+i\epsilon}\xi(E,\bm{q}).
\end{eqnarray}
%
\begin{figure}
\centering
\includegraphics[scale=1]{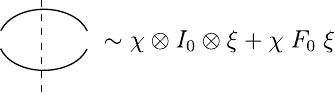}
    \caption{Diagrammatic representation of Eq.~\eqref{eq:sum-int-form} that will be used in Fig.~\ref{fig: CL splitting sum diagram} for evaluating $C_L$ in Eq.~\eqref{eq: two point correlation analytic 1}.
 \label{fig: sum-int-form diagram}}
\end{figure}
Since $\chi$ and $\xi$ are smooth functions, which is the case for interpolating fields and the kernels in Eq.~\eqref{eq: two point correlation analytic 1} below inelastic thresholds, similar to Eq.~\eqref{eq: example of exponential converging FV inf V integral}.
The only difference between the sum and the corresponding infinite-volume integral (beyond exponentially suppressed corrections) arise from singularities of $1/(E-\frac{{\bm q}^2}{M})$ function in $\bm{q}$.
Note that $i\epsilon$ is now redundant since the sum runs over discrete values of momenta. Subsequently, Eq.~\eqref{eq:I-convolution-2} can be written as
\begin{eqnarray}
	\chi \otimes I^V_0 \otimes \xi &=&
	\chi \otimes I_0 \otimes \xi + \chi \; F_0 \; \xi,
	\label{eq:sum-int-form}
\end{eqnarray}
where the first term, $\chi \otimes I_0 \otimes \xi$, is defined in the same way as the right-hand side of Eq.~\eqref{eq:I-convolution-1} except for the replacement $\frac{1}{L^3}\sum_{\bm q}\int\,\frac{dq_0}{2\pi} \to \int \frac{d^4q}{(2\pi)^4}$.
On the other hand, the second term, $\chi \; F_0 \; \xi$, can be interpreted as the ordinary product of three functions, that are in general matrices in the angular momentum basis.
In this basis, MEs of functions should be understood as projection of the original functions to a given partial-wave component with respect to the angular variable defined by $\bm{q}$.
Furthermore, the $F_0$ function, defined as the difference between the sum and integral, projects the functions adjacent to it to on-shell values of momentum, corresponding to $|\bm{q}| \to \sqrt{ME} \equiv p$.
A diagrammatic representation of Eq.~\eqref{eq:sum-int-form} is shown in Fig.~\ref{fig: sum-int-form diagram}.

For the s-wave projection that is a good approximation at low-energies, the finite-volume function $F_0$ can be written in a simple form
\begin{align}
	F_0(E) = \frac{1}{L^3}\sum_{\bm{q}}\hspace{-.5cm}\int \; \frac{1}{E-\frac{{\bm q}^2}{M}} 
	=\frac{M}{4\pi}\left[-4\pi \; c_{00}({\bm p}^2,L) +i p\right],
	\label{eq: F0 expression}
\end{align}
where the sum-integral difference notation is defined in Eq.~\eqref{eq: sumintegral definition}.
Furthermore, $c_{lm}$ is a function closely related to L\"uscher's $\mathcal{Z}$ function defined in Eq.~\eqref{eq: Luschers Z function full definition}:
\begin{align}
	c_{00}({\bm p}^2,L)=\frac{1}{\sqrt{4\pi^3}L} \mathcal{Z}_{00}\left[1;(pL/2\pi)^2\right],~\text{with}~~\mathcal{Z}_{00}[s;x^2]=\sum_{\bm n \in \mathbb{Z}^3}\frac{Y_{00}(\widehat{\bm n})}{\left(\bm{n}^2-x^2\right)^s},
    \label{eq: c00 definition}
\end{align}
where $\widehat{\bm{n}}$ denotes direction of $\bm{n}$.
Equation~\eqref{eq: F0 expression} is the special case of the general expression in Eq.~\eqref{eq: general F definition} with $s-$wave limit for single channel two-hadron system with identical particles. 
\begin{figure}
    \centering
    \includegraphics[scale=1]{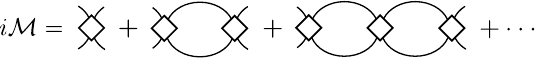}
    \caption{The full $NN\to NN$ scattering amplitude in the $^1S_0$ channel in infinite volume. Ellipsis denotes an expansion in arbitrary number of Bethe-Salpeter kernel, defined in Fig.~\ref{fig: two point correlation diagram}-d. A similar expansion can be formed for the scattering in $^3S_1$ channel with the corresponding Bethe-Salpeter kernel.}
    \label{fig: NN to NN full scattering amplitude}
\end{figure}

In the remainder of this thesis, the product of functions can be interpreted either as an ordinary product of scalar quantities in the s-wave approximation, or a matrix product of quantities expressed in the angular-momentum basis, in which case the generalized form of $F_0$ should be used, see Refs.~\cite{Luscher:1986pf,Luscher:1990ux,Kim:2005gf}.
For the spin-triplet channel, the spin quantum number must be taken into account, giving rise to the finite-volume function derived in Ref.~\cite{Briceno:2013lba}.
The s-wave approximation of this function, nonetheless, is the same at that presented in Eq.~(\ref{eq: F0 expression}).
Expressions for $F_0$ generalized to moving frames, hadrons with unequal masses, systems with relativistic kinematic, volumes with elongated sides, and general twisted BCs exist, see Refs.~\cite{Briceno:2017max} for a review.\footnote{Despite having used the s-wave approximation of finite-volume relations, the original ordering of the functions is preserved here so that generalization to higher partial waves could be achieved straightforwardly, i.e., by promoting the functions to their matrix form in the angular-momentum basis.}
\begin{figure}
    \centering
    \includegraphics[scale=1]{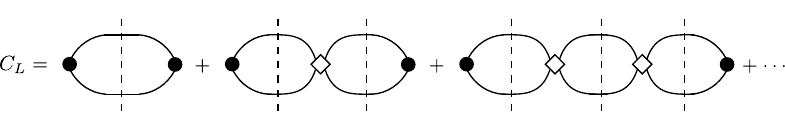}
    \caption{Diagrammatic representation of using Eq.~\eqref{eq:sum-int-form} iteratively in Eq.~\eqref{eq: two point correlation analytic 1} to obtain Eq.~\eqref{eq: two point correlation analytic 2}.
    The dotted line is defined in Fig.~\ref{fig: sum-int-form diagram}.
    }
    \label{fig: CL splitting sum diagram}
\end{figure}

Now using Eq.~(\ref{eq:sum-int-form}) to expand and rearrange Eq.~(\ref{eq: two point correlation analytic 1}), as shown diagrammatically in Fig.~\ref{fig: CL splitting sum diagram}, it is straightforward to see that $C_L(P)$ can be written as
\begin{equation}
	C_L (P) = C_{\infty}(P) + \mathcal{B} \; iF_0 \sum_{n=0}^{\infty} \left[ -\mathcal{M} F_0\right]^n\mathcal{B}^{\dagger}.
	\label{eq: two point correlation analytic 2}
\end{equation}
The first term is obtained by collecting all terms with $I_0$ in Eq.~\eqref{eq: two point correlation analytic 1} to produces the infinite-volume two-point correlation function.
All the functions in the second term are evaluated at the on-shell values of energy, i.e., they are solely a function of only one kinematic variable, $E$.
$\mathcal{B}$ and $\mathcal{B}^{\dagger}$ are the endcap functions which are built out of infinite-volume quantities.
Explicitly,
\begin{equation}
	\mathcal{B} \,=\, \bar{B} \sum_{n=0}^{\infty} \left[\otimes  \; I_0 \otimes \mathcal{K}\right]^n,
	~~ \mathcal{B}^{\dagger} \,=\, \sum_{n=0}^{\infty} \left[ \mathcal{K} \otimes I_0\; \otimes\right]^n\bar{B}^{\dagger}.
	\label{eq: in-and-out states in inf V}
\end{equation}
$i\mathcal{M}$ is the full on-shell $NN\to NN$ scattering amplitude in the spin-singlet channel, as shown in Fig.~\ref{fig: NN to NN full scattering amplitude}, and is given by
\begin{equation}
	i\mathcal{M} = -i\mathcal{K} \sum_{n=0}^{\infty} \left[\otimes \; I_0 \otimes \mathcal{K}\right]^n.
	\label{eq: NN to NN full amplitude}
\end{equation}
Given that the product of the functions in the second term of Eq.~\eqref{eq: two point correlation analytic 2} is now an ordinary (matrix) product, these terms can be summed to all orders in $F_0$:
\begin{align}
	C_L(P)&= C_{\infty}(P) \,+\, \mathcal{B}(E) \, i\mathcal{F}(E)\, \mathcal{B}^\dagger(E),
	\label{eq: two point correlation analytic 3}
\end{align}
where a new finite-volume function $\mathcal{F}$ is defined for convenience in later discussions:
\begin{align}
	\mathcal{F} & \equiv \frac{1}{F^{-1}_0+\mathcal{M}}.
	\label{eq: definition of mathcal F}
\end{align}

Equation~\eqref{eq: two point correlation analytic 3} has only singularities at interacting energies of two-nucleon system in the finite volume~\cite{Briceno:2015csa}, $E_n$. These arise from L\"uscher's `quantization condition': 
\begin{equation}
	F^{-1}_0(E)+\mathcal{M}(E)=0,~~\text{for}~~E=E_n.
	\label{eq: Luscher condition}
\end{equation}
The cubic volume does not respect the rotational symmetry, and as a result, the FV quantization conditions mix scattering amplitudes in all partial waves. 
However, at low energies the scattering amplitude is expected to be dominated by the $S$-wave interaction.
Ignoring the contribution from all higher-order partial waves, Eq.~\eqref{eq: Luscher condition} relates the $S$-wave phase shifts to a discrete set of FV energy eigenvalues, $E_n$.
For NN systems in the $^1S_0$ channel, it gives
\begin{equation}
    p_n\cot{\delta} = 4\pi c_{00}(p_n^2,L).
    \label{eq: quantization condition in phase shift}
\end{equation}

For a generic case with no s-wave approximation, the condition reads ${\rm det}\left[F^{-1}_0+\mathcal{M}\right]=0$, where determinant is taken in the angular-momentum space, as explained in Eq.~\eqref{eq: general F definition}.
In any case, a cutoff is required on the partial waves included, as otherwise the relation is not of practical use.

In summary, L\"uscher's quantization condition provides a constraint on the physical elastic amplitude of two nucleons at the finite-volume eigenenergies of the two-nucleon system, which are accessible from LQCD computations of Euclidean two-point correlation functions in a finite volume.
The condition is valid up to exponential corrections that go as $e^{-L/R}$, where $R \sim m_\pi^{-1}$ is the range of strong interactions.

Having arrived at the form in Eq.~\eqref{eq: two point correlation analytic 3} for $C_L(P)$, one can now perform integration over $E$ in Eq.~\eqref{eq: two point correlation(x-y) in C(P)} to obtain
\begin{equation}
	C_L(y_0-x_0,{\bm P}=0)=L^3\sum_{E_n}\;e^{-iE_n(y_0-x_0)}\;\mathcal{B}(E_n) \, \mathcal{R}(E_n)\, \mathcal{B}^\dagger(E_n),
	\label{eq: two point function integrated in p0}
\end{equation}
where the generalized Lellouch-L\"uscher residue matrix is given by:
\begin{equation}
	\mathcal{R}(E_n) \,=\, \lim_{E \to E_n} (E-E_n)\;\mathcal{F}(E) = \bigg[\frac{d\mathcal{F}^{-1}}{dE}\biggr|_{E=E_n}\bigg]^{-1} .
	\label{eq: definition of residue}
\end{equation}
Note that in order to arrive at this result in Minkowski space, the $i\epsilon$ prescription in the correlation functions must be recovered, i.e., $E \to E+i \epsilon$. Now comparing Eq.~(\ref{eq: two point function integrated in p0}) with Eq.~\eqref{eq: two point correlation: dispersion} for each value of $E_n$, one obtains the following matching condition:
\begin{equation}
	L^3\left[\langle 0|\,B(0) \vphantom{B^\dagger} \,| E_n;L\rangle\right]_L \left[\langle E_n;L|\,B^\dagger(0)\, |0\rangle\right]_L = \mathcal{B}(E_n) \, \mathcal{R}(E_n) \,\mathcal{B}^\dagger(E_n).
	\label{eq: matching condition for NN to NN}
\end{equation}
This relation, and its spin-singlet counterpart, will be used later in this thesis.

Going beyond two-hadron systems, the FV formalism has been extendded to three-hadron systems using various approaches~\cite{Polejaeva:2012ut,Briceno:2012rv,Hansen:2014eka,Hansen:2015zga,Hammer:2017uqm,Hammer:2017kms,Guo:2017ism,Mai:2017bge,Briceno:2017tce,Doring:2018xxx,Briceno:2018aml,Romero-Lopez:2019qrt,Jackura:2019bmu,Hansen:2020zhy}, see Ref.~\cite{Hansen:2019nir} for a recent review.
Some of those formalisms have been applied to constrain parameterizations of three-pion interactions from LQCD~\cite{Blanton:2019vdk,Mai:2018djl,Blanton:2019igq,Mai:2019fba,Culver:2019vvu} and other three-hadron systems~\cite{Kreuzer:2010ti}.
For more than three hadrons, the higher-body FV formalism extensions are not straightforward, and new ideas have been proposed that do not rely on L\"uscher's approach~\cite{Hansen:2017mnd,Bulava:2019kbi}.

For various electroweak processes involving up to two hadrons, extensions of FV formalism for electroweak currents have been worked out.
The next section reviews such extensions and provides a matching example for two-nucleon beta-decay amplitude.

\subsection{FV formalism for $2(J)\to2$ transition
\label{subsec: NN to NN  with 1 J}
}
Application of the FV formalism for matching the one-hadron to two-hadron nuclear MEs of external currents $(1(J)\to2)$ in a Euclidean space-time to the corresponding physical transition amplitude was first done by Lellouch and L\"uscher in Ref.~\cite{Lellouch:2000pv}.
Such matching requires the knowledge of FV energies and MEs along with the energy dependence of $2\to2$ scattering amplitude at those energies.
Since then, its generalizations to $0(J)\to2$, $1(J)\to2$ and $2(J)\to2$ processes for relativistic systems with generic currents have been performed~\cite{Briceno:2012yi, Meyer:2011um,Bernard:2012bi,Feng:2014gba,Briceno:2015tza,Briceno:2014uqa,Briceno:2015csa,Baroni:2018iau} and has been applied for numerical LQCD results~\cite{RBC:2015gro,Blum:2015ywa,Briceno:2016kkp}.
The main result for two-hadron transition via an electroweak current $\mathcal{J}(x)$ between the initial FV energy $E_{n_i}$ and final energy $E_{n_f}$ in a volume $L^3$  in the notations used in Sec.~\ref{sec: FV Formalism} is given by\footnote{The definition of $\mathcal{R}$ used here differs from the one in Ref.~\cite{Briceno:2015tza}. That is why the complex conjugate of $\mathcal{R}$ appears in Eq.~\eqref{eq: NN to NN 1J general result}.}~\cite{Briceno:2015tza}:
\begin{equation}
    {L^6}|\langle E_{n_f};L|\mathcal{J}(0)|E_{n_i}\rangle|^2={\rm Tr}\left[\mathcal{R}^{*}(E_{n_i})\mathcal{W}_{L,{\rm df}}(P_i,P_f,L)\mathcal{R}^{*}(E_{n_f})\mathcal{W}_{L,{\rm df}}(P_i,P_f,L)\right],
    \label{eq: NN to NN 1J general result}
\end{equation}
where $\mathcal{R}$ is defined in Eq.~\eqref{eq: definition of residue}, the trace is over the angular momentum and channel space, and the function $\mathcal{W}_{L,{\rm df}}(P_i,P_f,L)$ is defined as
\begin{equation}
    \mathcal{W}_{L,{\rm df}}(P_i,P_f,L) \equiv \mathcal{W}_{{\rm df}}(P_i,P_f) + \mathcal{M}(E_{n_f}) [G(L)\cdot w]\mathcal{M}(E_{n_i}),
    \label{eq: general definition of Wdf}
\end{equation}
where $P_{i(f)} = (E_{n_{i(f)}},{\bm p}_{i(f)})$, and $\mathcal{W}_{L,{\rm df}}$ is a divergence-free infinite volume transition amplitude where the on-shell divergences from $1(J) \to 1$ transitions on external-states hadrons are removed.
$\mathcal{M}(E_{n_i})$ and $\mathcal{M}(E_{n_f})$ are the two-hadron elastic scattering amplitudes in the initial and final state hadrons where they are assumed to be in the same channel.
$G(L)$ is a new FV function originating from the $s$-channel two-hadron loops with an insertion of the one-body current on the hadrons.
Its full definition is given in Ref.~\cite{Briceno:2015tza}.
$w$ is the $1(J) \to 1$ transition amplitude.

In the remaining section, we derive the matching relation for the two-nucleon $\beta-$decay transition in pionless EFT to set up the notation and introduce the methods used in Ch.~\ref{ch: DBD from LQCD}.
The final matching relation will be the appropriate limit of Eq.~\eqref{eq: NN to NN 1J general result}.

\subsubsection{Single-beta decay process 
\label{subsubsec: single beta decay from FV} 
}

Low-energy single-weak processes in the two-nucleon sector, including $pp$ fusion: $pp \to de^+\nu_e$, neutrino(anti-neutrino)-induced disintegration of the deuteron: $\nu(\bar{\nu}) d\to np\nu(\bar{\nu})$ and $\bar{\nu}d \to e^+nn$, and muon capture on the deuteron: $\mu d \to nn \nu_{\mu}$ have been studied in the past within the framework of pionless EFT~\cite{Kong:1999tw, Kong:2000px, Butler:2000zp, Chen:2002pv, Chen:2005ak, Acharya:2019fij}.
The Gamow-Teller ME in these processes is dominated by the single-nucleon contribution. At the LO, the ME is characterized by the nucleon's axial charge, $g_A$.
The two-nucleon contribution, characterized by the $L_{1,A}$ LEC in Eq.~\eqref{eq: vector axial-vector isovector 2 body}, is only a few percents of the full ME. It, nonetheless, constitutes the dominant source of theoretical uncertainty in the determinations of relevant cross sections~\cite{Adelberger:2010qa}.
A precise LQCD determination of the $L_{1,A}$ LEC at the physical values of the quark masses will be a critical goal of the next-generation nuclear LQCD studies in the coming years.
The two-nucleon contribution to Gamow-Teller ME can also be constrained from the half-life of the tritium~\cite{De-Leon:2016wyu, Baroni:2016xll}.
With the finite-volume LQCD technology developed in this work, simultaneous fits to single- and double-beta decay processes in the two-nucleon sector will be possible, which could provide better constraints on this unknown LEC, see also Refs.~\cite{Savage:2016kon,Shanahan:2017bgi,Tiburzi:2017iux}.
Alternatively, the constraint on $L_{1,A}$ from single-weak processes could be used to evaluate the significance of the higher-order two-weak two-nucleon contribution in the double-beta decay, hence testing the validity of the EFT power-counting in Ch.~\ref{ch: DBD from LQCD}.
Although the single-weak process $nn \to np({^3}S_1)$ does not occur in free space, its Gamow-Teller ME in the isospin-symmetric limit is the same as that in all the single-weak processes mentioned above.
Since this process constitutes a subprocess of the double-beta decay of the two-neutron system, the subscript $nn \to np$ is adopted for the amplitudes and correlation functions below, but the formalism is general for any $^3S_1 \to {^1}S_0$ or time-reversed transitions in two-nucleon systems.
Since these will be ingredients to the double-beta decay formalism in Ch.~\ref{ch: DBD from LQCD}, single-weak amplitudes and their matching to finite-volume correlation functions will be studied in detail in this section.
The derivation of the matching relation is new and was first given in Ref.~\cite{Davoudi:2020xdv}. 
It follows that of Ref.~\cite{Briceno:2015tza} for general $2(J) \to 2$ processes. The result obtained is in agreement with the earlier results on this problem presented in Refs.~\cite{Detmold:2004qn, Briceno:2012yi}.

\subsubsection{Physical single-beta decay amplitude
\label{subsubsec: singleIV}}
The focus of this section is on isovector transitions that lead to (double) $\beta$ decays.
Further, only low-energy processes are considered such that pionless EFT is a proper description.
As a result, the only relevant currents are the spatial component of the axial-vector isovector currents.
The momentum-independent scalar-isovector current in Eq.~\eqref{eq: scalar vector isovector 1 body} vanishes for beta-decay processes. 
Furthermore, the leading momentum-dependent vector-isovector current in Eq.~\eqref{eq: vector vector isovector 1 body} does not contribute to the ME in the limit where the current carries zero spatial momenta.
Finally, the leading momentum-dependent axial scalar-isovector current  Eq.~\eqref{eq: scalar axial-vector isovector 1 body} is proportional to $1/M$, and is therefore suppressed at the order of EFT considered here.
In fact, it is the nuclear ME of such currents, i.e., Gamow-Teller MEs, that are not constrained precisely phenomenologically and their determination will be the subject of a LQCD-EFT matching program.  At the LO in pionless EFT, the non-relativistic one-body operator with such quantum numbers can be formed from Eqs.~\eqref{eq: Full Hadronic current} and~\eqref{eq: vector axial-vector isovector 1 body}
\begin{figure}[t!]
    \centering
    \includegraphics[scale=1]{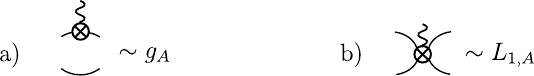}
    \caption{Diagrammatic representation of the a) single- and b) two-nucleon weak currents as defined in Eqs.~\eqref{eq: axial current : one body} and~\eqref{eq: axial current : two body}, respectively. The wavy line is a weak-current insertion.
    }
    \label{fig: gA and L1A diagram}
\end{figure}
%
\begin{equation}
	A_{k(1)}^{+} =\frac{g_{A}}{2} \, N^\dagger \tau _{+}\sigma_{k}N,
	\label{eq: axial current : one body}
\end{equation}
where $\tau_{+}=(\tau_{1}+i\tau_{2})/\sqrt{2}$ and $g_A$ is the nucleon axial charge.
$k=1,2,3$ is the spatial Lorentz index.
The momentum-independent two-body axial-vector/isovector current from Eq.~\eqref{eq: vector axial-vector isovector 2 body} is
\begin{equation}
	A_{k(2)}^{+}=L_{1,A}\big( N^{T}\widetilde{P}_{k}N\big) ^{\dagger }\big( N^{T}P_{+}N\big),
	\label{eq: axial current : two body}
\end{equation}
where $P_{+} = (P_{1}+iP_{2})/{\sqrt{2}}$.
This current contributes to the beta-decay amplitude at NLO, as shown below, leading to a renormalization group equation for $L_{1,A}$, obtained in terms of strong-interaction couplings in Eq.~\eqref{eq: EFT 2 nucleon Lagrangian} as well as the nucleon's axial charge, see Eqs.~\eqref{eq: L1A scale independent} and \eqref{eq: L1A tilde mu derivative}.
Diagrammatic representation of Eqs.~\eqref{eq: axial current : one body} and~\eqref{eq: axial current : two body} are shown in Fig.~\ref{fig: gA and L1A diagram}.

The nuclear weak-decay processes involve non-perturbative strong interactions between nucleons, governed by the terms in Eq.~\eqref{eq: EFT 2 nucleon Lagrangian}, and perturbative weak interactions between nucleons and leptons from $\mathcal{L}_{\rm CC}$ in Eq.~\eqref{eq: Charged current Lagrangian}.
The Feynman amplitude of a weak process involving $n$-lepton currents is obtained by considering the $n$th order term in the perturbative expansion of the S-matrix with respect to the CC interaction.
This amplitude can be decomposed into leptonic and hadronic parts as 
\begin{equation}
	i\mathcal{M}^{\rm full}_{X} = G_F^{n}\cdot i\mathcal{M}^{\rm lep.}_{X} \cdot i\mathcal{M}_{X},
\end{equation}
where $X$ denotes the transition considered, and all the amplitudes introduced have appropriate dependence on on-shell four-momenta of initial and final states that are suppressed for brevity.
The leptonic amplitude, $\mathcal{M}^{\rm lep.}_X$, involves the kinematics of the electron(s) and neutrino(s) and the hadronic amplitude, $\mathcal{M}_X$, is the amplitude involving initial and final nucleonic states. 

Given its non-perturbative nature, the hadronic amplitude $\mathcal{M}_{X}$ is obtained by solving the Schr\"{o}dinger's equation with the strong interaction potentials that act between nucleons in initial, final, and when relevant, intermediate nuclear states.\footnote{The potential language is appropriate as nucleons involved in low-energy processes are non-relativistic to a good approximation. Relativistic corrections can be included systematically in the EFT}
The beta-decay processes considered in this thesis involve two-nucleon initial and final states and Gamow-Teller transitions, and pionless EFT will be used to construct the amplitudes up to and including NLO.
Thus, as mentioned above, the only hadronic currents that contribute are those given in Eqs.~\eqref{eq: axial current : one body} and \eqref{eq: axial current : two body}. The Hamiltonian for these processes is given by 
\begin{equation}
	\hat{H} \, = \, \hat{H}_{S} + \hat{H}_{\rm CC}
	\label{eq: Full and Strong Hamiltonian}
\end{equation}
where $\hat{H}_S$ is the strong Hamiltonian defined in Eq.~\eqref{eq: Strong Hamiltonian}.
The weak-Hamiltonian operator $\hat{H}_{\rm CC}$ can be understood as the product of a leptonic part and a hadronic part, and the latter is the only relevant part when acting on the state of two nucleons.
In particular, assuming a final-state kinematic such that each hadronic current transfers zero spatial momenta (corresponding to pairs of electron and neutrino in a $n$-beta decay carrying away no spatial momenta), and at the EFT order considered here, the hadronic part, $\hat{A}_{\rm CC}$, can be written as
\begin{equation}
    \hat{A}_{\rm CC}=\hat{A}_{g_A}\bm{\otimes} \mathds{1} + \mathds{1} \bm{\otimes} \hat{A}_{g_A} + \hat{A}_{L_{1,A}} ,
    \label{eq: weak Hamiltonian HCC}
\end{equation}
where
\begin{equation}
    \hat{A}_{g_A}=\int \; d^3{\bm x} \; A_{k(1)}^{+}({x}), \quad
    \hat{A}_{L_{1,A}}=\int \; d^3 {\bm x}  \; A_{k(2)}^{+}({x}) ,
    \label{eq: CC hamiltonian, one-body and two-body}
\end{equation}
are constructed from the one-body and two-body current operators, respectively.
The one-body operator is extended to two-nucleon Hilbert space by taking a tensor product $(\bm{\otimes})$\footnote{Not to be mixed with the $\otimes$-product notation introduced in Sec.~\ref{sec: FV Formalism} to indicate sum or integral convolution of functions.} with the identity operator, $\mathds{1}$, in the non-participating nucleon space. For a given $NN\to NN$ process, Lorentz index $k$ determines the change in spin along the spin quantization axis.

The hadronic Feynman amplitude, $\mathcal{M}_{X}$, between the initial and final two-nucleon states with CM energies $E_i$ and $E_f$, respectively, is then given by
\begin{equation}
	i\mathcal{M}_{X} \;=\; -i \langle \phi(E_f^-),NN| \;J_{X}|\phi(E_i^+),NN \rangle ,
	\label{eq: Hadronic amplitude in terms of T-matrix}
\end{equation}
where $E_f^-$ is defined through an advanced Green's function, i.e., $i\epsilon  \to -i\epsilon$ in Eq.~\eqref{eq: Green's function: Free}, and $J_X$ is the hadronic part of the T-matrix for the weak interaction. For the $n$th order weak process, $J_X$ is constructed using $n$ insertions of $\hat{A}_{\rm CC}$ and strong retarded Green's function, $G_S (E^+)$, defined in Eq.~\eqref{eq: Green's function: Strong}, and the strong Hamiltonian eigenstates $|\phi(E^+),NN\rangle$ are defined in Eq.~\eqref{eq: Strong Hamiltonian Eigenstate}.

Equations~\eqref{eq: T-matrix for strong interaction},~\eqref{eq: Strong Hamiltonian Eigenstate}, and \eqref{eq: Hadronic amplitude in terms of T-matrix} imply that the hadronic Feynman amplitude can be expressed in terms of MEs of $\hat{V}_{S}$ and $\hat{A}_{\rm CC}$ between free eigenstates, defined in Eq.~\eqref{eq: two-nucleon eigenstates of free Hamiltonian}. 
Similar to Eqs.~\eqref{eq: Strong potential MEs I} and ~\eqref{eq: Strong potential MEs II}, one can obtain the MEs of the one-body and two-body weak hadronic Hamiltonian using Eqs.~\eqref{eq: axial current : one body}, \eqref{eq: axial current : two body}, and \eqref{eq: CC hamiltonian, one-body and two-body}, which are only nonvanishing when taken between the spin-singlet and spin-triplet states:
\begin{align}
    &\langle {\bm q}_1,{^1S_0}| \; \hat{A}_{g_A}\bm{\otimes} \mathds{1} \; | {\bm q}_2, ^3S_1 \rangle = 
    \frac{g_A}{2} \, (2\pi)^3\, \delta^3({\bm q}_1-{\bm q}_2),
    \label{eq: Weak Hamiltonian MEs I}
    \\
   & \langle {\bm q}_1,{^1S_0}| \; \hat{A}_{L_{1,A}}\; | {\bm q}_2, ^3S_1 \rangle =  L_{1,A} .
    \label{eq: Weak Hamiltonian MEs II}
\end{align}

Following the formalism presented in Sec.~\ref{sec: FV Formalism}, the single-beta decay amplitude will be constructed in this section at NLO in pionless EFT. 
The hadronic transition considered in Eq.~(\ref{eq: Hadronic amplitude in terms of T-matrix}) is $X \equiv nn\to np$.
For simplicity, we consider the initial two-nucleon state to be at rest with energy $E_i$ and further set the four-momentum carried away by the weak current to $(E,{\bf 0})$, so the two-nucleon system in the final state remains unboosted with energy $E_f$, where $E_f=E_i-E$.
The initial neutron-neutron state is a spin-singlet state while the final neutron-proton state is a spin-triplet state with $m=0,\pm1$, where $m$ is the eigenvalue of $\sigma_3$.
This state arise from the $(k=)m$th component of the currents in Eqs.~\eqref{eq: axial current : one body} and \eqref{eq: axial current : two body}.
Given the rotational symmetry, transition amplitudes with different $m$ values are equal.
Thus, we fix $m=0$ for the final state and omit the $m$ index from the notation.
\begin{figure}
    \centering
    \includegraphics[scale=1]{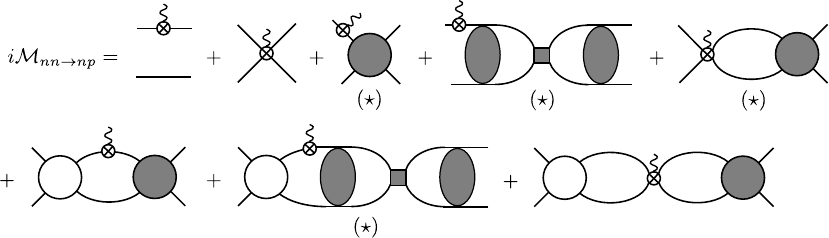}
    \caption{Diagrams contributing to the single-beta decay hadronic amplitude in Eq.~\eqref{eq: single beta decay full amplitude in inf V} at NLO in pionless EFT. All building blocks of this amplitude are introduced in Figs.~\ref{fig: C0 C2 Feynman diagram notation},~\ref{fig: LO and NLO NN in KSW}, and~\ref{fig: gA and L1A diagram}. The ($\star$) symbol under a given diagram indicates that a counterpart of the diagram must be also considered in which the diagram is reversed and the replacement ${^1}S_0 \leftrightarrow {^3}S_1$ is applied to all components except the initial and final two-nucleon states. For an on-shell amplitude, the external legs must be evaluated at on-shell kinematics.}
    \label{fig: single beta decay amplitude diagrams}
\end{figure}

\begin{figure}[t!]
    \centering
    \includegraphics[scale=1]{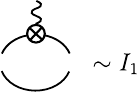}
    \caption{Diagrammatic representation of the $I_1$ loop defined in Eq.~\eqref{eq: I1 definition}.
    }
    \label{fig: I1 diagram}
\end{figure}

The hadronic part of the T-matrix for the $nn\to np$ transition, a first-order process in the weak Hamiltonian, is given by
\begin{equation}
	J_{nn\rightarrow np} \equiv \int d^3x \, \mathcal{J}_{nn \to pp}(x) \;=\; \hat{A}_{\rm CC},
	\label{eq: single beta decay T-matrix}
\end{equation}
where $\hat{A}_{\rm CC}$ is given in Eq.~\eqref{eq: weak Hamiltonian HCC}.
Substituting this in Eq.~\eqref{eq: Hadronic amplitude in terms of T-matrix} and using Eq.~\eqref{eq: Strong Hamiltonian Eigenstate}, one can decompose the amplitude into different terms depending on the number of $T_S$ insertions.
Diagrammatically, the contributions to the hadronic amplitude are shown in Fig.~\ref{fig: single beta decay amplitude diagrams}.
This amplitude can be computed up to NLO in strong interactions by inserting a complete set of hadronic states between the operators and using Eqs.~\eqref{eq: Green's function: Free}, \eqref{eq: ME of strong T-matrix I}, \eqref{eq: ME of strong T-matrix II}, \eqref{eq: Weak Hamiltonian MEs I}, and \eqref{eq: Weak Hamiltonian MEs II}.
A straightforward calculation gives
\begin{align}
    i\mathcal{M}_{nn\to np} (E_i,E_f) &= -ig_A (2\pi)^3 \delta^3(\bm{p}_i-\bm{p}_f)+ i\mathcal{M}^{\rm DF}_{nn\to np} (E_i,E_f)
    \nonumber\\
    &+ \frac{g_A}{E_i-E_f}\; \left[ i\mathcal{M}^{\rm (LO+NLO)}(E_i)-i\widetilde{\mathcal{M}}^{\rm (LO+NLO)}(E_f)\right],
    \label{eq: single beta decay full amplitude in inf V}
\end{align}
where $\bm{p}_{i(f)} = \sqrt{ME_{i(f)}}\;\widehat{\bm{p}}_{i(f)}$, with the wide hat used to denote the directionality of the three-vector.
The amplitude is conveniently split into three terms, the free-neutron decay contribution, a divergence-free (DF) term and a divergent term, with divergence referring to the behavior of the terms in the $E_i\to E_f$ limit.
Only the second term, the divergence-free amplitude, is relevant for the finite-volume calculation of this process, as shown in Sec.~\ref{subsubsec: singleV}.
It is given by
\begin{align}
    i\mathcal{M}^{\rm DF}_{nn\to np} &= i\widetilde{\mathcal{M}}^{\rm (LO+NLO)}(E_f) \; [\,i\,g_A\,I_1(E_f,E_i)\,] \; i\mathcal{M}^{\rm (LO+NLO)}(E_i)
    \nonumber\\
    & + i\widetilde{\mathcal{M}}^{\rm LO}(E_f)\; [i\widetilde{L}_{1,A}]\; i\mathcal{M}^{\rm LO}(E_i).
    \label{eq: single beta decay DF amplitude}
\end{align}
Here, $I_1$ is a loop introduced in Fig.~\ref{fig: I1 diagram} with three propagators, and is given by
\begin{align}
    I_1(E_1,E_2) &= \int \frac{d^3{q}}{(2\pi)^3} \frac{1}{E_1-\frac{{\bm q}^2}{M}+i\epsilon}\frac{1}{E_2-\frac{{\bm q}^2}{M}+i\epsilon}\nonumber\\
    &=\frac{1}{E_1-E_2}\,\left[I_0(E_2)-I_0(E_1)\right] \nonumber\\
    &=\frac{iM^{3/2}}{4\pi}\frac{1}{\sqrt{E_1}+\sqrt{E_2} }.
    \label{eq: I1 definition}
\end{align}
%
Furthermore, in arriving at Eq.~(\ref{eq: single beta decay DF amplitude}), the identity
\begin{align}
     \int \frac{d^3{q}}{(2\pi)^3} \frac{q^2}{(E_1-\frac{{\bm q}^2}{M}+i\epsilon)(E_2-\frac{{\bm q}^2}{M}+i\epsilon)}
     &=M\left[-I_0(E_1)+E_2I_1(E_1,E_2)\right]
     \nonumber\\
     &=M\left[-I_0(E_2)+E_1I_1(E_1,E_2)\right]
    \label{eq: I1 identity}
\end{align}
is used.
The $\widetilde{L}_{1,A}$ is a renormalization-scale independent combination of strong and weak couplings,\footnote{$\widetilde{L}_{1,A}$ is the one of the exceptions to the rule in this thesis that all quantities with a tilde over them correspond to the spin-triplet channel. The convention for this quantity is maintained to be compatible with the literature.}
\begin{equation}
    \widetilde{L}_{1,A} =
    \frac{L_{1,A}}{C_0\,\widetilde{C}_0}-\frac{g_A\,M}{2}\frac{(C_2+\widetilde{C}_2)}{C_0\,\widetilde{C}_0}.
    \label{eq: L1A scale independent}
\end{equation}
This is because, first of all, from Eqs.~\eqref{eq: I1 definition} and \eqref{eq: I0 definition}, $I_1(E_i,E_f)$ can be seen to be $\mu$ independent.
Secondly, divergent terms in Eq.~\eqref{eq: single beta decay full amplitude in inf V} are also renormalization-scale independent, as the $g_A$ coupling and the $NN \to NN$ elastic scattering amplitudes are scale independent.
Thus, for the physical amplitude in Eq.~\eqref{eq: single beta decay full amplitude in inf V} to be independent of the renormalization scale, $\widetilde{L}_{1,A}$ should satisfy
\begin{equation}
    \mu\frac{d}{d\mu} \widetilde{L}_{1,A} =0.
    \label{eq: L1A tilde mu derivative}
\end{equation}
This, along with Eq.~\eqref{eq: C0 scale dependence}, Eq.~\eqref{eq: C2 scale dependence}, and their spin-triplet counterparts, determines the $\mu$ dependence of $L_{1,A}$.
This result was first obtained in Refs.~\cite{Kong:2000px, Butler:2000zp}.
\subsubsection{Finite-volume correlation function
\label{subsubsec: singleV}
}
\begin{figure}
    \centering
    \includegraphics[scale=1]{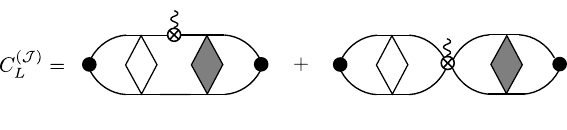}
    \caption{Diagrammatic representation of the finite-volume correlation function with a single insertion of the weak current corresponding to the expansion in Eq.~(\ref{eq:CLJ}). The black dots are the interpolating operators for the two-nucleon channels in spin-singlet or triplet states. All other components are introduced in Figs.~\ref{fig: two point correlation diagram} and~\ref{fig: gA and L1A diagram}. The loops are evaluated as a sum over discretized momenta, as discussed in the texts.}
    \label{fig: singleV finite diagrams}
\end{figure}

The first step in establishing a matching relation for the infinite-volume single-beta decay amplitude calculated in Sec.~\ref{subsubsec: singleIV} is to construct the momentum-space three-point function in a finite volume, $C_L^{(\mathcal{J})}(P_i,P_f)$. $P_{i(f)}$ denotes the four-momentum of the two-nucleon state in the spin-singlet (triplet) channel at rest with $P_{i(f)}=(E_{i(f)},\bm{P}_{i(f)}=\bm{0})$.
This correlation function can be expanded diagrammatically as shown in Fig.~\ref{fig: singleV finite diagrams}.
The result can be expressed as
\begin{equation}
    C_L^{(\mathcal{J})}(P_i,P_f) = \widetilde{\bar{B}}(P_f) \sum_{n=0}^{\infty} \big[\otimes \; I^V_0 \otimes \widetilde{\mathcal{K}} \big]^n \otimes \left[ig_A I^V_1+iL_{1,A}I^V_0\otimes I^V_0\right] \otimes \sum_{n=0}^{\infty} \left[ \mathcal{K} \otimes I^V_0 \otimes \right]^n \bar{B}^{\dagger}(P_i).
    \label{eq:CLJ}
\end{equation}
Here, $\bar{B}^{\dagger}(\widetilde{\bar{B}}^{\dagger})$ and $\bar{B}(\widetilde{\bar{B}})$ are two-nucleon source and sink interpolating operators for the spin-singlet (triplet) channel, respectively, all projected to zero spatial momentum.
$I_0^V$, $\mathcal{K}\;,\widetilde{\mathcal{K}}$, as well as $\otimes$ product notation, are defined in Sec.~\ref{sec: FV Formalism}.
$I_1^V$ is the finite-volume counterpart of the $I_1$ loop defined in Sec.~\ref{subsubsec: singleIV}, i.e., the replacement $\frac{d^3q}{(2\pi)^3} \to\frac{1}{L^3}\sum_{\bm q} $ must be applied.
It is important to expand the correlation function in Eq.~\eqref{eq:CLJ} in fixed orders in the EFT to ensure the renormalization-scale independence of physical observables extracted from the correlation function, including the energy eigenvalues and the finite-volume MEs.
For convenience, this simple form will be carried out for now but a pionless EFT expansion at NLO will be performed shortly.

The goal is to separate the infinite-volume correlation function and to re-arrange purely finite-volume contributions in a manner similar to that outlined in Sec.~\ref{sec: FV Formalism}.
Firstly as was shown in Eq.~\eqref{eq:sum-int-form}, the sum convolution of functions adjacent to $I_0^V$ can be separated into two contributions, the infinite-volume piece and a remnant finite-volume piece in which $\otimes$ products turn into ordinary (matrix) products of functions, and where the $F_0$ function puts any adjacent functions on shell.
The sum convolution of functions adjacent to $I_1^V$ for two generic left and right functions $\chi$ and $\xi$, defined as
\begin{equation}
    \chi \otimes I^V_1 \otimes \xi \equiv \frac{1}{L^3}\sum_{\bm{q}}
    \chi(\bm q)\frac{1}{E_i-\frac{{\bm q}^2}{M}}
    \frac{1}{E_f-\frac{{\bm q}^2}{M}}\xi(\bm{q}),
    \label{eq: I1 V LR definition}
\end{equation}
can proceed similarly, except now the left and right sides of $I_1^V$ have, in general, different on-shell kinematic. The general case of the convolution of functions with arbitrary momentum dependence adjacent to a (relativistic) $I_1^V$ is worked out in Ref.~\cite{Briceno:2015tza}.
Here, the situation is simpler, as at NLO in the pionless-EFT expansion of the three-point function, there are only two relevant scenarios to consider.
First, one encounters two contact (momentum-independent) kernels or current on both sides of the $I_1^V$ loop, in which case the convolution becomes trivially the ordinary product of (on-shell) functions.
The function $I_1^V$ in such ordinary products is that defined in Eq.~(\ref{eq: I1 definition}) upon replacements $\int \frac{d^3{q}}{(2\pi)^3} \to \frac{1}{L^3} \sum_{\bm q}$, $I_1 \to I_1^V$, and $I_0 \to I_0^V$,\footnote{Note that we have overloaded the $I_1^V$ symbol, used both as a sum with an arbitrary summand (made of left and right functions) times the corresponding propagators, and as the sum evaluated with just the propagators.
When $I_1^V$ is not convoluted with adjacent functions with the $\otimes$ product sign, it is meant to be a function on its own as defined in this paragraph.} with $I_0^V=I_0+F_0$, where $I_0$ and $F_0$ are defined in Eqs.~\eqref{eq: I0 definition} and \eqref{eq: F0 expression}, respectively.
It is evident that $ I^V_1$ is UV convergent and is hence scale independent.
Note also that this function is regular in the limit $E_i \to E_f$. The other possibility for the convolution of $I_1^V$ with two adjacent functions at NLO is when there is one momentum-dependent kernel $C_2 q^2$ (or $\widetilde{C}_2 q^2$) on either side of $I_1^V$ (with $\bm{q}$ being the summed momentum), in which case the application of the finite-volume counterpart of the identity in Eq.~\eqref{eq: I1 identity} leads to ordinary product of on-shell functions.

With these considerations, the expansion in Eq.~(\ref{eq:CLJ}) can be organized as follows.
First, collecting all terms with only infinite-volume loop contributions isolates the infinite-volume three-point function.
Second, one can collect all terms that contain at least one $F_0$ in the \[\sum_{n=0}^{\infty} \left[\otimes \; \widetilde{\mathcal{K}} \otimes I^V_0 \right]^n\] and \[\sum_{n=0}^{\infty} \left[ \mathcal{K} \otimes I^V_0 \otimes \right]^n\] factors.
These can then be expanded in a geometric sum to give $\widetilde{\mathcal{F}}(E_f)$ and $\mathcal{F}(E_i)$, respectively, as outlined in Sec.~\ref{sec: FV Formalism}. Now expanding only up to NLO in the pionless EFT as in Sec.~\ref{subsubsec: singleIV}, Eq.~\eqref{eq:CLJ} can be shown to reduce to
\begin{equation}
   C_L^{(\mathcal{J})}(P_i,P_f) = C_{\infty}^{(\mathcal{J})}(P_i,P_f) + 
   \widetilde{\mathcal{B}}(E_f)\,i\widetilde{\mathcal{F}}(E_f)
   \left[i\mathcal{M}^{\rm{DF},V}_{nn\to np} (E_i,E_f)\right]
   i\mathcal{F}(E_i)\,\mathcal{B}^\dagger(E_i)+\cdots.
   \label{eq: three-point correlation final}
\end{equation}
Here, ellipsis denotes all other terms in Eq.~(\ref{eq:CLJ}) that neither contribute to $C_{\infty}^{(\mathcal{J})}$ nor to the finite-volume terms in which factors of $\widetilde{\mathcal{F}}$ and $\mathcal{F}$ are both present.
The reason for such a rearrangement of the terms in the finite-volume correlation function will become evident shortly. $\mathcal{M}^{\rm{DF},V}_{nn\to np} $ is the finite-volume counterpart of the single-beta decay divergence-free amplitude $\mathcal{M}^{\rm DF}_{nn\to np} $ defined in Eq.~(\ref{eq: single beta decay DF amplitude}), in which $I_1$ is replaced with $I_1^V$. Explicitly,
\begin{eqnarray}
   i \mathcal{M}^{\rm{DF},V}_{nn\to np} (E_i,E_f)=
   i\mathcal{M}^{\rm DF}_{nn\to np} (E_i,E_f)+
   i\widetilde{\mathcal{M}}^{\rm (LO+NLO)}(E_f) \; \left[ig_A\,F_1(E_f,E_i)\right] \; i\mathcal{M}^{\rm (LO+NLO)}(E_i).
   \nonumber\\
   \label{eq: DF amplitude: finite V}
\end{eqnarray}
Here, $F_1=I_1^V-I_1$, or in terms of the sum-integral notation defined in Eq.~(\ref{eq: sumintegral definition}),
\begin{equation}
    F_1(E_f,E_i) = \frac{1}{L^3}\sum_{\bm{q}}\hspace{-.5cm}\int
    \frac{1}{E_i-\frac{{\bm q}^2}{M}}
    \frac{1}{E_f-\frac{{\bm q}^2}{M}}
    =\frac{1}{E_i-E_f}\,\left[F_0(E_f)-F_0(E_i)\right],
    \label{eq:F1def}
\end{equation}
where $F_0$ is defined in Eq.~(\ref{eq: F0 expression}). Finally, it should be noted that all terms in Eq.~(\ref{eq: three-point correlation final}) are evaluated at the on-shell kinematic, given their proximity to $F_0$ functions, and given the convolution rules with the $I_1^V$ loop explained above. Eq.~(\ref{eq: three-point correlation final}) will be used in the next subsection to derive the matching relation between the finite-volume ME and the physical hadronic amplitude for the single-beta decay process.

\subsubsection{The matching relation
\label{subsubsec: singleVIV}
}
To obtain the infinite-volume hadronic transition amplitude for the $nn\to np$ process from the corresponding ME in a finite volume, once again one could inspect the finite-volume correlation function. In momentum space, the correlation function is given by:
\begin{equation}
    C_L^{(\mathcal{J})} (P_i,P_f) =
    \int_Ld^3x\,d^3y \int dx_0\,dy_0
    \;e^{-iP_i\cdot x}\;e^{iP_f\cdot y}
    \left[\langle 0|\, T[\widetilde{B}(y)\,\mathcal{J}_{nn \to np}(0)\, B^\dagger(x)]\,|0\rangle \right]_L,
    \label{eq: three-point correlation (Pi,Pf)}
\end{equation}
while the momentum-time representation of this correlation functions can be written in different ways:
\begin{eqnarray}
    C_L^{(\mathcal{J})}(y_0-z_0,z_0-x_0)
    &\equiv& L^3 \int\, \frac{dE_i}{2\pi}\, \frac{dE_f}{2\pi} \; e^{-iE_i(z_0-x_0)} \; e^{-iE_f(y_0-z_0)}\;C_L
   ^{(\mathcal{J})}(P_i,P_f)
    \label{eq: three point correlation in diagrammatic definition}
    \\
    & = & \int_L d^3x\; d^3y\; d^3z\;
    \left[\langle 0|\, T[\widetilde{B}(y)\,\mathcal{J}_{nn\rightarrow np}(z)\,B^\dagger(x)]\, |0\rangle\right]_L
    \\
    &= &L^9\,\sum_{E_{n_i},E_{n_f}} e^{-iE_{n_i}(z_0-x_0)}\,e^{-iE_{n_f}(y_0-z_0)} \left[\langle 0|\,\widetilde{B}(0)\,| E_{n_f},L\rangle\right]_L
    \nonumber\\
    && \hspace{-0.1cm} \left[\vphantom{B^\dagger} \langle E_{n_f},L|\,\mathcal{J}_{nn\rightarrow np}(0)\,|E_{n_i},L\rangle\right]_L \left[\langle E_{n_i},L|\,B^\dagger(0)\, |0\rangle\right]_L.
    \label{eq: three-point correlation dispersive expression}
\end{eqnarray}
In the last equality $y_0>z_0>x_0$ is assumed, complete sets of finite-volume states are inserted between the operators, and the operators in the Heisenberg picture are expressed in the Schr\"odinger picture.
Now Eq.~(\ref{eq: three-point correlation final}) can be used in Eq.~(\ref{eq: three point correlation in diagrammatic definition}) to perform the energy integrations. This gives
\begin{align}
    &C_L^{(\mathcal{J})}(y_0-z_0,z_0-x_0)
    =L^3\,\sum_{E_{n_i},E_{n_f}}
    e^{-iE_{n_i}(z_0-x_0)}\,e^{-iE_{n_f}(y_0-z_0)}
    \nonumber\\
    &\hspace{4.5cm} \times 
    \widetilde{\mathcal{B}}(E_{n_f})\widetilde{\mathcal{R}}(E_{n_f})
    \left[i\mathcal{M}^{\rm{DF},V}_{nn\to np} (E_{n_i},E_{n_f})\right]
    \mathcal{R}(E_{n_i}) \mathcal{B}(E_{n_i}).
    \label{eq: three point correlation integrated in pi pf}
\end{align}
Here, the only contributions to the energy Fourier integrals arise from the poles of the $\widetilde{\mathcal{F}}$ and $\mathcal{F}$ functions, $E_{n_f}$ and $E_{n_i}$, which are the finite-volume energy eigenvalues of the spin-triplet and spin-singlet two-nucleon systems at rest, respectively, see Eq.~(\ref{eq: Luscher condition}).
$\mathcal{R}$  ($\widetilde{\mathcal{R}}$) is the residue of  $\mathcal{F}$ ($\widetilde{\mathcal{F}}$) at the corresponding finite-volume energies, as defined in Eq.~(\ref{eq: definition of residue}).
This step makes it clear why only the contributions with both factors of $\widetilde{\mathcal{F}}$ and $\mathcal{F}$ were collected in the correlation function explicitly, as the inverse Fourier integrals of the remaining contributions vanish.
For a comprehensive proof of the absence of any poles beside the finite-volume two-nucleon energies in the finite-volume correlation function, see Ref.~\cite{Briceno:2015tza}.

Having arrived at Eq.~\eqref{eq: three point correlation integrated in pi pf}, it can be compared with its equivalent form in Eq.~\eqref{eq: three-point correlation dispersive expression}. For each $E_{n_i}$ and $E_{n_f}$, one obtains:
\begin{align}
    &L^6\left[\langle 0|\,\widetilde{B}(0)\,| E_{n_f},L\rangle\right]_L \left[\vphantom{B^\dagger}\langle E_{n_f},L|\,\mathcal{J}_{nn\rightarrow np}(0)\,|E_{n_i},L\rangle\right]_L \left[\langle E_{n_i},L|\,B^\dagger(0)\, |0\rangle\right]_L
    \nonumber\\[10pt]
    &\hspace{4.5 cm} =\widetilde{\mathcal{B}}(E_{n_f})\widetilde{\mathcal{R}}(E_{n_f})
    \left[i\mathcal{M}^{\rm{DF},V}_{nn\to np} (E_{n_i},E_{n_f})\right]
    \mathcal{R}(E_{n_i}) \,\mathcal{B}^\dagger(E_{n_i}).
    \label{eq: matching relation for single 1}
\end{align}
Multiplying this equation by its complex conjugate and using Eq.~\eqref{eq: matching condition for NN to NN}, one finally arrives at the relation that connects the hadronic  ME in a finite volume with the divergence-free part of the physical amplitude:
\begin{align}
    L^6&\left|\left[\vphantom{B^\dagger}\langle E_{n_i},L|\,\mathcal{J}_{nn\rightarrow np}({0})\,|E_{n_f},L\rangle \right]_L\right|^2 = \left|\widetilde{\mathcal{R}}(E_{n_f})\right|\left|\mathcal{M}^{\rm{DF},V}_{nn\to np} (E_{n_i},E_{n_f})\right|^2  \left|\mathcal{R}(E_{n_i})\right|.
    \label{eq: matching relation for single 2}
\end{align}
This equation is the main result of this section. Note that the divergence-free part of the physical hadronic amplitude is embedded in $\mathcal{M}^{\rm{DF},V}_{nn\to np}$, see Eq.~(\ref{eq: DF amplitude: finite V}).
To construct the full hadronic transition amplitude, the divergent parts must be added to $\mathcal{M}^{\rm{DF}}_{nn\to np}$.
The divergent amplitudes, nonetheless, are comprised of contributions that can be determined from the nucleon single-beta decay amplitude ($g_A$ coupling), as well as the strong-interaction two-nucleon scattering amplitudes.
This relation, therefore, provides a means to constrain the $L_{1,A}$ LEC from a LQCD calculation of the finite-volume ME of the axial-vector current in the two-nucleon system, see Ref.~\cite{Savage:2016kon} for a first calculation along this direction.

Equation~\eqref{eq: matching relation for single 2} agrees with the general result in Eq.~\eqref{eq: NN to NN 1J general result}.
The divergence free amplitude $\mathcal{M}^{\rm{DF}}_{nn\to np}$ $(\mathcal{M}^{\rm{DF,V}}_{nn\to np})$ is equivalent to $\mathcal{W}_{L,{\rm df}}$ $(\mathcal{W}_{{\rm df}})$.
The function $F_1$ is the $s-$wave limit of the $G(L)$ function and $w$ in this case is just $g_A$.
The trace over angular momentum and channel space reduces to simple multiplication in the $s-$wave limit and $^1S_0\to {^3S_1}$ transition.

For non-local MEs derived from LQCD four-point functions, such as those relevant for rare kaon-decays and Compton amplitudes~\cite{Christ:2015pwa,Christ:2012se, Bai:2014cva, Christ:2016mmq, Bai:2017fkh, Bai:2018hqu,Christ:2019dxu,Briceno:2019opb}, and hadronic double-beta decays~\cite{Shanahan:2017bgi,Tiburzi:2017iux,Feng:2018pdq,Tuo:2019bue,Detmold:2020jqv,Feng:2020nqj,Davoudi:2020xdv,Davoudi:2020gxs}, in addition to identifying the volume dependence of MEs, another crucial step is involved.
For nuclear matrix elements of bi-local insertions of the electroweak current, the formalism outlined in this section has another complication from the multi-hadron states between two time-separated currents going on-shell.
This makes the mapping between the Minkowski and Euclidean space-time quantities non-trivial.
Dealing with these complications will be the central theme of the next chapter where the formalism developed in Ref.~\cite{Briceno:2019opb} for $1(2J) \to 1$ process for kinematics allowing only two-hadron on-shell intermediate states was extended.
In Sec.~\ref{sec: 2vbb}, formalism for obtaining the two-nucleon $2\nu\beta\beta$ decay amplitude, which is a $2(2J)\to 2$ transition amplitude, is given,~\cite{Davoudi:2020xdv}.
In Sec.~\ref{sec: 0vbb}, a similar formalism is presented, which further deals with the additional complication from an intermediate on-shell neutrino in the two-nucleon $0\nu\beta\beta$ decay amplitude~\cite{Davoudi:2020gxs}.

\section{Hamiltonian formulation of lattice gauge theories
\label{sec: Hamiltonian for LGT}
}
Despite the impressive achievements of LQCD in uncovering various aspects of strong interactions, there remain some unresolved issues.
LQCD is formulated in Euclidean space-time, which is not a problem for many physical quantities as they can be analytically continued back to Minkowski space-time.
However, there are some observables for which a mapping is not straightforward, as shown in  Sec.~\ref{sec: FV Formalism} and Ch.~\ref{ch: DBD from LQCD}, i.e., obtaining scattering amplitudes from LQCD calculation for non-threshold energies requires a non-trivial mapping because scattering S-matrix is defined by considering non-interacting external plane wave states in the Minkowski space-time.
No analytical continuation, in position or momentum space, between Minkowski-time and Euclidean-time correlation functions is possible above two-particle threshold~\cite{Maiani:1990ca,Hansen:2014wea}.
Furthermore, understanding dynamical properties of nuclear matter ranging from transport properties in the quark-gluon plasma to parton distribution functions of nucleons requires computation of real-time QCD correlations.
However, the path integral formalism for QCD action on a space-time lattice with Minkowski time does not have the probability distribution function interpretation, unlike the Euclidean time rotated action in Eq.~\eqref{eq: LQCD observable expectation value definition}. 
A similar problem occurs in Euclidean Monte Carlo simulations for simulating QCD in the presence of a non-zero chemical potential~\cite{Nagata:2021ugx}.
In this case, the presence of a non-zero fermionic chemical potential term in performing the fermionic integrals in Eq.~\eqref{eq: ZF definition} leads to a complex fermion propagator determinant which becomes a complex probability weight factor.
These are examples of the notorious ``sign problem" in QCD.

Sign problem can occur in Euclidean space-time even without the presence of fermions but from topological terms like the CP-violating term in Eq.~\eqref{eq: gauge interaction CP odd}, since it has only one time derivative that leads to a complex action after a Wick rotation.
Thus, the origin of the sign problem in QCD can be of fermionic or topological, or related to the signature of space-time.
Nonetheless, it undermines the whole principle of Monte Carlo simulations in LQCD.
There are methods to alleviate the sign problem in LQCD such as Taylor expansion, reweighting, Lefschetz thimbles, complex Langevin, dual variables, etc., that have been reviewed in a recent review~\cite{Berger:2019odf}, and will not be explored here any further.

The approach that this thesis is concerned with is the Hamiltonian method for solving LGTs.
In this approach, the space-time is kept as Minkowski and a time dependent observable $\hat{O}(t)$ for state $|\psi\rangle$ in the Hilbert space is calculated via the Hamiltonian of the underlying theory, $\hat{H}$, using
\begin{equation}
    \langle \psi |\; \hat{O}(t) \;|\psi\rangle = \langle \psi |\; e^{i\hat{H}\,t} \; \hat{O}(0) \; e^{-i\hat{H}\,t}\;|\psi\rangle.
    \label{eq: observable in Hamiltonian evolution}
\end{equation}
It is manifestly true that there is no sign problem in numerically evaluating Eq.~\eqref{eq: observable in Hamiltonian evolution} since it does not use the probability weights like in the path integral, but computes the matrix elements by performing operations on the states in the Hilbert space. 
This, however, has a problem as the 
Hilbert space grows exponentially fast with the system size, requiring an exponential scaling behavior of the conventional computational resources.
Nonetheless, there are tools that can address this problem and deal with this exponential scaling of the Hilbert space with polynomial scaling of resources with system size.
One such tool is the tensor network method that has been evolving progressively over the years and has gained a lot of attention in its applications to LGT problems.
The tensor-network ansatz tries to represent efficiently the physically relevant states of the system, and in many situations, this ansatz allows for an efficient solution of the problem leading to a polynomial scaling of resources with system size.
However, the ansatz is only efficient when dealing with states with bounded entanglement and can generally break down, e.g., in processes that are far from equilibrium.
This method will not be explored any further here and interested readers are referred to recent reviews~\cite{Banuls:2019rao,Meurice:2020pxc} on progresses in the tensor network method.

Another possible way is along the lines of an idea that Feynman formulated forty years ago~\cite{Feynman:1981tf}, which states that a way to study quantum-mechanical systems is to employ quantum mechanics directly. 
This can be achieved by engineering a physical 
system that mimics the dynamics of the considered theory as closely as possible, but it can also be realized experimentally, and can be controlled and measured. 
Alternatively, one can construct a more universal device, which we now call a quantum computer, that can be programmed to simulate the desired theory.
However, unlike a classical computer that is based on operations governed by the classical physics on deterministic states of its fundamental operational constituents, bits, a quantum computer operates according to the laws of quantum mechanics by applying (unitary) quantum operations on its fundamental operational constituent, quantum bits or qubits, that can be in any superposition of multiple states. 
This way, a quantum computer with $N$ qubits can encode information about $2^N$ states as a superposition.
This led to the field of quantum simulation, that is, using a controllable quantum system which is engineered to approximate the dynamics of the problem of interest.
A quantum simulation can be performed in a digital or analog way.
In the former case, a general algorithm for the discretized time evolution of a quantum state is given as a sequence of quantum gates (operations on qubits) that can be applied to a class of quantum computers~\cite{Lloyd:1996uni}.
While in the latter approach, the system is evolved continuously using a Hamiltonian that approximately or exactly maps on to the interactions of the model to be simulated~\cite{Jaksch_1998,Aidelsburger:2021mia,Zohar:2015hwa,Wiese:2013uua,Sch_fer_2020}.
Furthermore, there are also methods that use a hybrid approach by combining digital and analog quantum simulations~\cite{Davoudi:2021ney,Casanova:2012zz, Gonzalez-Cuadra:2022hxt, Babukhin_2020}.
Quantum simulation has made impressive technical and conceptual advances in the last decade, see recent reviews in Refs.~\cite{Preskill:2021apy, Georgescu:2013oza, Altman:2019vbv}, and aspects of its application for LGTs are discussed here.

Applications of quantum simulation tools for studying LGTs are active areas research, see reviews~\cite{Zohar:2021nyc, Banuls:2019bmf, Klco_2022, Bauer:2022hpo, Beck:2023xhh}.
As mentioned before, observables are calculated not by the path integral definition, but using the Hamiltonian of the underlying theory via Eq.~\eqref{eq: observable in Hamiltonian evolution}.
The aspects of its computational method can be understood by a careful inspection of Eq.~\eqref{eq: observable in Hamiltonian evolution}.
First, notice that the observable $\hat{O}$ and the exponentiated Hamiltonian $e^{\pm i \hat{H}\,t}$ are operators or mappings acting on the states in the Hilbert space.
This means computation of Eq.~\eqref{eq: observable in Hamiltonian evolution} using a quantum simulator involves encoding states in the Hilbert space onto the degrees of freedom of the quantum hardware in use, and manipulating these encoded states can be understood as actions of operators.
Second, the observables are computed on a specific state of the Hilbert space.
Thus, there needs to be a way to prepare a desired state in the encoded Hilbert space on the quantum device.
Finally, the projection of the evolved and operated state on the original state (or a different state when other state overlaps are sought) needs to be evaluated which amounts to measuring the final state and projecting it onto the initial state.
The probabilistic nature of quantum mechanics requires iterative measurements of the final state to calculate the observable.
This is also evident in the superposition nature of qubits which requires multiple measurements to know the probability of different configurations.

Each of those aspects are independently an active area of research, however, this thesis is concerned with a step even before the encoding of the underlying theory into a quantum simulator, and that is, its formulation in the Hamiltonian language suitable for encoding on quantum devices.
To elaborate, labeling states in the Hilbert space and constructing operations on them are not unique for a given theory.
Particularly in the case of LGTs, a local gauge symmetry leads to a highly constrained Hilbert space and a large number of redundant or unphysical states in the Hilbert space compared to the physical ones.
Different ways of constructing the Hilbert space and operators of a theory are referred to as different ``formulations" throughout this thesis.
Chapter~\ref{ch: LSH} proposes one such formulation of an SU(3) non-Abelian LGT in 1+1 dimension (1+1D), known as the loop-string-hadron (LSH) formulation, that could lead to a resource efficient encoding for quantum simulation of SU(3) LGT, and thus, potentially QCD.
The rest of this section provides the necessary background on the Hamiltonian formulation of LGTs in 1+1D, leads up to the case of LSH formulation for an SU(2) LGT in 1+1D, enlists its advantages over other equivalent formulations, and finally motivates constructing it for the SU(3) theory.

\subsection{Kogut-Susskind formulation of LGTs
\label{subsec: KS Hamiltonian formulation}
}
In his seminal paper titled `confinement of quarks'~\cite{Wilson:1974sk}, Wilson referred to the work by Kogut and Susskind~\cite{Kogut:1974ag} that also speculated on quark confinement but with a different approach than what Wilson proposed.
This Kogut-Susskind (KS) formulation in Ref.~\cite{Kogut:1974ag} dealt directly with the Hilbert space of an SU(2) non-Abelian LGT using its Hamiltonian formulation, in contrast to  Wilson's approach that formulated the LGT on a Euclidean space-time lattice.
This section reviews the KS formulation for 1+1D LGTs by building the Hamiltonian and discussing the construction of its Hilbert space.

The starting point of this formulation is setting up the discretization of the space-time and figuring out how to define fermions and gauge fields on such a lattice.
The KS Hamiltonian is formulated on a discrete space with a continuous Minkowski time direction.
For now, the lattice is considered with an infinite extent with lattice spacing $a$, and the lattice sites are labeled by their positions $n\in\mathds{Z}$.
To put a fermion field on this lattice, one has to revisit the fermion doubling problem mentioned in Sec.~\ref{sec: LQCD}.
The focus of this thesis is an $SU(3)$ LGT Hamiltonian in 1+1D considered in Ch.~\ref{ch: LSH}, and thus, the discussion here is restricted for the 1+1D case.
The naive discretization of a 1+1D fermion spinor field with mass $m$ at position $n$, $\Psi(n) = (\psi_1(n), \psi_2(n))^T$, where $(\psi_2)$ $\psi_1$ is the (anti-) particle component, leads to the Hamiltonian
\begin{equation}
     H_{\rm free} = a\sum_{n} \left[-i \overline{\Psi}(n)\gamma^1\frac{\Psi(n+1)-\Psi(n-1)}{2a} + m \overline{\Psi}(n)\Psi(n)\right],
     \label{eq: free fermion Hamiltonian before staggering}
\end{equation}
where $\overline{\Psi}=\Psi^\dagger\gamma^0$, $\gamma^0=\sigma_3$, and $\gamma^1 = -i\sigma_2$ with $\sigma_3=\big(\begin{smallmatrix}
  1 & 0\\
  0 & -1
\end{smallmatrix}\big)$ and $\sigma_2=\big(\begin{smallmatrix}
  0 & -i\\
  i & 0
\end{smallmatrix}\big)$ being two of the Pauli matrices.
The $a\sum_{n}$ factor comes from discretizing $\int dx$ in the continuum Hamiltonian definition.
The fermion doubling is seen from the fermion dispersion relation obtained from this Hamiltonian: $\omega = \pm(\sin{k})/a$ for $-\pi\leq k\leq\pi$, where $\omega$ is the frequency of free fermion plane waves and $k$ is the associated momentum.
The dispersion relation indicates that this lattice excites low-frequency modes near $\pm\pi$ that are not present in the continuum theory (which has the dispersion relation $\omega=k/a$).
These doublers are also present in the Wilson LQCD formulation where they are removed by adding a Wilson term as discussed in Sec.~\ref{sec: LQCD} or by other methods.
The Wilson fermion formulation can be used for the Hamiltonian simulation of LGTs, see Ref.~\cite{Zache:2018jbt}, however, in this thesis, the ``staggered" fermion solution to the fermion doubling problem introduced in Refs.~\cite{Kogut:1974ag,Susskind:1976jm,Banks:1975gq} is considered.
This solution reduces the Brillouin zone to half by deleting one branch of the dispersion relation, and it is achieved by restricting particle and anti-particle degrees of freedom to only even and odd sites, respectively, by setting the other mode to zero.
The equation of motion for $\psi_1$ and $\psi_2$ then assures that particle modes at odd site and anti-particle modes at even sites remain zero under the time evolution.
Effectively, this amounts to splitting each physical site $n$ into two staggered sites, denoted by label $r$ from here onward, such that the final form of the Hamiltonian is obtained by using just one field, $\psi(r)$, by setting $\psi(r)=\sqrt{a}\psi_1(n)$ at even $r$ sites and $\psi(r)=\sqrt{a}\psi_2(n)$ at odd $r$ sites.
The staggered fermion Hamiltonian, $H_{F}$, is then given by
\begin{equation}
     H_{F} = \frac{i}{2a}\sum_{r} \left[\psi^\dagger(r)\psi(r+1)-\psi^\dagger(r+1)\psi(r)\right] + \sum_{r}  m \; (-1)^r\; \psi^\dagger(r)\psi(r),
     \label{eq: staggered free fermion Hamiltonian}
\end{equation}
where the alternating sign in the mass term arises from $\gamma^0$ in the mass term in the original Hamiltonian in Eq.~\eqref{eq: free fermion Hamiltonian before staggering}, that gives the Dirac sea picture for the vacuum where all odd sites are filled with anti-particles.
The staggered fermion field obeys the fermionic anti-commutation relations given by
\begin{equation}
    \{\psi^\dagger(r),\psi^\dagger(r')\}=\{\psi(r),\psi(r')\}=0,
    \qquad \{\psi(r),\psi^\dagger(r')\}=\delta_{rr'}. 
    \label{eq: ferm_anticomm U(1)}
\end{equation}

Staggered fermions are used in the Wilson LQCD formulation as well.
Topics concerning generalization of staggered fermions to higher dimensions, their applications in LQCD calculation, and issues with taking their continuum limit are beyond the scope of the discussion here, and interested readers are referred to Refs.~\cite{Kilcup:1986dg,Kogut:1982ds,Golterman:2008gt,Gattringer:2010zz,Kronfeld:2007ek}.

Before moving on to the inclusion of a gauge field that interacts with the fermion field in this setup, let us take a moment to understand the Hamiltonian formulation of a pure gauge theory in the continuum space which will shed some light on the issue of gauge fixing and constraints on its Hilbert space.
For that purpose, a simpler case of the Abelian gauge theory in 1+1D is considered with the Lagrangian density
\begin{equation}
    \mathcal{L}_{\rm EM} = -\frac{1}{4}F_{\mu\nu}F^{\mu\nu} = \frac{E^2}{2} ,
    \label{eq: L QED in 1+1D}
\end{equation}
where the space-time coordinates of fields are suppressed for notational brevity.
Here, $F^{\mu\nu}=-F_{\mu\nu}=\big(\begin{smallmatrix}
  0 & E\\
  -E & 0
\end{smallmatrix}\big)$ with $E=-\partial_0 A^1-\partial_1A^0$ being the electric field.
Note, since there is only one spatial dimension, there is no magnetic field.

The canonical momentum $\Pi^i$ of the vector field $A_i$ is given by
\begin{equation}
    \Pi^i = \frac{\delta\mathcal{L}_{\rm EM}}{\delta(\partial_0A_i)} = \partial^iA^0-\partial^0A^i
    \label{eq: canonical conjugate}
\end{equation}
Thus, $\Pi^1 = E$ and $\Pi^0=0$, indicating that the $A_0$ is not a dynamical degree of freedom and it is essentially arbitrary.
The Hamiltonian is then given by
\begin{equation}
    H_{\rm EM} = \int\;dx \left[\frac{E^2(t,x)}{2}  - A_0(t,x) \;\partial_1E(t,x)\right],
    \label{eq: gauge Hamiltonian QED continuum}
\end{equation}
where $x$ $(t)$ denotes the space (time) coordinate, and the last term is obtained by performing integration by parts and assuming that fields vanish at infinity.
Equation~\eqref{eq: gauge Hamiltonian QED continuum} is a Hamiltonian with a Lagrange multiplier field $A_0$ that needs to be treated differently for quantizing the Hamiltonian~\cite{Blaschke:2020nsd}.
To see this, impose the equal time canonical quantization condition on conjugate variables pair $A_1$ and $\Pi_1$ as 
\begin{equation}
    \left[A_1(t,x),\Pi_1(t,y)\right] =  \left[A_1(t,x),E(t,y)\right] = i\;\delta(x-y).
    \label{eq: QED A E or A pi commutation}
\end{equation}
Clearly, such a quantization condition cannot be imposed on the conjugate variables pair $A_0$ and $\Pi_0$, since $\Pi_0=0$, and thus, the condition $\Pi_0=0$ needs to be imposed as a constraint.
However, using the Hamiltonian in Eq.~\eqref{eq: gauge Hamiltonian QED continuum}, one can obtain the Hamiltonian field equation for the field $\Pi_0$ as $\partial_0 \Pi_0 =\;\partial_1E$.
This implies that the canonical momentum of the $A_0$ field is dynamically generated by other fields and the operator level condition $\hat{\Pi}_0=0$ on the Hilbert space needs to be imposed during the time evolution.

Hamiltonian formulation with such quantization procedure is significantly more difficult, but it can be made easier by setting $A_0 = 0$.
This assures $\partial_0 \Pi_0 =0 $, and the dynamics can be contained to the $\hat{\Pi}_0=0$ sector.
The condition of restricting to a particular gauge is called \textit{gauge fixing}, and the choice $A_0=0$ is known as the temporal gauge.
Recall that the electromagnetic Lagrangian has a gauge symmetry $A_0\to A'_0 = A_0+\partial_0 \,\alpha(t,x)$, and fixing the temporal gauge, $A'_0(t,x) =0$, amounts to performing a gauge transformation using an $\alpha$ that is a solution to the equation $-\partial_0\,\alpha(t,x)=A_0(t,x)$.

This understanding of gauge fixing now allows one to revisit the discretized space.
In the temporal gauge, the gauge field Hamiltonian in Eq.~\eqref{eq: gauge Hamiltonian QED continuum} reduces to the electric field energy, $H_E$, and is given by
\begin{equation}
    H_E = a\sum_r \frac{E^2(r)}{2} ,
    \label{eq: Hamiltonian density QED lattice}
\end{equation}
where the time coordinate in $E$ is suppressed.
The gauge invariant gauge-matter interaction term on lattice can be constructed using the link variable $U(r)$ at link $r$ that connects lattice sites at $r$ and $r+1$.
The link variable is again as in Eq.~\eqref{eq: link variable in terms of gauge field} which is obtained by the path ordered exponential integral of the gauge field $A_\mu$ connecting $r$ and $r+1$.
For an Abelian theory, there is only one generator, $T=1$, and in the temporal gauge, there is only one non-trivial link operator which is along the space direction, $U(r) = e^{iagA_1(r)}$, where $g$ characterizes the interaction strength.
The link variable in the time direction are all identity elements of the gauge group due to the condition $A_0=0$.
The staggered matter then couples to the link operator in a similar manner as in Sec.~\ref{sec: LQCD}, and its Hamiltonian form is given by
\begin{equation}
    H_I = \frac{1}{2a}\sum_r\left[ \psi^\dagger(r) U(r) \psi (r+1) + {\rm H.c}\right].
    \label{eq: gauge-matter interaction HI for QED}
\end{equation}
Recall that electromagnetism is a $U(1)$ gauge theory, and thus, the gauge transformation in Eq.~\eqref{eq: guage rotation of quarks} for a staggered fermionic field is just a phase change with a local angle of rotation.
Upon expanding the link variable in the lattice spacing $a$, this interaction term leads to the covariant derivative up to $\mathcal{O}(a^2)$ corrections, and thus, it already contains the kinetic term for fermions.
Note that the factors $i$ and $-i$ in Eq.~\eqref{eq: staggered free fermion Hamiltonian} are absorbed via the field redefinition: $\psi(r) \to i^r\psi(r)$.

Putting everything together and restricting the lattice size to $N$ sites, the KS Hamiltonian for an U(1) gauge theory is given by~\cite{Susskind:1976jm,Banks:1975gq}
\begin{align}
    H_{U(1)} =  a\sum_{r=1}^{N'} \frac{E^2(r)}{2} +\frac{1}{2a}\sum_{r=1}^{N'}\left[ \psi^\dagger(r) U(r) \psi (r+1) + H.c\right] + \sum_{r=1}^{N}  m \; (-1)^r\; \psi^\dagger(r)\psi(r),
    \label{eq: QED Hamiltonian KS full}
\end{align}
where $N'=N-1$ $(N)$ for open (periodic) boundary conditions.

To construct the Hilbert space of this Hamiltonian, let us look at the (anti) commutation relations between the fields involved.
The fermion field obeys Eq.~\eqref{eq: ferm_anticomm U(1)}.
Thus, the local fermion states, $|f\rangle_r$, are labeled by the fermion occupation number $f=0,1$ to satisfy the Pauli's exclusion principle.
Note that, staggering implies $f=0$ (1) at even (odd) sites is the non-interacting vacuum that has the lowest energy due to the alternating sign in the mass term of Eq.~\eqref{eq: QED Hamiltonian KS full}, and $f=1$ (0) at even (odd) sites indicate a particle (anti-particle) excitation.

For the gauge fields, the discretized version of the equal time commutation relation in Eq.~\eqref{eq: QED A E or A pi commutation} is given by
\begin{equation}
    [A_1(r), E(r')] = \frac{i}{a}\delta_{rr'}.
    \label{eq: QED A E commutation discretized}
\end{equation}
Defining $\theta(r) = agA_1(r)$ such the $U(r)$ becomes a phase with angle $\theta(r)$, and rescaling the electric field $E_{\rm new}(r)\equiv E(r)/g$, one obtains
\begin{equation}
    [\theta, E_{\rm new}(r')] = i\,\delta_{rr'}.
    \label{eq: QED theta E commutation discretized}
\end{equation}
Dropping the subscript and relabeling $E_{\rm new}$ with $E$, the electric field energy becomes 
\begin{equation}
    H_E = \frac{g^2 a}{2}\sum_{r} E^2(r).
    \label{eq: HE rescaled QED}
\end{equation}

Since $\theta(r)$ is an angular variable with $0\leq\theta(r)\leq 2\pi$ such that $U(r)$ remains single valued, the commutation relation in Eq.~\eqref{eq: QED theta E commutation discretized} implies that the eigenvalues of $E(r)$ are quantized.
Labeling them with $j=0,\pm1,\pm2,\cdots $ and using
\begin{equation}
    [ E(r), U(r')] = i\,\delta_{rr'}\; U(r'),
    \label{eq: QED E U commutation discretized}
\end{equation}
which can be obtained from the definition $U(r) = e^{i\theta(r)}$ and Eq.~\eqref{eq: QED theta E commutation discretized}, one realizes that the link operator acts as a raising operator on the eigenstates of the electric field operator.
Thus, at each site $r$, the eigenstates of the electric field operator, $|j\rangle_r$, span the local Hilbert space of the gauge degrees of freedom, and the local Hilbert space of gauge and matter fields combined is spanned by states $|j\rangle_r\otimes|f\rangle_r$ with $j\in \mathds{Z}$ and $f\in \{0,1\}$.
A general state in the Hilbert space of the KS formulation of QED is then given by
\begin{equation}
    |\Psi\rangle^{(\rm KS)} = \goldieotimes_{r} |j\rangle_r\otimes|f\rangle_r,
    \label{eq: general KS state in QED}
\end{equation}
where $\goldieotimes_{r}$ denotes the tensor product over all local states labeled by $r$.

There is one final subtlety in the construction of the Hilbert space that comes from a remnant gauge symmetry.
This symmetry arises from time-independent gauge transformations, as the temporal gauge fixing still leaves $A_1(r) \to A_1(r) + \partial_1\alpha(r)$ in the continuous space where $\alpha$ only depends on the spatial coordinate.
Under these gauge transformations, $A_0(r)$ remain zero but $A_1(r)$ changes.
Since the Hamiltonian by construction is gauge-invariant, the generators of these time-independent gauge transformations naturally commute with the Hamiltonian.
In the pure gauge theory with no matter fields, these local symmetry generators are given by $\partial_1 E(r)$, as seen by the relation
\begin{equation}
    e^{i \int \,dr\, \alpha(r) \,\partial_1 E(r)} A_1(r) e^{-i \int\, dr \,\alpha(r) \, \partial_1 E(r)} = A_1(r) + \partial_1\alpha(r),
    \label{eq: generators of remnant gauge in pure EM theory}
\end{equation}
which can be obtained using Eq.~\eqref{eq: QED A E or A pi commutation}.
Thus, the states in the Hilbert space can be classified into separate sectors, corresponding to eigenvalues of these operators.
These sectors correspond to different static charge configurations.
The physical requirement that states that differ by time-independent gauge transformations be equivalent to each other leads to the demand that they should remain in the same static charge configurations.
This constraint is nothing but the Gauss's law which can be equivalently imposed on the Hilbert space of the LGT.
In that case, the Gauss's law for restricting to the zero static charge sector is given by
\begin{equation}
    G(r) |\Psi\rangle^{(\rm KS)} =0 \quad \forall r
    \label{eq: Gauss's law QED}
\end{equation}
where the generators $G(r)$ are given by the discretized $\partial_1E(r)$ along with the contribution from the local charge density of the fermion field configuration as~\cite{Banks:1975gq}:
\begin{equation}
    G(r) = E(r) - E(r-1) +\psi^\dagger(r)\psi(r) -\frac{1}{2}(1-(-1)^r)
    \label{eq: Gauss's law operator QED}
\end{equation}
where the last term ensures that traversing across a site to the right, the negatively charged particles at even sites lower the electric field by one unit if excited ($f=1$), and positively charged anti-particles at odd site raise the electric field by one unit if excited ($f=0$).

To summarize, the KS formulation of an Abelian LGT involves a gauge-invariant Hamiltonian (Eq.~\eqref{eq: QED Hamiltonian KS full}) and a Hilbert space constructed from the local quantum numbers (Eq.~\eqref{eq: general KS state in QED}) that is subjected to a Gauss's law constraint (Eq.~\eqref{eq: Gauss's law QED}) using a generator of time-independent gauge transformations (Eq.~\eqref{eq: Gauss's law operator QED}.
Extending this to  non-Abelian theories follows the same procedure.
The KS formulation of non-Abelian LGTs will be outlined here, and interested readers are referred to Refs.~\cite{Kogut:1974ag,Hamer:1981yq,Ryang:1983dd,Zohar:2014qma} for more details.

We consider here a non-Abelian gauge theory with gauge group SU(2) as an example.
Its KS formulation is based on the same assumptions as the Abelian gauge theory Hamiltonian: $1)$ a 1+1D system is considered with continuous time, $2)$ a spatial lattice with $N$ staggered lattice sites or equivalently $N/2$ physical sites is considered, and $3)$ a temporal gauge is used.
However, now for the non-Abelian gauge field, the gauge fixing involves setting the time component of every gauge field in the multiplet to zero, $A^a_0 = 0$, where `a' is the adjoint index that takes values from $1,2,3$.

In order to see what is different in the non-Abelian case compared to the Abelian one discussed so far, refer to the definition of the field strength tensor in Eq.~\eqref{eq: gluon field strength tensor definition} and gauge transformation in Eq,~\eqref{eq: transformation of gauge field}.
The non-commutative behavior leads to key changes in the Hamiltonian formulation.
To see this, let us look at the nature of the electric field and the link operator.
Because of the SU(2) group index $a$ on the gauge field $A^a_1(r)$, there are now three electric fields $E^a(r)$ with $a=1,2,3$.
From here onwards, these electric fields are referred to as ``chromoelectric" fields alluding to the fact that they carry an SU(2) group index, analogous to the color quantum number of the SU(3) group in QCD.
Similarly, in the definition of link operator, the exponentiated path ordered integral of the gauge field between lattice sites at $r$ and $r+1$ is now $U(r) = e^{iagA_1(r)}$ with $A_1(r) = \sum_a A^a_1(r)T^a$, where $T^{\rm a}=\frac{\sigma^{\rm a}}{2}$ are the generators of the SU(2) group.
Thus, the link variable $U(r)$ is now a unitary $2\times 2$ matrix defined on the link $r$ with components denoted by $U^\alpha{}_\beta(r)$, where $\alpha$, $\beta =$ 1, 2 are the SU(2) color indices.
Note that, the compact nature of the SU(2) group makes the link variable naturally compact, unlike the Abelian case where the angular nature of $\theta(r)$ variable was imposed by hand.
Thus, it is expected that a KS Hamiltonian gauge theory with a compact Lie group will naturally lead to the quantization in the electric field eigenvalues.

The main difference for the non-Abelian case arises from the relation between the chromoelectric field and the link operator.
For a general gauge theory with an Abelian or a non-Abelian group, the gauge transformations are internal rotations in the color space that are different at each lattice site.
This can be thought of as the color frame of references are oriented differently at each site, and a local gauge transformation re-orients (or rotates) the local frame.
The dimension of the color frame is determined by the number of generators of the gauge group.
For the U(1) group, the frame re-orientations are just changes in the phase since there is only one generator, however, for the non-Abelian case, they are rotations in the color space.
For constructing gauge invariant interactions between colored particles that can hop from one site to another, like the color-charged fermion matter field, there needs to be a connector that would rotate the hopping particles such that their color charges are re-assigned according to the orientation of the new local frame.
This is called a parallel transport, and it is precisely what the link operator does for the matter field, as seen in Eq.~\eqref{eq: gauge-matter interaction HI for QED}.

The situation is more involved for the chromoelectric fields.
Unlike the Abelian case where the electric field is gauge-invariant and does not carry the gauge-group index, the field stress tensor for a non-Abelian case transforms under gauge transformations that can be seen from Eq.~\eqref{eq: gluon field strength tensor definition}.
Thus, the chromoelectric fields also need to be parallel transported which indicates that the chromoelectric fields now live at the ends of each link.
Calling the ends as left ($L$) and right ($R$), the parallel transport relation between left and right chromoelectric fields, $E^{\rm a}(L,r)$ and $E^{\rm a}(R,r)$ along a link is given by
\begin{equation*}
    E^{\rm a}(R,r)T^{\rm a}=-U^\dagger(r)E^{\rm b}(L,r)T^{\rm b} U(r).
    \label{eq: E field parallel transport}
\end{equation*}
However, this operation preserves the gauge-invariant chromoelectric field energy which still has the same form as the electric field energy in U(1) theory.
It is given by the quadratic Casimirs on either side, $E^2(r)$,
\begin{equation}
    E^2(r)\equiv \sum_{\rm a} E^{\rm a}(L,r)E^{\rm a}(L,r)=\sum_{\rm a} E^{\rm a}(R,r)E^{\rm a}(R,r).
    \label{eq: electric field casimir constraint SU(2)}
\end{equation}
The equality of two sums on the right hand side of this equation can be clearly seen by squaring and taking the trace of Eq.~\eqref{eq: E field parallel transport}.

With this, the KS Hamiltonian for an $N$-site lattice can be constructed in the same manner as in Eq.~\eqref{eq: QED Hamiltonian KS full}:
\begin{align}
  H&= H_M+H_E+ H_{I} \nonumber\\ 
  &= m \sum_{r=1}^{N} (-1)^r \psi^\dagger_\alpha(r) \psi^\alpha (r) + \frac{g^2 a}{2}\sum_{r=1}^{N'}  E^2(r) \nonumber\\
  & \hspace{2 cm}+ \frac{1}{2a} \sum_{r=1}^{N'} \left[\psi^\dagger_\alpha(r)\, U^\alpha{}_{\beta}(r)\, \psi^\beta(r+1) + {\rm H.c.} \right] ,
  \label{eq: KS-ham SU(2)}
\end{align}
where $H_M$ is the matter self-energy, $H_E$ is the chromoelectric energy, and $H_{I}$ is the gauge-matter interaction. Here, $N'=N-1$ $(N)$ for open (periodic) boundary conditions.
The staggered fermionic matter is expressed by a fermion field in the fundamental representation $\psi_\alpha(r)$, where $\alpha$ is the SU(2) color index.
It obeys the fermionic anti-commutation relations given by
\begin{equation}
    \{\psi_\alpha^\dagger(r),\psi^\dagger_\beta(r')\}=\{\psi^\alpha(r),\psi^\beta(r')\}=0,
    \qquad \{\psi^\alpha(r),\psi^\dagger_\beta(r')\}=\delta^{\alpha}_\beta\,\delta_{rr'}. 
    \label{eq: ferm_anticomm SU(2)}
\end{equation}
This indicates that the local fermionic Hilbert space at site $r$ is spanned by states $|f_1,f_2\rangle$ where $f_i=0$, 1 is the fermion occupation number for color $i$.

The local Hilbert space of gauge degrees of freedom can be constructed by observing canonical  commutation relations between chromoelectric fields and the link operators:
\begin{align}
  [E^{\rm a}(L/R,r),E^{\rm b}(L/R,r')] &= \delta_{rr'} \sum_{{\rm c}=1}^3 i\,\epsilon^{\rm abc}\,E^{\rm c}(L/R,r),
  \label{eq: commutation-ELa-Elb SU(2)}\\
  [E^{\rm a}(L,r),E^{\rm b}(R,r')] &=0
  \label{eq: commutation-ELa-ERb SU(2)}
\end{align}
where $\epsilon^{\rm abc}$ are the Levi-Civita tensor that is the structure constants for SU(2), and $L/R$ indicates that the relation holds individually for both $L$ and $R$ sides of the link.
Equation~\eqref{eq: commutation-ELa-Elb SU(2)} denotes that the $L$ and $R$ chromoelectric play the role of the generators of SU(2) rotations in the analogy of gauge transformations as re-orienting of the local left and right color frames, respectively~\cite{Kogut:1974ag}. Equation~\eqref{eq: commutation-ELa-ERb SU(2)} indicates that these two spaces are independent.
So to span the Hilbert space of the gauge degrees of freedom using the eigenstates of chromoelectric field energy, just like in the U(1) case, one has to use the angular momentum representation states of the $SU(2)$ algebra for the $L$ and $R$ frame rotations independently.
This basis for link $r$ is given by $|j_L,m_L\rangle_r\otimes|j_R,m_R\rangle_r$, where $j_L(r)$ and $m_L(r)$ ($j_R(r)$ and $m_R(r)$) are the total angular momentum quantum number and the azimuthal quantum number for the color frame at the $L$ ($R$) end of the gauge link $r$, respectively.
The quantization of angular momentum implies $j_L(r)$, $j_R(r)$ = 0, $\frac{1}{2}$, 1, $\frac{3}{2}$, $\cdots$.
Furthermore, the allowed values for the azimuthal quantum number are $-j_L(r)\leq m_L(r) \leq j_L(r)$ and $-j_R(r)\leq m_R(r) \leq j_R(r)$.
However, $j_L(r)$ and  $j_R(r)$ cannot be independent since the quadratic Casimir operator of the $L$ and $R$ frames are related via Eq.~\eqref{eq: electric field casimir constraint SU(2)}, indicating $j_L(r)=j_R(r)$.
Thus, the full Hilbert space of the theory is spanned by states
\begin{equation}
    |\Psi\rangle^{(\rm KS)} = \goldieotimes_{r} |j_R,m_R\rangle_{r-1}\otimes|f_1,f_2\rangle_r \otimes |j_L,m_L\rangle_{r},
    \label{eq: general KS state in SU(2)}
\end{equation}
subjected to the constraint $j_L(r)=j_R(r)$.
The notation is chosen to show local Hilbert spaces around the lattice site at $r$ that has the right end of the link $r-1$ and left end of the link $r$.

The residual gauge redundancies from time-independent gauge transformations lead to a Gauss's law on these states.
However, unlike the U(1) case, the gauge regencies can be of three different kinds, corresponding to three different Euler angles for local gauge frame rotations.
The gauge invariance of Hamiltonian in Eq.~\eqref{eq: KS-ham SU(2)} means that it commutes with the generators of these residual gauge transformations, $G^{\rm a}(r)$, that are given by
\begin{equation}
  G^{\rm a}(r)= E^{\rm a}(L,r)+E^{\rm a}(R,r-1)+ \psi^\dagger_\alpha(r)\, \left(T^{\rm a}\right)^\alpha{}_\beta\, \psi^\beta(r).
  \label{eq: gauss-op SU(2)}
\end{equation}
The Hamiltonian generates the dynamics between states of the Hilbert space, $|\Psi\rangle$, and the gauge-invariance of observables implies that only the states $|\Psi\rangle^{(KS)}$ that satisfy Gauss's law constraints
\begin{equation}
  G^{\rm a}(r)|\Psi\rangle^{(KS)}=0 \quad \forall\, {\rm a}\,,r,
  \label{eq: gauss-law SU(2)}
\end{equation}
are the physical states in the Hilbert space.
An important point to note here is that the local Gauss's law generators given above do not commute among themselves.
Thus, the Hilbert space cannot be simultaneously diagonalized with respect to all local Gauss's law generators.

Finally, the action of the gauge-link variable $U^\alpha{}_\beta(r)$ on these states can be obtained from from the canonical commutation relations
\begin{align}
   \left[E^{\rm a}(L,r),U(r')\right] &= -\delta_{rr'}\,T^{\rm a} U(r)\,,\quad [E^{\rm a}(R,r),U(r')]= \delta_{rr'}\,U(r)T^{{\rm a}},
  \label{eq: commutation-ELa-U SU(2)} \\\left[U^\alpha{}_\beta(r),U^\gamma{}_\eta(r')\right] &= [U^\alpha{}_\beta(r),U^{\dagger\gamma}{}_\eta(r')]=0,
  \label{eq: commutation-U-U SU(2)}
\end{align}
which are analogous to Eq.~\eqref{eq: QED E U commutation discretized}.
Thus, the gauge-link operator raises or lower the left and right angular momentum values at the end of the link $r$ as
\begin{align}
    U^\alpha{}_\beta(r) |j,m_L\rangle_{r}\otimes|j,m_R\rangle_{r} &= \sum_{J=0,\frac{1}{2}, 1, \cdots} \sqrt{\frac{2j+1}{2J+1}}
    \nonumber\\
    &\hspace{-4 cm}\times\langle j,m_L;\frac{1}{2},\alpha|J,m_L+\alpha \rangle
    \langle j,m_R;\frac{1}{2},\beta|J,m_R+\beta \rangle \; |J,m_L+\alpha\rangle_{r}\otimes|J,m_R+\beta\rangle_{r}
    \label{eq: action of U on SU(2) KS}
\end{align}
where the Clebsch-Gordon coefficients denoted by state overlaps vanish for $J\neq j\pm\frac{1}{2}$, for an SU(2) group.
Here the fermion states are not shown as the link-operator does not act on them.
\begin{figure}[t]
	\centering
	\includegraphics[scale=1.1]{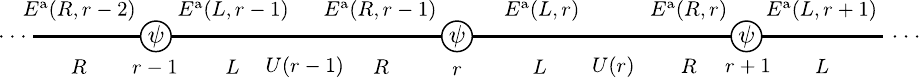}
	\caption{
    Three neighboring sites of the lattice located at positions $r-1$, $r$, and $r+1$ are shown in the KS formulation of an SU(2) LGT in 1+1D.
    They are connected by links with each link having a left end denoted by $L$ and a right end denoted by $R$. The chromoelectric field, $E^{\rm a}$, is identified by its location at either end of the link and its corresponding position on the link. The link operator, $U$, is defined at each position along the link, and the color-charged staggered matter, $\psi$, is defined at each lattice site represented by a circle in the diagram.
	\label{fig: KS dof SU(2)}}
\end{figure}

To summarize, the KS formulation of a non-Abelian gauge theory contains left and right chromoelectric fields, link operators and the staggered matter sites as degrees of freedom, as shown in Fig~\ref{fig: KS dof SU(2)}.
The chromoelectric fields across a gauge link are subjected to the constraint in Eq.~\eqref{eq: electric field casimir constraint SU(2)} that comes from their parallel transport in Eq.~\eqref{eq: E field parallel transport}.
The Hamiltonian is given by Eq.~\eqref{eq: KS-ham SU(2)} that acts on the states in the Hilbert space shown in Eq.~\eqref{eq: general KS state in SU(2)}.
The physical states in the Hilbert space obey Gauss's laws in Eq.~\eqref{eq: gauss-law SU(2)} for non-Abelian generators in Eq.~\eqref{eq: gauss-op SU(2)}.

The KS Hamiltonian further made it possible to perform perturbative calculations in the strong coupling limit, $g\to \infty$~\cite{Kogut:1974ag}.
It was shown that the continuum limit of this theory recovers the Yang-Mills theories coupled to matter, and this formulation is equivalent to the Wilson's path-integral formulation~\cite{Wilson:1974sk}.
For more details on the KS formulation of Abelian and non-Abelian theories, see Refs.~\cite{Kogut:1974ag,Banks:1975gq,Hamer:1981yq,Zohar:2021nyc}
The Hilbert space constructed here is appropriately called the angular momentum basis.
Note that, the SU(2) nature of the gauge-group was used only to construct the local Hilbert space of gauge degrees of freedom from the canonical commutation relations.
Thus, extending the KS formulation for a general SU(N) theory is straightforward.
It involves introducing appropriate changes in the color index of the matter field as well as the adjoint index of electric field and Gauss's law generators, making the link variable group elements of the corresponding gauge-group, and changing the local angular momentum basis to the irreducible representation (irrep) of the underlying gauge-group.
However, this construction of the Hilbert space is not unique.

The next subsection reviews alternate formulations of the KS Hamiltonian and/or Hilbert space for SU(2) gauge theory by discussing them briefly and comparing their advantages and disadvantages.
It motivates working with the Loop-String-Hadron (LSH) framework which implements non-Abelian Gauss’s laws using a complete set of gauge-invariant operators, and sets up the foundation for Ch.~\ref{ch: LSH} that develops the LSH formulation for an SU(3) LGT.

\subsection{Alternate formulations of KS Hamiltonian: A review
\label{subsec: KS comparison}
}
The KS formulation of the U(1) gauge theory in 1+1D, also known as the lattice Schwinger model, as well as non-Abelian gauge theories have been studied using tensor networks, see Ref.~\cite{Banuls:2019rao} for an extensive review.
Similarly, they have been employed in quantum simulation, see reviews in Refs.~\cite{Funcke:2023jbq,Banuls:2019bmf}.
However, to perform such calculations, the dimension of the Hilbert space of the theory needs to be finite in order to encode states on computing devices.
The Hilbert spaces spanned by the electric field basis of the lattice Schwinger model and the angular-momentum basis of the SU(2) LTG discussed in the previous subsection are obviously infinite dimensional.
Thus, the corresponding Hamiltonians are solved after truncating the Hilbert space to make them finite dimensional.
The simulated theory still gives viable results, if the truncation is performed such that the contribution of ignored states to the physics of interest is small.
The Hilbert space construction of the KS Hamiltonian provided in the previous section is not unique, and several different formulations have been proposed over the years, see Ref.~\cite{Banuls:2019bmf, Davoudi:2020yln,Bauer:2022hpo} for a review.
Therefore, a suitable truncation scheme depends on the formulation in use, and different formulations can lead to different sizes of the truncated Hilbert space while achieving the same precision on the calculated observables.
Furthermore, different formulations have different computational complexity in implementing the Hamiltonian, or imposing constraints on states, like the Gauss's law constraints or the equal angular momentum constraints on left and right ends of gauge links in the angular-momentum basis.
All these factors affect the computational cost, scaling of computational resources with system size, accuracy of imposing constraints, etc., when solving the Hamiltonian dynamics.
For these reasons, it is important to understand the advantages and drawbacks of a given formulation and identify computational resource efficient formulations of Hamiltonian LGT.

Recent work in Ref.~\cite{Davoudi:2020yln} performed a comparative study between different formulations of the KS framework of an SU(2) LGT in 1+1D coupled to matter to examine the classical computational-resource required for obtaining spectrum and dynamics from Hamiltonian computation. 
Along with the KS formulation presented in the previous subsection, the other formulations considered were the purely fermionic formulations~\cite{Hamer:1981yq,Banuls:2017ena,Sala:2018dui}, 
purely bosonic formulation~\cite{Zohar:2018cwb,Zohar:2019ygc}, and the loop-string-hadron (LSH) formulation~\cite{Raychowdhury:2018osk,Raychowdhury:2019iki}.
It found clear advantages in working with the LSH formulation over the others.
This formulation imposes the non-Abelian Gauss's law \textit{a priori} using gauge-invariant operators.
A brief overview of these formulations is provided here along with their advantages and drawbacks, with more emphasis on the LSH formulation.

In the KS formulation of an SU(2) LGT in 1+1D, the gauge degree of freedom is highly constrained from Gauss's laws, and the only dynamical gauge degree of freedom is the boundary mode in a lattice with periodic boundary condition.
Thus, for a lattice with an open boundary condition, any direct reference to gauge links and electric fields can be removed by performing a specific residual gauge transformation and using the Gauss's laws constraints%
\footnote{Derivation of purely fermionic formulation is performed in App.~\ref{app: purely fermionic formulation}. Although the theory considered there is an SU(3) LGT in 1+1D, the derivation is identical for any SU(N).}
~\cite{Hamer:1981yq,Banuls:2017ena,Sala:2018dui}.
This results in a formulation called the purely fermionic formulations since it is expressed entirely in terms of fermions.
All states in this formulation can be described using a full occupation-number basis of fermionic particles, and its Hamiltonian is equivalent to the KS Hamiltonian but only in the physical Hilbert space.
Thus, the Hilbert space of this formulation is finite dimensional with $4^N$ number of states for an $N$-site lattice and does not require any truncation for its simulation.
However, this comes at a cost of degenerate states arising from the fermion modes at the lattice endpoints.
Furthermore, the Hamiltonian is non-local and a fermion at each lattice site interacts with fermions at every other lattice sites.
Finally, the purely fermionic formulation is only feasible in 1+1D, as the constraints at each lattice site are inadequate in higher dimensions to eliminate the gauge degrees of freedom in all spatial dimensions.

Gauss's law can also be used to completely remove matter degrees of freedom instead of gauge variables.
In this purely bosonic formulation~\cite{Zohar:2018cwb,Zohar:2019ygc}, the local SU(2) gauge-group is enlarged to a U(2) gauge group to accommodate a sufficient number of constraints required to eliminate fermions, which can be generalized for higher dimensions~\cite{Pardo:2022hrp}.
In 1+1D, this procedure expands the Hilbert space of the SU(2) KS formulation by taking a tensor product with the Hilbert space of a newly added U(1) gauge-field, and an additional Gauss's law corresponding to this field needs to be imposed.
As a result, the infinite dimensional Hilbert space needs to be truncated for its computation.
However, in the open boundary condition where the KS Hilbert space is finite dimensional, the cutoff of the extra U(1) needs to be sufficiently high to accommodate the physical Hilbert space.
Furthermore, the extended Hamiltonian that includes the added gauge-field has nearest-link interactions originating from gauge-matter interaction term.
Even though the Hamiltonian remains local, the fermion sign from fermion anti-commutation relations needs to be embedded in gauge-boson interaction, which introduces additional complexity.

In all three formulations of SU(2) LGT in 1+1D discussed so far, that is the KS formulation in the angular-momentum basis, the purely fermionic formulation, and the purely bosonic formulation, the Gauss's laws played a central role in determining the physical Hilbert space from the much larger Hilbert space generated by the tensor product of independent gauge and matter Hilbert spaces.
This is a result of states in the larger Hilbert space transforming under the residual time-independent gauge transformations.
In the LSH formulation~\cite{Raychowdhury:2018osk,Raychowdhury:2019iki}, the Hilbert space is constructed not from two independent gauge and matter Hilbert spaces, but by building up from a vacuum operated with gauge-invariant combinations of local gauge and matter field operators such that the states are manifestly invariant under the residual gauge transformations.
The states are then labeled by a set of quantum numbers associated with these building block operators that are analogous (but not exact) to loops, strings and hadrons degrees of freedom in a gauge theory. 
The non-local features of the underlying gauge theory are imposed with a much simpler Abelian Gauss's law that appears as an algebraic relation on quantum numbers of neighboring sites reducing the complexity associated with finding physical states in the larger Hilbert space.
The Hamiltonian is then re-expressed in terms of these new quantum numbers, and it still remains local.
The LSH procedure for Hamiltonians in higher dimensions has been also worked out in Ref.~\cite{Raychowdhury:2019iki} for an SU(2) LGT.

Even though the 1+1D LSH Hilbert space is infinite dimensional for a periodic boundary condition, the cost of generating such a matrix is exponentially suppressed in the LSH formulation compared to others~\cite{Davoudi:2020yln}.
Furthermore, the terms in the LSH Hamiltonian map a given state to one and only one state as opposed to KS Hamiltonian that mixes the states in the angular-momentum basis via the operation in Eq.~\eqref{eq: action of U on SU(2) KS}.
From its comparison with other formulations mentioned here~\cite{Davoudi:2020yln} and a recent study on computational cost of its quantum simulation~\cite{Davoudi:2022xmb}, the LSH formulation appears to be the advantageous for the purpose of Hamiltonian simulation of the SU(2) LGT.
These encouraging results indicate that similar advantages can be seen for the Hamiltonian simulation of QCD with the LSH formulation.

As a step towards that direction, Ch.~\ref{ch: LSH} extends the LSH formulation to SU(3) LGT in 1+1D that is based on a recent work in Ref.~\cite{Kadam:2022ipf}.
In the rest of this section, an overview of constructing the LSH formulation of an SU(2) LGT in 1+1D is provided, which is based on the work in Ref.~\cite{Raychowdhury:2019iki}, and sets up the foundation for steps taken in Ch.~\ref{ch: LSH}.
The starting point of LSH formulation is the observation that the fermionic states in the angular-momentum basis in Eq.~\eqref{eq: general KS state in SU(2)}, are created using the fermion creation operators $\psi_\alpha^\dagger(r)$.
This operator transforms as the fundamental irrep under the SU(2) gauge transformations.
To construct the gauge-invariant creation operators for building the LSH Hilbert space, one needs to find bosonic creation operators that build up the angular momentum bases of left and right chromoelectric fields in Eq.~\eqref{eq: general KS state in SU(2)} by acting upon a vacuum state.
Such a construction of an SU(2) angular momentum states was given by Schwinger~\cite{osti_4389568}. This is known as the Schwinger boson or prepotential formulation, and it has been extended to other special unitary groups in Refs.~\cite{mukunda1965tensor,Chaturvedi:2002si,Mathur:2010wc,Mathur:2000sv,Mathur:2004kr,Mathur:2007nu,Anishetty:2014tta,Raychowdhury:2018tfj,Raychowdhury:2013rwa} for its applications in LGTs.

For the SU(2) irrep states $|j,m\rangle$ with angular momentum quantum number $j$ and azimuthal quantum number $m$, a set of two simple harmonic oscillators of type 1 and type 2 are considered along with their creation operators, $a^\dagger_1$ and $a^\dagger_2$, and annihilation operators, $a_1$ and $a_2$, such that they obey
\begin{equation}
    [a_i, a^\dagger_j]=\delta_{ij}, \quad [a^\dagger_i, a^\dagger_j]=[a_i, a_j]=0,
    \label{eq: SHO oscillator commutations}
\end{equation}
for $i,j$ =1, 2.
Then, a state 
\begin{equation}
    |n_1,n_2\rangle = \frac{(a^\dagger_1)^{n_1}(a^\dagger_2)^{n_2}}{\sqrt{n_1!}\sqrt{n_2!}}|0,0\rangle,
    \label{eq: n1 n2 state in Schwinger boson}
\end{equation}
where $|0,0\rangle$ is the vacuum defined by $a_i|0,0\rangle=0$ for $i=1$, 2, and $n_1$ and $n_2$ are positive integers, transforms as an SU(2) irrep with $j$ and $m$ under SU(2) transformations, where $j = (n_1+n_2)/2$ and $m = (n_1-n_2)/2$~\cite{osti_4389568,Sakurai:2011zz}.
Furthermore, the three angular momentum generators $J_x$, $J_y$, and $J_z$ can be expressed in terms of oscillators as
\begin{equation}
    J_+ = a^\dagger_1\,a_2, \quad J_- = a^\dagger_2\,a_1, \quad J_z = \frac{a^\dagger_1\,a_1 - a^\dagger_2\,a_2}{2},
    \label{eq: Js in terms of SHO oscillators}
\end{equation}
where $J_\pm = J_x\pm i J_y$.
Moreover, the multiplet $a\equiv(a_1, a_2)^T$ transforms under the SU(2) rotations generated by the three angular momentum generators, and the index 1 and 2 corresponds to the SU(2) group index.

With this, the left and right angular momentum states at site $r$ in Eq.~\eqref{eq: general KS state in SU(2)} can be constructed from a left and right set of local oscillators or Schwinger bosons, $a(L,r) = (a_1(L,r),$ $ a_2(L,r))^T$ and $a(R,r-1) = (a_1(R,r-1), a_2(R,r-1))^T$, respectively, from a vacuum that is annihilated by all annihilation type oscillators.
It then allows one to construct an alternative local basis by forming gauge-invariant operators using fermions and Schwinger bosons and acting on a vacuum, $|\Omega\rangle_r$, which is annihilated by $\psi_\alpha(r)$, $a_\alpha(L,r)$, and $a_\alpha(R, r-1)$, with $\alpha$ being the color or SU(2) group index.
This gives three conditions to find such building block operators: $1)$ They must be gauge-invariant under residual gauge transformations. $2)$ All operators must be creation-type operators. $3)$ They must not be null operators
\footnote{E.g.: $\psi^\dagger_\alpha(r) \psi^\dagger_\beta(r)\delta_{\alpha\beta}$, which a gauge-invariant operator of only creation-type operators, that is a null operator because of Eq.~\eqref{eq: ferm_anticomm SU(2)}.}.
The gauge-invariant operators are formed by contracting group indices using the gauge-invariant Kronecker delta, $\delta_{\alpha\beta}$, and the Levi-Civita tensor, $\epsilon_{\alpha\beta}$.
A complete set of SU(2) invariants has been obtained in Ref.~\cite{Raychowdhury:2019iki}, and the only four operators that satisfy these conditions are
\begin{align}
    \mathcal{L}^{++}(r) &= a^\dagger_\alpha(R,r-1)a^\dagger_\beta(L,r)\,\epsilon_{\alpha\beta},
    \label{eq: SU(2) LSH basis operators Lpp}\\
    \mathcal{S}^{++}_{\rm in}(r) &= a^\dagger_\alpha(R,r-1)\psi^\dagger_\beta(r) \,\epsilon_{\alpha\beta},
    \label{eq: SU(2) LSH basis operators Sin}\\
    \mathcal{S}^{++}_{\rm out}(r) &= a^\dagger_\alpha(L,r)\psi^\dagger_\beta(r)\,\epsilon_{\alpha\beta},
    \label{eq: SU(2) LSH basis operators Sou}\\
    \mathcal{H}^{++}(r) &= \frac{1}{2!}\psi^\dagger_\alpha(r)\psi^\dagger_\beta(r)\,\epsilon_{\alpha\beta}.
    \label{eq: SU(2) LSH basis operators Hadron}
\end{align}
$\mathcal{L}^{++}(r)$ has only boson type operators like loop degrees of freedom in gauge theories, $\mathcal{S}^{++}_{\rm in}$ and $\mathcal{S}^{++}_{\rm out}$ has a gauge and a matter type operator analogous to gauge-invariant strings in gauge theories, and finally, $\mathcal{H}^{++}$ has only fermion type operators like hadrons, thus the name loop-string-hadron formulation.

A local basis can be constructed from these operators that has two fermion like quantum numbers, $n_i$ and $n_o$, and a boson like quantum number $n_l$, and it is generated from the local vacuum $|\Omega\rangle_r$ as~\cite{Raychowdhury:2019iki}
\begin{align}
    |n_l; \; n_i=0,\,n_o=0\rangle_r &\propto (\mathcal{L}^{++}(r))^{n_l}|\Omega\rangle_r,
    \label{eq: SU(2) LSH basis ni 0 no 0}\\
    |n_l; \; n_i=1,\,n_o=0\rangle_r &\propto (\mathcal{L}^{++}(r))^{n_l}\mathcal{S}^{++}_{\rm in}(r)|\Omega\rangle_r,
    \label{eq: SU(2) LSH basis ni 1 no 0}\\
    |n_l; \; n_i=0,\,n_o=1\rangle_r &\propto (\mathcal{L}^{++}(r))^{n_l}\mathcal{S}^{++}_{\rm out}(r)|\Omega\rangle_r,
    \label{eq: SU(2) LSH basis ni 0 no 1}\\
    |n_l; \; n_i=1,\,n_o=1\rangle_r &\propto (\mathcal{L}^{++}(r))^{n_l}\mathcal{H}^{++}(r)|\Omega\rangle_r,
    \label{eq: SU(2) LSH basis ni 1 no 1}
\end{align}
where the operators appearing on the right-hand side are defined in Eqs.~\eqref{eq: SU(2) LSH basis operators Lpp}-\eqref{eq: SU(2) LSH basis operators Hadron}, and the proportionality constant is the normalization factor.
The states $|n_l; \; n_i,\,n_o\rangle_r$ span the local Hilbert space of combined gauge-matter degrees of freedom with $n_i\in\{0,1\}$, $n_o\in\{0,1\}$ and $n_l\in\{0,1,2,\cdots\}$.

A general SU(2) LSH state is then given by
\begin{equation}
    |\Psi\rangle^{(\rm LSH)} = \goldieotimes_{r} |n_l; \; n_i,\,n_o\rangle_r.
    \label{eq: general LSH state in SU(2)}
\end{equation}
This state is invariant under the residual gauge transformations by construction, that is, it satisfies $G^a(r)|\Psi\rangle^{(\rm LSH)}=0$ $\forall r$ for Gauss's law generators $G^a(r)$ in Eq.~\eqref{eq: gauss-op SU(2)}.
However, there is still an additional constraint that a physical state needs to satisfy which comes from the parallel transport of the chromoelectric field across the link.
This condition is the invariance of the quadratic Casimir across a link as given by Eq.~\eqref{eq: electric field casimir constraint SU(2)}, which reflects as the equality $j_L(r) = j_R(r)$.
A physical LSH state has to satisfy this constraint, which is translated into the LSH quantum numbers as~\cite{Raychowdhury:2019iki}
\begin{equation}
    [n_l+n_o(1-n_i)]_r = [n_l+n_i(1-n_o)]_{r+1} \quad \forall r
    \label{eq: LSH Abelian gausses law in SU(2)}
\end{equation}
Thus, among the three non-commuting Gauss's law constraints and the invariance of quadratic Casimir constraints in the KS formulation, only the latter one survives in the LSH formulation which appears as a simple relation between the LSH quantum numbers at neighboring sites.

Finally, the KS Hamiltonian in Eq.~\eqref{eq: KS-ham SU(2)} can be reformulated in terms of the number operators and raising and lowering operators of LSH quantum numbers.
This is done by expressing the link operator and chromoelectric fields in terms of Schwinger bosons and obtaining their action on LSH states.
The LSH Hamiltonian is not shown here, and interested readers are referred to Ref.~\cite{Raychowdhury:2019iki} for its detailed derivation and generalizations to higher dimensions.
Nonetheless, some prominent features of the LSH Hamiltonian are worth mentioning here: $1)$ The chromoelectric field energy and the fermion mass term remain diagonal in the LSH basis and dynamics is governed by the gauge-matter interaction term, just like the KS Hamiltonian. $2)$ The Hamiltonian has terms with only on-site interactions or nearest neighbor hopping interactions, and hence, remains local. $3)$ The raising and lowering operators for the bosonic LSH quantum number, $n_l$, in the LSH Hamiltonian appear as conditional operators on the fermionic quantum numbers, $n_i$ and $n_o$. That is, they are operated on a state depending on the values of $n_i$ and $n_o$.

The LSH formulation has been gaining attention in recent years, with proposals for its analog~\cite{Dasgupta:2020itb} and digital~\cite{Davoudi:2022xmb} quantum simulation developed, and implications of its global symmetries studied~\cite{Mathew:2022nep}.
Motivated by its advantageous features, Ch.~\ref{ch: LSH} extends the LSH formulation to an SU(3) LGT in 1+1D, which is a progression towards the goal of finding a computational-resource efficient formulation of QCD for its Hamiltonian simulation.

\renewcommand{\thechapter}{2}
\chapter{Obtaining double-beta Decay Amplitudes from Lattice QCD
\label{ch: DBD from LQCD}
}
\noindent
Double-beta decay is an isobaric transition between two nuclei.
The parent nucleus $(A,Z)$, where $A$ is the mass number and $Z$ is the atomic number, decays to a daughter nucleus $(A,Z+2)$ resulting from two neutrons getting converted into two protons.
Such a transition can be experimentally detected in isotopes that are even-even nuclei and lighter than the nearby odd-odd nucleus, for which the single-beta decay is forbidden or strongly suppressed.
There are two modes through which a double-beta decay can occur, $2\nu\beta\beta$ decay and $0\nu\beta\beta$ decay.
In the former case, the final state has two electrons and two electron-type antineutrinos: $(A,Z) \to (A, Z+2) + e e \bar{\nu}_e \bar{\nu}_e$.
Since the theoretical prediction of this process and the first estimation of its rate in 1930s~\cite{GoeppertMayer:1935qp}, such a decay has been observed conclusively in a dozen isotopes ranging from $^{48}$Ca to $^{238}U$~\cite{Barabash:2020nck, Saakyan:2013yna}.
This decay is one of the rarest SM processes in nature, with experimental half-lives ranging from $\sim 10^{19}$ to $\sim 10^{24}$ years~\cite{Barabash:2020nck, Saakyan:2013yna}.

The exotic BSM counterpart of this process, namely the neutrinoless mode $(A,Z) \to (A,Z+2)+ee$, remains at the center of vigorous theoretical studies and experimental searches~\cite{DellOro:2016tmg, Dolinski:2019nrj}, as its observation would unveil the nature of neutrinos and provide a mechanism for lepton-number violation, hence baryon-number violation, in the universe.
While several experiments searching for this decay have been performed in the past, such a transition has never been observed, and an extensive program of experiments continue to seek evidence for $0\nu\beta\beta$ decays~\cite{Biassoni:2020byh,Dolinski:2019nrj,Cappuzzello:2018wek,Cappuzzello:2016zlj,Bilenky:2014uka,Acharya:2023swl}.
Various BSM mechanisms can lead to $0\nu\beta\beta$ decays, which have been discussed in a number of reviews~\cite{Dolinski:2019nrj,DellOro:2016tmg, Vergados:2012xy,Avignone:2007fu,Rodejohann:2011mu}.
Regardless of the underlying mechanism, its observation would immediately imply that neutrinos have a Majorana mass component~\cite{Schechter:1981bd,Takasugi:1984xr}, i.e., they are their own anti-particles, unlike any other fermions in the SM.
The high stakes involved in the detection of $0\nu\beta\beta$ decay detection have made the associated theoretical nuclear physics calculations highly important, in particular the calculation of the involved nuclear ME that is related to its half-life, see Refs.~\cite{Cirigliano:2022oqy,Cirigliano:2022rmf,Acharya:2023swl}.

The $2\nu\beta\beta$ also continues to gain much attention, both theoretically and experimentally, for several reasons.
First, this decay mode is a dominant background for the much less probable $0\nu\beta\beta$ decay which could occur in the same isotopes. Therefore, accurate constraints on its decay rate, and on the spectral shape of electron energies emitted in the decay, are crucial for better understanding of the background in $0\nu\beta\beta$ decay searches~\cite{Saakyan:2013yna, Dolinski:2019nrj, Agostini:2017iyd}.
Second, assuming that only the SM weak interactions are in play, the measured decay rates can be converted to constraints on the corresponding nuclear MEs.
Subsequently, these MEs can be compared against theoretical determinations to test the validity of the nuclear-structure models used~\cite{Moreno:2008dz, Simkovic:2018rdz, NEMO-3:2019gwo, Azzolini:2019yib}.
This, along with theoretical calculations of the single-beta decay MEs, can refine these models, and inform the calculations of the exotic $0\nu\beta\beta$ decay process, for which the role of theory is crucial in constraining the new-physics mechanisms underlying the decay~\cite{DellOro:2016tmg, Dolinski:2019nrj}.
Finally, the increased sensitivity of $0\nu\beta\beta$ decay searches in recent years has led to a number of high-statistics measurements of the two-neutrino mode~\cite{KamLAND-Zen:2019imh, Argyriades:2009ph, Arnold:2016qyg, Arnold:2016ezh, Arnold:2018tmo, NEMO-3:2019gwo}, opening a window for probing potential BSM scenarios through this decay mode as well.
See for example, Ref.~\cite{Deppisch:2020mxv} for constraints on right-handed currents from the energy distribution and angular correlation of outgoing electrons in the $2\nu\beta\beta$ decay or probing the bosonic fraction of the neutrino wavefunction \cite{Dolgov:2005qi,Barabash:2007gb}.
Such investigations require accurate nuclear MEs to be computed within the SM, augmented with reliable uncertainties.

This chapter reviews the challenges in calculating double-beta decay nuclear ME and provides prospects that could mitigate errors in their calculations.
The nuclear ME is the crucial component in determining the half-life of the double-beta decay transition.
In the case of $2\nu\beta\beta$ decay, the nuclear ME $\mathcal{M}^{\rm full}_{2\nu}$ is related to its half-life $T_{2\nu}$ via~\cite{Saakyan:2013yna}
\begin{equation}
    (T_{2\nu})^{-1} = G_{2\nu} |\mathcal{M}^{\rm full}_{2\nu}|^2,
    \label{eq: 2vbb half life}
\end{equation}
where $G_{2\nu}$ is obtained by integrating over the phase space of the four leptons emitted in the decay.
Similarly, for the $0\nu\beta\beta$ decay, the nuclear ME $\mathcal{M}^{\rm full}_{0\nu}$ is related to its half-life $T_{0\nu}$ via
\begin{equation}
    (T_{0\nu})^{-1} = G_{0\nu} |\mathcal{M}^{\rm full}_{0\nu}|^2 \langle\eta\rangle^2,
    \label{eq: 0vbb half life}
\end{equation}
where $G_{0\nu}$ is a phase space factor from two electrons in the final state, and $\langle\eta\rangle^2$ is the lepton-number violating parameter that characterizes the beyond SM physics.
There are many BSM scenarios that can led to a $0\nu\beta\beta$ transition, in this thesis we consider the widely studied light neutrinos exchange scenario in which the SM neutrinos acquire a Majorana mass.
In this scenario the parameter $\langle\eta\rangle^2$ is given by the effective Majorana mass $m_{\beta\beta}^2$ that is defined and discussed in Sec~\ref{sec: 0vbb}.
The phase space factors $G_{2\nu}$ and $G_{0\nu}$ can be calculated very accurately.
Thus, in both cases, the respective nuclear MEs play a very important role in relating theory with experiments.
The role of $\mathcal{M}^{\rm full}_{0\nu}$ is even more important since its value is necessary in extracting the new physics parameter.
Unlike $\mathcal{M}^{\rm full}_{2\nu}$ which can be directly extracted from experiments, $\mathcal{M}^{\rm full}_{0\nu}$ is not related to any other observable to constrain its value from experiments, and one has to rely on nuclear models to do that.

Computing nuclear MEs for the double-beta decay process is a rather challenging task given the quantum many-body nature of the nuclear isotopes used, and thus, nuclear MEs are the main source of uncertainty in obtaining half-life predictions from theory.
Their determination involves two parts: $1)$ proper matching of the high-energy interactions to operators built from hadronic degrees of freedom and $2)$ nuclear structure calculations for the initial and final nuclear states.

For the first part, the interactions between quarks and leptons are considered as the high-energy interactions.
In the SM predictions of $\mathcal{M}^{\rm full}_{2\nu}$, these interactions are given by perturbative calculation of integrating out massive gauge bosons of electroweak theory to obtain the effective quark lepton interaction given in Eq.~\eqref{eq: Charged current Lagrangian}.
For the $\mathcal{M}^{\rm full}_{0\nu}$ predictions within the light neutrino exchange scenario, these interactions are obtained by considering SM as an EFT and writing down the lowest dimensional lepton number violating EFT operator, known as the Weinberg operator, that couples SM neutrinos with the SM Higgs doublet~\cite{Weinberg:1979sa}.
This leads to massive Majorana neutrinos that have the usual SM interactions with quarks and leptons.
Mapping these currents to the respective hadronic transition operators is the central topic of this chapter and is discussed in more details in the following sections.
For analysis involving higher dimension operators, see Ref.~\cite{Cirigliano:2018yza}.

For the second part of evaluating nuclear MEs, the \textit{ab initio} framework of nuclei allows building complex many-body correlations in the heavier isotopes using the few-nucleon interactions, see Refs.~\cite{Engel:2016xgb,Coraggio:2020iht} for recent reviews.
In this method, the nuclei are described as systems made of nucleons interacting via two- and three-body forces.
Their interactions with electroweak currents are obtained using many-body current operators like one- and two-body current operators discussed in Sec.~\ref{subsec: external currents in EFT}.
Due to their large computational cost, \textit{ab initio} calculations are generally performed in lighter nuclei, however, progress is made in applying them to experimentally relevant isotopes~\cite{Hergert:2015awm,Yao:2019rck,Wirth:2021pij,Belley:2020ejd,Jokiniemi:2021qqv}.
The other approaches involve approximated many-body methods to access the nuclei of experimental interest, which generally involves truncating the full Hilbert space of configurations and approximate the dynamics from the missing degrees of freedom through a renormalized effective Hamiltonian.
Nuclear models adopted to calculate $\mathcal{M}^{\rm full}_{0\nu}$ of nuclei of experimental interest are: the inretacting boson model~\cite{Barea:2012zz,Barea:2013bz,Barea:2009zza}, the quasiparticle random-phase approximation~\cite{Simkovic:2009pp,Fang:2011da,Neacsu:2015uja}, the energy density functional methods~\cite{Rodriguez:2010mn}, the covariant density functional theory~\cite{Jiao:2019thr,Jiao:2017dir,Jiao:2017opc,Yao:2018qjv} and the shell model~\cite{Coraggio:2018rou,Horoi:2013jx,Neacsu:2014bia,Brown:2015gsa}.
See a recent review for more details~\cite{Coraggio:2020iht}.
A comparison between different values of the nuclear MEs of $0\nu\beta\beta$ decay calculated with the different models is given in Fig. 5 of Ref.~\cite{Engel:2016xgb}. 
The values differ between different methods and the discrepancy is as large as a factor of three in values.

The source of these uncertainties can be traced to the two parts of evaluating the nuclear ME discussed above.
Here, the focus will be on the first part of evaluations.
This chapter illuminates on uncertainties in parameterizing two-nucleon nuclear MEs of electroweak currents in EFT framework for the $2\nu\beta\beta$ decay in Sec.~\ref{sec: 2vbb} and $0\nu\beta\beta$ decay in Sec.~\ref{sec: 0vbb}, and provides prescriptions for constraining them using LQCD calculations in the respective subsections.

\section{Two-nucleon \texorpdfstring{$2\nu\beta\beta$}{Bookmark Version} decay
\label{sec: 2vbb}
}
This section takes a bottom-up approach to the problem of calculating $2\nu\beta\beta$ nuclear ME and it is based on the work in Ref.~\cite{Davoudi:2020xdv}.
It lays out a possible framework for providing the SM input for the ME involved at the microscopic level, namely that in the $nn \to pp\,(ee\bar{\nu}_e\bar{\nu}_e)$ process, using a combined LQCD and EFT approach.
While such a process will not occur unless embedded in certain nuclear backgrounds, an apparent hierarchy of nuclear interactions and nuclear responses to external probes at low energies indicates that the most significant contribution to the nuclear decay may arise from single and correlated two-nucleon couplings, and corrections to this picture can be evaluated within an EFT power counting systematically.
The contributions to the ME at each order in the EFT depend upon the LECs for interactions and currents, which can ideally be constrained by matching to the SM determination of the same ME using LQCD methodology.
Setting up this step is the central subject of this section.
Upon successful completion of LQCD determinations of the double-weak two-nucleon ME~\cite{Shanahan:2017bgi,Tiburzi:2017iux}, the formalism presented can be employed to constrain EFTs that will be used in systematic \emph{ab initio} calculations of the decay rate in relevant isotopes.
The compatibility of the EFT description in the few-nucleon sector with \emph{ab initio} nuclear many-body methods, and the validity of the adopted EFT power counting when applied to a large nuclear isotope, will require further study and are not the focus of the current work.
The interested reader can refer to a recent review in Ref.~\cite{Tews:2020hgp} on the status and prospect of EFT-based studies of nuclei using \emph{ab initio} nuclear many-body methods.

For evaluating the decay $nn \to pp\,(ee\bar{\nu}_e\bar{\nu}_e)$, pionless EFT with KSW power counting has been employed, see Sec.~\ref{subsec: NN in pionless}.
As noted before, pionless EFT has shown notable success in studies of light nuclei and their electromagnetic and weak response.
Here, explicit calculation of the $nn \to pp\,(ee\bar{\nu}_e\bar{\nu}_e)$ amplitude in the pionless EFT with nucleonic degrees of freedom is provided.
The first attempt at constructing and constraining the nuclear ME for this process was presented in Refs.~\cite{Shanahan:2017bgi,Tiburzi:2017iux}, in which a dibaryon formulation of the pionless EFT~\cite{Beane:2000fi, Phillips:1999hh} was considered.
In this case, a contact LEC, called $h_{2,S}$ in this reference, was naturally introduced to account for the conversion of a spin-singlet dibaryon with $I_3=-1$ to that with $I_3=1$ at tree level, similar to the role of the so-called $l_{1,A}$ coupling of the dibaryon formalism that turns a spin-singlet dibaryon with $I_3=0$ to a spin-triplet dibaryon with $S_3=0$ at tree level. Here, $I_3$ ($S_3$) is the third component of the total isospin (spin) operator.
Importantly, the first LQCD computation of the relevant ME at a pion mass of $\approx 800~$MeV was performed at the same time, and led to a constraint on this new short-distance coupling, demonstrating its roughly equal contribution to the ME when compared with the $l_{1,A}$ effects in the deuteron-pole contribution. 
The effects of off-shell intermediate states had to be added to the tree-level coupling, leading to an effective short-distance coupling, called $\mathbb{H}_{2,S}$.
It was this effective coupling that was constrained with LQCD input on the ME. See Ref.~\cite{Tiburzi:2017iux} for details.

It is valuable to obtain an EFT amplitude within the nucleonic formulation, which is a more suitable framework in connecting to nuclear-structure calculations of the ME in larger isotopes.
Such an amplitude is calculated in the current work in Sec.~\ref{subsec: 2vbb amplitude in pionless EFT} for the $2\nu\beta\beta$ decay.
It is shown that, the axial coupling of the nucleon and the correlated coupling of the two-nucleon system to an axial-vector current are sufficient to obtain a renormalization-scale independent amplitude at NLO.
Such a conclusion for the power counting can be verified through simultaneous LQCD computations of both MEs.

Perhaps an even more significant component of the analysis presented here is to provide the framework for obtaining the hadronic amplitude for the two-nucleon double-beta decay from four-point correlation functions of two nucleons obtained with LQCD in a finite Euclidean spacetime.
As discussed in Sec.~\ref{sec: FV Formalism}, the multi-hadron amplitudes cannot be directly accessed from LQCD Euclidean correlation functions, and a mapping is required to constrain the amplitudes evaluated at eigenenergies of the system in a finite volume.
As reviewed in Sec~\ref{subsec: NN to NN  with 1 J}, formalism for constraining transition amplitudes with external currents requires FV MEs obtained from LQCD three-point functions and two nucleon FV energy eigenvalues are required.
For the non-local MEs like the double-beta decay process, due to the dependence of the time-ordered product of currents present in a four-point function on the spacetime signature, Euclidean and Minkowski MEs are fundamentally different when intermediate states with on-shell kinematics are present.
These must be subtracted from the Euclidean correlation function, and be separately constructed from the knowledge of appropriate two- and three-point functions.
These contributions must then be added back to the non-problematic contributions to obtain the complete Minkowski ME in a finite volume, which is used to construct the physical amplitude, see e.g., Ref.~\cite{Briceno:2019opb}.
Various components of the formalisms mentioned above, such as identifying volume corrections to external and intermediate two-hadron states, as well as reconstructing the Minkowski amplitude from the Euclidean four-point function, are relevant for the mapping of the nucleonic ME in the $2\nu\beta\beta$ transition as well.

The analysis of the single-beta decay in infinite and finite volume presented in Sec.~\ref{subsec: NN to NN  with 1 J} can now be extended to the case of the double-beta decay.
While the overall strategy is the same as that outlined in Secs.~\ref{subsubsec: singleV} and~\ref{subsubsec: singleVIV}, the  double-beta decay problem presents new features originating from the presence of two spacetime-displaced axial-vector currents in the MEs.
This, first of all, makes distinct contributions arising from different time ordering of the currents for general kinematics.
Second, and more significantly, it necessitates further steps to be taken to connect the Minkowski finite-volume ME to that calculated in Euclidean spacetime, which is the case with LQCD computations.
The first two subsections here follow the procedure of Sec.~\ref{subsubsec: singleV}, while the last subsection addresses this latter point.
It is found that no new contact two-nucleon two-current LEC is needed to renormalize the physical hadronic amplitude at NLO, therefore the only two-nucleon axial-current LEC that is to be constrained from the matching relation is $L_{1,A}$.
LQCD computations of this ME will be able to test the validity of this power counting.

\subsection{Two-nucleon \texorpdfstring{$2\nu\beta\beta$}{Bookmark Version} decay in pionless EFT
\label{subsec: 2vbb amplitude in pionless EFT}
}

\begin{figure}
    \centering
    \includegraphics[scale=0.75]{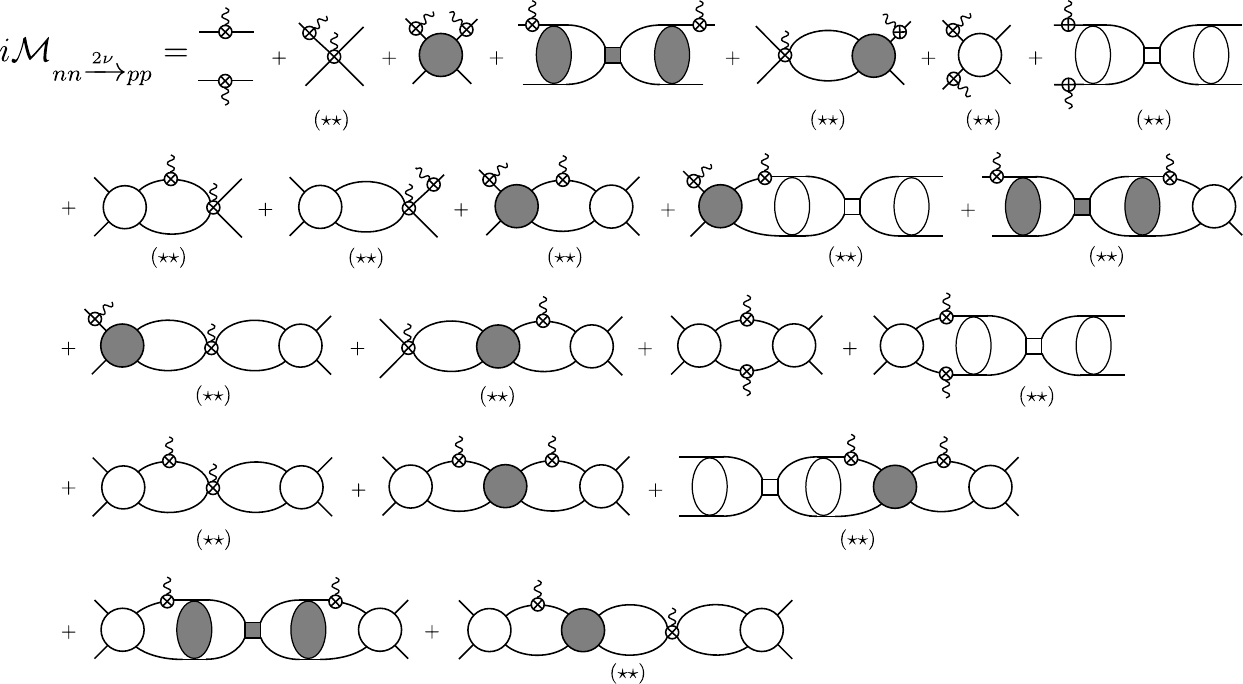}
    \caption{Diagrammatic representation of the hadronic transition amplitude contributing to the $nn \to pp$ process. All components of the diagrams are introduced in Figs.~\ref{fig: C0 C2 Feynman diagram notation},~\ref{fig: LO and NLO NN in KSW}, and~\ref{fig: gA and L1A diagram}.
    The external legs are evaluated at the on-shell kinematics, and the ($\star\star$) symbol under a given diagram indicates that the reversed diagrams must be included as well (without changing the initial and final states). An exact set of diagrams must be computed but with the first and the second currents reversed in order, which is equivalent to setting $E_1 \leftrightarrow E_2$ in these diagrams, where $E_1$ and $E_2$ are the energies carried out by each of the two currents, see the discussions in the text.}
    \label{fig: double weak infinite V amplitude diagrams}
\end{figure}
The hadronic amplitude for the double-beta decay is calculated here.
This is the amplitude given in Eq.~\eqref{eq: Hadronic amplitude in terms of T-matrix} for $X=nn\xrightarrow{2\nu} pp$.
The two currents are the same for this transition, and in the spin-isospin symmetric limit, all three spin components of the intermediate spin-triplet state contribute equally.
This leads to only Gamow-Teller type transitions for both currents.
This is because the intermediate state has an on-shell electron-antineutrino pair, indicating that both spin and isospin has to change by by one unit, but the Fermi type contribution only changes the isospin leaving the spin part unchanged. 

The calculation performed here considers only the amplitude for transitions with the $k=3$ component of the currents in Eq.~(\ref{eq: CC hamiltonian, one-body and two-body}), and in the full amplitude this contribution must be multiplied by a factor of 3. 
The currents are assumed to carry out no spatial momenta and their kinematics is given by $Q_1=(E_1,{\bm 0 })$ and $Q_2=(E_2,{\bm 0 })$.
The initial and final two-nucleon states are at rest, which constrains the intermediate states between the current to have zero spatial momentum. The hadronic contribution to the T-matrix is given by:
\begin{align}
	J_{nn\xrightarrow{2\nu} pp} \equiv \int d^3x \, \mathcal{J}_
	{nn  \xrightarrow{2\nu} pp}(x) = \hat{A}_{\rm CC}\,G_S\,\hat{A}_{\rm CC}.
	\label{eq: double beta T-matrix}
\end{align}
Here, $\hat{A}_{\rm CC}$ is the hadronic contribution to the weak Hamiltonian defined in Eq.~\eqref{eq: weak Hamiltonian HCC}, and $G_S$ is the strong retarded Green's function defined in Eq.~\eqref{eq: Green's function: Strong}.
There are two possibilities for the intermediate states, one with total energy $E_{*1}=E_i-E_1$ and one with total energy $E_{*2}=E_i-E_2$.
These should be considered separately in the time-independent Lippmann-Schwinger formalism that is adopted for computing the amplitudes.
The two contributions correspond to different time orderings of the currents in a time-dependent quantum field theory approach that is used in the definition of the finite-volume correlation functions.

$J_{nn \xrightarrow{2\nu} pp}$ can now be used in Eq.~\eqref{eq: Hadronic amplitude in terms of T-matrix} to obtain the hadronic amplitude up to NLO in pionless EFT.
The diagrammatic expansion of this amplitude is shown in Fig.~\ref{fig: double weak infinite V amplitude diagrams}.
Following the method described in Sec.~\ref{subsec: NN EFT}, one obtains
\begin{align}
    i\mathcal{M}_{nn\xrightarrow{2\nu} pp}& (E_i,E_1,E_f)=i\mathcal{M}^{\rm DF}_{nn\xrightarrow{2\nu} pp}(E_i,E_1,E_f)\nonumber\\
    & \hspace{-1.35cm}+
    i\,\frac{g_A^2}{4} (2\pi)^3 \delta^3(\bm{p}_i-\bm{p}_f)\left[ \frac{1}{E_1}+\frac{1}{E_2} \right]+\frac{g_A^2}{2(E_1+E_2)}\times
    \nonumber\\[10 pt]
    &\hspace{-1.35 cm} \left[\frac{i\mathcal{M}^{\rm (LO+NLO)}(E_i)-i\widetilde{\mathcal{M}}^{\rm (LO+NLO)}(E_{*1})}{E_2}-\frac{i\widetilde{\mathcal{M}}^{\rm (LO+NLO)}(E_{*1})-i\mathcal{M}^{\rm (LO+NLO)}(E_f)}{E_1}\right] 
    \nonumber\\
    & + \frac{g_A}{2} \left[\frac{i\mathcal{M}^{\rm DF}_{nn\to np} (E_i,E_{*1})}{E_2}- \frac{i\mathcal{M}^{\rm DF}_{np\to pp}(E_{*1},E_f)}{E_1} \right]+ (E_1 \leftrightarrow E_2).
    \label{eq: double amplitude full IV}
\end{align}
$(E_1 \leftrightarrow E_2)$ indicates that the counterpart of every term where $E_1$ is exchanged with $E_2$ must be included as well.
Similar to Eq.~\eqref{eq: single beta decay full amplitude in inf V}, terms in Eq.~\eqref{eq: double amplitude full IV} have been grouped into a divergence-free $nn\to pp$ amplitude, as well as divergent amplitudes which are singular in the limit $E_1,E_2\to0$.
The divergent part contains the divergence-free single-weak transition amplitude, $\mathcal{M}^{\rm DF}_{nn\to np}$ and $\mathcal{M}^{\rm DF}_{np\to pp}$, which are given by the expression in Eq.~\eqref{eq: single beta decay DF amplitude}.
Note that these amplitudes are the same in the isospin-symmetric limit.
Equation~\eqref{eq: double amplitude full IV} also contains $NN\to NN$ elastic scattering amplitudes for both spin-singlet and spin-triplet channels that are defined in Eq.~\eqref{eq: LO+NLO 2to2 Amplitude}.
\begin{figure}[t!]
    \centering
    \includegraphics[scale=1]{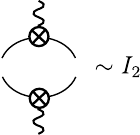}
    \caption{Diagrammatic representation of the $I_2$ loop defined in Eq.~\eqref{eq: I2 definition}.
    }
    \label{fig: I2 diagram}
\end{figure}

The divergence-free part of the $nn\xrightarrow{2\nu} pp$ amplitude, $\mathcal{M}^{\rm DF}_{nn\xrightarrow{2\nu} pp}$, is given by
\begin{align}
    \hspace{-3cm}i\mathcal{M}^{\rm DF}_{nn\xrightarrow{2\nu} pp}  (E_i,E_1,E_f) &=
    \frac{g_A}{2}\,i\mathcal{M}^{\rm (LO+NLO)}(E_f) 
    \nonumber\\
    &\hspace{-2.5 cm} 
    \times   \bigg[g_A\,i\,I_1(E_f,E_{*1})\,i\widetilde{\mathcal{M}}^{\rm (LO+NLO)}(E_{*1})
    \,i\,I_1(E_i,E_{*1})\,
    \nonumber\\
    &\hspace{-0.5cm}+ i\,\widetilde{L}_{1,A}\,i\widetilde{\mathcal{M}}^{\rm (LO+NLO)}(E_{*1}) \left[i\,I_1(E_i,E_{*1})+i\,I_1(E_{*1},E_f)\right]
    \nonumber\\
    &\hspace{0.5 cm} 
   +i\,g_A\,I_2(E_i,E_{*1},E_f) \bigg]
   \times i\mathcal{M}^{\rm (LO+NLO)}(E_i),
    \label{eq: double amplitude DF IV}
\end{align}
where $I_1$ is give by Eq.~\eqref{eq: I1 definition} and the new type of loop arising from four propagators and two weak currents, as shown in Fig.~\ref{fig: I2 diagram}-j, is given by
\begin{align}
    I_2(E_i,E_{*1},E_f) &= \int \frac{d^3{q}}{(2\pi)^3}
    \frac{1}{E_i-\frac{{\bm q}^2}{M}+i\epsilon}
    \frac{1}{E_{*1}-\frac{{\bm q}^2}{M}+i\epsilon}
    \frac{1}{E_f-\frac{{\bm q}^2}{M}+i\epsilon}
    \nonumber\\
    &=\frac{1}{E_i-E_f}\,[I_1(E_{*1},E_f)-I_1(E_i,E_{*1})],
    \label{eq: I2 definition}
\end{align}
and a useful identity is used to simplify the expressions:
\begin{align}
    \int \frac{d^3{q}}{(2\pi)^3}
    \frac{1}{E_i-\frac{{\bm q}^2}{M}+i\epsilon}
    \frac{\bm{q}^2}{E_{*1}-\frac{{\bm q}^2}{M}+i\epsilon}
    \frac{1}{E_f-\frac{{\bm q}^2}{M}+i\epsilon} 
    =-MI_1(E_{*1},E_f)+ME_iI_2(E_i,E_{*1},E_f).
    \label{eq: I2q2identity}
\end{align}
As the only new ingredient of the double-beta decay amplitude compared with the single-beta decay is $\mathcal{M}^{\rm DF}_{nn\xrightarrow{2\nu} pp}$, it is this quantity that is aimed to be constrained from a finite-volume matching relation in the next subsections.
At the order considered in the EFT, the physical hadronic amplitude in Eq.~\eqref{eq: double amplitude full IV} is evidently renormalization-scale independent (note that $I_2$ loop is UV convergent), and as a result, no new LEC beyond those present in the single-beta decay amplitude is needed to renormalize the amplitude, as stressed before.
A contact two-weak two-nucleon operator, in particular, should appear at next-to-NLO using a naive-dimensional analysis, and no new UV divergence at the previous order is observed to require its promotion to a lower order.


\subsection{Finite volume formalism for two-nucleon \texorpdfstring{$2\nu\beta\beta$}{Bookmark Version} decay in pionless EFT
\label{subsec: 2vbb matching}
}
\begin{figure}
    \centering
    \includegraphics[scale=0.85]{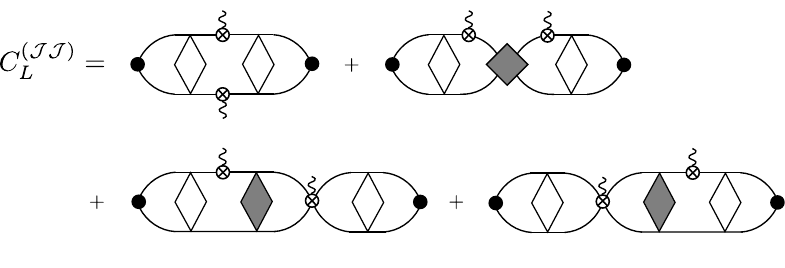}
    \caption{Diagrammatic representation of the finite-volume correlation function with two insertions of the weak current corresponding to the expansion in Eq.~(\ref{eq:CLJJ}). The black dots are the interpolating operators for the spin-singlet two-nucleon states.
    All other components are introduced in Figs.~\ref{fig: two point correlation diagram} and~\ref{fig: gA and L1A diagram}. The loops are evaluated as a sum over discretized momenta, as discussed in the texts.
    An exact set of diagrams must be computed but with the first and the second currents reversed in order, which is equivalent to setting $E_1 \leftrightarrow E_2$ in these diagrams, where $E_1$ and $E_2$ are the energies carried out by each of the two currents, see the discussions in the text.}
    \label{fig: double weak finite V correlation}
\end{figure}
The relevant finite-volume correlation function for the double-beta decay process in momentum space can be represented by diagrams shown in Fig.~\ref{fig: double weak finite V correlation}, where an expansion at NLO in the pionless EFT can be performed later.
These diagrams correspond to the following expansion
\begin{align}
    C_L^{(\mathcal{J}\mathcal{J})}(P_i,P_f) &= \bar{B}_{nn}(P_f) \sum_{n=0}^{\infty} \left[\otimes \, I^V_0 \otimes \mathcal{K} \right]^n \otimes
    \bigg\{\frac{g_A^2}{2}\,iI_2^V+i\frac{g_A^2}{2}\,I^V_1 \otimes \widetilde{\mathcal{K}} \sum_{n=0}^{\infty} \big[\otimes \, I^V_0 \otimes \widetilde{\mathcal{K}} \big]^n \otimes I_1^V
    \nonumber\\
    &+i\frac{g_A}{2}\,L_{1,A} I^V_1 \sum_{n=0}^{\infty} \big[\otimes \, \widetilde{\mathcal{K}} \otimes I^V_0 \big]^n \otimes I_0^V
    +i\frac{g_A}{2}\,L_{1,A} \,I_0^V \otimes \sum_{n=0}^{\infty} \big[ I^V_0 \otimes \widetilde{\mathcal{K}} \otimes \big]^n  I^V_1\bigg\}
    \nonumber\\
    &\hspace{7.25 cm}\otimes \sum_{n=0}^{\infty} \left[ \mathcal{K} \otimes I^V_0 \otimes \right]^n \bar{B}^{\dagger}_{pp}(P_i)+\cdots.
    \label{eq:CLJJ}
\end{align}
Here, the kinematic dependence of the functions is suppressed, except for the dependence of the interpolating functions on total initial and final momenta.
Ellipsis indicates that a similar expansion with $E_1 \leftrightarrow E_2$ must be included.
All ingredients in Eq.~\eqref{eq:CLJJ} are introduced in the previous sections, except for $\otimes I_2^V \otimes$, where $I_2^V$ is the finite-volume counterpart of the loop function with two insertions of the single-body weak current. For generic left and right functions $\chi$ and $\xi$, $\otimes$-product sign is defined as
\begin{equation}
    \chi \otimes I^V_2 \otimes \xi \equiv  \frac{1}{L^3}\sum_{\bm{q}}
    \chi(\bm q)\frac{1}{E_i-\frac{{\bm q}^2}{M}}
    \frac{1}{E_{*1}-\frac{{\bm q}^2}{M}}
    \frac{1}{E_f-\frac{{\bm q}^2}{M}}\xi(\bm{q}),
    \label{eq: I2 V definition II}
\end{equation}
and similarly for $E_{*2}$.
For the analysis of this work at NLO in the EFT, the left and right convoluting functions are either momentum independent kernels/current, in which case the $\otimes$ sign trivially becomes an ordinary product, or one of the convoluting functions is proportional to $\bm{q}^2$ and one is constant, in which case the finite-volume counterpart of the identity in Eq.~\eqref{eq: I2q2identity} (in which $\int \frac{d^3{q}}{(2\pi)^3} \to \frac{1}{L^3} \sum_{\bm q}$, $I_1 \to I_1^V$, and $I_2 \to I_2^V$) can be used to return to the ordinary product of functions.
When appearing in an ordinary product, $I_2^V$ is simply given by Eq.~\eqref{eq: I2 definition} upon replacements just mentioned.
Note that this function is regular in the limit $E_i \to E_f$. 

The expansion in Eq.~\eqref{eq:CLJJ} can now be turned into a useful form, in a similar fashion to Sec.~\ref{subsubsec: singleV}.
The idea is to first isolate the infinite-volume contribution, which arises from separating the loop functions to an infinite-volume integral and a sum-integral difference, and collecting all terms with exclusively infinite-volume loops.
Next, given the identity in Eq.~\eqref{eq:sum-int-form}, factors of $\mathcal{F}(E_f)$ and $\mathcal{F}(E_i)$ can be formed out of $\sum_{n=0}^{\infty} \left[\otimes \, \mathcal{K} \otimes I^V_0 \right]^n$ adjacent to $B$ and $B^\dagger$, up to additional contributions that do not concern us.
The $\mathcal{F}$ functions enforce on-shell kinematic on the adjacent function, which combined with identities on the $\otimes I_1^V \otimes$ and $\otimes I_2^V \otimes$ as described before, ensure that all functions in such a term are evaluated on-shell.
These features, along with straightforward algebra, lead to 
\begin{align}
   &C_L^{(\mathcal{J}\mathcal{J})}(P_i,Q_1,P_f) = C_{\infty}^{(\mathcal{J}\mathcal{J})}(P_i,Q_1,P_f) +
   \mathcal{B}_
   {nn}(E_f)\,i\mathcal{F}(E_f)\bigg[
   i\mathcal{M}^{\rm{DF},V}_{nn \xrightarrow{2\nu} pp} (E_i,E_1,E_f)
   \nonumber\\
   &\hspace{0.6 cm}+\frac{1}{2}\, i\mathcal{M}^{\rm{DF},V}_{np\to pp}(E_{*1},E_f)\,
   i\widetilde{\mathcal{F}}(E_{*1})\,
   i\mathcal{M}^{\rm{DF},V}_{nn\to np}(E_i,E_{*1}) + (E_1 \leftrightarrow E_2)\bigg]
   i\mathcal{F}(E_i)\,\mathcal{B}^\dagger_{pp}(E_i)+\cdots,
   \label{eq: four point correlation final}
\end{align}
where now ellipsis denotes any other terms left out while expanding Eq.~\eqref{eq:CLJJ} in terms of infinite- and finite-volume contributions.
The reason behind this peculiar rearrangement of the terms, as discussed in Sec.~\ref{subsubsec: singleVIV}, is to only keep contributions with poles in $E_i$ and $E_f$ (which are the finite-volume energy eigenvalues of two nucleons in the spin-singlet channel). 
This fact will be used in the matching relation that will be derived shortly. $\mathcal{M}^{\rm{DF},V}_{nn\to np}$ in Eq.~\eqref{eq: four point correlation final} is related to the physical divergence-free transition amplitude, $\mathcal{M}^{\rm{DF}}_{nn\to np}$, which are defined in Eqs.~\eqref{eq: DF amplitude: finite V} and~\eqref{eq: single beta decay DF amplitude}, respectively.
The only new quantity here that needs to be defined is $\mathcal{M}^{\rm{DF},V}_{nn \xrightarrow{2\nu} pp}$, which is the finite-volume counterpart of $\mathcal{M}^{\rm{DF}}_{nn \xrightarrow{2\nu} pp}$ in Eq.~\eqref{eq: double amplitude DF IV}, with replacements $I_1 \to I_1^V$ and $I_2 \to I_2^V$.
Explicitly,
\begin{align}
   &i\mathcal{M}^{\rm{DF},V}_{nn \xrightarrow{2\nu} pp}
   (E_i,E_1,E_f)
  =i\mathcal{M}^{\rm DF}_{nn\xrightarrow{2\nu} pp} (E_i,E_1,E_f)
  \nonumber\\
   &\hspace{0.2cm} + \frac{g_A}{2}\,i\mathcal{M}^{\rm (LO+NLO)}(E_f)\bigg[i\,g_A\,F_2(E_1,E_{*1},E_f)
   -g_A\,F_1(E_f,E_{*1})\,i\widetilde{\mathcal{M}}^{\rm (LO+NLO)}(E_{*1})\,I_1(E_i,E_{*1})
   \nonumber\\
   &\hspace{-0.1cm} - g_A\,I_1(E_f,E_{*1})\,i\widetilde{\mathcal{M}}^{\rm (LO+NLO)}(E_{*1})\,F_1(E_i,E_{*1})
   -g_A \,F_1(E_f,E_{*1}) \, i\widetilde{\mathcal{M}}^{\rm (LO+NLO)}(E_{*1}) \,F_1(E_i,E_{*1})
   \nonumber\\
& \hspace{3.0cm} -\widetilde{L}_{1,A}\,i\widetilde{\mathcal{M}}^{\rm (LO+NLO)}(E_{*1})
   \left[F_1(E_i,E_{*1})+ F_1(E_{*1},E_f)\right]\bigg] i\mathcal{M}^{\rm (LO+NLO)}(E_i).
   \label{eq: double DF in V}
\end{align}
Here, $F_2=I_2^V-I_2$, or in terms of the sum-integral notation defined in Eq.~\eqref{eq: sumintegral definition},
\begin{equation}
    F_2(E_i,E_{*1},E_f) =  \frac{1}{L^3}\sum_{\bm{q}}\hspace{-.5cm}\int
    \frac{1}{E_i-\frac{{\bm q}^2}{M}}
    \frac{1}{E_{*1}-\frac{{\bm q}^2}{M}}
    \frac{1}{E_f-\frac{{\bm q}^2}{M}}
    =\frac{1}{E_i-E_f}\,[F_1(E_{*1},E_f)-F_1(E_i,E_{*1})],
\end{equation}
where $F_1$ is defined in Eq.~\eqref{eq:F1def}. Equation~\eqref{eq: double DF in V} provides a direct relation between the finite-volume quantities and the physical divergence-free double-beta decay amplitude.

The momentum-space four-point correlation function in the finite volume can be written as
\begin{align}
    C_L^{(\mathcal{J}\mathcal{J})}(P_i,Q_1,P_f) &=
    \int_Ld^3x \, d^3y \, d^3z
    \int dx_0 \, dy_0 \, dz_0
    \;e^{-iP_i\cdot x}\;e^{iP_f\cdot y}\;e^{iQ_1\cdot z}
    \nonumber\\
    &\hspace{2cm}\times
    \bigg[\langle 0|\, T[B_{pp}(y)\,\mathcal{J}_{pp \to np}(z)\,\mathcal{J}_{pp \to np}(0)\, B_{nn}^\dagger(x)]\,|0\rangle \bigg]_L.
    \label{eq: four-point correlation (Pi,Pf,Q)}
\end{align}
The time-momentum representation of the same correlation function is 
\begin{align}
     C_L^{(\mathcal{J}\mathcal{J})}&(x_0,y_0,Q_1)
     \equiv \int \frac{dE_i}{2\pi}\,\frac{dE_f}{2\pi} \, e^{iE_i\cdot x_0} \, e^{-iE_f y_0}\,
    C_L^{(\mathcal{J}\mathcal{J})}(P_i,Q_1,P_f)
    \label{eq: four point correlation diagrammatic definition}
    \\
    & = \int_L d^3x\,d^3y\,
    \int  dz_0\int_L d^3z\, e^{iQ_1\cdot z}
    \bigg[\langle 0|\, T[B_{pp}(y)\,\mathcal{J}_{pp \to np}(z)\,\mathcal{J}_{pp \to np}(0)\,B^\dagger_{nn}(x)]\, |0\rangle\bigg]_L
    \label{eq: four point correlation dispersive definition 1}
    \\
    &=L^6\sum_{E_{n_i},E_{n_f}}
    e^{iE_{n_i}x_0}\,e^{-iE_{n_f}y_0}\,\int dz_0\int_L d^3z\, e^{iQ_1\cdot z}\left[\vphantom{B^\dagger}\langle 0|\,B_{pp}(0)\,| E_{n_f},L\rangle\right]_L
    \nonumber\\
    &\hspace{2.0cm}\times \, \left[\vphantom{B^\dagger}\langle E_{n_f},L|\,T[\mathcal{J}_{pp \to np}(z)\,\mathcal{J}_{pp \to np}(0)]\,|E_{n_i},L\rangle \right]_L \left[\langle E_{n_i},L|\,B^\dagger_{nn}(0)\, |0\rangle\right]_L,
    \label{eq: four point correlation dispersive expression 2}
\end{align}
written in different forms following similar steps as in the previous sections.
In the last equality, it is it assumed that $y_0>z_0>0>x_0$ or $y_0>0>z_0>x_0$.

The result in Eq.~\eqref{eq: four point correlation final} for $C_L^{(\mathcal{J}\mathcal{J})}$ can now be input in Eq.~\eqref{eq: four point correlation diagrammatic definition}.
Integration over $E_i$ and $E_f$ leads to non-vanishing contributions from the only poles of the function, which are the finite-volume eigenenergies in the spin-singlet channel, $E_{n_i}$ and $E_{n_f}$, respectively.
Note that these are solutions to quantization condition in Eq.~\eqref{eq: Luscher condition}.
This gives rise to
\begin{align}
    &C_L^{(\mathcal{J}\mathcal{J})}(x_0,y_0,Q_1)
    =\sum_{E_{n_i},E_{n_f}}e^{iE_{n_i}x_0} \, e^{-iE_{n_f}y_0}\,
    \mathcal{B}_{pp}(E_{n_f}) \mathcal{R}(E_{n_f})\bigg[i\mathcal{M}^{\rm{DF},V}_{nn\xrightarrow{2\nu} pp} (E_{n_i},E_1,E_{n_f})
    \nonumber\\
    &\hspace{1.0cm} +\frac{1}{2}\,i\mathcal{M}^{\rm{DF},V}_{nn\to np} (E_{n_i},E_{*1})\, i\widetilde{\mathcal{F}}(E_{*1})\, i\mathcal{M}^{\rm{DF},V}_{np\to pp} (E_{*1},E_{n_f}) + (E_1\leftrightarrow E_2)\bigg]
    \mathcal{R}(E_{n_i}) \,\mathcal{B}^\dagger_{nn}(E_{n_i}).
    \label{eq: four point correlation integrated}
\end{align}
Here, $\mathcal{R}$ is the residue of the finite-volume function $\mathcal{F}$ evaluated at the corresponding finite-volume energy, as defined in Eq.~\eqref{eq: definition of residue}.
Equating this for each $E_{n_i}$ and $E_{n_f}$ with Eq.~\eqref{eq: four point correlation dispersive expression 2} and using Eq.~\eqref{eq: matching condition for NN to NN}, one finally obtains the desired matching relation for the double-beta decay amplitude:
\begin{align}
    &L^{6}\left|\int dz_0\int_L d^3z\, e^{iQ_1\cdot z}
    \left[\vphantom{B^\dagger}\langle E_{n_f},L|\, T[\mathcal{J}_{pp \to np}(z)\,\mathcal{J}_{pp \to np}(0)]\, |E_{n_i},L\rangle\right]_L \right|^2 =\left|\mathcal{R}(E_{n_f})\right|\,\times
    \nonumber\\
    & \hspace{-0.25cm}\left|i\mathcal{M}^{\rm{DF},V}_{nn\xrightarrow{2\nu} pp} (E_{n_i},E_1,E_{n_f})
    +\frac{1}{2}i\mathcal{M}^{\rm{DF},V}_{nn\to np} (E_{n_i},E_{*1}) i\widetilde{\mathcal{F}}_{(np)}(E_{*1}) i\mathcal{M}^{\rm{DF},V}_{np\to pp} (E_{*1},E_{n_f}) + (E_1\leftrightarrow E_2) \right|^2
    \nonumber
   \\
   &\hspace{13 cm} \times \, \left|\mathcal{R}(E_{n_i})\right|,
    \label{eq: matching relation double}
\end{align}
with the finite-volume divergence-free amplitudes for single-beta and double-beta decays defined in Eqs.~\eqref{eq: DF amplitude: finite V} and~\eqref{eq: double DF in V}, respectively.
This relation is a main result of this work as it connects the (Minkowski) finite-volume ME of two time-ordered currents to the physical (Minkowski) amplitudes in infinite volume.
The only subtlety is that LQCD calculations do not directly have access to the former, i.e., the left-hand side of this equation. Instead, they provide the Euclidean counterpart of this ME.
It is, therefore, necessary to describe how to construct the Minkowski finite-volume ME from its Euclidean counterpart. Once this construction is carried out, Eq.~\eqref{eq: matching relation double} enables constraining the physical double-beta decay amplitude. 
\subsection{Minkowski to Euclidean matching
\label{subsec: doubleEM}
}
\noindent
The quantity derived in Eq.~\eqref{eq: matching relation double} in connection to the physical hadronic amplitude for double-beta decay is defined in Minkowski spacetime.
In particular, the ME on the left-hand side of Eq.~\eqref{eq: matching relation double} has the following spectral representation:
\begin{align}
    &\mathcal{T}_L^{(\rm M)} \equiv \int dz_0\int_L d^3z\, e^{iQ_1\cdot z}
    \left[\vphantom{B^\dagger}\langle E_f,L|\, T[\mathcal{J}_{pp \to np}(z)\,\mathcal{J}_{pp \to np}(0)]\, |E_i,L\rangle\right]_L =L^3\sum_{m=0}^{\infty}\int dz_0 \, \times
    \nonumber\\
   & \hspace{1.0 cm} \bigg[ e^{i(E_f+E_1-E_{*m}+i\epsilon) z_0}
   \left[\vphantom{B^\dagger}\langle E_f,L|\, \mathcal{J}_{pp \to np}(0)\,|E_{*m},L\rangle\right]_L\left[\vphantom{B^\dagger}\langle E_{*m},L |\, \mathcal{J}_{pp \to np}(0)\,|E_i,L\rangle\right]_L \theta(z_0)\,+
    \nonumber\\
    & \hspace{1.25 cm} e^{i(E_1-E_i+E_{*m}-i\epsilon) z_0}
   \left[\vphantom{B^\dagger}\langle E_f,L|\, \mathcal{J}_{pp \to np}(0)\,|E_{*m},L\rangle\right]_L\left[\vphantom{B^\dagger}\langle E_{*m},L |\, \mathcal{J}_{pp \to np}(0)\,|E_i,L\rangle\right]_L \theta(-z_0)\bigg]
    \nonumber\\
    & \hspace{0.75 cm} = iL^3\sum_{m=0}^{\infty} \left[ \frac{\left[\vphantom{B^\dagger}\langle E_f,L|\, \mathcal{J}_{pp \to np}(0)\,|E_{*m},L\rangle\right]_L\left[\vphantom{B^\dagger}\langle E_{*m},L |\, \mathcal{J}_{pp \to np}(0)\,|E_i,L\rangle\right]_L}{E_f+E_1-E_{*m}+i\epsilon} \right . 
    \nonumber\\
    &  \hspace{4.35 cm}  \left . - \frac{\left[\vphantom{B^\dagger}\langle E_f,L|\, \mathcal{J}_{pp \to np}(0)\,|E_{*m},L\rangle\right]_L\left[\vphantom{B^\dagger}\langle E_{*m},L |\, \mathcal{J}_{pp \to np}(0)\,|E_i,L\rangle\right]_L}{E_1-E_i+E_{*m}-i\epsilon} \right],
  \label{eq: 1}
\end{align}
where the kinematic dependence of $\mathcal{T}_L^{(\rm M)}$ on the left-hand side is left implicit.
$m$ is an index corresponding to the tower of finite-volume energy eigenstates with quantum numbers of $\mathcal{J}_{nn \to np}|E_i,L\rangle$ (or identically $\langle E_i,L|\mathcal{J}_{nn \to np}$).
In contrary, such a spectral decomposition is ill-defined in Euclidean spacetime as integration over the Euclidean time $\tau \equiv z_0^{(E)} = iz_0$ may diverge depending on the energy of intermediate states.
More explicitly, if there are intermediate states that could go on-shell given $E_i$, $E_n$, and $E_1$, the Minkowski and Euclidean correlation functions have fundamentally different analyticity properties.
Since a LQCD computation obtains the Euclidean finite-volume four-point correlation function, it is essential that the connection to this Minkowski relation is established.

Extending the procedure of Ref.~\cite{Briceno:2019opb} to non-local two-nucleon MEs of this work, one can start by defining the time-momentum representation of the Euclidean correlation function:
\begin{align}
   G_L(\tau)=\int_L d^3z \left[\langle E_f,L|\, T^{(\rm{E})}[\mathcal{J}^{(\rm E)}_{pp \to np}(\tau,\bm{z})\,\mathcal{J}^{(\rm E)}_{pp \to np}(0)]\, |E_i,L\rangle\right]_L,
  \label{eq:GL}
\end{align}
which can be directly accessed via a LQCD computation.
The initial and final states are at zero spatial momentum and hence the $\bm{p}_i$ and $\bm{p}_f$ dependence of the function is dropped. $T^{(\rm E)}$ means time ordering must be implemented with respect to the Euclidean-time variable.
Moreover, the Heisenberg-picture operator in Euclidean spacetime satisfies $\mathcal{J}^{(\rm E)}_{pp \to np}(\tau,\bm{z})=e^{\hat{P}_0\tau -i \hat{\bm{P}}\cdot \bm{z}} \, \mathcal{J}^{(\rm E)}_{pp \to np}(0) \,$ $ e^{-\hat{P}_0\tau+i \hat{\bm{P}}\cdot \bm{z}}$, where $\hat{P}_0$ and $\hat{\bm{P}}$ are energy (Hamiltonian) and momentum operators, respectively.
In the following, without loss of generality we assume that $\mathcal{J}^{(\rm E)}_{pp \to np}(0)$ is the same as its Minkowski counterpart, i.e., it has no phase relative to the Minkowski current at origin.
We thus drop the Euclidean superscript on the Schr\"odinger-picture currents.

Now assuming that there are $N$ lowest-lying intermediate states (indexed as $m= 0,\cdots,N-1$) with energy $E_{*m} \leq E_f+E_1$, and $N'$ lowest-lying intermediate states (indexed as $m= 0,\cdots,N'-1$) with energy $E_{*m} \leq E_i-E_1$, the contribution from all states except the lowest $N$ or $N'$ states in the theory can be directly analytically continued to Minkowski space.
With this observation, one can break the Minkowski  ME $\mathcal{T}_L^{(\rm M)}$ to two parts as following:
\begin{align}
    \mathcal{T}_L^{(\rm M)} =i\mathcal{T}_L^{(\rm{E})\,<}+i\mathcal{T}_L^{(\rm{E})\,{\geq}},
  \label{eq:TLMtwoparts}
\end{align}
where the first term, defined as
\begin{align}
    &\mathcal{T}_L^{(\rm{E})\,<} \equiv L^3 \left[\sum_{m=0}^{N-1} \frac{\left[\vphantom{B^\dagger}\langle E_f,L|\, \mathcal{J}_{pp \to np}(0)\,|E_{*m},L\rangle\right]_L\left[\vphantom{B^\dagger}\langle E_{*m},L |\, \mathcal{J}_{pp \to np}(0)\,|E_i,L\rangle\right]_L}{E_f+E_1-E_{*m}} - \right .
    \nonumber\\
    &  \hspace{2.175 cm} \left . \sum_{m=0}^{N'-1}  \frac{\left[\vphantom{B^\dagger}\langle E_f,L|\, \mathcal{J}_{pp \to np}(0)\,|E_{*m},L\rangle\right]_L\left[\vphantom{B^\dagger}\langle E_{*m},L |\, \mathcal{J}_{pp \to np}(0)\,|E_i,L\rangle\right]_L}{E_1-E_i+E_{*m}}\right],
  \label{eq: 2}
\end{align}
cannot be directly accessed via the Euclidean four-point correlation function, and can only be reconstructed from the knowledge of $N$ or $N'$ (whichever larger) finite-volume two-nucleon eigenenergies and the MEs of the weak current between initial/final two-nucleon states and the $N$ or $N'$ allowed intermediate states.
These must be evaluated using LQCD computations of two- and three-point correlation functions in a finite volume. The second term in Eq.~\eqref{eq:TLMtwoparts} needs the values of the four-point function evaluated with LQCD, as well as the two- and three-point functions that need to be evaluated in obtaining the first term. Explicitly,
\begin{align}
    \mathcal{T}_L^{(\rm{E})\,\geq} \equiv \int d\tau \,e^{E_1\tau} \left[ G_L(\tau)-G_L^{<}(\tau) \right],
  \label{eq: 3}
\end{align}
where $G_L$ is the Euclidean bi-local MEs defined in Eq.~\eqref{eq:GL} and $G_L^{<}$ is:
\begin{align}
   G_L^{<}(\tau)=\sum_{m=0}^{N} c_m \, \theta(\tau) \, e^{-(E_{*m}-E_f)|\tau|}+\sum_{m=0}^{N'} c_m \, \theta(-\tau) \, e^{-(E_{*m}-E_i)|\tau|},
  \label{eq:GLsmaller}
\end{align}
with the $c_m$ function defined in terms of finite-volume single-beta decay MEs  satisfying either $E_{*m} \leq E_f+E_1$ or $E_{*m} \leq E_i-E_1$, corresponding to the first and second sum in Eq.~\eqref{eq:GLsmaller}, respectively:
\begin{align}
   c_{m} \equiv L^3 \left[\vphantom{B^\dagger}\langle E_f,L|\, \mathcal{J}_{pp \to np}(0)\,|E_{*m},L\rangle\right]_L\left[\vphantom{B^\dagger}\langle E_{*m},L |\, \mathcal{J}_{pp \to np}(0)\,|E_i,L\rangle\right]_L.
  \label{eq: 4}
\end{align}

Given all these ingredients, the physical (infinite Minkowski spacetime) double-beta decay amplitude can be extracted from a (finite) set of relevant two-, three-, and four-point correlation functions in a Euclidean finite spacetime, quantities that are all accessible from LQCD.
In fact, as pointed out in Ref.~\cite{Briceno:2019opb}, it will be essential that the matching relation, including Minkowski spacetime construction and the finite-volume corrections to the ME, are all fit to LQCD data on the required $n$-point functions simultaneously to insure the exact cancellations among various contributions that lead to the correct analytic structure of the infinite-volume amplitude.

It should be noted that a similar method was used in Refs. ~\cite{Shanahan:2017bgi,Tiburzi:2017iux}, to extract the deuteron-pole contribution from the Euclidean four-point correlation function, resulting in the desired double-beta decay matrix element.
However, those studies used larger-than-physical values of the quark masses, which led to the initial and final states being bound nuclear states, and only one bound on-shell intermediate state (the deuteron) was considered.
As a result, a simpler formalism was used to extract the physical matrix element without the need for a finite-volume matching. In future lattice QCD calculations with lighter quark masses, where two-nucleon states are no longer bound, the complete framework presented in this work will be necessary.
The limitations of this formalism presented here are that it relies on a finite-order effective field theory expansion, which can be improved by including higher-order contributions, and it is only valid below three-hadron thresholds.
The latter limitation may be more challenging to overcome, but progress has been made in recent years in developing finite-volume technology for the three-hadron problem in lattice QCD, as discussed in Ref.~\cite{Hansen:2019nir}.

\subsection{Summary}
\noindent
The combination of lattice quantum chromodynamics with effective field theories has become an important tool for studying nuclear phenomena from first principles~\cite{Beane:2010em,Beane:2014oea,Briceno:2014tqa,Detmold:2019ghl,Cirigliano:2019jig,Kronfeld:2019nfb,Cirigliano:2020yhp,Drischler:2019xuo}.
This approach complements quantum many-body methods in nuclear physics.
The method presented here offers the basic principles for implementing this approach for the two-neutrino double-beta decay.
It presents the construction of the physical amplitude for the $nn \to pp , (ee \bar{\nu}_e\bar{\nu}_e)$ process within the framework of pionless effective field theory (EFT) with nucleonic degrees of freedom.
This EFT is a good description of the low-energy $nn \to pp$ process.
Although the $nn \to pp , (ee \bar{\nu}_e\bar{\nu}_e)$ process is not expected to occur in nature due to the short lifetime of the neutron compared to the inverse rate of the double-weak process, it is still an important process in isotopes undergoing ordinary double-beta decay.
Therefore, any constraint on the relevant nuclear matrix element at the two-nucleon level can serve as the microscopic input for quantum many-body calculations of the rate.
The analysis presented here investigated various contributions to the amplitude, including those with electron-neutrino emission on the external nucleons, and showed that they confirm the working principles of the pionless EFT as a renormalized approach to nuclear observables when applied to a double-weak amplitude.

It was shown that no additional low-energy constants are needed beyond the nucleon ($g_A$) and two-nucleon ($L_{1,A}$) axial-vector couplings to ensure a renormalization-scale independent amplitude at next-to-leading order.
This is different from the situation in pionless effective field theory with dibaryon degrees of freedom in Refs.~\cite{Shanahan:2017bgi,Tiburzi:2017iux}, where a new coupling ($h_{2,S}$) was introduced to account for short-distance two-nucleon two-weak current interactions.
The effect of $h_{2,S}$ was found to be comparable to the contribution from the $l_{1,A}$ coupling in the dibaryon formalism.
By examining the consistency of $L_{1,A}$ values when constrained from single- and double-weak processes independently, future analyses of lattice quantum chromodynamics results can determine the validity of the pionless EFT power counting with nucleonic degrees of freedom as presented in this thesis.
These analyses may also have implications for the "quenching" of $g_A$ in double-beta decays.

Furthermore, despite the lack of experimental constraints on the process, LQCD can still give access to the non-local correlation functions of two-nucleon systems with two insertions of the weak current, hence determining the relevant MEs contributing to the physical amplitude.
There is, however, a non-trivial procedure involved to perform this matching, as developed for the first time in this thesis for this problem.
The procedure builds upon the previously developed finite-volume formalisms for the local MEs of two nucleons and non-local MEs of a single hadron.
It requires the determination of the finite-volume two-nucleon spectrum in the spin-singlet and spin-triplet channels, and the three-point and four-point correlation functions corresponding to the single-beta and double-beta decay processes from LQCD for a range of kinematics
The reason is that the Euclidean and Minkowski correlation functions can not be mapped to each other straightforwardly when on-shell multi-hadron intermediate states are present, whose effects must be isolated and treated separately~\cite{Briceno:2019opb}.
Moreover, by assuming that only two-nucleon intermediate states can go on shell, power-law volume corrections to the amplitude are derived within the pionless-EFT approach at NLO, completing the matching relation.

This formalism is a step towards constraining the physical amplitude for the neutrinoless process $nn \to pp\,(ee)$ with the light Majorana exchange, or in turn the new short-distance LEC in pionless EFT that is present at the LO~\cite{Cirigliano:2018hja}.
Obtaining this constraint from a Euclidean four-point correlation function involves the convolution of the nuclear ME with the neutrino propagator, further complicating the relation to the Minkowski correlation function.
Nonetheless, upon slight modifications, identification of the finite-volume corrections to the reconstructed Minkowski amplitude follows the same procedure as outlined in this section.
These steps will be presented in in the next section.

\section{Two-nucleon \texorpdfstring{$0\nu\beta\beta$}{Bookmark Version} decay
\label{sec: 0vbb}
}
\noindent
The importance of $0\nu\beta\beta$ decay discovery was highlighted at the beginning of this chapter.
To re-emphasize, observation of such a transition would have profound consequences as: $1)$ It would demonstrate that lepton number is violated in nature.
This results in the violation of the only anomaly-free global symmetry of the SM, the difference between baryon number and lepton number $B-L$.
$2)$ It would also establish that neutrinos are Majorana particles.
This is important as neutrinos are the only fundamental particles in the SM that can be Majorana particles since all other fermions in the SM carry an electromagnetic charge, and as of now, the Majorana nature of neutrinos is only a theoretical possibility.
$3)$ The measured value of the new physics parameter $\langle\eta\rangle^2$ in Eq.~\eqref{eq: 0vbb half life}, will put a constraint neutrino masses and indicate the scale of the new physics.
Our knowledge of BSM mechanisms that may be responsible for this decay can only be enhanced by combining theoretical calculations of the rate, and other decay observables, with experimental findings~\cite{Dolinski:2019nrj, DellOro:2016tmg, Bilenky:2014uka}. Furthermore, planned experimental endeavors will crucially benefit from theoretical predictions of the expected rates in various isotopes given the BSM scenarios considered~\cite{Dolinski:2019nrj, Cappuzzello:2018wek, DellOro:2016tmg, Faessler:2012ku, Giuliani:2012zu}.

Among the many possible BSM scenarios that allow for this transition~\cite{Dolinski:2019nrj,DellOro:2016tmg, Bilenky:2014uka,Rodejohann:2011mu}, the light-neutrino exchange model has been the most espoused in the physics community.
It is because the light-neutrino exchange model is the minimal deviation scenario from the SM and the physics involved is the long-range light physics.
Most alternative mechanisms are short-range mechanisms with some exceptions, see a recent review~\cite{Giunti:2019aiy}, that too have been a subject of tight experimental constraints from recent results~\cite{STEREO:2022nzk}.
We thus restrict our discussion here to the light-neutrino exchange, in which case $\langle\eta\rangle^2$ is given by the square of effective Majorana mass $m_{\beta\beta}^2$ defined as
\begin{equation}
    m_{\beta\beta}= \left|\sum_i U_{ei}^2m_i\right|.
    \label{eq: effective Majorana mass definition}
\end{equation}
Here, $U_{ei}$ are the elements of the Pontecorvo-Maki-Nakagawa-Sato (PMNS) matrix~\cite{Pontecorvo:1957qd, Maki:1962mu}, and $m_i$ is the real and positive mass eigenvalue of the $i$th neutrino mass eigenstate with $i=1,2,3$.
It is important to note that the neutrino oscillations constraint only the differences between squares of neutrino masses, and the smallest neutrino mass is not currently known.
There are two options for neutrino mass ordering (or hierarchy): $m_3>m_2>m_1$ called normal ordering and $m_2>m_1>m_3$ known as inverted ordering.
The value of $m_{\beta\beta}$ depends on the type of hierarchy, and the next generation $0\nu\beta\beta$ experiments are expected to be sensitive enough~\cite{KATRIN:2021dfa} to probe the parameter space of $m_{\beta\beta}$ that could rule out inverted hierarchy, see Fig. 1 in Ref.~\cite{Engel:2016xgb}.
However, translating the constraints on half-life values to $m_{\beta\beta}$ requires precise calculation of the nuclear ME, as evident by Eq.~\eqref{eq: 0vbb half life} with $\langle\eta\rangle^2=m^2_{\beta\beta}$.

From the perspective that the SM is a low-energy EFT of $0\nu\beta\beta$ decay~\cite{Weinberg:1979sa, Babu:2001ex, Prezeau:2003xn, deGouvea:2007qla, Lehman:2014jma, Graesser:2016bpz, Cirigliano:2017djv, Cirigliano:2018yza}, $m_{\beta\beta}$ can be generated minimally at dimension five from the lepton-number violating Weinberg operator~\cite{Weinberg:1979sa} given by
\begin{equation}
    \mathcal{L}^{(\Delta L =2)}_{5} =\frac{C^{ll'}_5}{\Lambda_{LNV}} \left[\Phi \cdot \bar{L}^C_l \right] \left[ L_{l'} \cdot\Phi\right],
    \label{eq: Weinberg operator definition}
\end{equation}
where $C^{ll'}_5$ are the flavor-dependent Wilson coefficients, $\Lambda_{LNV}$ is the scale of lepton-number violating physics, $\Phi$ is the SM Higgs doublet, $L_l$ is the left-handed lepton doublet with flavor $l$, and $L^C_l=CL_l$ with $C=i\gamma_2\gamma_0$ denoting the charge conjugation matrix.
After the electroweak spontaneous symmetry breaking, where the Higgs fields gets a vacuum expectation value, the Weinberg operator generates the a Majorana mass term,
\begin{equation}
    \mathcal{L}^{(\Delta L = 2)}_\nu=-\frac{m_{\beta\beta}}{2} \nu_L^TC\nu_L+{\rm H.c.},
    \label{eq:dim-5}
\end{equation}
where $\nu_L$ is the left-handed (electron) neutrino field.
The contributions from higher-dimensional operators to the $0\nu\beta\beta$ decay can be also significant (considering the suppression from the small $m_{\beta\beta}$ factor form the lowest-dimensional Majorana mass operator).
Nonetheless, the scenario in which the the decay is induced by the operator in Eq.~\eqref{eq:dim-5} is the most natural possibility and will be the focus of the discussion here.

Equation~\eqref{eq:dim-5} indicates that the $0\nu\beta\beta$ transition with this operator only introduces the Majorana mass term, and thus, the currents involved in the transition is still the weak current of the SM.
In this situation, the anti-neutrino emitted during the first current interaction changes handedness under the neutrino mass term in Eq.~\eqref{eq:dim-5} and induces the second current interaction.
Given the estimated masses of the neutrinos in the sub-eV to a few-eV range~\cite{ParticleDataGroup:2022pth}, the corresponding nuclear matrix element is long range in nature and receives contributions from intermediate nuclear states.
Thus, the hadronic transition matrix between the initial state $|i\rangle$ and final state $|f\rangle$ of a nuclei with energies $E_i$ and $E_f$, respectively, is given by~\cite{Dolinski:2019nrj}
\begin{align}
    \int d^4x \;d^4y \; dq^0 \;e^{-iq\cdot(x-y)}\;&\langle f| T\{J^\mu(x)J^\nu(y)\}| i\rangle \nonumber\\
    &\propto \sum_m \frac{\langle f| J^\mu({\bm q})|n\rangle\langle n|J^\nu(-{\bm q})| i\rangle}{q(E_n + q + E_{e_1} -E_i)} + (e_1\leftrightarrow e_2, \mu \leftrightarrow \nu).
    \label{eq: anatomy of the 0vbb nuclear ME}
\end{align}
Here, $J^\mu(x) = \bar{u}\gamma^\mu(1-\gamma_5)d$ is the quark-level current at space-time coordinate $x$, and $T$ denotes time-ordered product.
The right hand side of the equation is obtained by taking $E_{e_{1(2)}}$ to be the energy emitted for the first (second) electrons, inserting a complete set of intermediate states $|n\rangle$ with corresponding energies $E_n$, expressing the currents in Heisenberg picture, and integrating over the leptonic part and space-time coordinates, see Ref.~\cite{Engel:2016xgb} for detailed derivation.
This resulted in the Fourier transformed $J^\mu(x)$ denoted by $J^\mu({\bm q})$ with ${\bm q}$ being the three momentum of the virtual neutrino.
In deriving this result, the three momenta of the emitted electrons and the small neutrino masses in the denominator of the neutrino propagator have been ignored.

Equation~\eqref{eq: anatomy of the 0vbb nuclear ME} shows some features of the transition that will be used in the discussion ahead. $1)$ Unlike the $2\nu\beta\beta$ decay where the electron electron-antineutrino pair is on-shell, the virtual neutrino in the intermediate state here allows for all intermediate states that are consistent with energy conservation to go on-shell.
$2)$ The transition amplitude receives significant contribution from the neutrino momentum transfer of the order of an average inverse spacing between nucleons, which is $q\sim 100$ MeV.
$3)$ Under the impulse approximation in which the MEs of current operator are taken as the sum over the individual free-nucleon MEs, i.e., only one nucleon experiences the weak decay without any interference from the surrounding nuclear medium, the transition amplitude receives contributions only from the Fermi and Gamow-Teller type form factors, and the tensor contribution are suppressed~\cite{Engel:2016xgb}.

Evaluating the transition amplitude between nuclei amounts to performing a nuclear structure calculation using the few-body processes, as explained earlier in this chapter.
Despite the long-range nature of the $0\nu\beta\beta$ process, recent nuclear EFT analyses of the elementary subprocess $nn \to pp\,(ee)$ have revealed a short-distance contribution to the amplitude at the LO, with a LEC of the corresponding isotensor contact operator that absorbs the UV scale dependence of the amplitude through Renormalization Group~\cite{Cirigliano:2018hja, Cirigliano:2019vdj}.
As such a subprocess cannot be observed in free space, and given the program that has been formed around the use of nuclear EFTs to systematically improve the \emph{ab initio} nuclear structure calculations of the nuclear matrix elements~\cite{Menendez:2011qq, Pastore:2017ofx, Basili:2019gvn, Yao:2019rck, Yao:2020olm} toward experimentally-relevant isotopes, the missing knowledge of the value of such a short-distance contribution appears to impede progress, and has promoted indirect but less certain estimations~\cite{Cirigliano:2019vdj,Cirigliano:2020dmx,Cirigliano:2021qko}.

Lattice QCD has the promise of reliably constraining the EFTs of $0\nu\beta\beta$ in the few-nucleon sector~\cite{Cirigliano:2020yhp, Davoudi:2020ngi, Cirigliano:2019jig}, and has already demonstrated its reach and capability in constraining pionic matrix elements for lepton-number violating processes $\pi^- \to \pi^+\,(ee)$ and $\pi^-\pi^- \to ee$ within the light-neutrino scenario~\cite{Feng:2018pdq, Tuo:2019bue, Detmold:2020jqv}, the $\pi^- \to \pi^+\,(ee)$ process within a heavy-scale scenario~\cite{Nicholson:2018mwc,Detmold:2022jwu}.
Lattice-QCD matrix elements for these processes, however, lack certain complexities compared with the desired $nn \to pp\,(ee)$ process with a light Majorana neutrino, whose determination is the key to matching to the EFT descriptions developed in recent years~\cite{Cirigliano:2018hja, Cirigliano:2019vdj, Cirigliano:2018yza}.
While the numerical evaluations of the matrix elements are underway, the interpretation of these matrix elements in terms of the physical amplitude, and their matching to EFTs have so far been missing from the course of developments.
In this section, which is based on the work in Ref.~\cite{Davoudi:2020gxs}, such a framework will be developed and presented, and the steps involved in the matching will be described.
This framework, along with a realistic example that will be outlined, therefore, demonstrate how the results of this section can be used in the ongoing and upcoming studies to obtain the short-distance LEC of the EFT from the lattice QCD output.

The matching framework presented here builds upon major developments in recent years in accessing local and non-local transition amplitudes in hadronic physics from the corresponding FV matrix elements in Euclidean spacetime obtained with lattice QCD~\cite{Lellouch:2000pv, Detmold:2004qn, Meyer:2011um, Briceno:2012yi, Bernard:2012bi, Briceno:2014uqa, Briceno:2015csa, Briceno:2015tza, Christ:2015pwa, Briceno:2019opb, Feng:2020nqj}, and in particular, a similar formalism for the two-neutrino process $nn \to pp\,(ee\bar{\nu}_e\bar{\nu}_e)$ shown in the previous section~\cite{Davoudi:2020xdv}.
Nonetheless, the neutrinoless process involves additional complexities due to a propagating neutrino in the intermediate state, requiring new components to be included in the matching condition.
The matching involves constructing the relation between (Minkowski and infinite-volume) physical amplitudes in the EFT and the Minkowski finite-volume matrix element in lattice QCD as the first ingredient, as well as the relation between the latter and its Euclidean counterpart, which is complicated by the presence of on-shell multi-particle intermediate states, as the second ingredient.
The Euclidean FV matrix element is what is provided by lattice QCD, and hence the connection to the physical amplitude is built with these two ingredients.

\subsection{\texorpdfstring{$0\nu\beta\beta$}{Bookmark Version} decay in pionless EFT
\label{subsec: 0vbb amplitude in pionless EFT}
}
While the $0\nu\beta\beta$ decay can only proceed in certain nuclear media, the subprocess to be studied is $nn \to pp\,(ee)$, in which the two $d$ quarks in the initial two-neutron state interact with the left-handed $W$ boson of the SM, the emitted antineutrino would turn into a neutrino through the operator in Eq.~\eqref{eq:dim-5} as it propagates, and gets absorbed by the two $u$ quarks mediated by another $W$ boson (and two electrons will be emitted in the final states).
To relate the rate of the decay in $nn \to pp\,(ee)$ to the underlying SM EFT, one needs to map this problem to a nuclear EFT, and constrain the EFT using a direct calculation of the matrix element with lattice QCD.

Given the scale of the neutrino momentum transfer, i.e. $q\sim100$ MeV, the energies involved are below $m_\pi$.
As discussed in Sec.~\ref{subsec: NN in pionless}, pionless EFT has shown to be a good description of nucleon physics below $m_\pi$ eneiges.
This subsection derives the LO $nn \to pp\,(ee)$ hadronic transition amplitude in pionless EFT using the interactions described in Sec.~\ref{subsec: NN in pionless} and~\ref{subsec: external currents in EFT}.
It will be shown that such an amplitude is divergent and requires introducing a new LEC at the LO to absorb the renormalization scale dependence of the regulated amplitude.
This derivation follows closely the Refs.~\cite{Cirigliano:2018hja, Cirigliano:2019vdj} that showed the need of such LEC for the first time.
In those works, the $nn \to pp\,(ee)$ is calculated in up to NLO in both chiral EFT and pionless EFT, and demonstrated that observables can only be made regulator-independent by inclusion of a LO short-range lepton number violating operator in the $s$-wave channel.
Furthermore, it was shown that no new LEC is required at the NLO in $s$-wave channels as the LECs appearing at that order are completely determined by a combination of known LECs and LECs appearing at the LO.
Moreover, higher partial waves channels also do not need a short-range operator.

In this section, only the $s$-wave transition will be considered at the LO in non-relativistic pionless EFT.
The pionless EFT that is discussed in Sec.~\ref{subsec: NN in pionless}, with Lagrangian given by Eqs.~\eqref{eq: EFT 1 nucleon Lagrangian} and~\eqref{eq: EFT 2 nucleon Lagrangian}.
Isospin symmetry will be assumed throughout as the isospin-breaking effects are small and contribute at higher orders than considered.
In this limit, both the Fermi and Gamow-Teller operators in Eqs.~\eqref{eq: scalar vector isovector 1 body}-\eqref{eq: vector axial-vector isovector 1 body} contribute.
However, for the LO amplitude, only Eqs.~\eqref{eq: scalar vector isovector 1 body} and~\eqref{eq: vector axial-vector isovector 1 body} current operators contribute as the others are suppressed by $1/M$.
Equations~\eqref{eq: scalar vector isovector 1 body} and~\eqref{eq: vector axial-vector isovector 1 body} can be combined together using the nucleon velocity and spin four-vectors given by $v=(1,\bm{0})$ and $\mathcal{S}=(0,\tfrac{\bm{\sigma}}{2})$, respectively, in the nucleon's rest frame.
Thus, the CC weak interaction in Eq.~\eqref{eq: Charged current Lagrangian} can be written as
\footnote{Note that factors here differ from that of in Ref.~\cite{Davoudi:2020gxs} because of the different definitions of $\tau^+$ and taking $V_{ud}\approx 1$.}
\begin{equation}
{\cal L}_{(\rm CC)}(x) =-G_{F} \big[\overline{e}_L\gamma ^{\mu }\nu_L\big]  \big[ N^\dagger \tau^{+}(v_\mu-2g_{A} \mathcal{S}_\mu)N\big] +{\rm H.c.} ,
\label{eq: CC Lagrangian for 0vbb}
\end{equation}
Here, the space-time coordinate for fields is not shown for brevity.
The leptonic current contains the left-handed electron, $e_L=\left(\frac{1-\gamma_5}{2}\right)e$, and neutrino, $\nu_L$, fields. 

The full amplitude for $nn \to pp\,(ee)$ transition between initial energy $E_i$ and final energy $E_f$, $i\mathcal{M}_{0\nu}^{\rm full}$, via two CC current interactions is given by
\begin{equation}
	i\mathcal{M}_{0\nu}^{\rm full} \;=\; \frac{-1}{2!} \int d^4x \;d^4y \;\langle e^-(E_1,{\bm k}_1) e^-(E_2,{\bm k}_2); \phi(E_f^-),{^1S_0}| T\{\mathcal{L}_{CC}(x)\mathcal{L}_{CC}(y)\}|\phi(E_i^+),{^1S_0} \rangle,
	\label{eq: 0vbb full amplitude initial setup}
\end{equation}
Here, $T$ is the time ordering operator.
The external electrons in the final state with energies (three-momenta) $E_1$ $({\bm k}_1)$ and $E_2$ $({\bm k}_2)$ are shown by the state $|e^-(E_1,{\bm k}_1) e^-(E_2,{\bm k}_2)\rangle$, and the strong Hamiltonian eigenstates $|\phi(E^+),NN\rangle$ are defined in Eq.~\eqref{eq: Strong Hamiltonian Eigenstate}.
The amplitude $i\mathcal{M}_{0\nu}^{\rm full}$ is not finite and needs additional contribution to absorb the regularization-scale dependence.
To see this, factorize $i\mathcal{M}_{0\nu}^{\rm full}$ by contracting the electron fields with external states.
Labeling this factor $\mathfrak{L}^{\mu,\nu}_{a,b}(x,y)$, where $\mu$ and $\nu$ are Lorentz indices, and $a$ and $b$ are Dirac indices one gets
\begin{align}
	i\mathcal{M}_{0\nu}^{\rm full} \;=\; \frac{-G^2_F}{2} \int d^4x \;d^4y \;& \mathfrak{L}^{\mu,\nu}_{a,b} (x,y)\;\langle 0| T\{\nu_a(x) (\nu^T C)_b(y)\}|0\rangle \nonumber\\
 &\hspace{-4.4 cm}\times\langle \phi(E_f^-),{^1S_0}| T\{\left[N^\dagger(x) \tau^{+}(v_\mu-2g_{A} \mathcal{S}_\mu)N(x)\right]\left[N^\dagger(y) \tau^{+}(v_\mu-2g_{A} \mathcal{S}_\mu)N(y)\right]\}|\phi(E_i^+),{^1S_0} \rangle,
	\label{eq: 0vbb full amplitude step 1}
\end{align}
where $|0\rangle$ is the fermionic vacuum, and $C$ is the charge conjugation matrix pulled out to define the Majorana neutrino propagator.
Converting $\nu_a$ to the mass eigenbasis, $\chi_i$, leads to the neutrino propagator
\begin{align}
    \langle 0| T\{\nu_a(x) (\nu^T C)_b(y)\}|0\rangle =  \langle 0| T\{U_{ei}\chi_i(x) U_{ej} \bar{\chi}_j(y)\}|0\rangle = U_{ei}U_{ej}\delta_{i,j} \int \frac{d^4q}{(2\pi)^4} \frac{i\;e^{-iq\cdot(x-y)}}{\slashed{q}-m_{i} +i\epsilon}.
    \label{eq: neutrino propagator definition}
\end{align}
Using the definition of $m_{\beta\beta}$ in Eq.~\eqref{eq: effective Majorana mass definition}, ignoring small masses $m_i$ in the denominator on the right hand side of Eq.~\eqref{eq: neutrino propagator definition}, and some straightforward manipulations leads to 
\begin{align}
	i\mathcal{M}_{0\nu}^{\rm full} & = \frac{-m_{\beta\beta}\;G^2_F}{2} \int d^4x \;d^4y \; \mathfrak{L}^{\mu,\nu}_{a,b} (x,y) \langle \phi(E_f^-),{^1S_0}| T\{\left[N^\dagger(x) \tau^{+}(v_\mu-2g_{A} \mathcal{S}_\mu)N(x)\right]\nonumber\\
 &\times \int \frac{d^4q}{(2\pi)^4} \frac{i\;e^{-iq\cdot(x-y)}}{q^2 +i\epsilon}\left[N^\dagger(y) \tau^{+}(v_\mu-2g_{A} \mathcal{S}_\mu)N(y)\right]\}|\phi(E_i^+),{^1S_0} \rangle.
	\label{eq: 0vbb full amplitude step 2}
\end{align}

The full transition amplitude for the $nn \to pp\,(ee)$ process is not separable to the hadronic and leptonic amplitudes as in the two-neutrino process given the presence of a neutrino that propagates between the two weak currents.
Nonetheless, the contribution from final-state electrons as well as constants proportional to $G_F$ can still be separated from a hadronic amplitude that includes the hadronic matrix element that is convoluted by the neutrino propagator.
This latter contribution is what one would evaluate in lattice QCD and match to nuclear EFTs.
A simple kinematic is assumed in which the total three-momenta of the system is zero, and the electrons are at rest, each having energy $E_1=E_2=m_e$, where $m_e$ is the electron's mass.

A careful look ~\cite{Cirigliano:2017tvr,Cirigliano:2017djv,Cirigliano:2018hja,Cirigliano:2019vdj,Cirigliano:2020dmx,Cirigliano:2021qko} at the amplitude in Eq.~\eqref{eq: 0vbb full amplitude step 1} through contributions from different energy scales of neutrino momentum $q$ concludes that, at the LO in the EFT where only $s$-wave interactions of the nucleons contribute, the ``radiation" region of the neutrino momentum, $q_0 \sim q << 100$ MeV, where the neutrino can be emitted and reabsorbed while the intermediate nuclear state propagates, are suppressed in EFT counting.
The most dominant contribution below $m_\pi$ energies come from the potential region, $q_0 \sim (100 {\text{ MeV}})^2/M$ and $q\sim 100$ MeV, where the neutrino exchange can be approximated with a static Coulomb-like potential, see~\cite{Cirigliano:2017tvr,Cirigliano:2017djv,Cirigliano:2018hja,Cirigliano:2019vdj} for a detailed discussion on the form of the potential.

\begin{figure}[t!]
    \centering
    \includegraphics[scale=1]{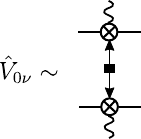}
    \caption{Diagrammatic representation of the static neutrino potential.
    The crossed circles in this case are the single nucleon currents with Fermi and Gamow-teller transitions defined in Eq.~\eqref{eq: CC Lagrangian for 0vbb}, and they are evaluated between the initial and final states as given in Eq.~\eqref{eq: neutrino potential ME definition}.
    }
    \label{fig: neutrino potential diagram}
\end{figure}

Here, only the hadronic amplitude mediated by the static neutrino potential, $\hat{V}_{0\nu}$, is considered, where
\begin{equation}
    \langle {\bm q}_1,{^1S_0}| \; \hat{V}_{0\nu} \; | {\bm q}_2, {^1S_0} \rangle =-m_{\beta\beta}\frac{(1+3g_A^2)}{({\bm q}_1-{\bm q}_2)^2}.
	\label{eq: neutrino potential ME definition}
\end{equation}
With this, the long-range component of the hadronic amplitude from the static potential, $i\mathcal{M}^{\rm long}_{0\nu}$, is given by
\begin{equation}
    i\mathcal{M}^{\rm long}_{0\nu}(E_i,E_f) =  -i \langle \phi(E_f^-),{^1S_0}| \;\hat{V}_{0\nu}|\phi(E_i^+),{^1S_0} \rangle.
	\label{eq: 0vbb long amplitude in terms of T-matrix}
\end{equation}
%
\begin{figure}
    \centering
    \includegraphics[scale=1]{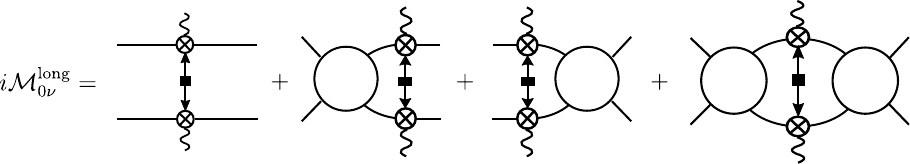}
    \caption{Diagrammatic representation of the long-range contribution to the hadronic transition amplitude of $nn \to pp(ee)$ process. All components of the diagrams are introduced in Figs.~\ref{fig: C0 C2 Feynman diagram notation},~\ref{fig: LO and NLO NN in KSW}, and~\ref{fig: neutrino potential diagram}.}
    \label{fig: M long diagram}
\end{figure}
Figure ~\ref{fig: neutrino potential diagram} shows a diagrammatic representation of Eq.~\eqref{eq: neutrino potential ME definition}.
With iterative application of Eq.~\eqref{eq: Strong Hamiltonian Eigenstate}, $i\mathcal{M}^{\rm long}_{0\nu}$ can be expressed diagrammatically at the LO as shown in Fig.~\ref{fig: M long diagram}, which evaluates to
\begin{align}
    i\mathcal{M}^{\rm long}_{0\nu}(E_i,E_f) =& -i \langle {\bm p}_f| \; \hat{V}_{0\nu}|{\bm p}_i \rangle \nonumber\\
    & \hspace{-2 cm}+ \int \frac{d^3{\bm q}_1}{(2\pi)^3} \frac{d^3{\bm q}_2}{(2\pi)^3}\langle {\bm p}_f| \; \hat{V}_{0\nu}|{\bm q}_1 \rangle \langle {\bm q}_f| \; G_0(E_i^+)|{\bm q}_2\rangle \langle{\bm q}_2|-iT^{(\rm LO)}_S(E_i^+) | {\bm p}_i\rangle \nonumber\\
    & \hspace{-2 cm} + \int \frac{d^3{\bm q}_1}{(2\pi)^3} \frac{d^3{\bm q}_2}{(2\pi)^3}  \langle{\bm p}_f|-iT^{(\rm LO)}_S(E_f^+) | {\bm q}_1\rangle \langle {\bm q}_1| \; G_0(E_f^+)|{\bm q}_2\rangle\langle {\bm q}_2| \; \hat{V}_{0\nu}|{\bm p}_i \rangle  \nonumber\\
    & \hspace{-2 cm} + \int \frac{d^3{\bm q}_1}{(2\pi)^3} \frac{d^3{\bm q}_2}{(2\pi)^3}\frac{d^3{\bm q}_3}{(2\pi)^3} \frac{d^3{\bm q}_3}{(2\pi)^3}  \langle{\bm p}_f|-iT^{(\rm LO)}_S(E_f^+) | {\bm q}_1\rangle \langle {\bm q}_1| \; G_0(E_f^+)|{\bm q}_2\rangle\langle {\bm q}_2| \; \hat{V}_{0\nu}|{\bm q}_3 \rangle\nonumber\\
    & \hspace{-1 cm} \times \langle {\bm q}_3| \; G_0(E_i^+)|{\bm q}_4\rangle \langle{\bm q}_4|T^{(\rm LO)}_S(E_i^+) | {\bm p}_i\rangle.
	\label{eq: 0vbb long amplitude step 2}
\end{align}
Using Eqs.~\eqref{eq: Strong Hamiltonian}-\eqref{eq: LO+NLO 2to2 Amplitude}, this simplifies to
\begin{align}
    \frac{i\mathcal{M}^{\rm long}_{0\nu}(E_i,E_f)}{m_{\beta\beta}} = &\frac{i(1+3g^2_A)}{2}\int d\cos{\theta}\frac{1}{p_i^2+p_f^2-2p_ip_f\cos{\theta}}\nonumber\\
    & - (1+3g_A^2)K(E_i,E_f)\,i\mathcal{M}^{(\rm LO)}(E_i)
    - (1+3g_A^2) \, i\mathcal{M}^{(\rm LO)}(E_f)\,K(E_i,E_f)\nonumber\\
    & - i(1+3g_A^2)\,i\mathcal{M}^{(\rm LO)}(E_f) \, J^{\infty}(E_i,E_f;\mu)\,i\mathcal{M}^{(\rm LO)}(E_i),
	\label{eq: 0vbb without gvNN}
\end{align}
where the first term comes from the projection of neutrino potential in the $^1S_0$ channel, and  $K(E_i,E_f)$ is given by~\cite{Cirigliano:2018hja,Cirigliano:2019vdj}
\begin{align}
    K(E_i,E_f) &= \int \frac{d^3{\bm q}}{(2\pi)^3} 
    \frac{1}{E-\frac{q^2}{M}+i\epsilon}\frac{1}{|\bm{q}_1-\bm{q}_2|^2}
    \label{eq: K definition}\\
    &=\frac{-M^2}{8\pi}\frac{i}{\sqrt{E_i}}\ln{\left(\frac{\sqrt{E_f}+\sqrt{E_i}}{\sqrt{E_f}-\sqrt{E_i}+i\epsilon}\right)}\nonumber,
\end{align}
which is a finite integral.
However, $J^{\infty}(E_i,E_f;\mu)$ defined as
\begin{eqnarray}
J^{\infty}(E_i,E_f;\mu)= \int \frac{d^3{\bm q}_1}{(2\pi)^3} \frac{d^3{\bm q}_2}{(2\pi)^3}
\frac{1}{E_i-\frac{q_1^2}{M}+i\epsilon}\frac{1}{E_f-\tfrac{q_2^2}{M}+i\epsilon}\frac{1}{|\bm{q}_1-\bm{q}_2|^2}
\label{eq:JinftyDef}
\end{eqnarray}
is a UV divergent integral that needs to be regularized.
Here, the superscript $\infty$ indicates infinite volume expression that will be a useful notation for the FV matching in the next subsection.
Expression for $J^{\infty}(E_i,E_f;\mu)$ in the PDS scheme is given by~\cite{Cirigliano:2018hja, Cirigliano:2019vdj}
\begin{equation}
J^{\infty}(E_i,E_f;\mu)=\frac{M^2}{32\pi^2} \bigg[-\gamma_E+\ln(4\pi)+L(E_i,E_f;\mu) \bigg],
\label{eq:Jinfty}
\end{equation}
with
\[L(E_i,E_f;\mu) \equiv \ln\left(\tfrac{\mu^2/M}{-(\sqrt{E_i}+\sqrt{E_f})^2-i\epsilon}\right)+1,\]
where $\mu$ is a UV renormalization scale.

The UV divergence of the loop function necessities the introduction of a counterterm at the same order that would absorb this divergence to make the LO amplitude regularization scale independent.
Such a term is given by a contact $\Delta L = 2$ four-nucleon-two-electron operator~\cite{Cirigliano:2017tvr, Cirigliano:2018hja, Cirigliano:2019vdj}
\begin{equation}
\mathcal{L}^{(\Delta L = 2)}_{NN} = G^2_{F}
m_{\beta\beta} \, g_\nu^{NN} \left[\bar{e}_LC\bar{e}_L^T\right]
 \big[(N^TP_-N)^{\dagger }(N^{T}P_+N)\big]+{\rm H.c.},
\label{eq:LNdeltaL2}
\end{equation}
where $g^{NN}_{\nu}$ is a new LEC that measures the strength of this interaction, which depends on $\mu$ and absorbs the divergent contribution in Eq.~\eqref{eq: 0vbb without gvNN}
\footnote{Factors here for ${\cal L}^{(\Delta L = 2)}_N$ term are different than in Ref.~\cite{Davoudi:2020gxs} to be consistent with the notation used in this thesis.}.
While the dimensional analysis suggests that this dimension nine operator must contribute at the NLO, RG considerations require promoting this operator to leading order, as derived in Refs.~\cite{Cirigliano:2018hja, Cirigliano:2019vdj}.
By matching the contribution from Eq.~\eqref{eq:LNdeltaL2} to the $nn \to pp\,(ee)$ transition amplitude with the form in Eq.~\eqref{eq: 0vbb full amplitude step 2}, the MEs of the contact potential, $\hat{V}_{g^{NN}_{\nu}}$, generated from Eq.~\eqref{eq:LNdeltaL2} to the hadronic amplitude is given by
\begin{figure}[t!]
    \centering
    \includegraphics[scale=1]{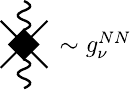}
    \caption{Diagrammatic representation of the lepton-number violating contact term defined in Eq.~\eqref{eq: gvNN potential ME definition}.
    }
    \label{fig: contact term potential diagram}
\end{figure}
%
\begin{equation}
    \langle {\bm q}_1,{^1S_0}| \; \hat{V}_{g^{NN}_{\nu}} \; | {\bm q}_2, {^1S_0} \rangle = 2 \; m_{\beta\beta}\;g^{NN}_{\nu},
	\label{eq: gvNN potential ME definition}
\end{equation}
which is shown as a diagram in Fig.~\ref{fig: contact term potential diagram}.
\begin{figure}
    \centering
    \includegraphics[scale=1]{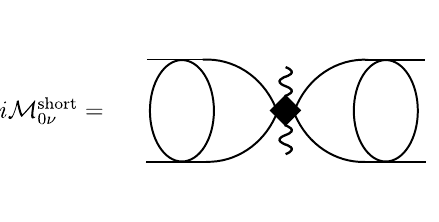}
    \caption{Diagrammatic representation of the short-range contribution to the hadronic transition amplitude of $nn \to pp(ee)$ process. All components of the diagrams are introduced in Figs.~\ref{fig: C0 C2 Feynman diagram notation},~\ref{fig: LO and NLO NN in KSW}, and~\ref{fig: contact term potential diagram}.}
    \label{fig: M short diagram}
\end{figure}

This contact term contribution gives the short-range part, $i\mathcal{M}^{\rm short}_{0\nu}$, of the full $nn \to pp\,(ee)$ hadronic amplitude whose diagrammatic expansion of is shown in Fig.~\ref{fig: M short diagram}, which evaluates to
\begin{equation}
    i\mathcal{M}^{\rm short}_{0\nu}(E_i,E_f) =  -i \langle \phi(E_f^-),{^1S_0}| \;\hat{V}_{g^{NN}_{\nu}}|\phi(E_i^+),{^1S_0} \rangle.
	\label{eq: 0vbb short amplitude in terms of T-matrix}
\end{equation}
A similar evaluation as Eq.~\eqref{eq: 0vbb long amplitude step 2} using the potential in Eq.~\eqref{eq: gvNN potential ME definition} gives
\begin{align}
    \frac{i\mathcal{M}^{\rm short}_{0\nu}(E_i,E_f)}{m_{\beta\beta}} =\,i\mathcal{M}^{(\rm LO)}(E_f) \, \left(\frac{2 g^{NN}_{\nu}}{C^2_0}\right)\,i\mathcal{M}^{(\rm LO)}(E_i),
	\label{eq: 0vbb gvNN amplitude}
\end{align}
which matches the form of the divergent term in the amplitude in Eq.~\eqref{eq: 0vbb without gvNN}.

The full $nn \to pp\,(ee)$ hadronic amplitude is then
\begin{equation}
    i\mathcal{M}_{nn\xrightarrow{0\nu}pp}(E_i,E_f)=i\mathcal{M}^{\rm long}_{0\nu}(E_i,E_f) + i\mathcal{M}^{\rm short}_{0\nu}(E_i,E_f),
    \label{eq: full 0vbb amplitude bothe parts combined}
\end{equation}
and demanding its renormalization scale independence leads to the RG evolution equation for the coupling $g^{NN}_{\nu}$, which is given by
\begin{equation}
    \mu\frac{d}{d\mu}\widetilde{g}_{\nu}^{NN} = \frac{1+3g_A^2}{2},
    \label{eq: RG equation for gvNN}
\end{equation}
where $\widetilde{g}_{\nu}^{NN}$ is a dimensionless parameter related to $g_{\nu}^{NN}$ via
\begin{equation}
\widetilde{g}_{\nu}^{NN}= \bigg(\frac{4\pi}{M C_0}\bigg)^2 g_{\nu}^{NN}\,.
\label{eq: gvNNtilde definition}
\end{equation}

The value of $g_{\nu}^{NN}$ is unknown since there is no experimental observation of $0\nu\beta\beta$ decay, and being the LEC of a lepton number violating term, it does not contribute any other known observable.
Nonetheless, Refs.~\cite{Cirigliano:2020dmx,Cirigliano:2021qko} constrained its value indirectly, but it has a large uncertainty from the phenomenological method used for obtaining those constraints.
A more precise and direct constraint on  $g_{\nu}^{NN}$ using LQCD will be desired, and a prescription for obtaining the $g_{\nu}^{NN}$ (or equivalently the $\widetilde{g}_{\nu}^{NN}$) value from a Euclidean four-point correlation function calculated using LQCD is shown in the next subsection.

\subsection{Finite volume formalism for two-nucleon \texorpdfstring{$0\nu\beta\beta$}{Bookmark Version} decay in pionless EFT
\label{subsec: 0vbb matching}
}
\begin{figure}[t!]
    \centering
    \includegraphics[scale=1]{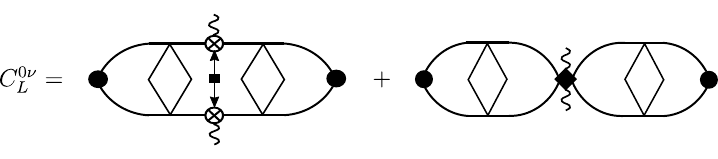}
    \caption{Diagrams representing the finite-volume correlation function defined in Eq.~\eqref{eq: 0vbb CL}. The black filled circles correspond to interpolating operators for the initial and final isotriplet states.
    All the ingredients have been defined in Figs.~\ref{fig: two point correlation diagram},~\ref{fig: neutrino potential diagram}, and~\ref{fig: contact term potential diagram}.
    } 
    \label{fig: 0vbb CL diagrams}
\end{figure}

Following closely the FV formalism used for the four-point function in the $2\nu\beta\beta$ case in Sec.~\ref{subsec: 2vbb matching}, the starting point of the FV matching for $0\nu\beta\beta$ case is the four-point function shown in Fig.~\ref{fig: 0vbb CL diagrams}.
Keeping the Minkowski signature of spacetime intact, and using the same derivation as obtaining Eq.\eqref{eq: four point correlation final}, the four-point function can be written as
\begin{align}
C_L(E_i,E_f) = & C_{\infty}(E_i,E_f) + 
\mathcal{B}_{pp}(E_f)\,i\mathcal{F}(E_f)
\nonumber\\
&\hspace{-2.2 cm}\times\bigg[i\mathcal{M}^{(\rm{Int.})}_{nn\xrightarrow{0\nu} pp} (E_i,E_f) + m_{\beta\beta}(1+3g_A^2) \, i\mathcal{M}(E_f)\,\delta J^V(E_f,E_i)\, i\mathcal{M}(E_i) \bigg] i\mathcal{F}(E_i) \mathcal{B}^\dagger_{nn}(E_i)+\cdots
\label{eq: 0vbb CL}
\end{align}
Here, $\mathcal{B}^\dagger_{nn}$ and $\mathcal{B}_{pp}$ are the matrix elements of initial- and final-state interpolating operators between vacuum and on-shell ``in'' and ``out'' two-nucleon states, respectively, as defined in Sec~\ref{subsec: NN from FV}.
The ellipsis denotes terms that will not matter for the matching relation below.
The $\mathcal{F}$ is a finite-volume function defined in Eq.~\eqref{eq: definition of mathcal F}.
$\mathcal{M}^{(\rm{Int.})}_{nn\xrightarrow{0\nu} pp}$ denotes contributions to the infinite volume hadronic transition in which the neutrino propagates between two nucleons dressed by strong interactions on both sides as well as the full contribution from the short-range amplitude that depends upon the short-distance LEC $g_{\nu}^{NN}$.
Diagrammatically, $\mathcal{M}^{(\rm{Int.})}_{nn\xrightarrow{0\nu} pp}$ is given by contributions shown in Fig.~\ref{fig: M int diagram} that evaluates to
\begin{equation}
    \mathcal{M}^{(\rm{Int.})}_{nn\xrightarrow{0\nu} pp} (E_i,E_f)=m_{\beta\beta}\;\mathcal{M}(E_f) \bigg [ -(1+3g_A^2) J^{\infty}(E_i,E_f;\mu)+\frac{2g_\nu^{NN}(\mu)}{C_0^2(\mu)}\bigg]
    \mathcal{M}(E_i).
    \label{eq:MICnnpp}
\end{equation}
Figure~\ref{fig: M int diagram} implies that $\mathcal{M}^{(\rm{Int.})}_{nn\xrightarrow{0\nu} pp}$ contains $\mathcal{M}^{\rm short}_{0\nu}(E_i,E_f)$ and a part of $\mathcal{M}^{\rm long}_{0\nu}(E_i,E_f)$ that comes from the last diagram in Fig.~\ref{fig: M long diagram}.
Furthermore, $\mathcal{M}^{(\rm{Int.})}_{nn\xrightarrow{0\nu} pp}$ is related to the full hadronic transition amplitude via
\begin{equation}
    i\mathcal{M}_{nn\xrightarrow{0\nu} pp} (E_i,E_f) = i\mathcal{M}^{(\rm{Ext.})}_{nn \xrightarrow{0\nu} pp} (E_i,E_f) \; +\; i\mathcal{M}^{(\rm{Int.})}_{nn \xrightarrow{0\nu} pp} (E_i,E_f),
    \label{eq: Mfull in terms of Mint and Mext}
\end{equation}
where $\mathcal{M}^{(\rm{Ext.})}_{nn\xrightarrow{0\nu} pp}$ is the remaining part of $\mathcal{M}^{\rm long}_{0\nu}(E_i,E_f)$ consisting of the first three diagrams appearing in the sum in Fig.~\ref{fig: M long diagram}, i.e., contributions in which the neutrino propagates between two external nucleons.
These contributions will not matter for matching to the lattice-QCD matrix element, however, their expression can be easily read off using Eqs.~\eqref{eq: 0vbb without gvNN},~\eqref{eq: 0vbb gvNN amplitude},~\eqref{eq: full 0vbb amplitude bothe parts combined} and~\eqref{eq:MICnnpp}.

Finally, a new finite-volume function $\delta J^V$, corresponding to the two-loop diagram with the exchanged neutrino propagator needs to be evaluated:
\begin{figure}
    \centering
    \includegraphics[scale=1]{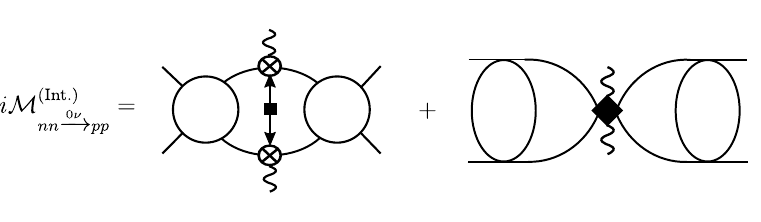}
    \caption{Diagrammatic representation of the long-range contribution to the hadronic transition amplitude of $nn \to pp(ee)$ process. All components of the diagrams are introduced in Figs.~\ref{fig: C0 C2 Feynman diagram notation},~\ref{fig: LO and NLO NN in KSW}, and~\ref{fig: neutrino potential diagram}.}
    \label{fig: M int diagram}
\end{figure}
%
\begin{equation}
\delta J^V(E_i,E_1,E_f)=
\bigg[\frac{1}{{L^6}}\sum_{\substack{\bm{k}_1,\bm{k}_2 \\ \bm{k}_1 \neq \bm{k}_2}}-
\int \frac{d^3k_1}{(2\pi)^3}\frac{d^3k_2}{(2\pi)^3}
\bigg]
\frac{1}{E_i-\tfrac{\bm{k}_1^2}{M}+i\epsilon} \frac{1}{E_f-\tfrac{\bm{k}_2^2}{M}+i\epsilon}\frac{1}{|\bm{k}_1-\bm{k}_2|^2},
\label{eq:deltaJV}
\end{equation}
where in the summations, $\bm{k}_1,\bm{k}_2  \in \frac{2\pi}{L}\mathbb{Z}^3$.
This sum-integral difference can be evaluated numerically for given values of $E_i$ and $E_f$, the detail of which is presented in App.~\ref{app: Two-loop sum-integral difference}.
The requirement $\bm{k}_1 \neq \bm{k}_2$ removes the zero spatial-momentum mode of the neutrino in the loop to render the finite-volume sum finite.
Correspondingly, the finite-volume correlation function in lattice QCD will need to implement a zero-mode regulated neutrino propagator to match to this expression.
Such a treatment of the infrared singularities in a finite volume is customary in the lattice QCD+QED studies of hadronic masses~\cite{Hayakawa:2008an, Borsanyi:2014jba, Davoudi:2018qpl}, decay amplitudes~\cite{Lubicz:2016xro, Carrasco:2015xwa}, and two-hadron scattering~\cite{Beane:2014qha, NPLQCD:2020ozd}.
Other prescriptions have been proposed and implemented for lattice-QCD calculation of the $\pi^- \to \pi^+\;(ee)$ amplitude~\cite{Tuo:2019bue, Detmold:2020jqv}. These can be generalized to be applicable to the $nn \to pp\;(ee)$ process following similar steps given in there.

To proceed with finding the matching relation, one notes that the finite-volume correlation function in Eq.~\ref{eq: 0vbb CL} has the same general structure as that for the two-neutrino process obtained in Sec.~\ref{subsec: 2vbb matching}.
As a result, all steps introduced in Sec.~\ref{subsec: 2vbb matching} can be closely followed to obtain the matching relation between finite and infinite-volume matrix elements.
In particular, upon Fourier transforming Eq.~\ref{eq: 0vbb CL} with $E_i$ and $E_f$ to form the correlation function in the mixed time-momentum representation, and comparing it against the same correlation function that is obtained from a direct four-point function upon inserting complete sets of intermediate finite-volume states between the currents, one arrives at
\begin{equation}
    L^6\;\bigg|\mathcal{T}^{(\rm M)}_L \bigg|^2=\bigg|\mathcal{R}(E_{n_f})\bigg |\;\bigg|\mathcal{M}^{0\nu,V}_{nn \xrightarrow{0\nu} pp} (E_{i},E_{f})\bigg|^2 \bigg|\mathcal{R}(E_{n_i})\bigg|,
    \label{eq: 0vbb IVFVmatching}
\end{equation}

where $\mathcal{R}(E)$ is defined in Eq.~\eqref{eq: definition of residue}, $E_{n_{i(f)}}$ denotes a finite-volume energy of the initial (final) two-nucleon state, and 
\begin{equation}
    \mathcal{M}^{0\nu,V}_{nn \xrightarrow{0\nu} pp} (E_{i},E_{f}) = \mathcal{M}^{(\rm{Int.})}_{nn\xrightarrow{0\nu} pp} (E_{i},E_{f})-m_{\beta\beta}(1+3g_A^2)\mathcal{M}^{(\rm LO)}(E_i)\delta J^V(E_{i},E_{f}) \mathcal{M}^{(\rm LO)}(E_f).
    \label{eq: Finite volume amplitude 0vbb}
\end{equation}
$\mathcal{T}^{(\rm M)}_L$ denotes the Minkowski finite-volume matrix element defined as
\begin{equation}
    \mathcal{T}^{(\rm M)}_L \equiv \int dz_0\, e^{iE_1 z_0}\int_L d^3z\big[\langle E_{n_f},L|\, T[\mathcal{J}(z_0,\bm{z})\,S_\nu(z_0,\bm{z})
    \mathcal{J}(0)]\, |E_{n_i},L\rangle\big]_L.
    \label{eq: 0vbb TLM}
\end{equation}
Here, $\mathcal{J}=\bar{q} \tau _+\gamma_\mu(1-\gamma_5)q$ with $q=\begin{psmallmatrix} u \\ d \end{psmallmatrix}$, which can be implemented in lattice-QCD calculations.
At the hadronic level, it matches to $N^\dagger \tau _{+}(v_\mu-2g_{A} \mathcal{S}_\mu)N$ in Eq.~\eqref{eq: CC Lagrangian for 0vbb}.
Nonetheless, being a quark-level current means that $\mathcal{T}^{(\rm M)}_L$ also incorporates the contact $\Delta L=2$ interaction in Eq.~\eqref{eq:LNdeltaL2}.
$S_\nu(z_0,\bm{z})$ denotes the Minkowski finite-volume propagator of a Majorana neutrino, with its zero spatial-momentum mode removed:
\begin{equation}
    S_\nu(z_0,\bm{z}) = \frac{1}{L^3}\sum_{\bm{q}  \in \frac{2\pi}{L}\mathbb{Z}^3\neq \bm{0}} \int \frac{dq_0}{2\pi}e^{i\bm{q} \cdot \bm{z}-iq_0z_0} \frac{-im_{\beta\beta}}{q_0^2-|\bm{q}|^2+i\epsilon}.
\end{equation}
\subsection{Minkowski to Euclidean matching: Neutrino propagator case
\label{subsec: 0vbb minkowski Euclidean matching}
}
The quantity $\mathcal{T}^{(\rm M)}_L$ in Eq.~\eqref{eq: 0vbb TLM}, whose connection to the physical amplitude was established in Eq.~\eqref{eq: 0vbb IVFVmatching}, is defined with a Minkowski signature.
On the other hand, with lattice QCD only Euclidean correlation functions can be evaluated.
Unfortunately in the case of non-local matrix elements, generally one cannot obtain the former from the latter upon an analytical continuation, as discussed in Sec.~\ref{subsec: doubleEM}.
To appreciate the subtlety involved, and to introduce a procedure that, nonetheless, allows constructing the Minkowski matrix element from its counterpart in Euclidean spacetime, one may consider the following correlation function:
\begin{equation}
    G_L^{(\rm E)}(\tau)=\int_L d^3z \; \big [\langle E_f,L|T^{(\rm E)} [\mathcal{J}^{(\rm E)}(\tau,\bm{z})S_\nu^{(\rm E)}(\tau,\bm{z})\mathcal{J}^{(\rm E)}(0)] |E_i,L \rangle \big]_L,
    \label{eq: 0vbb GLE}
\end{equation}
which can be computed directly with lattice QCD.\footnote{It is assumed that the continuum limit of lattice-QCD correlation functions is taken prior to the matching to infinite volume. Therefore, the discretized nature of lattice-QCD quantities will not affect the framework presented.} $\tau \equiv iz_0$ is the Euclidean time, and the subscript (E) on quantities is introduced to distinguish Euclidean forms from Minkowski counterparts.
In particular, $S_\nu^{(\rm E)}$ is the Euclidean neutrino propagator in a finite volume with its zero spatial-momentum mode removed,
\begin{align}
    S_\nu^{(\rm E)}(\tau,\bm{z}) &=  \frac{1}{L^3}\sum_{\bm{q}  \in \frac{2\pi}{L}\mathbb{Z}^3\neq \bm{0}} \int \frac{dq_0^{\rm (E)}}{2\pi}e^{i\bm{q} \cdot \bm{z}-iq_0^{\rm (E)}\tau} \frac{m_{\beta\beta}}{q_0^{\rm (E)2}+|\bm{q}|^2}
    \nonumber\\
    &=\frac{m_{\beta\beta}}{2L^3}\sum_{\bm{q}  \in \frac{2\pi}{L}\mathbb{Z}^3\neq \bm{0}}\frac{e^{i\bm{q} \cdot \bm{z}}}{|\bm{q}|}\bigg[\theta(\tau)e^{-|\bm{q}| \tau}+\theta(-\tau)e^{|\bm{q}| \tau} \bigg].
\end{align}
It is now clear that simply integrating over the Euclidean time with weight $e^{E_1 \tau}$ can be problematic if on-shell intermediate states are allowed. Here, $E_1$ is the energy of the first or the second electron depending on the time ordering.
This can be seen by expressing the Heisenberg-picture operator in Euclidean spacetime as $\mathcal{J}^{(\rm E)}(\tau,\bm{z})=e^{\hat{P}_0\tau -i \hat{\bm{P}}\cdot \bm{z}} \, \mathcal{J}^{(\rm E)}(0) \,$ $ e^{-\hat{P}_0\tau+i \hat{\bm{P}}\cdot \bm{z}}$, where $\hat{P}_0$ and $\hat{\bm{P}}$ are energy (Hamiltonian) and momentum operators, respectively, and upon inserting a complete set of single- and multi-particle states in the volume between the two currents.
Without loss of generality, we assume that $\mathcal{J}^{(\rm E)}(0)$ is the same as its Minkowski counterpart, i.e., it has no phase relative to the Minkowski current at the origin.
The Euclidean superscript of the Schr\"odinger-picture currents will therefore be dropped. It then becomes clear that for those values of intermediate-states energies and momenta such that 
\begin{align}
i) \; |\bm{P}_{*m}|+E_{*m} \leq E_f+E_1\quad \text{or}\quad ii)\; |\bm{P}_{*m}|+E_{*m} \leq E_i-E_1, 
\label{eq:on-shell}
\end{align}
the integration over Euclidean time with $e^{E_1 \tau}$ will be divergent.
Here, $E_{*m}$ are the finite-volume energy eigenvalues of the intermediate spin-triplet two-nucleon state with total momentum ${\bm P}_{*m}$, and we assume that three-particle intermediate states with on-shell kinematics are not possible given the initial-state energy.
The problematic contributions satisfying conditions $i$ and $ii$ can be subtracted from Eq.~\eqref{eq: 0vbb GLE}, leaving the rest to read
\begin{align}
\mathcal{T}_L^{(\rm{E})\,\geq} \equiv \int d\tau \,e^{E_1\tau} \left[ G_L^{(\rm E)}(\tau)-G_L^{(\rm{E})<}(\tau) \right].
\label{eq: 0vbb TLElarger}
\end{align}
The spectral decomposition of $\mathcal{T}_L^{(\rm{E})\,\geq}$ has, therefore, exactly the same form as the Minkowski counterpart upon an overall $i$ factor.
Here,
\begin{equation}
G_L^{(\rm{E})<}(\tau) \equiv \sum_{m=0}^{N-1} \theta(\tau) \, c_m \, e^{-(|\bm{P}_{*m}|+E_{*m}-E_f)|\tau|} + \sum_{m=0}^{N'-1} \theta(-\tau) \, c_{m} \, e^{-(|\bm{P}_{*m}|+E_{*m}-E_i)|\tau|},
\label{eq: 0vbb GLlesser}
\end{equation}
where it is assumed that there are $N~(N')$ states satisfying condition $i~(ii)$ above, and
\begin{equation}
c_m \equiv \frac{m_{\beta\beta}}{2|\bm{P}_{*m}|}  \big [\langle E_f,L|\mathcal{J}(0)| E_{*m},L \rangle \langle E_{*m},L|\mathcal{J}(0) |E_i,L \rangle \big]_L.
\label{eq: 0vbb cm}
\end{equation}

The remaining contributions arising from on-shell intermediate states, called $\mathcal{T}_L^{(\rm{E})\,<}$, can be formed separately with the knowledge of the single-current matrix elements in a finite volume between the initial (final) and intermediate states:
\begin{equation}
    \mathcal{T}_L^{(\rm{E})\,<} \equiv \sum_{m=0}^{N-1} \frac{c_m}{|\bm{P}_{*m}|+E_{*m}-E_f-E_1} + \sum_{m=0}^{N'-1} \frac{c_{m}}{|\bm{P}_{*m}|+E_{*m}-E_i+E_1}.
    \label{eq: 0vbb TLEsmaller}
\end{equation}

Eqs.~\eqref{eq: 0vbb TLEsmaller} and \eqref{eq: 0vbb TLElarger}) can now be combined to construct the desired Minkowski quantity $\mathcal{T}_L^{(\rm M)}$,
\begin{equation}
\mathcal{T}_L^{(\rm M)} =i\mathcal{T}_L^{(\rm{E})\,<}+i\mathcal{T}_L^{(\rm{E})\,{\geq}},
\label{eq: 0vbb TLM realted to Euclidean}
\end{equation}
whose relation to the physical $nn \to pp \;(ee)$ amplitude at the LO in the EFT was already established in Eq.~\eqref{eq: 0vbb IVFVmatching}.
This completes the matching  framework that relates $\mathcal{M}^{(\rm{Int.})}_{nn\xrightarrow{0\nu} pp}$ in Eq.~\eqref{eq:MICnnpp}, and hence the new short-distance LEC $g_\nu^{NN}$, to the lattice-QCD correlation function $G_L^{(\rm E)}(\tau)$ in Eq.~\eqref{eq: 0vbb GLE}.
It should be noted that the single-current matrix elements required for this matching relation, i.e., those appearing in Eq.~\eqref{eq: 0vbb cm}, can themselves be evaluated with lattice QCD, and can be matched to the physical amplitude for the single-beta transition amplitude~\cite{Briceno:2012yi, Davoudi:2020xdv}.

\subsection{Summary
\label{subsec: 0vbb summary}
}
Given significant progress in lattice QCD studies of nuclear matrix elements in recent years~\cite{Beane:2015yha, Savage:2016kon, Shanahan:2017bgi, Winter:2017bfs, Chang:2017eiq, Detmold:2020snb, Davoudi:2020ngi}, albeit yet with unphysical quark masses, it is expected that lattice QCD will be able to evaluate the four-point correlation function in Eq.~\eqref{eq: 0vbb GLE} at the physical values of the quark masses, along with the required two- and three-point functions that allow the construction of the finite-volume Minkowski amplitude in Eq.~\eqref{eq: 0vbb TLM realted to Euclidean}.
This can then be used in Eq.~\eqref{eq: 0vbb IVFVmatching}) to constrain the physical EFT amplitude, hence the unknown short-distance contribution.
The practicality of the method, however, relies on the presence of only a finite (and few) number of on-shell intermediate states that are composed of no more than two hadrons, such that the Minkowski amplitude can be constructed.
One can estimate the expected nature and the number of intermediate states by examining a plausible example.
Let us take $L=8$ fm to ensure the validity of the finite-volume formalism used with physical quark masses, up to exponentially suppressed contributions~\cite{Sato:2007ms, Briceno:2013bda}.
The finite-volume spectrum of the two-nucleon isotriplet channel at rest arising from singularities of the function in Eq.~\eqref{eq: definition of mathcal F} can be determined using the experimentally known phase shifts~\cite{Stoks:1994wp}, giving the ground-state energy $E_{n_i}=-2.6482$ MeV (which polynomially approaches zero as $L \to \infty$).\footnote{A more detailed discussion on obtaining numerical results for all values in this section is done in Sec.~\ref{sec: DBD sensitivity analysis}.}
A simple kinematic can be considered for the transition amplitude such that $E_{i}(=E_{n_i})=E_f(=E_{n_f})$, and where the currents carry zero energy and momentum so that the final-state two-nucleon system remains at rest.
Given the available total energy, and the quantum numbers of the currents, the only allowed intermediate state is the two-nucleon isotriplet channel at rest, whose low-lying spectrum in this volume is $\widetilde{E}_{*m}=\{-5.5774,10.5308,\cdots\}$ MeV.
While it may appear that the ground state of this system constitutes an on-shell intermediate state, requiring construction of the Minkowski amplitude through an evaluation of the isosinglet to isotriplet matrix element, one must note that since the zero spatial momentum is not allowed for the neutrino propagation in the finite volume, none of the on-shell conditions in Eq.~\eqref{eq:on-shell} can be satisfied with the kinematics considered (noting that the minimum allowed energy of an on-shell neutrino in this volume is $|\bm{P_*}|=2\pi/L=154.9802$ MeV).
As a result, $\mathcal{T}_L^{(\rm M)} =i\mathcal{T}_L^{(\rm{E})}=i\int d\tau \,e^{E_1\tau} G_L^{(\rm E)}(\tau)$, and Eq.~\eqref{eq: 0vbb IVFVmatching} can be readily used to obtain the physical amplitude from the lattice-QCD four-point function $G_L^{(\rm E)}(\tau)$. 

The example described demonstrates that obtaining the physical amplitude of the $nn \to pp\;(ee)$ process from lattice QCD is even more straightforward than its two-neutrino counterpart, as in the latter there is a larger kinematic phase space allowed for on-shell intermediate states---a feature that is shared with the nuclear structure calculations of the corresponding nuclear matrix elements~\cite{Engel:2016xgb}.
Without experimental input, such nuclear many-body calculations, however, are unable to isolate and quantify the short-distance contribution to the matrix element, motivating the need for first-principles determination of the matrix elements in the few-nucleon sector, and a thorough matching to nuclear EFTs. The current framework, therefore, takes an essential step in enabling such a matching program in the upcoming years.
Besides its application in the $0\nu\beta\beta$ process, the formalism outlined here will find its use in a range of hadronic processes that consist of single- or two-hadron initial, intermediate, and final states, and where a light lepton (or photon) propagator is present, such as in the semi-leptonic rare decays of the kaon~\cite{Christ:2020hwe}.

\section{Extraction of \texorpdfstring{$L_{1,A}$ and $g_\nu^{NN}$}{Bookmark Version} from Lattice QCD: A Sensitivity Analysis
\label{sec: DBD sensitivity analysis}
}
Nuclear reactions mediated by weak interactions are central to a variety of frontier problems in nuclear and astrophysics as well as high-energy physics.
Single-weak-current processes like $pp$-fusion and (anti)neutrino-deuteron scattering are two prominent examples.
The former is a critical process in understanding the energy production mechanism in a range of stars~\cite{Adelberger:2010qa}, and the latter is used to probe properties of neutrinos in several neutrino experiments~\cite{Aharmim:2011vm,Fukuda:2001nj,Fukuda:2002pe}.
At the next order in weak currents, double-beta decay transitions are of major importance.
Two important modes of this transition are $2\nu\beta\beta$ decay and $0\nu\beta\beta$ decay discussed in Ch.~\ref{ch: DBD from LQCD}, where it was argued that a major source of uncertainty in calculating the decay rate of weak processes is the MEs of weak currents between the initial and final nuclear states.
The double-beta decays nuclear-ME calculations suffer from uncertainties that stem from both approximations in quantum many-body methods as well as uncertainties in (multi)nucleon interactions and weak currents~\cite{Engel:2016xgb,Giuliani:2012zu}. The latter can be mitigated by improving the accuracy of MEs in the two-nucleon (NN) sector using an effective Lagrangian along with a power-counting scheme, and then using them as an input in an \textit{ab initio} framework to calculate the many-body MEs for larger nuclei~\cite{Coraggio:2020iht,Engel:2016xgb}.

For SM processes involving more than two nucleons, the nuclear MEs of isovector axial-vector currents corresponding to Gamow-Teller transitions are not constrained precisely in pionless EFT, as discussed in Ch.~\ref{ch: DBD from LQCD}.
This is in part due to a large uncertainty on the renormalization-scale ($\mu$) dependent LEC $L_{1,A}$ that contributes at NLO and determines the strength of the momentum independent isovector axial-vector two-body current~\cite{Kong:1999tw,Butler:1999sv,Butler:2000zp}.
While constituting only a few percent of the total amplitude, the contribution to the Gamow-Teller transitions from the $L_{1,A}$ term remains the dominant source of uncertainty in determining the decay rate of processes such as $pp$ fusion in Sun and similar stars~\cite{Adelberger:2010qa}.
The value of $L_{1,A}$ determined from experimental data has improved over the years~\cite{Chen:2002pv, Butler:2002cw, Butler:2000zp, Chen:2005ak, De-Leon:2016wyu}, with the most recent constraint given by\footnote{Throughout this work, values of $\mu$-dependent LECs are given at $\mu=m_{\pi}$.} $L_{1,A}= 4.9 ^{+1.9}_{-1.5}\text{ fm}^3$~\cite{Acharya:2019fij}, which has a significant uncertainty.
On the other hand, no experimental constraint exists on the nuclear ME of $0\nu\beta\beta$ decay transition due to lack of observation, and it the light neutrino exchange scenario of the hadronic amplitude of $0\nu\beta\beta$ decay transition in the two-nucleon sector, i.e. $nn\xrightarrow{0\nu} pp$, has an unknown LEC, $g_\nu^{NN}$, at the LO in pionless EFT~\cite{Cirigliano:2017tvr,Cirigliano:2018hja,Cirigliano:2019vdj}.
An indirect estimate of $g_\nu^{NN}$ was obtained in Refs.~\cite{Cirigliano:2020dmx,Cirigliano:2021qko}: $\widetilde{g}_{\nu}^{NN} = 1.3 \pm 0.6$,
where $\widetilde{g}_{\nu}^{NN}$ is a dimensionless parameter related to $g_{\nu}^{NN}$ via Eq.~\eqref{eq: gvNNtilde definition}.
Subsequent analyses using this value showed that the missing short-range contribution to the nuclear ME of various candidate nuclei is comparable to the rest of the contributions~\cite{Wirth:2021pij,Jokiniemi:2021qqv}.
This indicates the importance of improving the constraint on $g_\nu^{NN}$, preferably using a  first-principles approach such as LQCD.

A direct way of constraining nuclear MEs is to solve the underlying short-distance theory of QCD numerically using the technique of LQCD~\cite{Davoudi:2020ngi,Briceno:2014tqa,Cirigliano:2019jig,Kronfeld:2019nfb,Drischler:2019xuo,Cirigliano:2020yhp}.
LQCD was in fact used  in Ref.~\cite{Savage:2016kon} to constrain $L_{1,A}$ from the relevant LQCD three-point correlation functions albeit at unphysical quark masses corresponding to $m_{\pi}\approx 806$ MeV, see also Ref.~\cite{Detmold:2021oro}.
The obtained value of $L_{1,A}=3.9(1.4) \text{ fm}^3$ at the physical quark masses required an uncertain quark-mass extrapolations but found to be comparable to experimental constraints with similar uncertainties.
On the other hand, no LQCD determination of the $g_\nu^{NN}$ coupling is yet reported although progress in simpler $0\nu\beta\beta$ processes in the pion sector is being made in recent years~\cite{Feng:2018pdq, Tuo:2019bue, Detmold:2020jqv, Nicholson:2018mwc}.

The formalism for obtaining single-beta and double-beta decays in the NN sector to obtain the needed matching relations that constrain the $L_{1,A}$ and $g_\nu^{NN}$ LECs from the LQCD was discussed in Ch.~\ref{ch: DBD from LQCD}.
Given the complexity of the matching relations involved, it is not immediately obvious what the precision requirement of the upcoming LQCD studies at the physical quark masses should be to reach the precision goal of the LECs, that is to be compatible or superior to phenomenological constraints.
In particular, it is important to ask if anticipated uncertainties on the lowest-lying FV energies and on the MEs, as well as achievable physical volumes in LQCD, will guarantee precise determinations of LECs such as $L_{1,A}$ and $g_\nu^{NN}$.
As a result, in this chapter we embark on an investigation based on synthetic data to determine the sensitivity of the output of the matching relations (hence the LECs) along with their uncertainties on the values and uncertainties of the input to these relations, namely the LQCD energies and MEs.
This also allows determining the range of volumes which leads to better constraints, hence guiding future LQCD calculations on their resource planning.
This follows the spirit of Ref.~\cite{Briceno:2013bda} which demonstrated that a precise determination of the small S-D mixing parameter in the deuteron channel from LQCD is achievable in future LQCD calculations of the lowest-lying spectra of NN systems in boosted frames.
This investigation, furthermore, aligns with recent valuable analyses of the sensitivity of nuclear spectra and MEs to the uncertainties in the input LECs of interactions and currents, when those uncertainties are propagated through \emph{ab initio} many-body calculations~\cite{Ekstrom:2019lss}.
\begin{figure}[t!]
    \centering
    \includegraphics[scale=1]{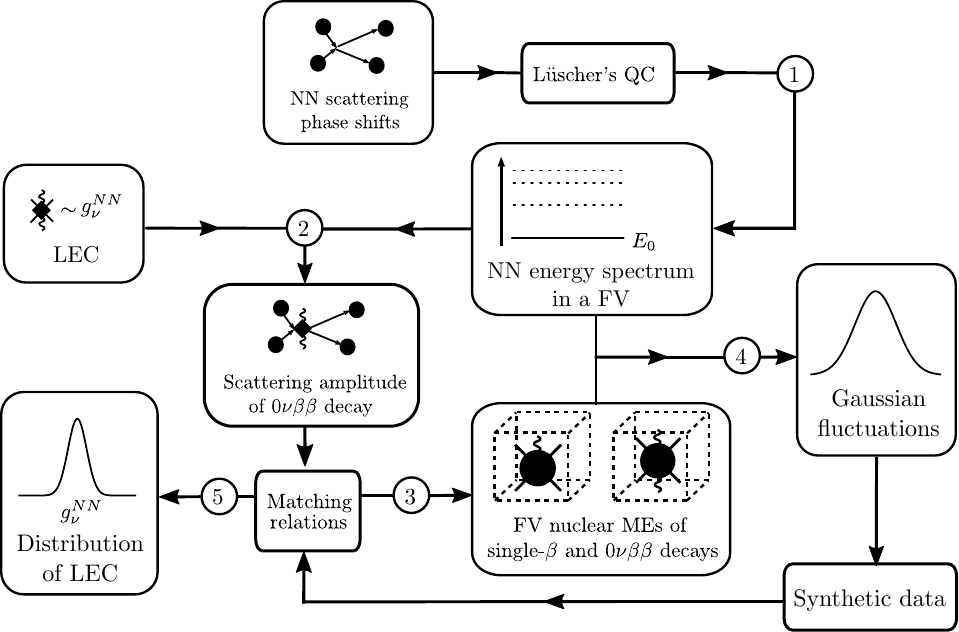}
    \caption{The procedure used to perform the sensitivity analysis of $L_{1,A}$ and $g_\nu^{NN}$. The sequence of steps followed is indicated by the numbers enclosed in the circles. The LECs $L_{1,A}$ and $g_\nu^{NN}$ are represented by a crossed circle and a solid diamond, respectively. The wavy line denotes external leptons from a single weak-current insertion. A nucleon is denoted by the small solid circle in the diagrams for NN processes in infinite volume. Dotted lines in the NN energy spectrum in a finite volume are the excited-state energies, and the ground state energy, $E_0$ ($\widetilde{E}_0$), in the spin-singlet (spin-triplet) channel is denoted by the solid line. The FV nuclear MEs for the decay transitions are represented by large solid circles enclosed in dotted cubes with one and two weak-current insertions, respectively. The solid line denotes the FV nucleon state. The simulation of LQCD uncertainties using Gaussian fluctuations and uncertainty analysis of LECs from the synthetic data is discussed in Secs.~\ref{subsec: L1A} and~\ref{subsec: gvNN}. \label{fig: flowchart}}
\end{figure}

Explicitly, we consider the single-beta decay (Sec.~\ref{subsec: L1A}) and $0\nu\beta\beta$ decay (Sec.~\ref{subsec: gvNN}) transitions in the two-nucleon sector: $nn \to np e^-\bar{\nu}_e$ and $nn \to pp e^-e^-$, respectively.
First using the L\"uscher's quantization condition (QC), the low-energy spectra of NN systems in a range of spatial cubic volumes with periodic boundary conditions (PBCs) are calculated using the phase shifts reported in the experimental NN scattering database~\cite{NNonline}.
These spectra are expected to be the same as those calculated from the two-point function with LQCD at the physical quark masses, up to exponential corrections in $m_\pi$.
Second, the central values of $L_{1,A}$ and $g_{\nu}^{NN}$ from Ref.~\cite{Acharya:2019fij} and Refs.~\cite{Cirigliano:2020dmx,Cirigliano:2021qko} are used to evaluate the physical transition amplitudes for single-beta and (neutrinoless) double-beta decay processes with initial and final energies set to the lowest energy eigenvalues obtained in the first step.
These scattering amplitudes are then used in matching relations of Secs.~\ref{sec: 2vbb} and~\ref{sec: 2vbb}, respectively, to obtain a reasonable guess for the central values of the corresponding FV three- and four-point functions.
Next, Gaussian fluctuations are introduced to the quantities that are expected to be extracted from LQCD, namely the FV energy eigenvalues and the three- and four-point functions, to generate a set of synthetic data for performing the sensitivity analysis.
This introduces uncertainties in the supposedly LQCD ingredients.
Finally, matching relations are used once again to obtain $L_{1,A}$ and $g_{\nu}^{NN}$ from the synthetic dataset, along with their uncertainties.
Figure~\ref{fig: flowchart} summarizes the procedure used for performing the sensitivity analysis of this work.

A detailed account of findings of this chapter is provided in Sec.~\ref{subsec: sensitivity analysis summary}.
To summarize, achieving small uncertainties in $L_{1,A}$ is found to be more challenging than $g_{\nu}^{NN}$, and demands (sub)percent-level precision in the two-nucleon spectra and the ME to supersede the current phenomenological constraints.
On the other hand, the short-distance coupling of the neutrinoless double-beta decay, $g_{\nu}^{NN}$, turns out to be less sensitive to uncertainties on both LQCD energies and the ME, and promises competitive precision compared with the current indirect estimates, even with few-percent uncertainties on LQCD energies and MEs.
The volume requirements are moderate and for ground-state to ground-state transitions, smaller volumes are shown to lead to more precise extractions. 
\subsection{Luscher's formalism for two-nucleons: A numerical analysis
\label{subsec: Luscher numerical formalism}
}
\noindent
All the FV ingredients required to perform the sensitivity analyses of Secs.~\ref{subsec: L1A} and~\ref{subsec: gvNN} have been introduced in Ch.~\ref{ch: Intro} and~\ref{ch: DBD from LQCD}.
The NN scattering amplitudes and contact LECs introduced in Sec.~\ref{subsec: NN in pionless} are needed in matching relations for $L_{1,A}$ and $g_{\nu}^{NN}$ in Eq.~\eqref{eq: matching relation for L1A} and~\eqref{eq: matching relation gvNN}, respectively.
Nonetheless, this section recaps and highlights key results, and performs a numerical analysis of the  L\"uscher's quantization condition that relates the FV energy eigenvalues to the physical two-nucleon scattering amplitudes.

In LQCD, the $n$-point correlation functions are computed on a finite Euclidean spacetime lattice.
Assuming the continuum limit for a hypercubic lattice with periodic boundary conditions, L\"uscher's quantization condition gives a direct relation between the FV energy eigenvalues of two hadrons obtained from LQCD and the corresponding scattering amplitudes.
The mapping is valid up to exponentially suppressed corrections governed by the range of the interactions.
For the low-energy NN systems, the interaction range is set by the Compton wavelength of the pion.
The quantization conditions are then valid up to $\mathcal{O}(e^{-m_\pi L})$ corrections, where $L$ denotes the spatial extent of the volume. 
\begin{figure}[t]
    \centering
    \includegraphics[scale=1]{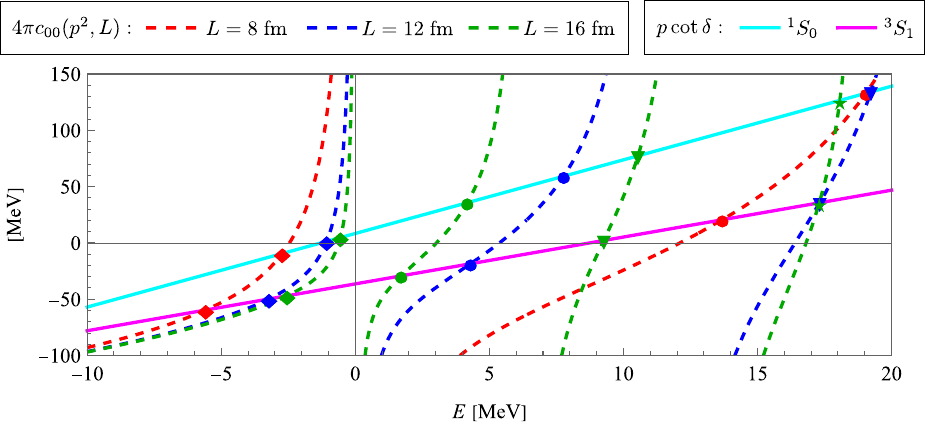}
    \caption{The effective-range function (solid lines) and L\"uscher's function (dotted lines) in Eq.~\eqref{eq: quantization condition in phase shift again} are plotted independently against the CM energy of NN systems.
    Equation~\eqref{eq: ERE definition} is used for the effective-range function with the effective-range expansion parameters given in Eq.~\eqref{eq: ERE values} for the two channels, $^1S_0$ (cyan) and $^3S_1$ (magenta).
    The function $4\pi c_{00}(p^2,L)$ is plotted for three different volumes with $L=8\;{\rm fm}$ (red), $L=12\;{\rm fm}$ (blue) and $L=16\;{\rm fm}$ (green).
    The diamonds, circles, triangles, and stars denote, respectively, the location of energy eigenvalues of the ground, first, second, and third excited states in each volume, and satisfy the quantization condition in Eq.~\eqref{eq: quantization condition in phase shift again} (and its counterpart for the $^3S_1$ channel).
    The numerical values associated with this figure are provided in Appendix~\ref{app: numerical tables for sensitivity analysis}.
    \label{fig: QC plot}}
\end{figure}
\begin{figure}[t]
    \centering
    \includegraphics[scale=1]{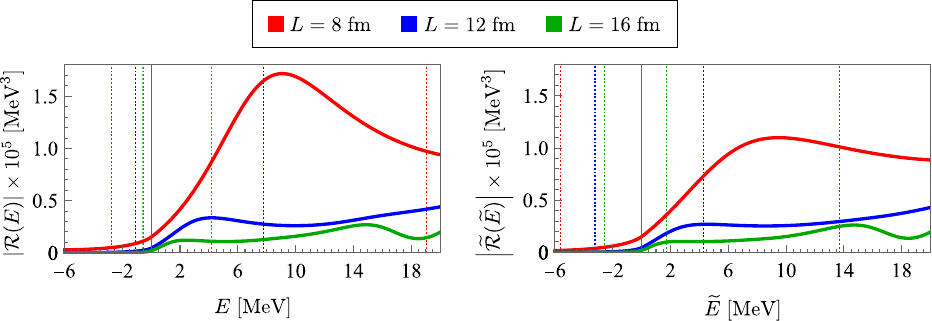}
    \caption{The absolute values of the LL residue function in the $^1S_0$ (left) and $^3S_1$ (right) channels is plotted against the CM energy for three different volumes with $L=8\;{\rm fm}$ (red), $L=12\;{\rm fm}$ (blue), and $L=16\;{\rm fm}$ (green).
    Dashed lines indicate energy eigenvalues in the respective volumes. The numerical values of $|\mathcal{R}|$ and $|\widetilde{\mathcal{R}}|$ evaluated at the FV ground- and first excited-state energies in the corresponding volumes are provided in Appendix \ref{app: numerical tables for sensitivity analysis}.
     \label{fig: R plot}}
\end{figure}

The cubic volume does not respect the rotational symmetry, and as a result, the FV quantization conditions mix scattering amplitudes in all partial waves.
However, at low energies the scattering amplitude is expected to be dominated by the $S$-wave interaction.
Ignoring the contribution from all higher-order partial waves, the FV quantization condition relates the $S$-wave phase shifts to a discrete set of FV energy eigenvalues, $E_n$.
For NN systems in the $^1S_0$ channel, the quantization condition has already been discussed in Sec~\ref{subsec: NN from FV}, which is given by Eq.~\eqref{eq: quantization condition in phase shift}.
This equation is at the center of the discussion in this section, and thus, it reiterated here:
\begin{equation}
    p_n\cot{\delta} = 4\pi c_{00}(p_n^2,L).
    \label{eq: quantization condition in phase shift again}
\end{equation}
Here, the phase shift $\delta$ is related to the scattering amplitude $\mathcal{M}$ via Eq.~\eqref{eq: M in pcot del deifinition}, and $C_{00}$ is defined in Eq.~\eqref{eq: c00 definition}.
Another useful quantity, which appears in the matching relations in Eqs.~\eqref{eq: matching relation for L1A} and~\eqref{eq: matching relation gvNN}, is the generalized Lellouch-L\"uscher (LL) residue matrix, $\mathcal{R}$, which is the residue of the FV function $\mathcal{F}$ at FV energies $E_n$, and is defined in Eq.\eqref{eq: definition of residue}
In the limit where higher partial waves are ignored, Eq.~\eqref{eq: quantization condition in phase shift again} is also valid for the $^3S_1$ channel after replacing $\delta$ with $\widetilde{\delta}$.
Similarly, the replacement $\mathcal{M} \to \widetilde{\mathcal{M}}$ in Eq.~\eqref{eq: definition of mathcal F} defines $\mathcal{F}\to\widetilde{\mathcal{F}}$, which leads to the FV residue function $\widetilde{\mathcal{R}}$ for the $^3S_1$ channel.\par
\begin{figure}[t]
    \centering
    \includegraphics[scale=1]{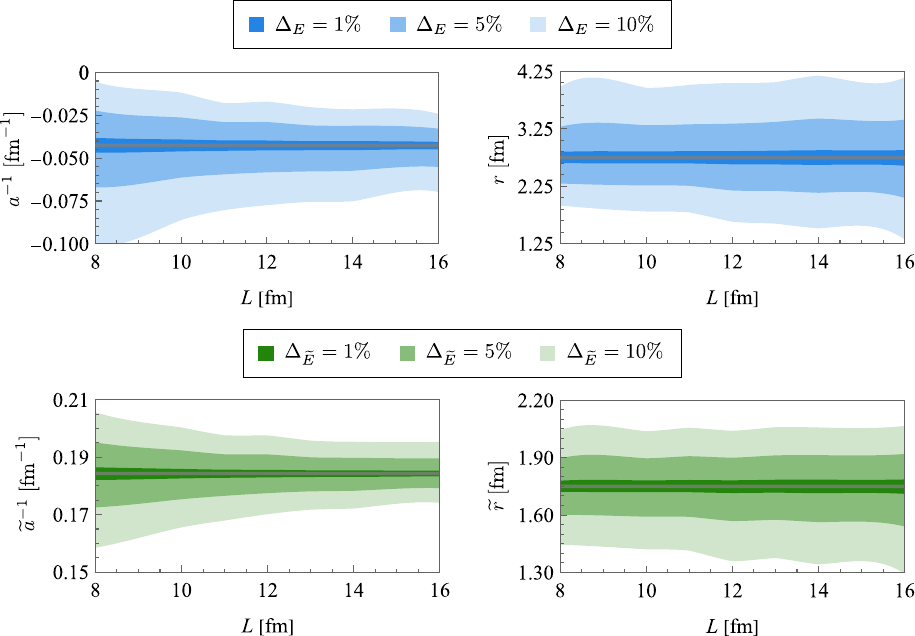}
    \caption{The inverse scattering length (left column) and the effective range (left column) for the $^1S_0$ (top row) and $^3S_1$ (bottom row) channels obtained from synthetic data with $\Delta_{E},\Delta_{\widetilde{E}}=10\%$, $5\%$, and $1\%$, from lighter to darker bands, respectively, are shown as a function of $L$. The bands indicate mid-$68\%$ uncertainty on the parameters from synthetic data, whereas gray thin bands denote the corresponding experimental values.
    Selected numerical values associated with this figure are provided in Appendix~\ref{app: numerical tables for sensitivity analysis}. 
    \label{fig: band ERE}}
\end{figure}
In the following sections, the sensitivity of constraining LECs $L_{1,A}$ and $g^{NN}_{\nu}$ to LQCD inputs from future LQCD calculations of the corresponding three- and four-point correlation functions at physical quark masses is investigated.
The lowest-lying FV energy eigenvalues in each of the NN channels enter the necessary matching relations and these energies will be evaluated \emph{ab initio} from LQCD FV two-point correlation functions.
As no LQCD determination of the FV spectrum at the physical quark masses exist to date, one can estimate the expected energies for given volumes by solving L\"uscher's quantization condition in Eq.~\eqref{eq: quantization condition in phase shift again} using experimental input for scattering amplitudes, as illustrated in Fig.~\ref{fig: QC plot}.
Here, the function $p\cot \delta$ ($p\cot \widetilde{\delta}$) on the left-hand side of Eq.~\eqref{eq: quantization condition in phase shift again} is given by the effective-range expansion defined in Eq.~\eqref{eq: ERE definition} (and its counterpart for the $^3S_1$ channel).
The effective-range expansion parameters are obtained using NN phase shifts for $S$-wave scattering generated by the Nijmegen phenomenological NN potential~\cite{Stoks:1994wp}, that are the result of fits to NN scattering data in Ref.~\cite{NNonline}.
These values are shown in Eq.~\eqref{eq: 1S0 ERE values} and~\eqref{eq: 3S1 ERE values} for $^1S_0$ and $^3S_1$ channels, respectively, and reiterated here for emphasis:
\begin{equation}
    \begin{split}
        a & =-23.5\;[\text{fm}], \hspace{2cm} r = 2.75\;[\text{fm}],
        \\
        \widetilde{a} & =5.42\;[\text{fm}], \hspace{2.35cm} \widetilde{r} = 1.75\;[\text{fm}],
    \end{split}
    \label{eq: ERE values}
\end{equation}
The ground-state energies of the NN systems in the $^1S_0$ ($^3S_1$) channel with the CM energy $E_0$ ($\widetilde{E}_0$) for the volumes shown are negatively shifted compared with the threshold as noted in Fig.~\ref{fig: QC plot}, and asymptote polynomially (exponentially) to zero (to -2.2245 MeV) in the infinite-volume limit.
Additionally, the absolute values of the LL residue functions are plotted in Fig.~\ref{fig: R plot} as a function of energy for $L=8,12$ and $16$~fm. Note that only the absolute values of these functions appear in the matching relations for the matrix elements.

The small uncertainties on the scattering parameters from experiment are ignored as the goal is to obtain central values of FV energies.
For the sensitivity analyses of the upcoming sections, uncertainties need to be artificially introduced on these energies in generating synthetic data to mimic the expected LQCD uncertainties on energy extractions.
This indicates that the scattering parameters associated with these energies will become uncertain too. Since the scattering parameters enter the LO and NLO NN scattering amplitudes, and hence impact the matching relations of the next sections, the subsequent uncertainty on scattering parameters must be taken into account.
Uncertainties on the first two lowest-lying energies in each channel (which is a minimal set in a single volume to constrain the scattering length and effective range) can be introduced through a randomly-generated Gaussian distribution of energies with central values equal to $E_0$ and $E_1$ ($\widetilde{E}_0$ and $\widetilde{E}_1$) and the width equal to $\Delta_{E_0} \times |E_0|$ and $\Delta_{E_1} \times |E_1|$ ($\Delta_{\widetilde{E}_0} \times |\widetilde{E}_0|$ and $\Delta_{\widetilde{E}_1} \times |\widetilde{E}_1|$) for the ground- and first excited-state energies of the NN systems in the $^1S_0$ ($^3S_1$) channels, respectively.
The scattering length and effective range corresponding to each channel for the choices of $\Delta_{E}\equiv\Delta_{E_0}=\Delta_{E_1}=10\%,~5\%$, and $1\%$ and $\Delta_{\widetilde{E}}\equiv\Delta_{\widetilde{E}_0}=\Delta_{\widetilde{E}_1}=10\%,~5\%$, and $1\%$ are then obtained by solving the quantization condition in Eq.~\eqref{eq: quantization condition in phase shift again}, resulting in uncertainties in the scattering parameters as shown in Fig.~\ref{fig: band ERE}.
A similar analysis was performed in Ref.~\cite{Briceno:2013bda} in the isosinglet channel to study the viability of the extraction of the S-D mixing parameter from the upcoming LQCD calculations. 

Constraints on more than two energies, including in more than one volume and with various different boost vectors, will improve uncertainties on the extracted scattering parameters, possibilities that are not considered in this initial analysis.
More radically, one may attempt to input the experimental determination of the scattering parameters (and hence the energy eigenvalues derived using quantization conditions) to avoid an uncertainty introduced in both quantities in costly LQCD calculations.
This can reduce the uncertainty on the extracted LECs, as the only LQCD input will be matrix elements that are unknown experimentally. Nonetheless, the upcoming LQCD calculations will first evaluate these matrix elements at the isospin-symmetric limit where quantum electrodynamics (QED) effects and the non-vanishing mass difference among the light quarks are ignored.
This means that for consistency, one needs to input the scattering parameters associated with the $^1S_0$ and $^3S_1$ channels in such a limit. As obtaining the isospin-symmetric parameters from experimental data involves model/EFT uncertainties, it is preferred that all inputs to the quantization and matching conditions are evaluated from first-principles LQCD calculations consistently.
That is the strategy adopted in this synthetic data analysis. While the experimental parameters are used to obtain the central values of the FV energies, the subsequent analysis assumes energies and hence the scattering parameters are obtained directly from LQCD and hence involve likely sizable uncertainties in early calculations.\footnote{The inaccuracy in the central values of the FV energies compared to what is expected at the isospin symmetric limit will have minimal impact in the conclusions reached in the upcoming sections, as we have verified by slightly changing the central values of the synthetic data in our analysis and observed no significant sensitivity in achieved uncertainties on the LECs.}


\subsection{Sensitivity analysis for \texorpdfstring{$L_{1,A}$}{Bookmark Version}
\label{subsec: L1A}
}
\noindent
The two-body axial-vector current in Eq.~\eqref{eq: vector axial-vector isovector 2 body} contributes to single- and double-weak processes, including $pp$ fusion, neutrino(antineutrino)-induced disintegration of the deuteron, and muon capture on the deuteron~\cite{Butler:1999sv,Butler:2000zp,Davoudi:2020xdv}, at the NLO in the pionless EFT, and its strength is characterized by the LEC $L_{1,A}$.
Constraints on $L_{1,A}$ were obtained using elastic and inelastic (anti)neutrino-deuteron scattering data from nuclear reactors: $L_{1,A}=3.6 \pm 5.5\text{ fm}^3$~\cite{Chen:1999tn}, as well as from Sudbury Neutrino Observatory~\cite{Bellerive:2016byv} and Super-K~\cite{Fukuda:2001nj,Fukuda:2002pe} experiments: $L_{1,A}=4.0 \pm 6.3\text{ fm}^3$~\cite{Butler:2002cw}.
A more precise constraint was obtained in Ref.~\cite{Acharya:2019fij} where improved low-energy chiral EFT results of inelastic (anti)neutrino-deuteron scattering amplitude were matched to those of pionless EFT, resulting in: $L_{1,A}= 4.9 ^{+1.9}_{-1.5}\text{ fm}^3$.
It is expected that the uncertainty in $L_{1,A}$ will be reduced to $\sim 1.25 \text{ fm}^3$ from the precise measurement of reaction rate of muon capture on the deuteron that is underway in the MuSun experiment~\cite{Andreev:2010wd}. 

Furthermore, a constraint on $L_{1,A}$ has been obtained from a LQCD study of the $pp$-fusion process in Ref.~\cite{Savage:2016kon} giving the value $L_{1,A}=3.9(0.2)(1.4) \text{ fm}^3$.
Even though the statistical uncertainty shown in the first parentheses is small, the overall uncertainty is similar to the experimental constraints due to the large systematic uncertainty indicated in the second parentheses.
The major source of uncertainty is the extrapolation to the physical quark masses as the correlation function for the $pp$-fusion process was calculated at larger quark masses corresponding to $m_\pi\approx 806$ MeV.
Thus, it is expected that this uncertainty will improve in future LQCD calculations at lighter quark masses.
However, the extraction of this LEC at such a large pion mass did not require the involved matching relation that will be presented shortly, as the NN states appeared deeply bound.
Furthermore, achieving the quoted statistical uncertainty with quark masses near the physical values will be challenging.
The question that will be addressed here is whether these features will limit the precise extraction of $L_{1,A}$ at the physical values of the quark masses.

This section investigates the accuracy with which $L_{1,A}$ can be obtained from future LQCD calculations performed at the physical pion mass.
The matching relation by Eq.~\eqref{eq: matching relation for single 2}, which relates the hadronic scattering amplitude for the $nn \to npe^-\bar{\nu}_e$ decay (or alternatively the $pp$ fusion process $pp \to npe^+\nu_e$) to the corresponding nuclear ME calculated using LQCD, will be at the center of this analysis, and thus briefly summarized here.

Consider the single-beta decay transition $nn \to npe^-\bar{\nu}_e$ with the kinematics chosen such that the total three-momentum of the electron and anti-neutrino is zero, and the NN systems are unboosted in the initial and final states.
Such a process was studied in detail in Sec.~\ref{subsec: NN to NN  with 1 J}.
The hadronic amplitude was shown to receive non-vanishing contribution from the Gamow-Teller-type transitions mediated by one-body and two-body axial-current operators in Eqs.~\eqref{eq: vector axial-vector isovector 1 body} and~\eqref{eq: vector axial-vector isovector 2 body}, respectively.
A matching relation given by Eq.~\eqref{eq: matching relation for single 2} was obtained in Sec.~\ref{subsubsec: single beta decay from FV} that related the infinite volume amplitude in the spin-isospin symmetric limit, to the FV nuclear ME of the weak current between the ground states of the NN system with energies $E_0$ and $\widetilde{E}_0$ corresponding to the $^1S_0$ and $^3S_1$ channels, respectively.
Recall from Ch.~\ref{ch: Intro} that $L_{1,A}$ characterizes the strength of two-body axial-current operators in Eq.~\eqref{eq: vector axial-vector isovector 2 body}, and the matching relation for this LEC is given by~\cite{Davoudi:2020xdv, Briceno:2012yi}
\begin{align}
    L^6&\left|\left[\vphantom{B^\dagger}\langle E_{0},L|\,\mathcal{J}({0})\,|\widetilde{E}_{0},L\rangle \right]_L\right|^2 = \left|\widetilde{\mathcal{R}}(\widetilde{E}_{0})\right| \left|\mathcal{M}^{\rm{DF},V}_{nn\to np} (E_{0},\widetilde{E}_{0})\right|^2  \left|\mathcal{R}(E_{0}) \vphantom{\widetilde{\mathcal{R}}(\widetilde{E}_{0})} \right|,
    \label{eq: matching relation for L1A}
\end{align}
where all components of this equation and the equation itself have been derived or defined in Ch.~\ref{ch: Intro} (see the discussion around Eq.~\eqref{eq: matching relation for single 2} for more details).
Furthermore, the quantity, $\mathcal{M}^{\rm{DF},V}_{nn\to np}$, is related to infinite-volume amplitude via Eq.~\eqref{eq: DF amplitude: finite V}, which depends on the LEC $L_{1,A}$ via the amplitude $\mathcal{M}^{\rm DF}_{nn\to np}$ defined in Eq.~\eqref{eq: single beta decay DF amplitude}.

Future constraints on $L_{1,A}$ from LQCD calculations at the physical quark masses will depend on LQCD determinations of the low-lying FV energy eigenvalues of the NN systems in the $^1S_0$ and $^3S_1$ channels, as well as the nuclear MEs of the axial-vector current between these states, as is clear from the ingredients of Eq.~\eqref{eq: matching relation for L1A}.
Furthermore, the matching relation depends upon the LO and NLO NN scattering amplitudes in both the $^1S_0$ and $^3S_1$ channels, as well as the derivative of scattering amplitudes with respect to energy that enters the LL residue function in Eq.~\eqref{eq: definition of residue}, requiring the values of the scattering length and effective range in the two NN channels.
These are obtained from the knowledge of at least two energy levels in the spectrum, i.e., the ground and the first excited states, as outlined in Sec.~\ref{subsec: Luscher numerical formalism}.
The precision with which $L_{1,A}$ can be obtained depends on the precision and correlation of these ingredients.
In order to quantify the uncertainty on $L_{1,A}$ extracted from a future LQCD calculation performed in a given volume, one can introduce percent precision with which the nuclear ME of a single axial-vector current and the NN ground- (and first excited-) state energies are expected to reach, to be denoted by $\Delta_{\beta}$ and $\Delta_{E(\widetilde{E})}$, respectively.
A sample set of these ingredients is then generated from a Gaussian distribution with the mean represented by the value obtained from (the central values of) the phenomenological constraint for the quantity, and the  precision level multiplied by the mean for its standard deviation.

The mean values for the expected ground- and first excited-state energies of the NN channels are obtained using the quantization condition in Eq.~\eqref{eq: quantization condition in phase shift again} with the NN phase shifts for the $S$-channel from Ref.~\cite{NNonline}, as was already discussed in Sec.~\ref{subsec: Luscher numerical formalism} and demonstrated in Fig.~\ref{fig: QC plot}.
The mean value of the expected FV ME is obtained by using the matching relation in Eq.~\eqref{eq: matching relation for L1A} with the FV energies being the mean values discussed above, the experimental value of $g_A\approx1.27$, and the central value of the $L_{1,A}$ (or and its scale-independent counterpart) from a recent phenomenological determination~\cite{Acharya:2019fij}
\begin{equation}
   L_{1,A}= 4.9 ^{+1.9}_{-1.5}\;\text{ fm}^3~~\text{or}~~\widetilde{L}_{1,A}= -449.7^{+19.5}_{-15.4}\;\text{ fm}^3.
    \label{eq: L1A value}
\end{equation}
\begin{figure}[t]
    \centering
    \includegraphics[scale=1]{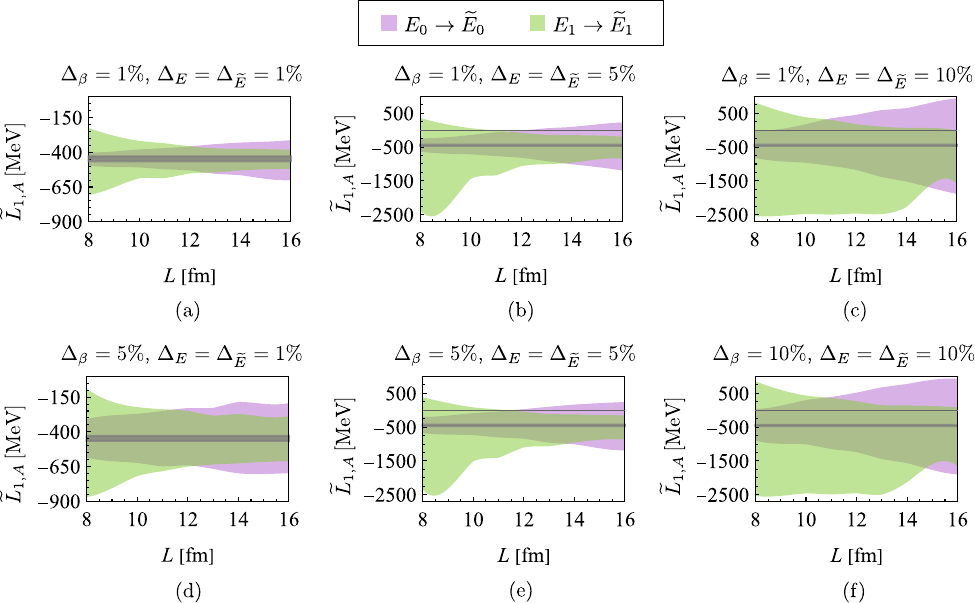}
    \caption{The value of $\widetilde{L}_{1,A}$ as a function of $L$ for the $^1S_0 \to {^3}S_1$ transition obtained from synthetic data with various combinations of $\Delta_{E}=\Delta_{\widetilde{E}}$ and $\Delta_{\beta}$ values.
    The gray horizontal band denotes the experimental value, whereas the colored bands indicate mid-$68\%$ uncertainty on extracted $L_{1,A}$ for the ground-state to ground-state  (purple) and first excited-state to first excited-state (green) transitions.
    Note the smaller range of the $\widetilde{L}_{1,A}$-axes in the most-left plots compared to the rest. Selected numerical values associated with this figure are provided in Appendix~\ref{app: numerical tables for sensitivity analysis}.
    \label{fig: band L1A}}
\end{figure}

The expected mean values are then used to generate the Gaussian samples for the FV energies and FV ME. With the samples generated, the matching relation in Eq.~\eqref{eq: matching relation for L1A} is used once again to solve for the $L_{1,A}$ values associated with each set of energies and MEs, leading to a distribution for the expected $L_{1,A}$ values.
In the following, the scale-dependent quantity $\widetilde{L}_{1,A}$ is used but it can be converted to $L_{1,A}$ values give the values of the NN LECs evaluated at the corresponding values of the scattering length and effective range.
Note that since the scattering parameters are obtained \emph{ab initio} from LQCD, the uncertainties in energies impact their precision, as discussed in Sec.~\ref{subsec: Luscher numerical formalism}.

The effect of $\Delta_{\beta}$ and $\Delta_E$ on determining $\widetilde{L}_{1,A}$ is illustrated in Fig.~\ref{fig: band L1A}, where the volume dependence of $L_{1,A}$ values obtained from the sample sets for various combinations of $\Delta_E$ and $\Delta_{\beta}$ values is shown.
In all cases, the uncertainty on $L_{1,A}$ (determined from the mid-$68\%$ of the sample) increases with increasing $\Delta_E$, $\Delta_{\widetilde{E}}$, and $\Delta_\beta$. Only the most precisely determined sample set and at volumes with $L \approx 8\;{\rm fm}$, constraints on $L_{1,A}$ become comparable in precision to that in Eq.~\eqref{eq: L1A value}.
Thus, future LQCD calculations at the physical quark masses need to determine the NN ground and first excited-state energies and the FV MEs with below percent-level precision to supersede the current phenomenological constraints.
The situation is likely alleviated in the actual LQCD calculations where energy and ME extractions are partially correlated, and where the NN scattering amplitude can be determined more precisely with a larger set of precise FV energies.

Since LQCD can, in principle, obtain FV MEs for transitions involving excited states, one may wonder if constraining $L_{1,A}$ through the first excited-state to the first excited-state transition will be more beneficial and relaxes the precision requirements on the FV energies and ME above.
The green bands in Fig.~\ref{fig: band L1A} denote the $\widetilde{L}_{1,A}$ values and uncertainties obtained from the first excited-state to the first excited-state transition.
It is clear that the ground-state to ground-state transition leads to better constraints at smaller volumes---volumes that are more readily accessibly to upcoming LQCD calculations at the physical pion mass, but for larger volumes with $L \gtrsim 14\;{\rm fm}$, the constraints from the first excited-state to the first excited-state transition become comparable or more precise.
The reverse trend in uncertainties as a function of volume between the two cases is a consequence of different behavior of the LL residue functions near negative and positive CM energies, as illustrated in Fig.~\ref{fig: R plot}.
One cautionary note is the loss of accuracy in using the effective-range expansion and the associated LO and NLO NN scattering amplitudes in the pionless EFT near the first excited-state energies.
However, at  large volume where the $\widetilde{L}_{1,A}$ constraints from excited-state transition become more precise, the FV energies tend to their asymptotic value of zero and are therefore near or within the t-channel cut. On the other hand, at such large volumes, the density of states in the spectrum increases, and the identification of excited states with current methods may present a challenge.
Variational techniques such as those developed in Refs.~\cite{Horz:2020zvv, Green:2021qol, Amarasinghe:2021lqa} will likely constrain the lowest-lying levels with comparable precision to the ground state.
\subsection{Sensitivity analysis for $g_{\nu}^{NN}$
\label{subsec: gvNN}
}
\noindent
Section~\ref{subsec: 0vbb amplitude in pionless EFT} showed that the light neutrino exchange model of the low-energy $nn \to pp e^-e^-$ decay, there exists an undetermined LEC, $g_{\nu}^{NN}$, at the LO the pionless EFT, which is introduced to absorb the UV scale dependence of the amplitude through renormalization group~\cite{Cirigliano:2017tvr,Cirigliano:2018hja,Cirigliano:2019vdj}.
The contact interaction associated with $g_{\nu}^{NN}$ is given in Eq.~\eqref{eq:LNdeltaL2}, however, its contribution to the short distance part of the hadronic amplitude of $nn \to pp e^-e^-$ transition is unknown at the LO due to the undetermined LEC $g_{\nu}^{NN}$.
Nonetheless, Refs.~\cite{Cirigliano:2020dmx,Cirigliano:2021qko} gave a constraint on the $g_{\nu}^{NN}$ value by expressing the  $nn \to pp e^-e^-$ decay amplitude as a product of momentum integral of the Majorana neutrino propagator and the generalized forward Compton scattering amplitude, in analogy to the
Cottingham formula~\cite{Cottingham:1963zz,Harari:1966mu} for the electromagnetic contribution to hadron masses.
A model-independent representation of the integrand using the chiral EFT and operator product expansion can then be obtained, and the missing parts of the full amplitude can be filled by interpolating between the known regions using nucleon form factors for the weak current and information on NN scattering.
The constraint on $g_{\nu}^{NN}$ via this method is:
\begin{equation}
    \widetilde{g}_{\nu}^{NN} = 1.3 \pm 0.6,
    \label{eq: gvNNtilde value}
\end{equation}
where $\widetilde{g}_{\nu}^{NN}$ is a dimensionless parameter related to $g_{\nu}^{NN}$ and the momentum-independent NN LEC in Eq.~\eqref{eq: EFT 2 nucleon Lagrangian} via Eq.~\eqref{eq: gvNNtilde definition}.
The value in Eq.~\eqref{eq: gvNNtilde value} has a large uncertainty, and a more precise and direct constraint on  $g_{\nu}^{NN}$ using LQCD will be desired.
As shown in Sec.~\ref{sec: 0vbb}, a prescription exists for obtaining the $g_{\nu}^{NN}$ (or equivalently the $\widetilde{g}_{\nu}^{NN}$) value from a Euclidean four-point correlation function calculated using LQCD.
With LQCD calculations of these correlation functions underway, it would be useful to know the precision with which one can constrain the  $\widetilde{g}_{\nu}^{NN}$ value for a given LQCD setup.

This section performs the sensitivity analysis of constraining  $\widetilde{g}_{\nu}^{NN}$ by estimating the uncertainty on  $\widetilde{g}_{\nu}^{NN}$ from a synthetic data representing a future LQCD calculation of the four-point correlation function at the physical quark masses.
The starting point of this analysis is to consider the transition $nn \to pp e^-e^-$ in the spin-isospin symmetric limit with simple kinematics, where the currents carry zero energy and momentum such that the initial CM energy, $E_i \equiv E_{CM}$, remains unchanged.
The Euclidean four-point function for this process, which is accessible via LQCD methods, can be analytically continued to Minkowski spacetime to obtain $\mathcal{T}^{(\rm M)}_L$ defined in Eq.~\eqref{eq: 0vbb TLM} using the procedure described in Sec.~\ref{subsec: 0vbb minkowski Euclidean matching}.
Recall that in Eq.~\eqref{eq: 0vbb TLM} the neutrino four-momentum is given by $(k_0,\bm{k})$ with quantized spatial momenta $\bm{k}$, contributions from the small non-zero neutrino mass in the denominator of the neutrino propagator are ignored at the LO in the EFT power counting, and the infrared divergence is regulated by removing the zero-momentum mode of the neutrino.
It is important to note that for a ground-state to ground-state transitions at low energies corresponding to the FV energy eigenvalues in the range of volumes studied, no intermediate single-neutrino-two-nucleon state can go on shell and the analytic continuation from the Euclidean correlation function of LQCD to the Minkowski counterpart in Eq.~\eqref{eq: 0vbb TLM} is straightforward.
With on-shell intermediate states, the complete formalism of Sec.~\ref{subsec: 0vbb minkowski Euclidean matching} needs to be implemented but this will not be necessary in the upcoming LQCD calculations given realistic volumes and energies.

$\mathcal{T}^{(\rm M)}_L$ is related to the physical decay amplitude through the matching relation in Eq.~\eqref{eq: 0vbb IVFVmatching} which is the focus of the study here.
Thus, it is re-written here for emphasis.
\begin{equation}
    L^6\;\bigg|\mathcal{T}^{(\rm M)}_L(E_i,E_f) \bigg|^2=\bigg|\mathcal{R}(E_i)\bigg | \, \bigg|\mathcal{M}^{0\nu,V}_{nn\xrightarrow{0\nu} pp} (E_i,E_f) \bigg|^2 \bigg|\mathcal{R}(E_f)\bigg |,
    \label{eq: matching relation gvNN}
\end{equation}
where the right-hand side of Eq.~\eqref{eq: matching relation gvNN} contains the LL residue matrix, $\mathcal{R}$, defined in Eq.~\eqref{eq: definition of residue}, and the FV quantity $\mathcal{M}^{0\nu,V}_{nn\xrightarrow{0\nu} pp}$ which is related to the physical scattering amplitude of the $0\nu\beta\beta$ decay with the initial (final) CM energy $E_i$ $(E_f)$, as defined in Eq.~\eqref{eq: Finite volume amplitude 0vbb}.
Recall that $\mathcal{M}^{(\rm{Int.})}$ in Eq.~\eqref{eq: Finite volume amplitude 0vbb} is the infinite-volume decay amplitude evaluated in the pionless EFT after removing the contributions from the diagrams in which the neutrino propagates between two external nucleons, and its expression is given in Eq.~\eqref{eq:MICnnpp}.

\begin{figure}[t]
    \centering
    \includegraphics[scale=1]{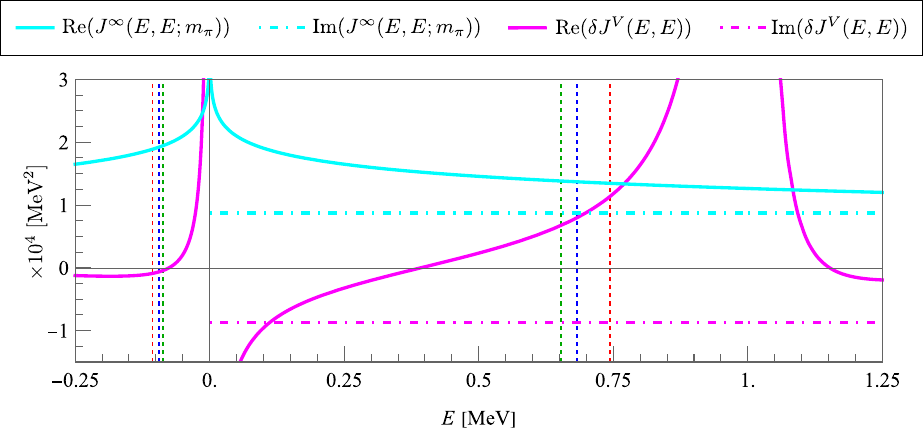}
    \caption{The real (solid cyan) and imaginary (dotted dashed cyan) parts of $J^{\infty}(E_i,E_f;\mu=m_\pi)$ defined in Eq.~\eqref{eq:Jinfty}), as well as real (solid magenta) and imaginary (dotted dashed magenta) parts of $\delta J^V(E_i,E_f)$ defined in Eq.~\eqref{eq:deltaJV}, both evaluated at $E_i=E_f\equiv E$. The red, blue, and green dashed lines denote the FV ground-state energy eigenvalues with $L=8$, $12$, and $16\;{\rm fm}$, respectively, obtained from the quantization condition in Eq.~(\ref{eq: quantization condition in phase shift again}) (as plotted in Fig.~\ref{fig: QC plot}). These are the values at which the LQCD four-point function will be evaluated in the future studies at the physical quark masses.  Selected numerical values for the functions shown are provided in Appendix~\ref{app: numerical tables for sensitivity analysis}.
    \label{fig:delJ}
    }
\end{figure}

$J^{\infty}(E_i,E_f;\mu)$ in Eq.~\eqref{eq:MICnnpp} is a function arising from the s-channel two-loop diagram with an exchanged Majorana neutrino, and its expression is given in Eq.~\eqref{eq:Jinfty}.
Similarly, the definition of $\mathcal{M}^{0\nu,V}_{nn\xrightarrow{0\nu} pp}$ in Eq.~\eqref{eq: Finite volume amplitude 0vbb} contains the FV function originating from $J^{\infty}(E_i,E_f;\mu)$, $\delta J^V(E_i,E_f)$, that is defined in Eq.~\eqref{eq:deltaJV}.
This sum-integral difference is calculated numerically using the technique presented in App.\ref{app: Two-loop sum-integral difference}. The real and imaginary parts of $J^{\infty}$ and $\delta J^V$ are depicted in Fig.~\ref{fig:delJ} for a range of negative and positive $E_i=E_f\equiv E$ values.

\begin{figure}[t]
    \centering
    \includegraphics[scale=1]{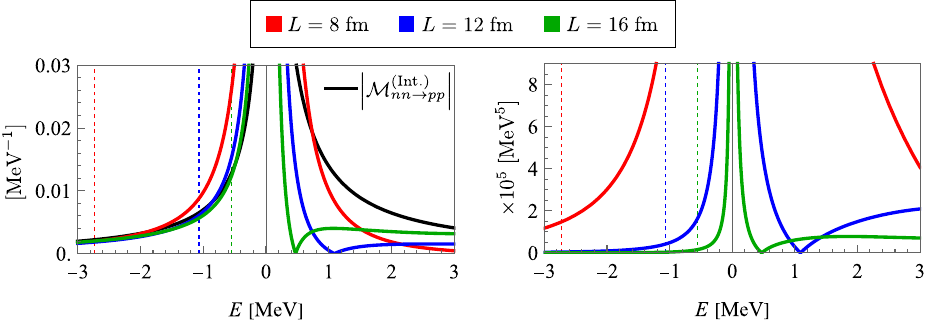}
    \caption{$\big| \mathcal{M}^{0\nu,V({\rm Int.})}_{nn\to pp}\big|$ (left) and $\big|\mathcal{T}_L^{(M)}\big|$ (right) functions defined in Eqs.~\eqref{eq:MICnnpp} and~\eqref{eq: 0vbb TLM}, that is evaluated using Eq.~\eqref{eq: matching relation gvNN} with $L=8\;{\rm fm}$ (red), $L=12\;{\rm fm}$ (blue), and $L=16\;{\rm fm}$ (green) are plotted against the CM energy of the NN state, considering the kinematics $E_i=E_f \equiv E$.
    The effective neutrino mass $m_{\beta\beta}$ is set to $1\;{\rm MeV}$.
    The dashed lines in both panels denote the ground-state energy eigenvalues in the corresponding volumes obtained from the quantization condition in Eq.~(\ref{eq: quantization condition in phase shift again}) (as plotted in Fig.~\ref{fig: QC plot}).
    Selected numerical values for the functions shown are provided in Appendix~\ref{app: numerical tables for sensitivity analysis}.
    \label{fig: TLMMV vs ECM}
    }
\end{figure}
The absolute value of the FV amplitude $\mathcal{M}^{0\nu,V}_{nn\xrightarrow{0\nu} pp}$ for the kinematics $E_i=E_f \equiv E$ is plotted against the CM energy in the left panel of Fig.~\ref{fig: TLMMV vs ECM} along with $|\mathcal{M}^{\rm (Int.)}_{nn \xrightarrow{0\nu} pp}|$, using the value of $g_{\nu}^{NN}$ obtained from the central value of the constraint in Eq.~\eqref{eq: gvNNtilde value}.
The dependence of the $|\mathcal{T}_L^{(M)}|$ on the CM energy of the NN system in different volumes is shown in the right panel of Fig.~\ref{fig: TLMMV vs ECM}(b) using the matching relation in Eq.~\eqref{eq: matching relation gvNN}. 

Equation~\eqref{eq: matching relation gvNN} indicates that the precision with which $g_{\nu}^{NN}$, and thus $\widetilde{g}_{\nu}^{NN}$, can be obtained from LQCD depends on the precision with which the FV ground-state energy in a given volume, $E_0$, and the FV ME are obtained from the LQCD calculations of the corresponding two- and four-point functions, respectively.
Furthermore, the matching relation depends upon the LO NN scattering amplitude in the $^1S_0$ channel as well as the derivative of the NLO+LO scattering amplitude with respect to energy that enters the LL residue function in Eq.~\eqref{eq: definition of residue}, requiring the values of the scattering length and effective range in the $^1S_0$ channel.
These depend on the central value and the uncertainty of at least two energy levels in the spectrum, e.g., the ground and the first excited states, as outlined in Sec.~\ref{subsec: Luscher numerical formalism}.
This section investigates the uncertainty on $\widetilde{g}_{\nu}^{NN}$ from the precision levels with which these LQCD inputs are obtained in future LQCD calculations at the physical quark masses.

The expected value of $E_0$ for a given volume is calculated using L\"uscher's quantization condition in Eq.~\eqref{eq: quantization condition in phase shift again} and NN phase shifts in the $^1S_0$ channel obtained from Ref.~\cite{NNonline}.
This expected value of $E_0$ and the central value of the constraint on $\widetilde{g}_{\nu}^{NN}$ given in Eq.~\eqref{eq: gvNNtilde value} are then used to obtain an estimate on the expected value of $\mathcal{T}^{(\rm M)}_L$ with the use of Eqs.~\eqref{eq: matching relation gvNN} and~\eqref{eq:MICnnpp}.
Note that even though the expected value of $\mathcal{T}^{(\rm M)}_L$ from Eqs.~\eqref{eq:MICnnpp}-\eqref{eq: 0vbb TLM} is dependent on $m_{\beta\beta}$, the mean value and the uncertainty on $\widetilde{g}_{\nu}^{NN}$ obtained from synthetic data using Eq.~\eqref{eq: matching relation gvNN} is independent of $m_{\beta\beta}$.

\begin{figure}[t]
    \centering
    \includegraphics[scale=1]{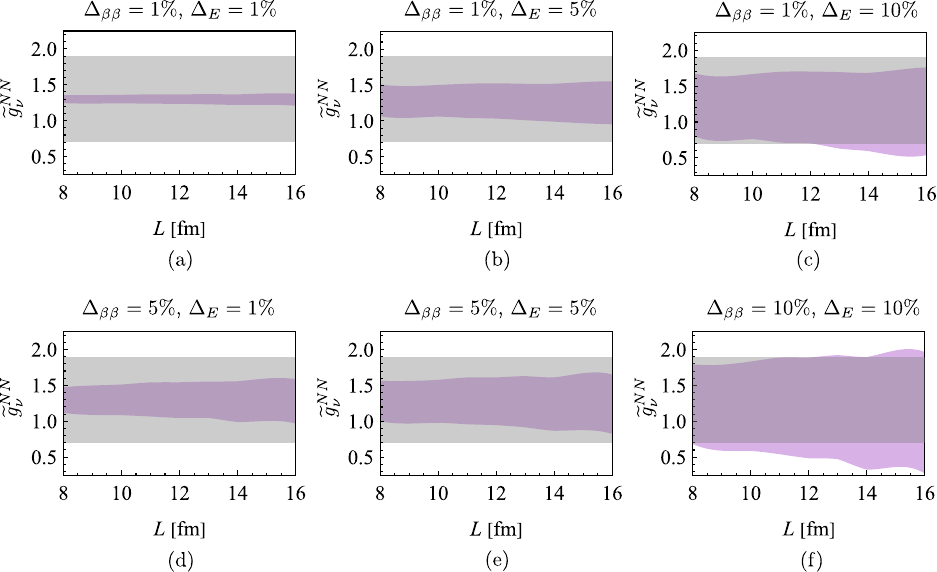}
    \caption{ The value of $\widetilde{g}_{\nu}^{NN}$ obtained from the synthetic data is plotted against $L$ for different combinations of $\Delta_{\beta\beta}$ and $\Delta_E$. The gray band denotes the uncertainty in the value of $\widetilde{g}_{\nu}^{NN}$ from Eq.~\eqref{eq: gvNNtilde value} from the indirect determination of Ref.~\cite{Cirigliano:2020dmx}. The corresponding central value is used to obtain the expected values of $\mathcal{T}^{(\rm M)}_L$, which enables this sensitivity analysis.  The purple band is the mid-$68\%$ uncertainty band corresponding to the sample sets with uncorrelated fluctuations. Selected numerical values associated with this figure are provided in Appendix~\ref{app: numerical tables for sensitivity analysis}.
    \label{fig: band gvNN}}
\end{figure}
The percent precision on $E_0$ (and $E_1$) is denoted by $\Delta_E$, whereas the percent precision on $\mathcal{T}^{(\rm M)}_L$ is denoted by $\Delta_{\beta\beta}$.
Similar to Sec.~\ref{subsec: L1A}, the uncertainty on $\widetilde{g}_{\nu}^{NN}$ is taken as the mid-$68\%$ of the ensemble of $\widetilde{g}_{\nu}^{NN}$ values obtained from synthetic data that incorporates uncertainties on $E_{0(1)}$ and $\mathcal{T}^{(\rm M)}_L$ as Gaussian fluctuations.
The precision levels, $\Delta_E$ and $\Delta_{\beta\beta}$, are incorporated in this synthetic data by making the standard deviation of the fluctuations equal to the expected values of the quantities multiplied by the corresponding percent precision.
The scattering length and effective range in the $^1S_0$ channel are obtained by solving L\"uscher's quantization condition in Eq.~\eqref{eq: quantization condition in phase shift again} for the generated ensembles of the ground- and the first excited-state energies, as outlined in Sec.~\ref{subsec: Luscher numerical formalism}.

The $\widetilde{g}_{\nu}^{NN}$ values obtained for various combinations of $\Delta_{\beta\beta}$ and $\Delta_E$ are plotted against $L$ in Fig.~\ref{fig: band gvNN}.
The LQCD constraints on $\widetilde{g}_{\nu}^{NN}$ are almost always more precise than the constraint of Ref.~\cite{Cirigliano:2020dmx} for input uncertainties below $\sim10\%$ level, which indicates that future LQCD calculations can confidently improve the current constraint, especially for smaller volumes, provided that $\Delta_{\beta\beta}$ and $\Delta_E$ are a few percents.
This situation is more promising than the case of $L_{1,A}$, where (sub)precent-level uncertainties appear to be the requirement. As the LQCD input for energies and the ME will be partially correlated, the constraint on $\widetilde{g}_{\nu}^{NN}$ will likely be further improved.
\subsection{Summary
\label{subsec: sensitivity analysis summary}
}
\noindent
This chapter presents an analysis of the effect of uncertainties in the future lattice quantum chromodynamics calculations at the physical quark masses on the accuracy with which the hadronic amplitudes of $\beta$ decays can be constrained in the two-nucleon sector.
The nuclear matrix elements of the single-beta decay and the neutrinoless double-beta decay within the light neutrino exchange scenario are studied for this purpose, and the precision with which the low-energy constants $L_{1,A}$ and $\tilde{g}_{\nu}^{NN}$, corresponding to the respective two-body isovector and isotensor operators, can be obtained from future calculations was deduced from a synthetic data analysis.

For processes that are studied here, matching relations exist that relate the three- and four-point functions of LQCD evaluated in a finite Euclidean spacetime to their respective physical scattering amplitudes~\cite{Briceno:2012yi,Detmold:2004qn,Briceno:2015tza,Davoudi:2020xdv,Davoudi:2020gxs}.
The LQCD inputs that go into these matching relations involve the lowest-lying two-nucleon energy spectra for a given volume, the matrix elements of a single axial-vector weak current (for the single-beta decay), and of two axial-vector weak currents along with a Majorana neutrino propagator (for the $0\nu\beta\beta$ decay) between appropriate two-nucleon states.
Using these matching relations, constraints were obtained on $L_{1,A}$ and $\tilde{g}_{\nu}^{NN}$ from the synthetic data of the relevant LQCD ingredients.
In order to synthesize this data to represent the underlying LQCD uncertainties, Gaussian fluctuations were introduced on the supposedly LQCD ingredients that go into these matching relations. 

The precision with which $L_{1,A}$ and $\tilde{g}_{\nu}^{NN}$ can be obtained from the synthetic data was obtained for a range of input uncertainties at or below $\sim10\%$ level.
The uncertainty on the LECs grows with volume in both cases assuming ground-state to ground-state transitions, and so smaller volumes that are more feasible computationally appear to be more advantageous.
The constraints from LQCD studies on $L_{1,A}$ will likely be worse than the current experimental constraints for the range of volumes and plausible input uncertainties considered here, and may require (sub)percent-level precision on the finite-volume energies and matrix element.
The situation may be alleviated in actual LQCD calculations where the uncertainties in the inputs to the matching relations are (partially) correlated.
Furthermore, one may imagine inputting the precise experimental parameters and associated FV energies in those analyses, rather than obtaining them directly from LQCD calculations, to decrease the uncertainty in the extraction of the unknown LECs.
Nonetheless, such an approach will not be \emph{ab initio}, particularly since the early calculations will take place at the isospin-symmetric limit and excluding QED, and for consistency and model independency, scattering parameters need to be evaluated directly from LQCD. 

Finally, for precision levels on the LQCD energies and the ME below $10\%$, the constraint on $\tilde{g}_{\nu}^{NN}$ will likely improve the existing constraint, and will therefore provide a direct precise determination arising from first-principles calculations rooted in QCD.
As a result, the present study further motivates future studies of the $nn \to ppee$ process within the light Majorana exchange scenario from LQCD at or near the physical values of the quark masses.

\renewcommand{\thechapter}{3}

\chapter{Reformulating Hamiltonian Lattice Gauge Theory with Loops, Strings, and Hadrons
\label{ch: LSH}
}
\noindent
Quantum technology is a rapidly growing field that has numerous applications in physics, including the quantum simulation of quantum mechanical systems~\cite{Feynman:1981tf,Lloyd:1996uni}.
Although simulating quantum-mechanical systems is not a recent development, a significant amount of computational power is currently devoted to classical supercomputers that simulate chemical and physical systems where quantum phenomena play a crucial role.
The problem is that simulations on classical machines often face limitations in simulating certain systems~\cite{Troyer:2004ge} or with scaling up the problem size.
Lattice quantum field theory, specifically lattice quantum chromodynamics (QCD), is the area of high-energy physics where simulations are most prevalent, and where there is a possibility for a quantum boost~\cite{Bauer:2022hpo}.
While classical simulation methods exist, they are not directly applicable to quantum hardware.
As a result, the emergence of the so called ``second quantum revolution"~\cite{Dowling:2002jjw} has started a race to understand how to design quantum simulation protocols in theory and to subsequently bring theoretical proposals into alignment with experimental reality.

Full-blown lattice QCD is marvelously rich in technical features, including a multitude of quark flavors and masses, three spatial dimensions, confinement of quarks, and spontaneous breaking of chiral symmetry, but perhaps the single most defining feature of QCD is its non-Abelian, SU(3) gauge symmetry.
Understanding how to implement SU(3)-symmetric matter and interactions, therefore, will be one of the essential milestones on the path to quantumly simulating QCD.
The goal of this chapter to advance our understanding of how SU(3) gauge theories can be formulated for quantum computers.

Some of the most promising opportunities for a quantum boost in lattice field theory include calculations whose path integrals suffer from so-called ``sign problems.''
While sign problem solutions may be found for special cases~\cite{Grabowska:2012ik}, or in particular parameter regimes~\cite{deForcrand:2002hgr}, a general-purpose solution to a wide variety of problems of interest has not been found and may be non-existent~\cite{Troyer:2004ge}.
Scenarios in which sign problems arise include non-zero chemical potential, topological terms in the Lagrangian, and far-from-equilibrium and real-time dynamics.
Traditional lattice (gauge) field theory~\cite{Wilson:1974sk} has been based on importance sampling of path integrals in imaginary time.
But the Monte Carlo importance sampling approach breaks down when the Euclidean action acquires imaginary contributions or when considering finite intervals of real time.
In contrast, the Hamiltonian formalism, expressed in terms of vector spaces, linear operators, time evolution in a Schr\"{o}dinger (or Heisenberg or Dirac) picture, operations on states, and measurement probabilities does not obviously suffer from a sign problem.
Instead, it suffers from the exponential growth of the Hilbert space, regardless of the nature of terms present in the Hamiltonian.
Quantum simulators, themselves featuring exponentially large Hilbert spaces and being naturally expressed in the Hamiltonian framework, may be the key to opening up this method of calculation.
There have been continuous efforts over the past decade~\cite{Banerjee:2012pg, Banerjee:2012xg, Zohar:2012xf, Tagliacozzo:2012df, Zohar:2015hwa, Tagliacozzo:2012vg, Zohar:2011cw,Aidelsburger:2021mia, Mazza:2011kf, Gonzalez-Cuadra:2017lvz, Zohar:2016iic, Kasper:2016mzj, Muschik:2016tws,Atas:2022dqm, Farrell:2022wyt, Farrell:2022vyh, Armon:2021uqr, Carena:2022kpg, Martinez:2016yna, Davoudi:2019bhy, Klco:2018kyo, Klco:2019evd,Davoudi:2021ney,Raychowdhury:2018osk, Dasgupta:2020itb, Mil:2019pbt, Semeghini:2021wls, Yang:2020yer, Atas:2021ext, Ciavarella:2021nmj, Paulson:2020zjd,Riechert:2021ink, Halimeh:2019svu, Halimeh:2021vzf, Zhou:2021kdl,Stryker:2021asy,Mendicelli:2022ntz,Yamamoto:2022jnn,Kane:2022ejm}
toward this research goal. 

Quantum simulation of field theory Hamiltonians raises important and interesting questions surrounding truncation~\cite{Kessler:2003cv,Tong:2021rfv} and field discretization~\cite{Klco:2018zqz}, but for gauge theories in particular there has been a flurry of investigations into different formulations of a given model~\cite{Davoudi:2020yln}.
These formulations include the original Kogut-Susskind formulation~\cite{Kogut:1974ag}, dual or magnetic variables~\cite{Horn:1979fy,Ukawa:1979yv,Kaplan:2018vnj,Haase:2020kaj,Bauer:2021gek}, purely fermionic formulations (in 1D space)~\cite{Hamer:1981yq,Banuls:2017ena,Farrell:2022wyt,Atas:2022dqm}, local-multiplet bases~\cite{Banuls:2017ena,Klco:2019evd,Ciavarella:2021nmj}, quantum link models~\cite{Chandrasekharan:1996ih}, tensor formulations~\cite{Bazavov:2015kka,Meurice:2021bcz,Meurice:2020pxc}, prepotential (Schwinger boson) formulations~\cite{Mathur:2004kr,Mathur:2007nu,Anishetty:2009ai,Anishetty:2009nh,Mathur:2010wc,Raychowdhury:2013rwa,Raychowdhury:2018tfj}, the loop-string-hadron (LSH) formulation~\cite{Raychowdhury:2019iki,Raychowdhury:2018osk} derived from prepotentials, and more~\cite{Fontana:2020xzp}.
The lattice community benefits from having various formulations, as some formulations can be more efficient for different parameter regimes~\cite{Haase:2020kaj,Bauer:2021gek}, and alternative formulations can serve as a way to ensure consistency in the post-classical computing era.
Therefore, it is crucial to understand the advantages and drawbacks of a given formulation.

Much can be learned about the challenges that will be encountered in QCD by first considering gauge groups simpler than SU(3), namely U(1) and SU(2).
Understanding SU(2) is especially crucial as it features the non-Abelian interactions that are so characteristic of QCD, and the representation theory is non-trivial.
In the case of SU(2), the recently proposed LSH formulation was suggested to have several attractive features from the perspective of quantum simulation.
The LSH formulation is derived from the Schwinger boson or prepotential formulation, and in certain respects represents the culmination of the work that has gone into the latter.
The LSH Hilbert space and Hamiltonian have been developed at length for the SU(2) gauge group, as well as matter in the form of one flavor of staggered fermions.
Benefits of the LSH framework have been established~\cite{Davoudi:2020yln} and discovered~\cite{Davoudi:2022xmb}.
Nonetheless, there is much work left to do, especially when dealing with a multidimensional space, where it is already known that the qubit savings could break down~\cite{Raychowdhury:2018osk}.
Still, these preliminary investigations encourage the continued exploration into the LSH framework.

One of the exciting possibilities is that the LSH approach will generalize to SU(3) and continue to offer computational advantages, perhaps even yielding greater savings.
The original LSH-formulated theory was derived from the earlier Schwinger boson or prepotential formulations of SU(2) LGTs, as mentioned in Sec.~\ref{subsec: KS comparison}.
The transition from SU(2) Schwinger bosons to SU(3) comes with considerable new technical challenges.
However, it will be shown here that an SU(3) LSH framework can be constructed using essentially the same building blocks found in SU(2) LGT in 1+1D coupled to one flavor of staggered quarks.
A recent algorithmic study of the SU(2) LSH confirms cost saving advantages with the LSH formulation~\cite{Davoudi:2022xmb}, that could also hold for its SU(3) counterpart presented here.

Although progress has been made in creating simulation procedures for SU(3) gauge theories using different formulations, the subject is still in its early stages.
The first study to ever explore algorithms for U(1), SU(2), and SU(3) lattice gauge theories was Ref.~\cite{Byrnes:2005qx}, while a more contemporary and exhaustive approach has been worked out more recently in Ref.~\cite{Kan:2021xfc}.
In recent studies, the emphasis has been on near-term applications, and small SU(3) systems have been implemented on current hardware~\cite{Ciavarella:2021nmj,Atas:2022dqm,Farrell:2022vyh,Farrell:2022wyt}. 
However, most of these applications take advantage of the capability to entirely eliminate gauge fields in 1+1D.
The LSH formulation of SU(3) LGT with staggered quarks in 1+1D presented here retains the gauge fields that are needed in higher dimensions while also postponing the complications of chromomagnetic interactions to future work.

Section~\ref{sec: SU(3) prepotentials} presents the prepotential formulation of the KS Hamiltonian from which the LSH formulation is derived.
The actual derivation of the LSH Hilbert space and Hamiltonian from SU(3) Schwinger bosons is deferred to Sec.~\ref{sec: LSH-framework}, for it turns out the calculations involved can be long and technical yet they have little bearing on applying the resultant LSH framework.
A numerical confirmation of its ability to produce the spectrum expected from an alternate but equivalent purely fermionic formulation is given in Sec.~\ref{subsec: result-numerics}.
Our conclusions in Sec.~\ref{sec: LSH conclusions} summarize the SU(3) LSH formulation and provide perspectives on how this work contributes to the program of developing quantum-technological applications for lattice QCD.

\section{Prepotential formulation of SU(3) LGT in 1+1D
\label{sec: SU(3) prepotentials}
}
This section begins with revisiting the KS formulation in Sec.~\ref{subsec: KS Hamiltonian formulation}.
It summarizes the key relations but for an 1+1D SU(3) LGT by making appropriate changes.
The equivalent prepotential formulation for describing the gauge bosons is then provided.
Later in Sec.~\ref{subsec: prepotential-with-matter}, the prepotential degrees of freedom are coupled to matter fields in the form of one flavor of staggered fermions.

\begin{figure}
    \centering
    \begin{subfigure}{\textwidth}
        \captionsetup{justification=centering}
        \includegraphics[width=\textwidth]{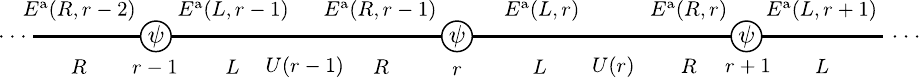}
        \caption{Kogut-Susskind variables}
        \label{subfig: KS variables}
    \end{subfigure}
    \par\bigskip
    \begin{subfigure}{\textwidth}
        \captionsetup{justification=centering}
        \includegraphics[width=\textwidth]{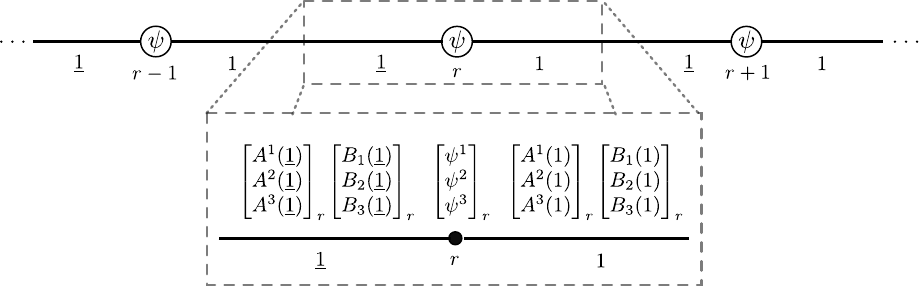}
        \caption{Prepotential variables.}
        \label{subfig: prepotential variables}
    \end{subfigure}
\caption{(a) Three adjacent lattice sites at $r-1$, $r$, and $r+1$, and their links with Kogut-Susskind degrees of freedom. The left and right ends of a link are denoted by $L$ and $R$, respectively. The chromoelectric field, $E^{\rm a}$ is labeled by a side, $L$ or $R$, denoting the end of the link at which it resides, and a position argument indicating the corresponding link. The link operator, $U$, is defined on each link as denoted by its position argument, and the staggered matter, $\psi$, is defined at each lattice sites denoted by a circle in this figure. (b) The prepotential degrees of freedom for the same lattice as in Fig.~\ref{subfig: KS variables}. 
The prepotential formulation of the same theory introduces four independent sets of SU(3) irreducible Schwinger bosons for each site $r$: $A^\alpha(1,r)$ and $B_\alpha(1,r)$ residing on the $L$ side of link $r$, and $A^\alpha(\obar,r)$ and $B_\alpha(\obar,r)$ residing on the $R$ side of link $r-1$, such that degrees of freedom are essentially site-localized.
In tandem with the emphasis on site-local degrees of freedom, the conventional $L$ and $R$ labels for left and right ends of links are replaced by $1$ and $\obar$, respectively.
Additionally, for brevity in this figure, we have used the shorthand notation for expressing the position labels of $A$, $B$ and $\psi$ by denoting the position label $r$ as $[]_r$.
}
\end{figure}

The Kogut-Susskind Hamiltonian for the 1+1 dimensional SU(3) lattice gauge theory with staggered quarks is defined on a spatial lattice with $N$ staggered lattice sites (or $N/2$ physical sites) and continuous time.
Lattice sites are referred by their position label $r=1,2,\cdots,N$ and the link joining sites $r$ and $r+1$ by `$r$' as well.
The gauge degrees of freedom in this formulation are described by the canonically conjugate variables on each link, namely, the chromoelectric fields and the holonomy or link operators that remain after choosing the temporal gauge.
The chromoelectric fields reside at the left, $L$, and the right, $R$, ends of each link $r$, and they are denoted by $E^{\rm a}(L,r)$ and $E^{\rm a}(R,r)$, respectively, where `a' is the adjoint index that takes values from ${\rm a}=1,\cdots,\,8$.
Meanwhile, the link operators are unitary $3\times 3$ matrices defined on each link $r$, with components denoted by $U^\alpha{}_\beta(r)$, where $\alpha$, $\beta =$ 1, 2, 3 are the color indices.
The electric fields and link operators are illustrated together in Fig.~\ref{subfig: KS variables}.
The canonical commutation relations are given by
\begin{eqnarray}
  [E^{\rm a}(L/R,r),E^{\rm b}(L/R,r')] &=& \delta_{rr'} \sum_{{\rm c}=1}^8 i\,f^{\rm abc}\,E^{\rm c}(L/R,r),
  \label{eq: commutation-ELa-Elb}\\
  \left[E^{\rm a}(L,r),U(r')\right] &=& -\delta_{rr'}\,T^{\rm a} U(r)\,,\quad [E^{\rm a}(R,r),U(r')]= \delta_{rr'}\,U(r)T^{{\rm a}},
  \label{eq: commutation-ELa-U}\\
  \left[U^\alpha{}_\beta(r),U^\gamma{}_\eta(r')\right] &=& [U^\alpha{}_\beta(r),U^{\dagger\gamma}{}_\eta(r')]=0,
  \label{eq: commutation-U-U}
\end{eqnarray}
where $f^{\rm abc}$ are the structure constants for SU(3), $T^{\rm a}=\frac{\lambda^{\rm a}}{2}$ with $\lambda^{\rm a}$ being the Gell-Mann matrices, and $L/R$ indicates that the relation holds individually for both $L$ and $R$ sides of the link.
The left and right chromoelectric fields on a link are connected by parallel transport, that is they are related to each other via the relation
\begin{equation*}
    E^{\rm a}(R,r)T^{\rm a}=-U^\dagger(r)E^{\rm b}(L,r)T^{\rm b} U(r).
\end{equation*}
As a consequence, the quadratic Casimirs on either side, $E^2(r)$, must be equal:
\begin{equation}
    E^2(r)\equiv \sum_{\rm a} E^{\rm a}(L,r)E^{\rm a}(L,r)=\sum_{\rm a} E^{\rm a}(R,r)E^{\rm a}(R,r).
    \label{eq: electric field casimir constraint}
\end{equation}
The one-flavor fermionic matter is expressed by a staggered fermionic field at each lattice site $r$, $\psi_\alpha(r)$, where $\alpha$ is again the color index. It obeys the fermionic anti-commutation relations given by
\begin{equation}
    \{\psi_\alpha^\dagger(r),\psi^\dagger_\beta(r')\}=\{\psi^\alpha(r),\psi^\beta(r')\}=0,
    \qquad \{\psi^\alpha(r),\psi^\dagger_\beta(r')\}=\delta^{\alpha}_\beta\,\delta_{rr'}. 
    \label{eq: ferm_anticomm}
\end{equation}
The KS Hamiltonian for an $N$-site lattice is then given by
\begin{align}
  H&= H_M+H_E+ H_{I} \nonumber\\ 
  &= \mu \sum_{r=1}^{N} (-1)^r \psi^\dagger_\alpha(r) \psi^\alpha (r) + \sum_{r=1}^{N'}  E^2(r) + x \sum_{r=1}^{N'} \left[\psi^\dagger_\alpha(r)\, U^\alpha{}_{\beta}(r)\, \psi^\beta(r+1) + {\rm H.c.} \right] ,
  \label{eq: KS-ham}
\end{align}
where $H_M$ is the matter self-energy, $H_E$ is the chromoelectric energy, and $H_{I}$ is the gauge-matter interaction.
Here, repeated indices are summed over, and $N'=N-1$ $(N)$ for open (periodic) boundary conditions. 
The form presented here differs from Eq.~\eqref{eq: KS-ham SU(2)} as it is obtained by rescaling the Hamiltonian in Eq.~\eqref{eq: KS-ham SU(2)} by $\frac{2}{g^2a}$ such that the dimensionless couplings $\mu = \frac{2m}{g^2a}$ and $x=\frac{1}{(ga)^2}$ are related to the fermion mass $m$ and the coupling constant $x$, respectively; see Ref.~\cite{Hamer:1981yq} for details.

The Hamiltonian in Eq.~\eqref{eq: KS-ham} is accompanied by a constraint structure, known as the
Gauss's law constraints, given by
\begin{equation}
  G^{\rm a}(r)|\Psi\rangle=0 \quad \forall\, {\rm a}\,,r.
  \label{eq: gauss-law}
\end{equation}
where $G^{\rm a}(r)$ are the full generators of the local gauge transformations:
\begin{equation}
  G^{\rm a}(r)= E^{\rm a}(L,r)+E^{\rm a}(R,r-1)+ \psi^\dagger_\alpha(r)\, \left(T^{\rm a}\right)^\alpha{}_\beta\, \psi^\beta(r).
  \label{eq: gauss-op}
\end{equation}
The gauge invariance of Hamiltonian means that it commutes with $G^{\rm a}(r)$ for all ${\rm a}$, $r$. The Hamiltonian generates the dynamics between states of the Hilbert space, $|\Psi\rangle$, that satisfy the Gauss's laws.

The convenient choice of basis for Kogut-Susskind formulation is the strong coupling basis, where the gauge-invariant ground state is defined as having zero electric flux, no particles, and no antiparticles. The gauge-matter interaction Hamiltonian $H_{I}$, when applied to this state, builds up the physical Hilbert space of the theory.
However, the physical Hilbert space is only a really small subspace of the full gauge theory Hilbert space that is created by applying individual link operators on the strong coupling vacuum. The gauge-invariant Hilbert space contains non-local Wilson loops, non-local strings, and hadrons.
Such a space of states, conventionally known as the ``loop space,'' albeit gauge invariant, is spanned by an over-complete and non-local basis that is not that useful for practical computation~\cite{Mathur:2007nu}.
However, the reformulation of the theory in terms of prepotentials leads to a local and complete basis for the physical Hilbert space of the theory \cite{Mathur:2004kr}.
\subsection{Prepotential formulation
\label{subsec: prepotential}
}
\noindent
The prepotential formulation of gauge theories, that started developing more than a decade ago for SU(2), gives a reformulation of the same Kogut-Susskind theory allowing one to construct a local and complete gauge invariant basis.
Generalization of the prepotential formulation for SU(3) as well as for arbitrary SU(N) is an involved process~\cite{Anishetty:2009ai, Anishetty:2009nh, Raychowdhury:2013rwa}.
This subsection briefly summarize the key results for SU(3).

Recall that prepotentials are the Schwinger boson (or harmonic oscillator) operators, in the fundamental representation of the gauge group.
For the SU(2) gauge group, which is a rank one group with a two dimensional fundamental representation, the set of prepotential operators consists of one Schwinger boson doublet at the $L$ and $R$ ends of each link.
Similarly, the prepotential formulation of the SU(3) gauge group, which is a rank-two group with a three-dimensional fundamental representation, requires two independent Schwinger boson triplets at both ends of each link.
They are denoted by $a^\dagger_\alpha(L,r)$ and $b^{\dagger\alpha}(L,r)$ for Schwinger bosons on the $L$ end of link $r$, and $a^\dagger_\alpha(R,r)$ and $b^{\dagger\alpha}(R,r)$ for Schwinger bosons on the $R$ end of the link $r-1$.
Here, $\alpha$ and $\beta$ are the color indices that take integer values from 1 to 3. Note that, whereas bosonic variables of the KS formulation are identified by their links, the Schwinger boson definitions are focused on sites.
The SU(3) Schwinger bosons obey the ordinary bosonic commutation relations:
\begin{align}
    \big[a^\alpha(s,r),a^{\alpha'}(s',r')\big]&=\big[b_\alpha(s,r),b_{\alpha'}(s',r')\big]=0,
    \label{eq: SU3 schwinger boson aa bb commutation}\\
    \big[a^\alpha(s,r),b_{\alpha'}(s',r')\big]&=\big[a^\alpha(s,r),b^{\dagger\alpha'}(s',r')\big]=0,
    \label{eq: SU3 schwinger boson ab commutation}\\
    \big[a^\alpha(s,r),a^\dagger_{\alpha'}(s',r')\big]&=\big[b_{\alpha'}(s,r),b^{\dagger\alpha}(s',r')\big]=\delta_\alpha^{\alpha'}\delta_{ss'}\delta_{rr'}
    \label{eq: SU3 schwinger boson non-zero commutation},
\end{align}
where $s$ and $s'$ can each take values $L$ or $R$.

Following the Schwinger boson construction for SU(3) generators, the left and right chromoelectric fields on a link could be defined in terms of the prepotentials as
\begin{subequations}
\label{eq: E in terms of a and b}
\begin{align}
    E^{\rm a}(L,r)&= a^\dagger_\alpha(L,r)\,\big(T^{\rm a}\big)^\alpha{}_\beta \, a^\beta(L,r) - b^{\dagger\alpha}(L,r)\,\big(T^{*\rm a}\big)_\alpha{}^\beta\, b_\beta(L,r) , \\
    E^{\rm a}(R,r-1)&= a^\dagger_\alpha(R,r)\,\big(T^{\rm a}\big)^\alpha{}_\beta \, a^\beta(R,r) - b^{\dagger\alpha}(R,r)\,\big(T^{*\rm a}\big)_\alpha{}^\beta\, b_\beta(R,r) ,
\end{align}
\end{subequations}
with $T^{\rm a}$ and the $L/R$ notation being the same as in Eqs.~\eqref{eq: commutation-ELa-Elb} and~\eqref{eq: commutation-ELa-U}.
It was shown in~\cite{Anishetty:2009ai,Anishetty:2009nh} that this particular set of Schwinger bosons are not suitable for the construction of the gauge theory Hilbert space.
To appreciate this fact, it is instructive to look at a simpler example of SU(2) gauge bosons.

In the case of the SU(2) gauge group, only one doublet of Schwinger bosons, $a^\dagger_\alpha$ with $\alpha=1,2$ is needed to represent the SU(2) irreps, as mentioned before.
An SU(2) irrep corresponding to angular momentum $j$ is constructed in terms of these Schwinger bosons as
\begin{equation}
    |j\rangle_{\vec{\alpha}} = \mathcal N a^\dagger_{\alpha_1}\ldots a^\dagger_{\alpha_{2j}}|0\rangle,
    \label{eq: SU2-j-state}
\end{equation}
where $\vec{\alpha} = (\alpha_1,\cdots,\alpha_{2j})$, $\mathcal N$ is the normalization factor, and the $j=0$ state, $|0\rangle$, is defined by the condition $a_1 |0\rangle=a_2 |0\rangle=0$.
The indices $\alpha_{i}$ for $i=1,2, \dots, 2j$ can take any value between 1 and 2, and their values determine the azimuthal quantum number $m$ of that irrep. (We have left the azimuthal quantum number in Eq.~\eqref{eq: SU2-j-state} implicit for brevity.)
A similar monomial of the Schwinger bosons $a^\dagger_{\alpha}$ and $b^{\dagger\beta}$ will likewise create representations of SU(3), however they are generally not irreducible~\cite{mukunda1965tensor,Chaturvedi:2002si}.
The simplest case of the $(1,1)$ irrep of SU(3) is naively constructed as $a^\dagger_{\alpha} b^{\dagger\beta}|0,0\rangle$, however its proper construction is instead
\begin{equation}
   |1,1\rangle_\alpha^\beta= a^\dagger_{\alpha} b^{\dagger\beta}|0,0\rangle - \frac{\delta_{\alpha}^{\beta}}{3} (a^\dagger \cdot b^\dagger)|0,0\rangle,  \label{eq: SU3-11-irrep}
\end{equation}
such that the irrep is traceless, which is a fundamental property of any irrep. Above, $\alpha\,,\beta=1, 2, 3$ are the color indices,  $a^\dagger\cdot b^\dagger \equiv \sum_{\gamma=1}^3 a^\dagger_\gamma b^{\dagger\gamma}$, and the $(0,0)$ irrep state, $|0,0\rangle$, satisfies $a^\alpha|0,0\rangle = b_{\alpha}|0,0\rangle=0$ for all values of $\alpha$.
For a general irrep $(P,Q)$, one has to extract out all the traces from the monomial state $a^\dagger_{\alpha_1}\cdots a^\dagger_{\alpha_P} b^{\dagger\beta_1}\cdots b^{\dagger\beta_Q}|0,0\rangle $ to satisfy the tracelessness condition (see, for example, equation (35) of Ref.~\cite{Mathur:2000sv}).
Such a traceless construction is cumbersome to use, and on top of that there is also a multiplicity problem because $(a^\dagger\cdot b^{\dagger})^{\rho}\displaystyle|P,Q\rangle_{\vec{\alpha}}^{\vec{\beta}}$ for any positive integer $\rho$ transforms in the same way under SU(3) as the $(P,Q)$ irrep~\cite{Chaturvedi:2002si}.
A novel solution that uses the naive Schwinger bosons discussed thus far to construct a set of modified Schwinger bosons, called irreducible Schwinger bosons (ISBs), has previously been proposed in the Ref.~\cite{Anishetty:2009ai}, providing a monomial construction of traceless SU(3) irreps that solves the multiplicity problem.

The ISBs are constructed in terms of the naive SU(3) Schwinger bosons as
\begin{eqnarray}
    A^{\dagger}_{\alpha}&\equiv& a^{\dagger}_{\alpha}-\frac{1}{\hat{N}_a+ \hat{N}_b+1}(a^\dagger\cdot b^\dagger)b_{\alpha},
    \label{eq: Adagg-def} \\
    B^{\dagger\alpha}&\equiv& b^{\dagger\alpha}-\frac{1}{\hat{N}_a+\hat{N}_b+1}(a^\dagger\cdot b^\dagger)a^{\alpha}.
    \label{eq: Bdagg-def}
\end{eqnarray}
Above,
\begin{equation}
    \hat{N}_a \equiv a^{\dagger}\cdot a = \sum_{\alpha=1}^{3} a^{\dagger}_\alpha\,a^{\alpha} \quad\text{and}\quad \hat{N}_b \equiv b^{\dagger}\cdot b = \sum_{\beta=1}^{3} b^{\dagger\beta}\,b_{\beta} 
    \label{eq: prepotential number operators}
\end{equation}
are the total number operators for $a$-type and $b$-type Schwinger bosons, respectively.
Using ISBs, the state in Eq.~\eqref{eq: SU3-11-irrep} may now be expressed as a monomial state: $|1,1\rangle^\beta_\alpha= A^\dagger_{\alpha} B^{\dagger\beta}|0,0\rangle$.
In general, the monomial states constructed out of the ISBs in Eq.~\eqref{eq: Adagg-def} and~\eqref{eq: Bdagg-def} as
\begin{equation}
    |P,Q\rangle_{\vec{\alpha}}^{\vec{\beta}}= \mathcal{N} A^\dagger_{\alpha_1}\ldots A^\dagger_{\alpha_P} B^{\dagger\beta_1}\ldots B^{\dagger\beta_Q}|0\rangle
    \label{eq: SU3-irrep}
\end{equation}
are indeed the traceless SU(3) irreps.
Here $\mathcal N$ corresponds to the normalization factor, the indices $\alpha_{i}$ with $i=1, \cdots, P$ and  $\beta_{j}$ with $j=1, \cdots, Q$ can take integer values between 1 to 3, and $\alpha_{i}$ and $\beta_{j}$ determine the isospin and hypercharge of the irrep state~\cite{mukunda1965tensor,Chaturvedi:2002si}.
The monomial states constructed in Eq.~\eqref{eq: SU3-irrep} also solve the multiplicity problem as the states satisfy 
\begin{equation}
    A^\dagger\cdot B^\dagger\, |P,Q\rangle_{\vec{\alpha}}^{\vec{\beta}} =0  \quad\text{and}\quad A\cdot B\, |P,Q\rangle_{\vec{\alpha}}^{\vec{\beta}} =0,
    \label{eq: ISB multiplicity constraint}
\end{equation}
which ensures that they are restricted to the $\rho=0$ subspace or equivalently the kernel of the operator $a\cdot b$, see Ref.~\cite{Anishetty:2009nh} for details.
This implies that the operators $A^\dagger\cdot B^\dagger$ and $A\cdot B$ are effectively the null operators within the Hilbert space spanned by states in Eq.~\eqref{eq: SU3-irrep}.
This leads to the following modified commutation relations for ISBs:
\begin{eqnarray}
    &&[A^\alpha, A^\dagger_\beta] \simeq \left( \delta^\alpha_{\beta}-\frac{1}{\hat{N}_a+\hat{N}_b+2}B^{\dagger\alpha}B_\beta \right),
    \label{eq: AAdagg-commutator}\\
    &&[B_\alpha, B^{\dagger\beta}] \simeq  \left( \delta_\alpha^{\beta}-\frac{1}{\hat{N}_a+\hat{N}_b+2}A^{\dagger}_{\alpha}A^\beta \right),
    \label{eq: BBdagg-commutator}\\
    && [A^\alpha,B^{\dagger\beta}] \simeq   -\frac{1}{\hat{N}_a+\hat{N}_b+2}B^{\dagger\alpha}A^\beta, 
    \label{eq: ABdagg-commutator}\\
    && [B_\alpha,A^\dagger_\beta] \simeq   -\frac{1}{\hat{N}_a+\hat{N}_b+2}A^{\dagger}_\alpha B_\beta,
    \label{eq: AdaggB-commutator}
\end{eqnarray}
along with
\begin{eqnarray}
    [ A^\dagger_\alpha, A^\dagger_{\beta}]=[ A^\alpha, A^{\beta}]=[B_\alpha,B_\beta]=[ B^{\dagger\alpha}, B^{\dagger\beta}]=[A^\alpha,B_\beta]=[A^\dagger_\alpha, B^{\dagger\beta} ]=0,
    \label{eq: zero-commutators}
\end{eqnarray}
where $\simeq$ indicates that the above set of commutation relations are valid within the vector subspace spanned by SU(3) irreps defined in Eq.~\eqref{eq: SU3-irrep}.
Note that, in the same subspace, the number operators for $A$-type and $B-$type ISBs satisfy
\begin{equation}
    \hat{N}_A \equiv A^\dagger\cdot A \simeq \hat{N}_a \quad \text{and} \quad \hat{N}_B \equiv B^\dagger\cdot B \simeq \hat{N}_b\,.
    \label{eq: NA Na and NB Nb equivalence}
\end{equation}

One can then proceed towards the prepotential framework for SU(3) lattice gauge theory as a straightforward extension of the SU(2) case using the SU(3) ISBs given in Eqs.~\eqref{eq: Adagg-def} and~\eqref{eq: Bdagg-def}.
Sets of ISB prepotential operators  $A^\dagger_\alpha(L,r)$ and $B^{\dagger\alpha}(L,r)$ are introduced for the $L$ end of link $r$, and $A^\dagger_\alpha(R,r)$ and $B^{\dagger\alpha}(R,r)$ are introduced for the $R$ end of link $r-1$.
These ISBs with different arguments commute, and they obey the commutation relations in Eqs.~\eqref{eq: AAdagg-commutator}-~\eqref{eq: zero-commutators} only if they have the same direction and position arguments. 
The electric fields in terms of ISBs follow the same format that was used in the naive definitions of Eqs.~\eqref{eq: E in terms of a and b}:
\begin{subequations}
\begin{align}
    E^{\rm a}(L,r)= A^\dagger_\alpha(L,r)\,\big(T^{\rm a}\big)^\alpha{}_\beta \, A^\beta(L,r) - B^{\dagger\alpha}(L,r)\big(T^{*\rm a}\big)_\alpha{}^\beta B_\beta(L,r). \\
    E^{\rm a}(R,r-1)= A^\dagger_\alpha(R,r)\,\big(T^{\rm a}\big)^\alpha{}_\beta \, A^\beta(R,r) - B^{\dagger\alpha}(R,r)\big(T^{*\rm a}\big)_\alpha{}^\beta B_\beta(R,r).
\end{align}
\label{eq: E in terms of A and B}
\end{subequations}
In order for this redefinition to satisfy Eq.~\eqref{eq: electric field casimir constraint}, the prepotential number operators~\eqref{eq: NA Na and NB Nb equivalence} defined at each end of the link must be related to each other by either $(\hat{N}_A(L,r) - \hat{N}_A(R,r+1)) \ket{\Psi} = (\hat{N}_B(L,r) - \hat{N}_B(R,r+1)) \ket{\Psi} = 0$, or
\begin{align}
    (\hat{N}_A(L,r) - \hat{N}_B(R,r+1)) \ket{\Psi} =
    (\hat{N}_B(L,r) - \hat{N}_A(R,r+1)) \ket{\Psi} = 0
    \label{eq: AGL in ISB}
\end{align}
for any state $\ket{\Psi}$ belonging to the Hilbert space created by ISBs.
However, the fact that the gauge theory Hilbert space is built up by the action of link operators $U^\alpha{}_\beta$, transforming as in Eq.~\eqref{eq: commutation-ELa-U} is only consistent with the second choice of the constraint.
Hence, the ISB construction of the link operator is defined to satisfy
\begin{align}
    [ U^\alpha{}_\beta(r) , \hat{N}_A(L,r) - \hat{N}_B(R,r+1)] = [ U^\alpha{}_\beta(r) , \hat{N}_B(L,r) - \hat{N}_A(R,r+1)] &= 0 .
    \label{eq: link operator AGL}
\end{align}
In the prepotential formulation of gauge theories, the constraints in Eq.~\eqref{eq: AGL in ISB} are known as the Abelian Gauss's law constraints of the theory, and they are responsible for the non-locality of the gauge-invariant (or physical) degrees of freedom of the theory. 

The link operator satisfying Eq.~\eqref{eq: link operator AGL} can be expressed in terms of the prepotential bosons as~\cite{Anishetty:2009nh}
\begin{align}
     U^\alpha{}_{\beta}(r)&= B^{\dagger\alpha}(L,r) \,\eta(r) \, A^{\dagger}_{\beta}(R,r+1) + A^{\alpha}(L,r)\, \theta(r)\, B_{\beta}(R,r+1)\nonumber\\
     &\quad + \bigl(A^{\dagger}(L,r) \wedge B(L,r)\bigl)^\alpha \, \delta(r) \, \bigl(B^{\dagger}(R,r+1)\wedge A(R,r+1)\bigr)_{\beta},
    \label{eq: link operator in prepotential}
\end{align}
where $(A^\dagger\wedge B)^\alpha \equiv \epsilon^{\alpha\gamma\delta}A^\dagger_{\gamma}B_{\delta}$ and $(B^\dagger\wedge A)_\beta \equiv \epsilon_{\beta\gamma\delta}B^{\dagger\gamma}A^{\delta}$, and the coefficients $\eta(r)$, $\theta(r)$, $\delta(r)$ are fixed by the unitarity and unit-determinant conditions on $U(r)$ given by $U^\dagger(r) U(r)= \mathds{1}_{3\times 3}$ and $\det U(r)=\mathds{1}$, respectively.
It is important to note that just like in the prepotential formulation of the SU(2) gauge group, the link operator matrix $U(r)$ for the SU(3) gauge group can be written as a product of two SU(3) matrices defined in terms of the $(L,r)$ and $(R,r+1)$ prepotential operators~\cite{Anishetty:2009ai,Anishetty:2009nh}. 

The motivation behind the prepotential formulation of lattice gauge theories is to be able to construct a local and minimal set of gauge-invariant operators along with an orthonormal Hilbert space for the pure Yang-Mills theory.
For the one-dimensional lattice, the gauge theory becomes dynamical in the presence of matter fields.
The next subsection will discuss the coupling of prepotentials to staggered fermions.
\subsection{Irreducible prepotentials for SU(3) coupled to staggered matter
\label{subsec: prepotential-with-matter}
}
\begin{table}[t]
    \renewcommand{\arraystretch}{1.8}
    \centering
    \begin{tabular}{C{3 cm} | C{3 cm}}
        \hline
        Fundamental & Anti-fundamental \\
        $(1,0)$ or \textbf{3} & $(0,1)$ or \textbf{3*}\\
        \hline
        \hline
        $A^\dagger_{\alpha}(\obar,r)$ & 
        $A^\alpha(\obar,r)$\\
        $A^\dagger_{\alpha}(1,r)$ &
        $A^\alpha(1,r)$ \\
        $B_{\alpha}(\obar,r)$ & 
        $B^{\dagger\alpha}(\obar,r)$\\
        $B_{\alpha}(1,r)$ &
        $B^{\dagger\alpha}(1,r)$\\
        $\psi^\dagger_\alpha(r)$ &
        $\psi^{\alpha}(r)$ \\
        \hline
    \end{tabular}
    \caption{Prepotential variables and their irreps under the SU(3) gauge group at any site $r$. Equation~\eqref{eq: SU3-irrep} defines the irreps of $A^\dagger_\alpha$ and $B^{\dagger\alpha}$ ISBs. The choice of matter field irreps to be fundamental or anti-fundamental is arbitrary and it only affects the construction of gauge invariant singlets in Sec.~\ref{subsubsec: LSH-operators}; we have chosen the $\psi^\dagger(r)$ triplet to transform as the fundamental irrep under the SU(3) gauge transformations.\label{tab: prepotential irreps}}
\end{table}
To couple the prepotential operators with fermionic matter, we consider the KS staggered fermionic matter field, $\psi^\alpha(r)$, at each lattice site $r$.
It satisfies the anticommutation relations given in Eq.~\eqref{eq: ferm_anticomm} and transforms as a triplet under the SU(3) gauge group.
The matter field $\psi^\alpha(r)$ along with the prepotential operators, $A^\alpha(L,r)$, $B_{\alpha}(L,r)$, $A^\alpha(R,r)$ and $B_{\alpha}(R,r)$, forms the set of local prepotential variables that completely describe a one-dimensional SU(3) lattice gauge theory. 
Given the site-centric labeling of ISB degrees of freedom, and in anticipation of higher-dimensional generalizations, we relabel the $L$ and $R$ ends of links with $1$ and $\obar$, respectively.
This choice of labels is not standard, so to familiarize readers with it, we contrast it against the conventional notation in Fig.~\ref{subfig: prepotential variables} by illustrating the prepotential variables with $1$ and $\obar$ labels next to KS variables with $L$ and $R$ labels.

The prepotential variables transform as triplets under the SU(3) gauge group either as fundamental, $(1,0)$, or anti-fundamental, $(0,1)$, irreps. Equation~\eqref{eq: SU3-irrep} implies that oscillators of type $A^\dagger_\alpha$ transform as the $(1,0)$ irrep and the oscillators of type $B^{\dagger\alpha}$ transform as $(0,1)$ irrep, while their conjugates transform as $(0,1)$ and $(1,0)$ irreps, respectively.
For the fermionic matter field, we have taken, without loss of generality,  $\psi^\dagger_\alpha(r)$ and $\psi^\alpha(r)$ to transform as $(1,0)$ and $(0,1)$ irreps, respectively.
Thus, the fermionic matter at each lattice site is expressed in terms of fundamental, and not anti-fundamental, matter fields.
The complete set of fundamental and anti-fundamental prepotential variables present at a lattice site are tabulated in Table~\ref{tab: prepotential irreps}.
One can use the various fundamental and anti-fundamental fields to construct a complete set of local gauge singlets via symmetric or antisymmetric contractions with the Kronecker delta, $\delta_\alpha^\beta$, or Levi-Civita symbols, $\epsilon_{\alpha\beta\gamma}$ or $\epsilon^{\alpha\beta\gamma}$, respectively.
We focus on a subset of these local singlets that acts as the on-site building blocks of the Kogut-Susskind Hamiltonian in Eq.~\eqref{eq: KS-ham} re-written in terms of prepotentials variables.

The mass Hamiltonian, $H_M$, in the staggered formulation as given in Eq.~\eqref{eq: KS-ham} is carried over unchanged to the prepotential formulation.
The electric Hamiltonian, $H_E$, consisting of the quadratic Casimir operator defined at each end of a link is given by replacing chromoelectric fields in Eq.~\eqref{eq: KS-ham} with their corresponding prepotential definition given in Eqs.~\eqref{eq: E in terms of A and B}.
This leads to
\begin{align}
    H_E &= \sum_{r=1}^{N'} H_E(r) \nonumber\\
    &= \sum_{r=1}^{N'} \bigg[ \frac{1}{3} \left( \hat{N}_A(1,r)^2 + \hat{N}_B(1,r)^2 + \hat{N}_A(1,r) \hat{N}_B(1,r) \right) + \hat{N}_A(1,r) + \hat{N}_B(1,r) \bigg].
    \label{eq: HE in prepotential}
\end{align}
Note that, because of the equality of Casimirs in Eq.~\eqref{eq: electric field casimir constraint}, it is arbitrary whether $H_E$ is defined in terms of $1$-side or $\obar$-side ISBs.

The dynamical part of the Hamiltonian is governed by the interaction between fermions and gauge fields denoted by $H_I$ in Eq.~\eqref{eq: KS-ham}.
It contains local gauge singlet terms involving the gauge link operator $U(r)$, and the staggered matter field $\psi(r)$.
The prepotential reformulation of $H_I$ is obtained by replacing $U(r)$ in the term $\psi^\dagger_\alpha(r)\, U^\alpha{}_{\beta}(r)\, \psi^\beta(r+1)$ with its expression in terms of prepotential variables given in Eq.~\eqref{eq: link operator in prepotential}.
This leads to
\begin{align}
    \psi^\dagger_\alpha(r)\, U^\alpha{}_{\beta}(r)\, \psi^\beta(r+1) &=
    \left[\psi^\dagger_\alpha\, B^{\dagger\alpha}(1) \; \eta(1) \right]_r  \left[ \eta(\obar) \; \psi^\beta \, A^\dagger_\beta(\obar) \right]_{r+1} \nonumber\\
    & \quad + \,\left[\psi^\dagger_\alpha \, A^\alpha(1) \; \theta(1) \right]_r  \left[ \theta(\obar) \; \psi^\beta \, B_\beta(\obar) \right]_{r+1}\nonumber\\
    & \quad + \,\left[\psi^\dagger_\alpha \, (A^\dagger(1)\wedge B(1))^\alpha  \; \delta(1) \right]_r \left[ \delta(\obar) \; \psi^\beta \, (B^{\dagger}(\obar) \wedge A(\obar))_\beta\right]_{r+1},
    \label{eq: HI local term in prepotential}
\end{align}
where we have used the shorthand notation $[ \ ]_r$ for the position labels for brevity.
Above, each of the $\eta(r)$, $\theta(r)$ and $\delta(r)$ in Eq.~\eqref{eq: link operator in prepotential} have been factored into a product of two operators, one for each end of the link, as
\begin{equation}
    \eta(r) = \eta(1,r)\; \eta(\obar,r+1), \quad \theta(r) = \theta(1,r)\; \theta(\obar,r+1), \quad \text{and}\quad \delta(r) = \delta(1,r)\;\delta(\obar,r+1).
    \label{eq: eta theta delta factorization}
\end{equation}
The conditions $U^\dagger(r) U(r)= \mathds{1}_{3\times 3}$ and $\det U(r)=\mathds{1}$ on $U(r)$ in Eq.~\eqref{eq: link operator in prepotential} determine the form of these decomposed operators to be
\begin{align}
    \eta(1,r) &= \frac{1}{\sqrt{B(1,r)\cdot B^\dagger(1,r)}},  \quad\quad \eta(\obar,r) = \frac{1}{\sqrt{A^\dagger(\obar,r)\cdot A(\obar,r)}},
    \label{eq: eta 1 and 1 bar expressions}\\
    \theta(1,r) &= \frac{1}{\sqrt{A^\dagger(1,r)\cdot A(1,r)}},  \quad\quad \theta(\obar,r) = \frac{1}{\sqrt{B(\obar,r)\cdot B^\dagger(\obar,r)}},
    \label{eq: theta 1 and 1 bar expressions}\\
   \delta(1/\obar,r) &= \frac{1}{\sqrt{\big(A^\dagger(1/\obar,r)\cdot A(1/\obar,r) +2 \big)\;B^\dagger(1/\obar,r)\cdot B(1/\obar,r)}}.
    \label{eq: delta 1 and 1 bar expressions}
\end{align}
where $1/\obar$ indicates that the expression holds individually for both $1$ and $\obar$ direction arguments.

Putting everything together, the prepotential reformulation of the KS Hamiltonian in Eq.~\eqref{eq: KS-ham} is given by
\begin{align}
    H&= H_M+H_E+ H_{I} \nonumber\\ 
    &=\mu \sum_{r=1}^{N} (-1)^r \psi^\dagger(r) \cdot \psi(r) \nonumber \\
    &\quad + \sum_{r=1}^{N'} \left( \frac{1}{3} \Big( \hat{N}_A(1,r)^2 + \hat{N}_B(1,r)^2 + \hat{N}_A(1,r) \hat{N}_B(1,r) \Big) + \hat{N}_A(1,r) + \hat{N}_B(1,r) \right) \nonumber\\
    &\quad + x\sum_{r=1}^{N'} \Bigl(
    \left[\psi^\dagger\cdot B^{\dagger}(1) \; \eta(1) \right]_r  \left[ \eta(\obar) \; \psi \cdot A^\dagger(\obar) \right]_{r+1} \nonumber  + \,\left[\psi^\dagger \cdot A(1) \; \theta(1) \right]_r  \Bigl[ \theta(\obar) \; \psi \cdot B(\obar) \Bigr]_{r+1}\nonumber\\
    &\quad \quad + \,\left[\psi^\dagger \cdot A^\dagger(1) \wedge B(1)  \; \delta(1) \right]_r \left[ \delta(\obar) \; \psi \cdot B^{\dagger}(\obar) \wedge A(\obar)\right]_{r+1} + \mathrm{H.c.}\Bigl).
    \label{eq: KS ham in prepotential final}
\end{align}
Importantly, the above Hamiltonian is written explicitly in terms of SU(3)-invariant contractions of site-local triplets.
It is also easy to see that it conserves the Abelian Gauss's laws.
In the next section, we will define a suite of SU(3)-invariant loop-string-hadron operators to ultimately derive the LSH Hamiltonian corresponding to Eq.~\eqref{eq: KS ham in prepotential final} in terms of them.

\section{SU(3) LSH framework derived from prepotentials
\label{sec: LSH-framework}
}

With all of the machinery established for both SU(3) irreducible Schwinger bosons, a staggered fermion field, and a Hamiltonian expressed in terms of manifestly SU(3)-invariant contractions at sites, the groundwork now exists to derive the loop-string-hadron formulation of the same theory.
The LSH approach ultimately replaces the SU(3)-covariant fields used in the ISB formulation with fields that are intrinsically SU(3)-invariant.
This reformulation has the distinct advantage of having only local Abelian constraints to deal with (preserving the locality of the Hamiltonian that is otherwise lost when total gauge-fixing is applied), while substantially reducing the number of bosonic degrees of freedom.

Section~\ref{subsubsec: LSH-operators} provides a heuristic discussion of gauge-invariant operators and excitations, which is intended to motivate the local basis that is established in Sec.~\ref{subsec: LSH-basis}.
In Sec.~\ref{subsubsec: LSH operator factorization}, with the local basis in hand, the operator factorizations are obtained that make matrix elements of SU(3)-singlet operators explicit.
Sections~\ref{subsubsec: LSH Ham construction} and~\ref{subsubsec: LSH operators in terms of basis and AGL} provide a summary of the LSH formulation suitable for the whole lattice, including the Hamiltonian and a global-basis implementation of the operators that appear in the Hamiltonian.
\subsection{Site-local SU(3) singlets and LSH excitations
\label{subsubsec: LSH-operators}
}

An essential feature of the Hamiltonian for ISBs plus fermions is its construction in terms of site-local, SU(3)-invariant singlets, formed by suitable contractions of the ISBs and/or quark fields with the invariant tensors $\delta$ and $\epsilon$.
From now until Sec.~\ref{subsubsec: LSH Ham construction}, the focus of the discussion will be on operators all living at a given site $r$, and hence, the position arguments are suppressed for brevity.

Mass terms in Eq.~\eqref{eq: KS ham in prepotential final} are expressed entirely in terms of the SU(3)-invariant total fermion numbers, $ \psi^\dagger \cdot \psi$.
Similarly, the Casimir operators to either side of a site given in Eq.~\eqref{eq: electric field casimir constraint}, which are necessary for the electric Hamiltonian, are expressed entirely in terms of the SU(3)-invariant Schwinger boson occupation numbers: $\hat{N}_A(1)$, $\hat{N}_B(1)$, $\hat{N}_A(\obar)$, and $\hat{N}_B(\obar)$.
Together, the five operators
\begin{align}
    \hat{N}_{\psi} \equiv \psi^\dagger \cdot \psi, \ \ \ \hat{P}(\obar) \equiv  \hat{N}_A(\obar) ,\ \ \ \hat{Q}(\obar)\equiv\hat{N}_B(\obar) , \ \ \ \hat{P}(1) \equiv \hat{N}_B(1)  , \ \ \ \hat{Q}(1) \equiv \hat{N}_A(1) , 
    \label{eq: irrep quantum numbers}
\end{align}
furnish a local complete set of commuting observables (CSCO) for the SU(3)-invariant subspace of the ISBs plus fermions.
Note that the definitions of $\hat{P}$ and $\hat{Q}$ in terms of $A$- and $B$-type number operators are interchanged for the $1$ and $\obar$ sides of a site.
The reason for this choice will become apparent later.

The gauge-matter interaction, $H_I$, is distinct because it couples the fermions and gauge bosons, making it responsible for generating the Hilbert space of gauge-invariant states and for the dynamics among them.
In addition to the CSCO identified above, hopping terms in $H_I$ explicitly depend on the local SU(3) singlets
\begin{align}
    & \psi^\dagger \cdot B^\dagger(1) , \ \psi^\dagger \cdot A(1) , \ \psi^\dagger \cdot A^\dagger(1) \wedge B(1) , \  \psi^\dagger \cdot B^\dagger(\obar) , \ \psi^\dagger \cdot A(\obar) , \ \psi^\dagger \cdot A^\dagger (\obar) \wedge B(\obar) , \text{ and H.c.}
    \label{eq: Hinter singlets}
\end{align}
Local, observable configurations of quarks and gauge flux can be reached from the local fermionic and bosonic vacuum, a subspace on which all operators in the local CSCO evaluate to zero.
In other words, all site-local configurations can be generated by acting on the local vacuum with the SU(3) singlets from which $H$ is constructed.
It will be shown that all local configurations can be characterized in terms of two distinct SU(3)-invariant bosonic modes and three distinct SU(3)-invariant fermionic modes.
The rest of this section aims to provide an intuitive explanation for the links between different operators that are invariant under SU(3), and different fermionic and bosonic excitations that are also invariant under SU(3).
This will establish the context for a future basis ansatz that can truly encompass all potential excitations at a site.

In the Kogut-Susskind formulation and the ISB formulation with fermions, there are three quark modes per site, denoted by $\psi^\dagger_\alpha$ with $\alpha=1,2,3$ for the three quark colors.
By definition, however, all of the $\psi^\dagger_\alpha$ are gauge covariant rather than gauge invariant.
Still, the total quark number $\psi^\dagger \cdot \psi$ is a gauge-invariant quantum number and can take values from 0 to 3, suggesting that there should exist three SU(3)-invariant quark modes.
The LSH approach seeks to identify and use such excitations as elementary degrees of freedom.
The first task, then, is to identify the three SU(3)-invariant fermionic excitations.

Two important singlets that are clearly needed are those in Eq.~\eqref{eq: Hinter singlets} that involve creation operators only, namely, $\psi^\dagger \cdot B^\dagger(1)$ and $\psi^\dagger \cdot B^\dagger(\obar)$, because these can create non-trivial states when applied to the local vacuum.
Taking these to correspond with distinct modes, one may associate a fermionic creation operator $\chi_1^\dagger$ ($\chi_{\obar}^\dagger$) with $\psi^\dagger \cdot B^\dagger(1)$ ($\psi^\dagger \cdot B^\dagger(\obar)$).
One then searches for a third independent mode that can be excited from the local vacuum. Such an operator must (i) be gauge invariant, (ii) contain only creation operators, and (iii) have exactly one fermionic creation operator.
From Table~\ref{tab: prepotential irreps}, it can be deduced that $\psi^\dagger \cdot A^\dagger(1) \wedge A^\dagger(\obar)$ is the only such SU(3) singlet not yet enumerated.
To see that the $\psi^\dagger \cdot A^\dagger(1) \wedge A^\dagger(\obar)$ excitation is indeed required by the Hamiltonian, one first notes that
\begin{align}
    \{ \psi^\dagger \cdot A^\dagger(1) \wedge B(1) \, , \, \psi^\dagger \cdot B^\dagger(1) \} &= - \psi^\dagger \cdot \psi^\dagger \wedge A^\dagger(1) ,
\end{align}
implying the necessity of a two-quark excitation $\psi^\dagger \cdot \psi^\dagger \wedge A^\dagger(1)$, and then that
\begin{align}
    [ \psi^\dagger \cdot \psi^\dagger \wedge A^\dagger(1) \, , \, \psi \cdot A^\dagger(\obar) ] &= \psi^\dagger \cdot A^\dagger(\obar) \wedge A^\dagger(1) ,
\end{align}
where $\psi \cdot A^\dagger(\obar)$ is considered because it is one of the singlets identified in Eq.~\eqref{eq: Hinter singlets}.
Returning to the original set of singlets from Eq.~\eqref{eq: Hinter singlets}, the operators $\psi^\dagger \cdot A(1)$ and $\psi^\dagger \cdot A(\obar)$
are associated with $\hat{\chi}_{\obar}^\dagger$ and $\hat{\chi}_{1}^\dagger$, respectively, while both of $\psi^\dagger \cdot A^\dagger(1) \wedge B(1)$ and $\psi^\dagger \cdot A^\dagger(\obar) \wedge B(\obar)$ are associated with $\hat{\chi}_0^\dagger$;
these associations will be validated later, in Sec.~\ref{subsubsec: LSH operator factorization}.

In addition to the fermionic modes, there must be two bosonic excitations of the form $A^\dagger(\obar) \cdot B^\dagger(1)$ and $B^\dagger(\obar) \cdot A^\dagger(1)$ because of the relations
\begin{align}
    \{ \psi^\dagger \cdot B^\dagger(1) , \psi \cdot A^\dagger(\obar) \} &= A^\dagger(\obar) \cdot B^\dagger(1) , \\
    \{ \psi^\dagger \cdot B^\dagger(\obar) , \psi \cdot A^\dagger (1)\} &= B^\dagger(\obar) \cdot A^\dagger(1) .
\end{align}
Note that $A^\dagger(\obar) \cdot B^\dagger(1)$ [$B^\dagger (\obar)\cdot A^\dagger(1)$] increases $\hat{P}(\obar)$ and $\hat{P}(1)$ [$\hat{Q}(\obar)$ and $\hat{Q}(1)$] by one unit, and will hence be associated with a bosonic degree of freedom $\hat{n}_P$ [$\hat{n}_Q$].
Beyond these two excitations, there are no other independent, purely bosonic SU(3)-invariant excitations.
To support this claim, one may use counting arguments to deduce there are at most two unconstrained bosonic degrees of freedom:
(i) In the Kogut-Susskind formulation, the states of the adjacent link ends can each be characterized in terms of two Casimirs and three additional quantum numbers (hypercharge, isospin, and third component of isospin), giving a total of ten variables that are then subject to eight non-Abelian Gauss's law constraints.
(ii) In the prepotential plus fermions formulation, the two triplets on either side of a site add up to 12 bosonic occupation numbers, which are then subject to
eight components of Gauss's law and the two additional constraints $A^\dagger(1) \cdot B^\dagger(1) \simeq 0$ and $A^\dagger(\obar) \cdot B^\dagger(\obar) \simeq 0$.
At a more practical level, and in contrast to the counting arguments, one may alternatively note that there are only four possible bilinear contractions that can be made out of the bosonic creation operators $A^\dagger_\alpha(1)$, $A^\dagger_\alpha(\obar)$, $B^{\dagger \, \alpha}(1)$, and $B^{\dagger \, \alpha}(\obar)$.
Two of these bilinears are the bosonic excitations identified above, while the other two, $A^\dagger(1) \cdot B^\dagger(1)$ and $A^\dagger (\obar)\cdot B^\dagger(\obar)$, are equivalent to zero with the irreducible Schwinger boson construction.
One could also imagine constructing trilinears by contractions with the $\epsilon$ tensor, but any non-trivial contraction would have to involve at least one annihilation triplet.

To summarize, the operators appearing directly in the Hamiltonian can be associated with a variety of purely-creation SU(3) singlets with the following associations:
\begin{align*}
    \psi^\dagger \cdot B^\dagger(1) , \ \psi^\dagger \cdot A(\obar) \ &: \quad \text{apply } \hat{\chi}_1^\dagger , \\
    \psi^\dagger \cdot B^\dagger(\obar) , \ \psi^\dagger \cdot A(1) \ &: \quad \text{apply } \hat{\chi}_{\obar}^\dagger , \\
    \psi^\dagger \cdot A^\dagger (1)\wedge B(1) , \ \psi^\dagger \cdot A^\dagger(\obar) \wedge B(\obar) , \ \psi^\dagger \cdot A^\dagger(\obar) \wedge A^\dagger(1) \ &: \quad \text{apply } \hat{\chi}_0^\dagger , \\
    A^\dagger(\obar) \cdot B^\dagger(1) \ &: \quad \text{raise $\hat{n}_P$ by one} , \\
    B^\dagger (\obar)\cdot A^\dagger(1) \ &: \quad \text{raise $\hat{n}_Q$ by one} .
\end{align*}
In addition to these, there are also multi-quark excitations that can be regarded as composite:
\begin{align*}
    \psi^\dagger \cdot B^\dagger(\obar) \, \psi^\dagger \cdot B(1) \ &: \quad \text{apply } \hat{\chi}_{\obar}^\dagger \chi_1^\dagger \\ 
    \psi^\dagger \cdot \psi^\dagger \wedge A^\dagger(1) \ &: \quad \text{apply } \hat{\chi}_0^\dagger \chi_1^\dagger \\ 
    \psi^\dagger \cdot \psi^\dagger \wedge A^\dagger(\obar) \ &: \quad \text{apply } \hat{\chi}_{\obar}^\dagger \chi_{0}^\dagger \\
    \psi^\dagger \cdot \psi^\dagger \wedge \psi^\dagger \ &: \quad \text{apply $\hat{\chi}_{\obar}^\dagger \hat{\chi}_{0}^\dagger \chi_1^\dagger$}
\end{align*}
Note that, one may take a different perspective on how to identify the set of local excitations, which proceeds by taking all available triplets at the site and listing all non-trivial bilinear and trilinear contractions of purely creation operators.
One will ultimately arrive at the same set of creation operators that cover both ``elementary'' degrees of freedom and composite excitations.

The following sections will serve to derive LSH operator factorizations for the above operators by considering a local basis and demonstrating that the factorizations behave in all ways identically to the SU(3)-invariant contractions, at least within the space of excitations that are dynamically relevant.

\subsection{LSH basis: Single site
\label{subsec: LSH-basis}
}
Above, the SU(3)-singlet operators relevant to expressing the Hamiltonian have been identified and arguments have been made for the existence of exactly two SU(3)-invariant bosonic excitations and three SU(3)-invariant fermionic excitations. 
This section makes this concrete by again restricting the discussion to one site, writing down a local-basis ansatz in terms of the two bosonic and three fermionic quantum numbers, and showing that the Hilbert space spans all possible local configurations that can be generated by Hamiltonian.
With a valid local basis identified, one can use it to convert the SU(3)-invariant contractions involving ISBs into factorized expressions that only make reference to the SU(3)-invariant degrees of freedom, following what was done for the SU(2) LSH formulation~\cite{Raychowdhury:2019iki}.
To carry out these calculations, one must make repeatedly apply the (anti)commutation rules that are obeyed by the colored fermion components and the ISBS.
The non-trivial bosonic commutation relations obeyed by ISBs, which are necessary for SU(3), make the calculations substantially more involved than the case of SU(2).
Once the factorized expressions are obtained, it is possible to express the entire formulation without any reference whatsoever to gauge-covariant quarks or ISBs.
\begin{figure}[t!]
    \centering
    \includegraphics[scale=1]{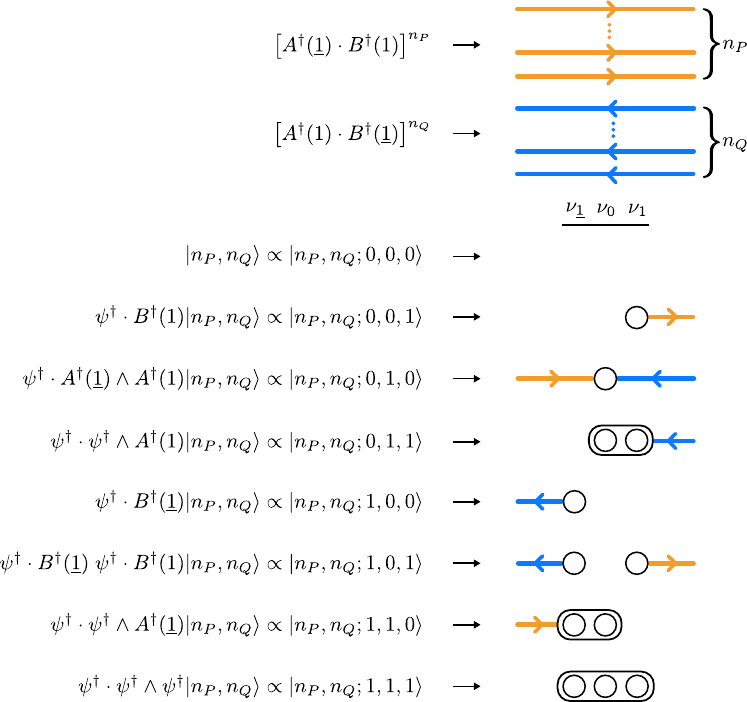}
    \caption{Pictorial representation of single-site SU(3) loop string-hadron states defined in Eqs.~\eqref{eq: LSH states unnormalized def} and \eqref{eq: Unnormalized LSH-basis-state}. At each site, electric flux of two types can pass through the site, illustrated by rightward (or orange) arrows counted by $n_P$, and leftward (blue) arrows counted by $n_Q$.
    Additionally there may exist three gauge singlet gauge-matter excitations, characterized by three fermionic occupation numbers $\nui,\num,\nuo$.
    For states with $\sum_f \nu_f = 1$, the quarks either source a unit of flux outward to one side, or they sink two incoming units of flux, one unit per side.
    For states with $\sum_f \nu_f = 2$, the quarks either source two units of outward flux, one unit per side, or they sink a single incoming unit of flux from one side.
    The $\sum_f \nu_f = 3$ configuration is the on-site baryon, which neither sources nor sinks any gauge flux.
    \label{fig: pict-LSH}}
\end{figure}

A basis ansatz is now proposed for site-local excitations that is characterized in terms of a set of gauge-invariant quantum numbers, i.e., the loop-string-hadron quantum numbers in the strong coupling basis.
The first object needed is the local normalized vacuum ket, denoted $\ket{\Omega}$, defined by the simultaneous conditions
\begin{align}
    \hat{P}(1) \ket{\Omega} = \hat{Q}(1) \ket{\Omega} = \hat{P}(\obar) \ket{\Omega} = \hat{Q}(\obar) \ket{\Omega} = \hat{N}_\psi \ket{\Omega} &= 0, \\
    \braket{\Omega | \Omega} &= 1.
\end{align}
Next, following the discussion of the preceding section, an ansatz for eight orthogonal local matter configurations $\kket{0 \, 0 \, ; \nu_{\obar} , \nu_0 , \nu_1}$ can be defined in terms of three bits $\nu_{\obar}$, $\nu_0$, $\nu_1$ as
\begin{subequations}
\begin{align}
    \kket{0 \, 0 ; 0 \, 0 \, 0} &\equiv \ket{\Omega} , \\
    \kket{0 \, 0 ; 0 \, 0 \, 1} &\equiv
    \psi^\dagger \cdot B^\dagger(1) \ket{\Omega} , \\
    \kket{0 \, 0 ; 1 \, 0 \, 0} &\equiv
    \psi^\dagger \cdot B^\dagger(\obar) \ket{\Omega} , \\
    \kket{0 \, 0 ; 0 \, 1 \, 0} &\equiv
    \psi^\dagger \cdot A^\dagger(\obar) \wedge A^\dagger(1) \ket{\Omega} , \\
    \kket{0 \, 0 ; 1 \, 0 \, 1} &\equiv
    \psi^\dagger \cdot B^\dagger(\obar) \ \psi^\dagger \cdot B^\dagger(1) \ket{\Omega} , \\
    \kket{0 \, 0 ; 0 \, 1 \, 1} &\equiv
    \tfrac{1}{2} \psi^\dagger \cdot \psi^\dagger \wedge A^\dagger(1) \ket{\Omega} , \\
    \kket{0 \, 0 ; 1 \, 1 \, 0} &\equiv 
    \tfrac{1}{2} \psi^\dagger \cdot \psi^\dagger \wedge A^\dagger(\obar) \ket{\Omega} , \\
    \kket{0 \, 0 ; 1 \, 1 \, 1} &\equiv \tfrac{1}{6} \psi^\dagger \cdot \psi^\dagger \wedge \psi^\dagger \ket{\Omega} ,
\end{align}
\label{eq: LSH states unnormalized def}
\end{subequations}
where the double-bar ket $\kket{\ }$ is used to denote that the above kets are not necessarily normalized.
The infinite towers of bosonic excitations are accommodated via 
\begin{equation}
    \kket{n_P,n_Q\,;\,\nu_{\obar},\nu_0,\nu_1} \equiv \bigl(A^\dagger(\obar)\cdot B^\dagger(1)\bigr)^{n_P} \bigl(B^\dagger(\obar)\cdot A^\dagger(1)\bigr)^{n_Q}\kket{0,0\,;\,\nu_{\obar},\nu_0,\nu_1} ,
    \label{eq: Unnormalized LSH-basis-state}
\end{equation}
where $n_P$ and $n_Q$ are non-negative integers and the double-bar notation serves the same purpose as in Eqs.~\eqref{eq: LSH states unnormalized def}.

The above states can be represented pictorially as shown in Fig.~\ref{fig: pict-LSH}. Such a pictorial representation is often useful for obtaining an intuitive understanding of the LSH degrees of freedom, which can be described as follows:
\begin{itemize}
    \item $n_P$ and $n_Q$ are associated with two independent types of bosonic flux running through the site, without originating from or terminating on any quarks. Here $n_P$ is chosen to count units of ``rightward'' (or orange) flux running from $\obar\to 1$, and $n_Q$ to count units of ``leftward'' (or blue) flux running from $1\to\obar$.
    \item A lone excitation of $\nu_{1}$ ($\nu_{\obar}$) sources one unit of outward flux, pointing in the $1$ (\obar) direction.
    \item The $\nu_{1}$ and $\nu_{\obar}$ excitations can be present simultaneously, with each sourcing an outward flux unit as described above.
    \item A lone excitation of $\nu_0$ sinks two units of flux, with one unit pointing inward from both sides of the site.
    \item The excitations of $\nu_0$ paired with $\nu_{1}$ ($\nu_{\obar}$) acts like an antiquark by sinking one unit of flux, pointing inward from the $1$ ($\obar$) direction.
    \item The excitation of all $\nu_{\obar}$, $\nu_0$, and $\nu_1$ modes together is an on-site baryon, which neither sources nor sinks any gauge flux (but flux units can flow through it).
\end{itemize}

Later, in Sec.~\ref{subsubsec: LSH operators in terms of basis and AGL}, it will be discussed how the states in the physical Hilbert space of the whole lattice are constructed from the tensor product of local LSH states in Eq.~\eqref{eq: LSH states unnormalized def} subjected to the Abelian Gauss's law constraints (equivalent to Eq.~\eqref{eq: AGL in ISB}).
These constraints are simple to visualize in terms of the pictorial representation: they translate into conservation of rightward and leftward gauge fluxes along any given link.

Finally, the above states yield an orthonormal basis given by
\begin{equation}
    \ket{n_P,n_Q\,;\,\nu_{\obar},\nu_0,\nu_1} =\mathcal N^{n_P,n_Q}_{\nu_{\obar},\nu_0,\nu_1} \bigl(A^\dagger(\obar)\cdot B^\dagger(1)\bigr)^{n_P} \bigl(B^\dagger(\obar)\cdot A^\dagger(1)\bigr)^{n_Q}\kket{0,0\,;\,\nu_{\obar},\nu_0,\nu_1} , 
    \label{eq: LSH-basis-state}
\end{equation}
where $\mathcal N^{n_P,n_Q}_{\nu_{\obar},\nu_0,\nu_1}$ are the normalization factors obtained by solving the condition
\begin{equation}
    \braket{n_P',n_Q'\,;\,\nu'_{\obar},\nu'_{0},\nu'_1|\,n_P,n_Q\,;\,\nu_{\obar},\nu_{0},\nu_1} = \delta_{n_P}^{n_P'}\; \delta_{n_Q}^{n_Q'}\; \delta_{\nu_{\obar}}^{\nu_{\obar}'}\; \delta_{\num}^{\num'}\; \delta_{\nu_1}^{\nu_1'},
    \label{eq: normalization-cond}
\end{equation}
and using the algebraic relations in Eqs.~\eqref{eq: ferm_anticomm}, and~\eqref{eq: AAdagg-commutator}-\eqref{eq: zero-commutators}.
The coefficients' expression in terms of the LSH occupation numbers turns out to be
\begin{equation}
    \mathcal {N}^{\,n_P,n_Q}_{\,\nu_{\obar},\num,\nu_1} =  \left[\frac{1}{2} (n_P+n_Q+3-\delta_{\num\,\nu_{\obar}}\delta_{\num\,\nu_{1}}) (n_P+2-\delta_{\num\,\nu_{1}})! \, (n_Q+2-\delta_{\num\,\nu_{\obar}})! \, n_P! \, n_Q! \right]^{-1/2} .
    \label{eq: normalization factor expression}
\end{equation}

With the basis ansatz above, the next step is to confirm that all possible states that are dynamically connected to the local vacuum are indeed spanned by it.
To do this, it is sufficient to confirm that the states are closed under the application of each and every SU(3)-singlet operator appearing in the Hamiltonian.
The simplest place to start is with the operators appearing in $H_M$ and $H_E$.
It is easy to see that $\hat{N}_\psi$ is diagonalized by this basis, with
\begin{align}
    \hat{N}_\psi \ket{n_P , n_Q ; \nu_{\obar}, \nu_0, \nu_1} &= (\nu_{\obar} + \nu_0 + \nu_1) \ket{n_P,n_Q\,;\,\nu_{\obar},\nu_{0},\nu_1} .
\end{align}
As for the total ISB occupation numbers to either side, $\hat{P}(1/\obar)$ and $\hat{Q}(1/\obar)$, these are also diagonalized.
Note that each factor of $A^\dagger(\obar)\cdot B^\dagger(1)$ increases $\hat{P}(1)$ and $\hat{P}(\obar)$ by one unit,
while each factor of $B^\dagger(\obar)\cdot A^\dagger(1)$ increases $\hat{Q}(1)$ and $\hat{Q}(\obar)$ by one unit.
It is for this reason that the definitions of $\hat{P}$ and $\hat{Q}$ were swapped at one end of a link relative to the other in Eq.~\eqref{eq: irrep quantum numbers} and that the symbols $n_P$ and $n_Q$ were chosen.
Careful inspection of the matter configurations reveals that the ISB occupation numbers may be increased by one additional unit, depending on the specific combination of $\nu_f$'s.
These observations are encapsulated by
\begin{subequations}
    \label{eq: P Q numbers on LSH states}
\begin{align}
    \hat{P}(1) \ket{n_P,n_Q\,;\,\nu_{\obar},\nu_{0},\nu_1} &= ( n_P +  \nu_1\left(1-\nu_0  \right) ) \ket{n_P,n_Q\,;\,\nu_{\obar},\nu_{0},\nu_1} , \\
    \hat{Q}(1) \ket{n_P,n_Q\,;\,\nu_{\obar},\nu_{0},\nu_1} &= ( n_Q +  \nu_0 (1-\nu_{\obar}  ) ) \ket{n_P,n_Q\,;\,\nu_{\obar},\nu_{0},\nu_1} , \\
    \hat{Q}(\obar) \ket{n_P,n_Q\,;\,\nu_{\obar},\nu_{0},\nu_1} &= ( n_Q +  \nu_{\obar}\left(1-\nu_0  \right) ) \ket{n_P,n_Q\,;\,\nu_{\obar},\nu_{0},\nu_1} , \\
    \hat{P}(\obar) \ket{n_P,n_Q\,;\,\nu_{\obar},\nu_{0},\nu_1} &= ( n_P +  \nu_0\left(1-\nu_1  \right) ) \ket{n_P,n_Q\,;\,\nu_{\obar},\nu_{0},\nu_1} .
\end{align}
\end{subequations}
Pictorially, the total types of each flux flowing to either side of the site are counted according to Eqs.~\eqref{eq: P Q numbers on LSH states} as follows:
\begin{itemize}
    \item $P(1)$ counts total units of flux pointing out from the $1$ side, with $n_P$ units coming from pure gauge flux, plus an additional unit if $\nu_1$ is excited, unless $\nu_0$ is also excited.
    \item $Q(\obar)$ counts total units of flux pointing out from the $\obar$ side, with $n_Q$ units coming from pure gauge flux, plus an additional unit if $\nu_{\obar}$ is excited, unless $\nu_0$ is also excited.
    \item $Q(1)$ counts total units of flux pointing in from the $1$ side, with $n_Q$ units coming from pure gauge flux, plus an additional unit if $\nu_0$ is excited, unless $\nu_{\obar}$ is also excited.
    \item $P(\obar)$ counts total units of flux pointing in from the $\obar$ side, with $n_P$ units coming from pure gauge flux, plus an additional unit if $\nu_0$ is excited, unless $\nu_{1}$ is also excited.
\end{itemize}

Turning to $H_I$, every term is strictly off-diagonal, so the real validation of the basis ansatz lies in its closure under the application of local singlets in Eq.~\eqref{eq: Hinter singlets}.
For example, in the case of $\psi^\dagger \cdot B^\dagger(1)$, one can show using Eqs.~\eqref{eq: ferm_anticomm},~\eqref{eq: AAdagg-commutator}-\eqref{eq: zero-commutators},~\eqref{eq: LSH states unnormalized def}, and \eqref{eq: normalization factor expression} that
\begin{subequations}
\begin{align}
    \label{eq: example-action-first}
    \psi^\dagger \cdot B^\dagger(1) \ket{n_P,n_Q;0 \, 0 \, 0} &= \sqrt{n_P+2} \sqrt{\frac{n_P+n_Q+3}{n_P+n_Q+2}} \ket{n_P\,n_Q;001} , \\
    \psi^\dagger \cdot B^\dagger(1) \ket{n_P,n_Q;0 \, 1 \, 0} &= - \sqrt{n_P+1} \sqrt{\frac{n_P+n_Q+4}{n_P+n_Q+3}} \ket{n_P+1,n_Q;011} , \\
    \psi^\dagger \cdot B^\dagger(1) \ket{n_P,n_Q;1 \, 0 \, 0} &= - \sqrt{n_P+2} \ket{n_P\,n_Q;101} , \\
    \psi^\dagger \cdot B^\dagger(1) \ket{n_P,n_Q;1 \, 1 \, 0} &= \sqrt{n_P+1} \ket{n_P+1,n_Q;111} , \\
    \psi^\dagger \cdot B^\dagger(1) \ket{n_P,n_Q; \nu_{\obar} \, \num \, 1} &= 0 \quad \forall\; \nu_{\obar},\,\num \, .
\end{align}
\label{eq: 1 psi-dagg B-dagg operator actions}
\end{subequations}
For its Hermitian conjugate, one has
\begin{subequations}
\begin{align}
    \psi \cdot B(1) \ket{n_P,n_Q;0\, 0\, 1} &=  \sqrt{n_P+2} \sqrt{\frac{n_P+n_Q+3}{n_P+n_Q+2}} \ket{n_P\,n_Q;000} , \\
    \psi \cdot B(1) \ket{n_P,n_Q;0\, 1\, 1} &= -\sqrt{n_P}   \sqrt{\frac{n_P+n_Q+3}{n_P+n_Q+2}} \ket{n_P-1,n_Q;010} , \\
    \psi \cdot B(1) \ket{n_P,n_Q;1\, 0\, 1} &= -\sqrt{n_P+2} \ket{n_P\,n_Q;100} , \\
    \psi \cdot B(1) \ket{n_P,n_Q;1\, 1\, 1} &=  \sqrt{n_P}   \ket{n_P-1,n_Q;110} , \\
    \psi \cdot B(1) \ket{n_P,n_Q; \nu_{\obar} \, \num \, 0} &= 0 \quad \forall\; \nu_{\obar},\,\num \, .
\end{align}
\label{eq: 2 psi B operator actions}
\end{subequations}
Hence, the basis is closed under applications of $\psi^\dagger \cdot B^\dagger(1)$ and $\psi \cdot B(1)$; symmetry arguments imply the same is true for applications of $\psi^\dagger \cdot B^\dagger(\obar)$ and $\psi \cdot B(\obar)$.
One now proceeds systematically through the remaining SU(3) singlets.
For example,
\begin{subequations}
\begin{align}
    \psi^\dagger \cdot A(1) \ket{n_P,n_Q;0 \, 0 \, 0} &= \sqrt{n_Q} \ket{n_P,n_Q-1;100} , \\
    \psi^\dagger \cdot A(1) \ket{n_P,n_Q;0 \, 0 \, 1} &= \sqrt{n_Q} \sqrt{\frac{n_P+n_Q+3}{n_P+n_Q+2}} \ket{n_P,n_Q-1;101} , \\
    \psi^\dagger \cdot A(1) \ket{n_P,n_Q;0 \, 1 \, 0} &= \sqrt{n_Q+2} \ket{n_P\,n_Q;110} , \\
    \psi^\dagger \cdot A(1) \ket{n_P,n_Q;0 \, 1 \, 1} &= \sqrt{n_Q+2} \sqrt{\frac{n_P+n_Q+3}{n_P+n_Q+2}} \ket{n_P\,n_Q;111} , \\
    \psi^\dagger \cdot A(1) \ket{n_P,n_Q;1 \, \num \, \nu_1} &= 0 \quad \forall\; \num, \, \nu_{1} ,
\end{align}
\label{eq: 3 psi-dagg A operator actions}
\end{subequations}
showing that the basis is closed under applications of $\psi^\dagger \cdot A(1)$ and, by extension, $\psi^\dagger \cdot A(\obar)$.
The same is true for their adjoints, though we omit the specific formulas.
As a final explicit example, one can apply $\psi^\dagger \cdot A^\dagger(1) \wedge B(1)$ to the basis with the following results:
\begin{subequations}
\begin{align}
    \psi^\dagger \cdot A^\dagger(1) \wedge B(1) \ket{n_P,n_Q;0 \, 0 \, 0} &= - \sqrt{n_P} \sqrt{n_Q+2} \ket{n_P-1,n_Q;010} , \\
    \psi^\dagger \cdot A^\dagger(1) \wedge B(1) \ket{n_P,n_Q;0 \, 0 \, 1} &= - \sqrt{n_P+2} \sqrt{n_Q+2} \ket{n_P\,n_Q;011} , \\
    \psi^\dagger \cdot A^\dagger(1) \wedge B(1) \ket{n_P,n_Q;1 \, 0 \, 0} &= \sqrt{n_P} \sqrt{n_Q+1} \ket{n_P-1,n_Q+1;110} , \\
    \psi^\dagger \cdot A^\dagger(1) \wedge B(1) \ket{n_P,n_Q;1 \, 0 \, 1} &= \sqrt{n_P+2} \sqrt{n_Q+1} \ket{n_P,n_Q+1;111} , \\
    \psi^\dagger \cdot A^\dagger(1) \wedge B(1) \ket{n_P,n_Q;\nu_{\obar} \, 1 \, \nu_1} &= 0 \quad \forall\; \nu_{\obar},\,\nu_{1} .
    \label{eq: example-action-last}
\end{align}
\label{eq: 4 psi-dagg A-dagg B operator actions}
\end{subequations}
The above four explicit examples, together with the analogous calculations for the operators' adjoints and exchange of $1\leftrightarrow \obar$, confirms that the basis ansatz does in fact capture all local states excited by $H_I$.
\begin{figure}[t]
    \includegraphics[scale=1]{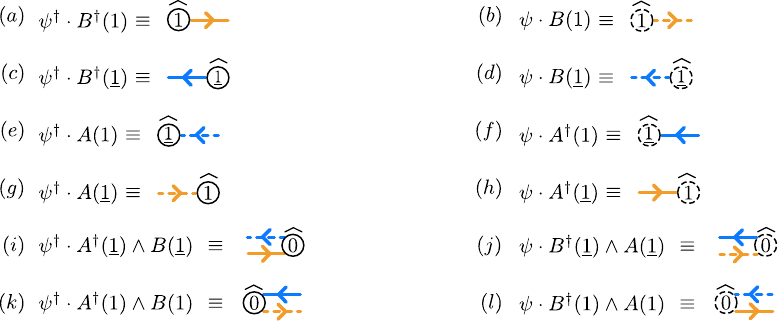}
    \caption{Pictorial representations of the LSH operators in Eq.~\eqref{eq: HI local term in prepotential} that act as local building blocks of the interaction Hamiltonian of the theory. 
    The fermion circles and arrows with lines collectively describe the creation (when solid) or annihilation (when dashed) of on-site quarks and directional gauge flux units to either side of the site.
    For a more complete explanation of this pictorial scheme, see the text in Sec.~\ref{subsec: LSH-basis}}
    \label{fig: operator actions}
\end{figure}

In Fig.~\ref{fig: pict-LSH}, a pictorial interpretation was provided for basis states of a single site.
One can also introduce a pictorial scheme for on-site SU(3) singlet operators as shown for the most important singlet operators, those appearing directly in $H_I$, in Fig.~\ref{fig: operator actions}. The rules for interpreting these pictorial representations are as following:
\begin{enumerate}
    \item A factor of ${\psi}^\dagger$ (${\psi}$) is illustrated by a solid-line (dashed-line) circle with a label $f\in \{\obar,0,1\}$ inside, corresponding to the $\nu_f$ mode it fills (empties).
    \item A factor of $A^\dagger(1)$ ($A(1)$) 
    is illustrated by a solid (dashed) arrow pointing inward from the 1 side of the site, corresponding to the creation (annihilation) of a unit of leftward gauge flux to that side.
    \item A factor of $A^\dagger(\obar)$ ($A(\obar)$) is illustrated by a solid (dashed) arrow pointing inward from the $\obar$ side of the site, corresponding to the creation (annihilation) of a unit of rightward gauge flux to that side.
    \item A factor of $B^\dagger(1)$ ($B(1)$) is illustrated by a solid (dashed) arrow pointing outward to the 1 side of the site, corresponding to the creation (annihilation) of a unit of rightward gauge flux to that side.
    \item A factor of $B^\dagger(\obar)$ ($B(\obar)$) is illustrated by a solid (dashed) arrow pointing outward to the $\obar$ side of the site, corresponding to the creation (annihilation) of a unit of leftward gauge flux to that side.
    \item A hat may be added to the drawing to distinguish it as an operator rather than a state configuration.
\end{enumerate}
Later, when multiple lattice sites are discussed, operator cartoons from different sites may be stitched together to represent more complex SU(3)-invariant operators that also manifestly preserve (or break) the Abelian Gauss's laws along links.

\subsection{LSH operator factorizations: Single site
\label{subsubsec: LSH operator factorization}
}

Above, a local on-site basis characterized by strictly SU(3)-invariant quantum numbers has been given and we have argued that they span the space relevant to gauge-invariant time evolution.
In the process, a number of matrix elements for the various SU(3)-singlet contractions are computed by applying the singlets of Eq.~\eqref{eq: Hinter singlets} to basis kets and using the algebra of SU(3) ISBs and the colored fermion components.
To generate these results and depart from their expression in terms of SU(3) ISBs, it is now appropriate to introduce, following the original SU(2) LSH construction~\cite{Raychowdhury:2019iki}, the normalized ladder operators and diagonal functions that are sufficient to express LSH dynamics without any reference to underlying degrees of freedom that transform under SU(3).

The first and most obvious objects are number operators for the LSH occupation numbers, $\hat{n}_l$ and $\hat{\nu}_f$ for the bosonic modes $l\in \{P,Q\}$ and the fermionic modes $f\in\{\obar,0,1\}$. Being diagonal in the LSH occupation number basis, these satisfy
\begin{align}
    \hat{n}_l \ket{n_P , n_Q ; \nu_{\obar}, \nu_0, \nu_1} &= n_l \ket{n_P , n_Q ; \nu_{\obar}, \nu_0, \nu_1}, \\
    \hat{\nu}_f \ket{n_P , n_Q ; \nu_{\obar}, \nu_0, \nu_1} &= \nu_f \ket{n_P , n_Q ; \nu_{\obar}, \nu_0, \nu_1}.
\end{align}
Next are the SU(3)-invariant quark modes, $\hat{\chi}_f$.
These are defined as canonical fermionic operators satisfying
\begin{align}
    \bigl\{ \hat{\chi}_{f}, \hat{\chi}_{f'} \bigr\} = \bigl\{ \hat{\chi}_{f}^{\dagger}, \hat{\chi}_{f'}^{\dagger} \bigr\} &= 0\,, \\
    \bigl\{ \hat{\chi}_{f}, \hat{\chi}_{f'}^{\dagger} \bigr\} &= \delta_{f \, f'}\,, \\
    \hat{\chi}_f \ket{n_P , n_Q \,;\, 0, 0, 0} &= 0\,.
\end{align}
They also raise or lower the $\hat{\nu}_f$ quantum numbers, as expressed by
\begin{align}
    [\hat{\nu}_{f}, \hat{\chi}_{f'}^\dagger] &= \delta_{f \, f'} \hat{\chi}_{f}^\dagger\,, \\
    [\hat{\nu}_{f}, \hat{\chi}_{f'}] &= - \delta_{f \, f'} \hat{\chi}_{f}\,.
\end{align}
Equivalently, $\hat{\nu}_f = \hat{\chi}_f^\dagger \hat{\chi}_f$.

Then there are normalized bosonic ladder operators $\hat{\ladder}_l$,
where $\hat{\ladder}_l$ ($\hat{\ladder}_l^\dagger$) lowers (raises) the bosonic quantum number $n_l$ by one unit, without rescaling a state's coefficients, aside from $\hat{\ladder}_l$ destroying the basis kets with $n_l=0$.
Algebraically,
\begin{align}
    [\hat{n}_{l} , \hat{\ladder}_{l'}^\dagger ] &= \delta_{l \, l'} \hat{\ladder}_{l}^\dagger\,,
\end{align}
\begin{align}
    [\hat{n}_{l} , \hat{\ladder}_{l'} ] &= - \delta_{l \, l'} \hat{\ladder}_{l} \,,\\
    [\hat{\ladder}_{l} , \hat{\ladder}_{l'}^\dagger ] &= \delta_{l \, l'} \int_{-\pi}^{\pi}  \frac{d\phi}{2\pi} e^{i \phi \hat{n}_{l} } \,.
\end{align}
Above, $\int_{-\pi}^{\pi}  \frac{d\phi}{2\pi} e^{i \phi \hat{n}_{l}}$ is used to represent a projection operator onto the local bosonic $l$-mode vacuum. The bosonic ladder operators also do not ``talk'' to the fermionic modes whatsoever:
\begin{align}
    [ \hat{\nu}_{f} , \hat{\ladder}_{l} ] = [ \hat{\chi}_{f} , \hat{\ladder}_{l} ] = [ \hat{\chi}_{f}^\dagger , \hat{\ladder}_{l} ] &= 0 \,.
\end{align}
Equipped with the normalized fermionic and bosonic ladder operators, one can construct the local orthonormal basis by applying the raising operators on the bosonic and fermionic vacuum as
\begin{align}
    \ket{n_P, n_Q\,;\,\nu_{\obar},{\nu_0},\nu_1} &\quad\mapsto\quad ( \hat{\ladder}_{P}^{\dagger} ) ^{n_P}
    ( \hat{\ladder}_{Q}^{\dagger} ) ^{n_Q}
    ( \hat{\chi   }_{\obar}^{\dagger} ) ^{\nu_{\obar}}
    ( \hat{\chi   }_{0}^{\dagger} ) ^{\nu_0}
    ( \hat{\chi   }_{1}^{\dagger} ) ^{\nu_1} \ket{0,0\,;\,0,0,0} .
    \label{eq: basis in term of creation operators}
\end{align}
This local-basis definition involves choosing an order for the fermionic creation operators, although the final results of this section (the operator factorizations) are independent of this choice.

The last structures that are useful to define are those of diagonal functions and conditional (bosonic) ladder operators.
Diagonal functions are any operators that can be written in closed form in terms of only LSH number operators.
For example, the diagonal function $\sqrt{\hat{n}_l + \hat{\nu}_f}$ is defined by the matrix elements
\begin{align*}
    \bra{n_P',n_Q'\,;\,\nu'_{\obar},\nu'_{0},\nu'_1} \sqrt{\hat{n}_l + \hat{\nu}_f} \ket{n_P , n_Q ; \nu_{\obar}, \nu_0, \nu_1} &= \sqrt{n_l + \nu_f} \ \delta_{n_P}^{n_P'} \, \delta_{n_Q}^{n_Q'} \, \delta_{\nu_{\obar}}^{\nu_{\obar}'} \, \delta_{\num}^{\num'} \, \delta_{\nu_1}^{\nu_1'}  \,.
\end{align*}
The conditional ladder operators are essentially controlled operations.
For example, the expression $( \hat{\ladder}_l )^{  \hat{\nu}_f} $ is used for an operator that, when applied to kets, lowers $n_l$ if $\nu_f=1$ and does nothing if $\nu_f=0$.
Formulas for the various conditional ladder operators that arise are as follows:
\begin{alignat}{4}
    ( \hat{\ladder}_l )^{  \hat{\nu}_f} &\equiv \hat{\ladder}_l \, \hat{\nu}_f + ( 1 - \hat{\nu}_f ), \qquad& ( \hat{\ladder}_l^{\dagger} )^{  \hat{\nu}_f} &\equiv \hat{\ladder}_l^{\dagger} \, \hat{\nu}_f + ( 1 - \hat{\nu}_f ), \\
    ( \hat{\ladder}_l )^{ 1 - \hat{\nu}_f} &\equiv \hat{\ladder}_l \, ( 1 - \hat{\nu}_f )  + \hat{\nu}_f, & ( \hat{\ladder}_l^{\dagger} )^{ 1 - \hat{\nu}_f} &\equiv \hat{\ladder}_l^{\dagger} \, ( 1 - \hat{\nu}_f ) + \hat{\nu}_f .
\end{alignat}
All of the necessary ingredients to express the LSH formulation are now available to give operator factorizations for the SU(3) singlets of Eq.~\eqref{eq: Hinter singlets} as follows:
\begin{subequations}
\label{eq: Hinter opfacs}
\begin{align}
    \psi^\dagger \cdot B^\dagger(1) &\quad\mapsto\quad \hat{\chi}_1^\dagger ( \hat{\ladder}_P^\dagger )^{\hnum}
    \sqrt{\hat{n}_P+2 - \hnum} \sqrt{\frac{\hat{n}_P+\hat{n}_Q+3+\hnum}{\hat{n}_P+\hat{n}_Q+2+\hnui+\hnum}} \\
    \psi \cdot B(1) &\quad\mapsto\quad \hat{\chi}_1 ( \hat{\ladder}_P )^{\hnum}
    \sqrt{\hat{n}_P+2 (1- \hnum)} \sqrt{\frac{\hat{n}_P+\hat{n}_Q+3}{\hat{n}_P+\hat{n}_Q+2+\hnui}} \\
    \psi^\dagger \cdot B^\dagger(\obar) &\quad\mapsto\quad \hat{\chi}_{\obar}^\dagger ( \hat{\ladder}_Q^\dagger )^{\hnum}
    \sqrt{\hat{n}_Q+2 - \hnum} \sqrt{\frac{\hat{n}_P+\hat{n}_Q+3+\hnum}{\hat{n}_P+\hat{n}_Q+2+\hat{\nu}_1+\hnum}} \\
    \psi \cdot B(\obar) &\quad\mapsto\quad \hat{\chi}_{\obar} ( \hat{\ladder}_Q )^{\hnum}
    \sqrt{\hat{n}_Q+2 (1- \hnum)} \sqrt{\frac{\hat{n}_P+\hat{n}_Q+3}{\hat{n}_P+\hat{n}_Q+2+\hat{\nu}_1}}\\
    \psi^\dagger \cdot A(1)         &\quad\mapsto\quad \hat{\chi}_{\obar}^\dagger ( \hat{\ladder}_Q )^{1-\hnum}
    \sqrt{\hat{n}_Q+2 \,\hnum} \sqrt{\frac{\hat{n}_P+\hat{n}_Q+2+\hat{\nu}_1}{\hat{n}_P+\hat{n}_Q+2}} \\
    \psi \cdot A^\dagger(1) &\quad\mapsto\quad \hat{\chi}_{\obar} ( \hat{\ladder}_Q^\dagger )^{1-\hnum}
    \sqrt{\hat{n}_Q+1+\hnum} \sqrt{\frac{\hat{n}_P+\hat{n}_Q+3-\hnum+\hat{\nu}_1}{\hat{n}_P+\hat{n}_Q+3-\hnum}}\\
    \psi^\dagger \cdot A(\obar)         &\quad\mapsto\quad \hat{\chi}_1^\dagger ( \hat{\ladder}_P )^{1-\hnum}
    \sqrt{\hat{n}_P+2 \,\hnum} \sqrt{\frac{\hat{n}_P+\hat{n}_Q+2+\hnui}{\hat{n}_P+\hat{n}_Q+2}} \\
    \psi \cdot A^\dagger(\obar) &\quad\mapsto\quad \hat{\chi}_1 ( \hat{\ladder}_P^\dagger )^{1-\hnum}
    \sqrt{\hat{n}_P+1+\hnum} \sqrt{\frac{\hat{n}_P+\hat{n}_Q+3-\hnum+\hnui}{\hat{n}_P+\hat{n}_Q+3-\hnum}}\\
    \psi^\dagger \cdot A^\dagger(1) \wedge B(1) &\quad\mapsto\quad - \hat{\chi}_0^\dagger ( \hat{\ladder}_P )^{1-\hat{\nu}_1} ( \hat{\ladder}_Q^\dagger )^{\hnui}
    \sqrt{\hat{n}_P+2 \,\hat{\nu}_1} \sqrt{\hat{n}_Q+2 - \hnui}\\
    \psi \cdot B^\dagger(1) \wedge A(1) &\quad\mapsto\quad \hat{\chi}_0 ( \hat{\ladder}_P^\dagger )^{1-\hat{\nu}_1} ( \hat{\ladder}_Q )^{\hnui}\sqrt{\hat{n}_P+1+\hat{\nu}_1} \sqrt{\hat{n}_Q+2(1-\hnui)}
\end{align}
\begin{align}
    \psi^\dagger \cdot A^\dagger(\obar) \wedge B(\obar) &\quad\mapsto\quad \hat{\chi}_0^\dagger ( \hat{\ladder}_P^\dagger )^{\hat{\nu}_1} ( \hat{\ladder}_Q )^{1-\hnui}
    \sqrt{\hat{n}_P+2 - \hat{\nu}_1} \sqrt{\hat{n}_Q+2 \,\hnui} \\
     \psi \cdot B^\dagger(\obar) \wedge A(\obar) &\quad\mapsto\quad - \hat{\chi}_0 ( \hat{\ladder}_P )^{\hat{\nu}_1} ( \hat{\ladder}_Q^\dagger )^{1-\hnui}
    \sqrt{\hat{n}_P+2(1-\hat{\nu}_1)} \sqrt{\hat{n}_Q+1+\hnui}
\end{align}
\label{eq: on site operator factorinzation}
\end{subequations}
To obtain the set of equations in Eq.~\eqref{eq: on site operator factorinzation}, we performed a similar analysis as Eqs.~\eqref{eq: 1 psi-dagg B-dagg operator actions}-\eqref{eq: 4 psi-dagg A-dagg B operator actions} for all operators in Eq.~\eqref{eq: Hinter singlets}, and assumed the basis could be written in the form of Eq.~\eqref{eq: basis in term of creation operators}.
Note that a different fermionic ordering in Eq.~\eqref{eq: basis in term of creation operators} would lead to equivalent operator factorizations as in Eq.~\eqref{eq: Hinter opfacs} but they must satisfy all the same algebraic relations.
We stress that it is the algebraic relations of the operators and the characterization of the local vacuum that are the defining features of the LSH formulation.
For completeness, analogous factorizations for the other SU(3) singlets used to construct the local Hilbert space as in Eqs.~\eqref{eq: LSH states unnormalized def}-\eqref{eq: Unnormalized LSH-basis-state} can be found in Appendix~\ref{app: extra factorizations}.

\subsection{Complete LSH formulation including Hamiltonian and Abelian symmetries
\label{subsubsec: LSH Ham construction}
}
The construction of the LSH operators and the LSH basis has thus far been focused on a single site.
To complete the formulation, the framework will be adapted to the whole lattice and the Hamiltonian will be expressed.

The bosonic occupation numbers $\hat{n}_l$, bosonic lowering operators $\hat{\ladder}_l$, fermionic occupation numbers $\hat{\nu}_f$, and fermionic annihilation operators $\hat{\chi}_f$ (along with their conjugates) are elevated to site-dependent fields:
\begin{align}
\hat{n}_l \to \hat{n}_l(r), \quad  \hat{\nu}_f \to \hat{\nu}_f(r), \quad \hat{\ladder}_l \to \hat{\ladder}_l(r), \quad \hat{\chi}_f \to \hat{\chi}_f(r), \qquad (r=1,2,\cdots,N) .
\end{align}
Operators belonging to the same site obey the commutation and anti-commutation relations as they were given for a single site in Sec.~\ref{subsubsec: LSH operator factorization}.
Operators belonging to different sites, however, have vanishing commutators or anti-commutators as appropriate for their statistics.
The algebra of LSH occupation numbers and ladder operators across the entire lattice is summarized by
\begin{align}
    [\hat{\nu}_{f} (r), \hat{\nu}_{f'} (r')] &= [\hat{n}_{l} (r), \hat{n}_{l'} (r')] = [\hat{\nu}_{f} (r), \hat{n}_{l} (r')] = 0 , \\
    [\hat{\chi}_{f} (r) , \hat{n}_{l} (r')] &= [\hat{\nu}_{f} (r), \hat{\ladder}_{l} (r')] = [\hat{\chi}_{f} (r), \hat{\ladder}_{l} (r')] = [\hat{\chi}_{f}^\dagger (r), \hat{\ladder}_{l} (r')] = 0 , \\
    \bigl\{ \hat{\chi}_{f} (r), \hat{\chi}_{f'} (r') \bigr\} &= 0 , \\
    \bigl\{ \hat{\chi}_{f} (r), \hat{\chi}_{f'}^{\dagger} (r') \bigr\} &= \delta_{f \, f'} \delta_{r \, r'} , \\
    [\hat{\nu}_{f} (r), \hat{\chi}_{f'} (r')] &= - \delta_{f \, f'} \delta_{r \, r'} \hat{\chi}_{f} (r) , \\
    [\hat{n}_{l} (r), \hat{\ladder}_{l'} (r')] &= - \delta_{l \, l'} \delta_{r \, r'} \hat{\ladder}_{l} (r) , \\
    [\hat{\ladder}_{l} (r), \hat{\ladder}_{l'}^\dagger (r')] &= \delta_{l \, l'} \delta_{r \, r'} \int_{-\pi}^{\pi}  \frac{d\phi}{2\pi} e^{i \phi \hat{n}_{l}(r) } ,
\end{align}
where $r$ and $r'$ take values from 1 to $N$.
Above, $\int_{-\pi}^{\pi}  \frac{d\phi}{2\pi} e^{i \phi \hat{n}_{l}(r)}$ is used to represent a projection operator onto the local bosonic vacuum at site $r$, which is a subspace of the lattice's Hilbert space.
Next, the fermionic modes, normalized ladder operators, and diagonal functions may be put together to express the Hamiltonian.
One should first note that the SU(3)-invariant quantum numbers of the underlying prepotentials plus staggered fermions formulation (Eq.~\ref{eq: irrep quantum numbers}) are translated as
\begin{align}
    \hat{N}_\psi (r) \mapsto &\ \hnui(r) + \hnum(r) + \hnuo(r) ,
\end{align}
\begin{align}
    \label{eq: P Q 1bar side in LSH variables}
    \hat{P}(\obar,r) \mapsto \hat{n}_P(r) + \hnum(r) (1 - \hnuo(r)), &\quad \hat{Q}(\obar,r) \mapsto \hat{n}_Q(r) + \hnui(r) (1 - \hnum(r)) , \\
    \label{eq: P Q 1 side in LSH variables}
    \hat{P}(1,r) \mapsto \hat{n}_P(r) + \hnuo(r) (1 - \hnum(r)), &\quad \hat{Q}(1,r) \mapsto \hat{n}_Q(r) + \hnum(r) (1 - \hnui(r)) .
\end{align}
Similarly, in light of Eqs.~\eqref{eq: on site operator factorinzation} the operators in Eq.~\eqref{eq: Hinter singlets} for a lattice site $r$ have the following operator factorizations:
\begin{subequations}
\begin{align}
    \psi^\dagger(r) \cdot B^\dagger(1,r) &= \left[\hat{\chi}_1^\dagger ( \hat{\ladder}_P^\dagger )^{\hnum}
    \sqrt{\hat{n}_P+2 - \hnum} \sqrt{\frac{\hat{n}_P+\hat{n}_Q+3+\hnum}{\hat{n}_P+\hat{n}_Q+2+\hnui+\hnum}}\ \right]_r \,,\\
    \psi(r) \cdot B(1,r) &= \left[\hat{\chi}_1 ( \hat{\ladder}_P )^{\hnum}
    \sqrt{\hat{n}_P+2 (1- \hnum)} \sqrt{\frac{\hat{n}_P+\hat{n}_Q+3}{\hat{n}_P+\hat{n}_Q+2+\hnui}}\ \right]_r \,, \\
    \psi^\dagger(r) \cdot A(1,r) &= \left[\hat{\chi}_{\obar}^\dagger ( \hat{\ladder}_Q )^{1-\hnum}
    \sqrt{\hat{n}_Q+2 \,\hnum} \sqrt{\frac{\hat{n}_P+\hat{n}_Q+2+\hat{\nu}_1}{\hat{n}_P+\hat{n}_Q+2}}\ \right]_r \,,\\
    \psi(r) \cdot A^\dagger(1,r) &= \left[\hat{\chi}_{\obar} ( \hat{\ladder}_Q^\dagger )^{1-\hnum}
    \sqrt{\hat{n}_Q+1+\hnum} \sqrt{\frac{\hat{n}_P+\hat{n}_Q+3-\hnum+\hat{\nu}_1}{\hat{n}_P+\hat{n}_Q+3-\hnum}}\ \right]_r \,,\\
    \psi^\dagger(r) \cdot A^\dagger(1,r) \wedge B(1,r) &= -\left[ \hat{\chi}_0^\dagger ( \hat{\ladder}_P )^{1-\hat{\nu}_1} ( \hat{\ladder}_Q^\dagger )^{\hnui}
    \sqrt{\hat{n}_P+2 \,\hat{\nu}_1} \sqrt{\hat{n}_Q+2 - \hnui} \right]_r,\\
    \psi(r) \cdot B^\dagger(1,r) \wedge A(1,r) &= \left[\hat{\chi}_0 ( \hat{\ladder}_P^\dagger )^{1-\hat{\nu}_1} ( \hat{\ladder}_Q )^{\hnui}
    \sqrt{\hat{n}_P+1+\hat{\nu}_1} \sqrt{\hat{n}_Q+2(1-\hnui)} \right]_r,
\end{align}
\end{subequations}
The factorizations for $\psi^\dagger(r) \cdot B^\dagger(\obar,r)$ and $\psi^\dagger(r) \cdot A(\obar,r)$ are obtained from their 1-side counterparts by interchanging the labels $P\leftrightarrow Q$ and $1\leftrightarrow \obar$.
The same is true for obtaining $\psi^\dagger(r) \cdot A^\dagger(\obar,r) \wedge B(\obar,r)$ and $\psi(r) \cdot B^\dagger(\obar,r) \wedge A(\obar,r)$, except one must also apply an overall minus sign in both cases.

Noting that the hopping Hamiltonian of Eq.~(\ref{eq: KS ham in prepotential final}) also involves the $\eta(1/\obar)$, $\theta(1/\obar)$, and $\delta(1/\obar)$ operators, which are diagonal functions in the sense defined above.
These can be translated from Eqs.~\eqref{eq: eta 1 and 1 bar expressions}-\eqref{eq: delta 1 and 1 bar expressions} into diagonal functions of $\hat{P}$ and $\hat{Q}$ operators via Eq.~\eqref{eq: irrep quantum numbers}, resulting in
\begin{alignat}{4}
    \eta(1,r) &= \frac{1}{\sqrt{\hat{P}(1,r)+2}}\sqrt{\frac{\hat{P}(1,r)+\hat{Q}(1,r)+3}{\hat{P}(1,r)+\hat{Q}(1,r)+2}}, & \hspace{18pt} \eta(\obar,r) &= \frac{1}{\sqrt{\hat{P}(\obar,r)}}, \\
    \theta(\obar,r) &= \frac{1}{\sqrt{\hat{Q}(\obar,r)+2}}\sqrt{\frac{\hat{P}(\obar,r)+\hat{Q}(\obar,r)+3}{\hat{P}(\obar,r)+\hat{Q}(\obar,r)+2}}, & \theta(1,r) &= \frac{1}{\sqrt{\hat{Q}(1,r)}}, \\
    \delta(1,r) &= \frac{1}{\sqrt{(\hat{Q}(1,r)+2)\hat{P}(1,r)}}, & \delta(\obar,r), &= \frac{1}{\sqrt{(\hat{P}(\obar,r)+2)\hat{Q}(\obar,r)}} .
\end{alignat}
Note that the $\hat{P}$ or $\hat{Q}$ operators technically have zero in their spectrum,
making their inverse square roots formally undefined.
However, they cannot actually encounter singularities because the link operator in Eq.~\eqref{eq: link operator in prepotential} that defines the $\eta(1/\obar)$, $\theta(1/\obar)$, and $\delta(1/\obar)$ operators is non-singular.

Once the $\eta$, $\theta$, and $\delta$ operators are multiplied with the SU(3) singlets of Eq.~\eqref{eq: Hinter singlets} as dictated by Eq.~\eqref{eq: KS ham in prepotential final}, the full LSH Hamiltonian can be expressed as $H=H_M+H_E+H_I$, with
\begin{align}
    H_M &= \sum_{r=1}^{N} H_M(r) \equiv \mu \sum_{r=1}^{N} (-1)^r (\hnui(r) + \hnum(r) + \hnuo(r)) , 
    \label{eq: HM in LSH operators}\\
    H_E &= \sum_{r=1}^{N'} H_E(r) \equiv \sum_{r=1}^{N'} \frac{1}{3} \left( \hat{P}(1,r)^2 + \hat{Q}(1,r)^2 + \hat{P}(1,r) \hat{Q}(1,r) \right) + \hat{P}(1,r) + \hat{Q}(1,r) , 
    \label{eq: HE in LSH operators}
\end{align}
\begin{align}
    H_I = \sum_{r=1}^{N'} H_I(r) &\equiv \sum_{r} x \left[ 
    \hat{\chi}_{1}^\dagger ( \hat{\ladder}_P^\dagger )^{\hnum}
    \sqrt{ 1  - \hnum/(\hat{n}_P + 2)} \sqrt{ 1 - \hnui/(\hat{n}_P+\hat{n}_Q+3)}
    \ \right]_{r} \nonumber \\
    & \hspace{1.2cm} \otimes
    \left[
    \sqrt{ 1  + \hnum/(\hat{n}_P + 1)} \sqrt{ 1 + \hnui/(\hat{n}_P+\hat{n}_Q+2)} \,
    \hat{\chi}_{1} ( \hat{\ladder}_P^\dagger )^{1-\hnum}
    \right]_{r+1} \nonumber \\
    &\quad + x \left[
    \hat{\chi}_{\obar}^\dagger ( \hat{\ladder}_Q )^{1-\hnum}
    \sqrt{ 1 + \hnum/(\hat{n}_Q + 1)}\sqrt{ 1 + \hat{\nu}_{1}/(\hat{n}_P+\hat{n}_Q+2)}
    \ \right]_{r} \nonumber \\
    &\hspace{1.2cm} \otimes \left[
    \sqrt{ 1 - \hnum/(\hat{n}_Q + 2)} \sqrt{ 1 - \hat{\nu}_{1}/(\hat{n}_P+\hat{n}_Q+3)} \, \hat{\chi}_{\obar} ( \hat{\ladder}_Q )^{\hnum}
    \right]_{r+1} \nonumber \\
    &\quad + x \left[
    \hat{\chi}_0^\dagger ( \hat{\ladder}_P )^{1-\hat{\nu}_{1}} ( \hat{\ladder}_Q^\dagger )^{\hnui}
    \sqrt{ 1 + \hat{\nu}_{1}/({\hat{n}_P+1}}) \sqrt{ 1 - {\hnui}/({\hat{n}_Q+2})}
    \ \right]_{r} \nonumber \\
    & \hspace{1.2cm} \otimes \left[
    \sqrt{ 1 - {\hat{\nu}_{1}}/({\hat{n}_P+2}}) \sqrt{ 1 + {\hnui}/({\hat{n}_Q+1}}) \,
    \hat{\chi}_0 ( \hat{\ladder}_P )^{\hat{\nu}_{1}} ( \hat{\ladder}_Q^\dagger )^{1-\hnui}
    \right]_{r+1} + \mathrm{H.c.} 
    \label{eq: HI in LSH operators}
\end{align}
This is the final Hamiltonian operator for the LSH formulation of (1+1)-dimensional SU(3) lattice gauge theory with staggered fermions.

The LSH Hamiltonian, while having SU(3) symmetry intrinsically built into every constituent operator, features several remnant Abelian symmetries.
First there are three global U(1) symmetries generated by $\sum_r \hat{\nu}_f(r)$ for each type $f$, implying conservation of total fermions of each type.
These quantum numbers will be discussed in more detail in Sec.~\ref{subsec: super-selection}.
Less obviously, the Hamiltonian commutes with all required Abelian Gauss's law constraints:
\begin{align}
    \label{eq: LSH Hamiltonian commutes with AGL}
    [ H \, , \, \hat{P}(1,r)-\hat{P}(\obar,r+1) ] &= [ H \, , \, \hat{Q}(1,r)-\hat{Q}(\obar,r+1) ] =  0
\end{align}
for each link $r$, where $\hat{P}$ and $\hat{Q}$ operators are as defined in Eqs.~\eqref{eq: P Q 1bar side in LSH variables}-\eqref{eq: P Q 1 side in LSH variables}.
Physical states are restricted to the simultaneous kernel of all Abelian Gauss's law generators.

\subsection{Explicit basis implementation of the LSH Hamiltonian
\label{subsubsec: LSH operators in terms of basis and AGL}
}
\begin{figure}[t]
    \includegraphics[scale=0.99]{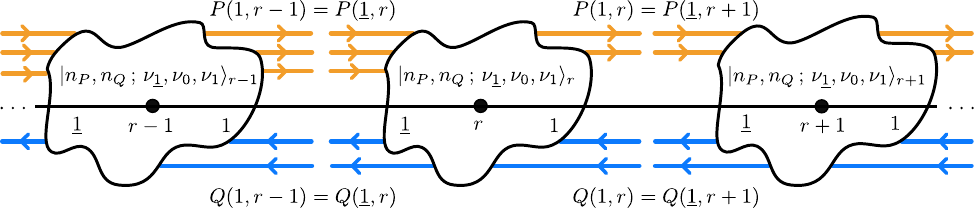}
    \caption{
    Schematic representation of LSH degrees of freedom in an occupation number basis, with on-site states that satisfy the Abelian Gauss's laws.
    Here, filled circles labeled by $r-1$, $r$, and $r+1$ depict the lattice sites at those positions. The left and right directions from each lattice site are denoted by $\obar$ and $1$, respectively. The orange and blue lines denote the $P$-type and $Q$-type gauge flux units, respectively, and they are related to the on-site LSH quantum numbers at each site through Eqs.~\eqref{eq: P Q 1bar side in LSH variables} and~\eqref{eq: P Q 1 side in LSH variables}. The arrows on them indicate the direction in which the corresponding gauge flux passes through a site, rightward $(\obar \rightarrow 1)$ of $P$-type and leftward $(1 \rightarrow \obar)$ for $Q$-type gauge flux. The LSH basis states satisfy the Abelian Gauss's law constraints defined in Eq.~\eqref{eq: AGL} which requires that the number of flux units of each $P$- and $Q$-type of gauge flux to be conserved over each link connecting two sites, as shown in the figure.\label{fig: AGL_pict}}
\end{figure}
For classical computation or digital quantum computation, it is helpful to have an explicit basis representation of the Hilbert space and Hamiltonian.
The Hilbert space can be constructed as a tensor product space of local Hilbert spaces all in an occupation number basis:
\begin{align}
    \label{eq: LSH global basis ket}
    \otimes_{r} \ket{n_P(r), n_Q(r) \,;\; \nui(r), \num(r) , \nuo(r)} &\equiv \otimes_{r} \ket{n_P, n_Q \,;\; \nui, \num , \nuo}_r
\end{align}
where, for brevity, the $r$-dependences of the $n_l$ and $\nu_f$ quantum numbers have been collected into the single subscript $r$ on the ket.
The above states constitute what can be referred to as the ``full'' Hilbert space in the LSH formulation, i.e., the Hilbert space that contains physical and unphysical states.
The subspace relevant to physical dynamics, per Eq.~\eqref{eq: LSH Hamiltonian commutes with AGL}, is restricted to those satisfying
\begin{align}
    n_P(r) +  \nu_1(r)\left(1-\nu_0 (r) \right) &= n_P(r+1) +  \nu_0(r+1) \left(1-\nu_1 (r+1) \right) , \\
    n_Q(r) +  \nu_0(r)\left(1-\nu_{\obar} (r) \right) &= n_Q(r+1) +  \nu_{\obar}(r+1)\left(1-\nu_0 (r+1) \right),
\end{align}
or more compactfully
\begin{align}
    P(1,r) = P(\obar,r+1)  \quad\text{and}\quad
    Q(1,r) = Q(\obar,r+1),
    \label{eq: AGL}
\end{align}
where $P$s and $Q$s are define in Eqs.~\eqref{eq: P Q 1bar side in LSH variables} and~\eqref{eq: P Q 1 side in LSH variables}.

This constraint is a realization of the Abelian Gauss's law constraints originally given for ISBs in Eq.~\eqref{eq: AGL in ISB}.
The Abelian Gauss's laws may be stated as: the total number of flux units flowing in either direction (rightward or leftward) is preserved from one side of a link to the other.
This is illustrated in Fig.~\ref{fig: AGL_pict}.

In practice, for numerical calculations a truncation may be necessary for finite dimensionality of the Hilbert space, for which one may introduce a cutoff $\Lambda$ on all bosonic $n_l$ quantum numbers.
In the case of open boundary conditions, $P(\obar,1)$ and $Q(\obar,1)$ are static, and the Abelian Gauss's laws constrain $P(1,N)\leq P(\obar,1) + N$ and $Q(1,N)\leq Q(\obar,1) + N$ such that it is possible to truncate the basis without necessarily truncating the dynamically relevant Hilbert space.

In the occupation number basis above, number operators are explicitly given by
\begin{subequations}
\begin{align}
    \hat{n}_l (r) &= \sum_{n_P, n_Q,\nu_{\obar} , \nu_0 , \nu_1} n_l(r) \ket{n_P,n_Q\,;\,\nu_{\obar},\nu_0,\nu_1}\bra{n_P,n_Q\,;\,\nu_{\obar},\nu_0,\nu_1}_r,
\end{align}
\begin{align}
    \hat{\nu}_f (r) &= \sum_{n_P, n_Q,\nu_{\obar} , \nu_0 , \nu_1} \nu_{f}(r) \ket{n_P,n_Q\,;\,\nu_{\obar},\nu_0,\nu_1} \bra{n_P,n_Q\,;\,\nu_{\obar},\nu_0,\nu_1}_r,
\end{align}
\end{subequations}
The bosonic lowering operators are defined by
\begin{subequations}
\begin{align}
    \hat{\ladder}^{}_{P} (r) &= \sum_{n_P=1}^\infty \ \sum_{n_Q,\nu_{\obar} , \nu_0 , \nu_1} \ket{n_P-1, n_Q\,;\,\nu_{\obar},\nu_0,\nu_1} \bra{n_P,n_Q\,;\,\nu_{\obar},\nu_0,\nu_1}_r, \\
    \hat{\ladder}^{}_{Q} (r) &= \sum_{n_Q=1}^\infty \ \sum_{n_P,\nu_{\obar} , \nu_0 , \nu_1} \ket{n_P,n_Q-1\,;\,\nu_{\obar},\nu_0,\nu_1} \bra{n_P,n_Q\,;\,\nu_{\obar},\nu_0,\nu_1}_r ,
\end{align}
\end{subequations}
with the corresponding raising operators $\hat{\ladder}_P^\dagger(r)$, $\hat{\ladder}_Q^\dagger(r)$ being their conjugates.
When it comes to the fermions, some care is needed in order to ensure the proper realization of fermionic statistics, i.e., an ordering of creation operators acting on the global fermionic vacuum has to be prescribed for any populated modes.
A simple choice would be to order sites $1$ to $N$ from left to right, such that quarks at site $N$ are created first, then site $N-1$, and so on down to site 1, and within each lattice site $r$ the creation operators are ordered as $(\chi^\dagger_{\obar})^{\nu_{\obar}}(\chi^\dagger_{0})^{\nu_{0}}(\chi^\dagger_{1})^{\nu_{1}}$, like in Eq.~\eqref{eq: basis in term of creation operators}.
The appropriate definitions of the fermionic annihilation operators are then
\begin{subequations}
\begin{align}
    \hat{\chi}_{\obar}(r) &= \left( \prod_{r'=1}^{r-1} (-1)^{\hat{\nu}_{\obar}(r')+\hat{\nu}_0(r')+\hat{\nu}_1(r')} \right) \sum_{n_P , n_Q , \nu_0 , \nu_1 } \ket{n_P,n_Q\,;\,0 ,\nu_0 ,\nu_1} \bra{n_P,n_Q\,;\,1 ,\nu_0 ,\nu_1}_r, \label{eq: chi-1bar-global} \\
    \hat{\chi}_0 (r) &= \left( \prod_{r'=1}^{r-1} (-1)^{\hat{\nu}_{\obar}(r')+\hat{\nu}_0(r')+\hat{\nu}_1(r')} \right) \sum_{n_P , n_Q , \nu_{\obar} , \nu_1 } \ket{n_P,n_Q\,;\,\nu_{\obar} ,0 ,\nu_1} \bra{n_P,n_Q\,;\,\nu_{\obar} ,1 ,\nu_1}_r (-1)^{\nu_{\obar}}, \label{eq: chi-o-global} \\
    \hat{\chi}_{1}(r) &= \left( \prod_{r'=1}^{r-1} (-1)^{\hat{\nu}_{\obar}(r')+\hat{\nu}_0(r')+\hat{\nu}_1(r')} \right) \sum_{n_P , n_Q , \nu_{\obar} , \nu_0 } \ket{n_P,n_Q\,;\,\nu_{\obar} ,\nu_0 ,0} \bra{n_P,n_Q\,;\,\nu_{\obar} ,\nu_0 ,1}_r (-1)^{\nu_{\obar} + \nu_0}, \label{eq: chi_1_global} 
\end{align}
\end{subequations}
where $(-1)^{\hat{\nu}_f}$ is a conditional phase that can be equivalently expressed as $(-1)^{\hat{\nu}_f} = 1 - 2 \hat{\nu}_f$.
The creation operators $\hat{\chi}_{\obar}^\dagger(r)$,  $\hat{\chi}_0^\dagger(r)$, and  $\hat{\chi}_{1}^\dagger(r)$ are obtained by Hermitian conjugation.
The above definitions, when used in Eqs.~\eqref{eq: HM in LSH operators}-\eqref{eq: HI in LSH operators}, along with the diagonal functions appearing in $H_I$ (such as $\sqrt{1+\hnuo/(\hat{n}_P+1)}$) being defined as explained in the previous sections, give a complete description of the Hamiltonian matrix with respect to the LSH occupation-number basis.

Beyond the essential Abelian Gauss's law constraints, there are also superselection sectors corresponding to the different possible global quantum numbers, which will be discussed in Sec.~\ref{subsec: super-selection}.

\section{Global symmetry sectors of the SU(3) LSH formulation
\label{subsec: super-selection}
}
It is important to note the global symmetries of the LSH formulation that lead to a block-diagonalized structure of the LSH Hamiltonian. The gauge-invariant dynamics then remains confined into each block providing significant computational benefit.
To identify these global symmetries, we notice that mass and electric Hamiltonians in Eqs.~\eqref{eq: HM in LSH operators} and~\eqref{eq: HE in LSH operators} are already diagonal in the LSH basis.
On the other hand, the interaction Hamiltonian in Eq.~\eqref{eq: HI in LSH operators}, although being non-diagonal, conserves the total number of fermions of each fermion type $f$.
The conserved Abelian symmetries are manifest from the couplings of the form $\hat{\chi}_f^\dagger(r) \hat{\chi}_f(r+1)$ in Eq.~\eqref{eq: HI in LSH operators} directly in the interaction Hamiltonian:
global rotations of the form $\hat{\chi}_f(r) \to e^{i \phi} \hat{\chi}_f(r)$ leave the Hamiltonian invariant.
These U(1) transformations consequently preserve the following global fermionic quantum numbers:
\begin{equation}
    \sum_{r=1}^{N} \nu_{\obar}(r) ~,~ \sum_{r=1}^{N} \num(r) ~,~ \sum_{r=1}^{N} \nu_{1}(r).
    \label{eq: global conserved charges LSH}
\end{equation}
The global symmetries of the Hamiltonian as characterized above are generated by the associated total fermion number operators, $\sum_r \hat{\nu}_f(r)$, which all mutually commute with each other and independently generate their own Abelian symmetry.

More insightful linear combinations of these three global U(1) conserved charges in Eq.~\eqref{eq: global conserved charges LSH} are given by 
\begin{align}
    \mathcal{F} &= \sum_{r=1}^{N} \big( \nu_{\obar}(r) +  \num(r) + \nu_{1}(r)\big) ,
    \label{eq: conserved total fermi number}\\
    \Delta\mathcal{P} &= \sum_{r=1}^{N} \big(\nu_1(r)-\nu_0 (r) \big) ,
    \label{eq: conserved total P}\\
    \Delta\mathcal{Q} &= \sum_{r=1}^{N} \big(\num(r)-\nu_{\obar} (r) \big),
    \label{eq: conserved total Q}
\end{align}
where $\mathcal{F}$ is the total fermion number that can take integer values from 0 to $3N$.
The charges $\Delta\mathcal{P}$ and $\Delta\mathcal{Q}$ denote the net bosonic flux of $P$-type and $Q$-type that the system sources or sinks when traversed from left to right.
Thus, for given (background) gauge fluxes fixed at the first lattice site, $(P(\obar,1),Q(\obar,1)) = (\mathcal{P}_0,\mathcal{Q}_0)$, the total gauge flux at the last site is then given by $(P(1,N),Q(1,N))=(\mathcal{P}_0+\Delta\mathcal{P}, \mathcal{Q}_0+\Delta\mathcal{Q})$.
Equations~\eqref{eq: P Q 1bar side in LSH variables} and~\eqref{eq: P Q 1 side in LSH variables} along with the Gauss's law conditions in Eq.~\eqref{eq: AGL} imply that each lattice site can source and/or sink up to one unit of gauge flux of either type, indicating that $\Delta \mathcal{P}$ and $\Delta\mathcal{Q}$ can both take integer values from $-N$ to $N$ as elaborated in Sec.~\ref{subsec: LSH-basis} in conjunction with a pictorial illustration in Fig.~\ref{fig: pict-LSH}.

An important subspace of the Hilbert space is the space that contains the strong-coupling vacuum, i.e., the ground state of the theory in the limit of $x\to 0$.
This state is characterized as the state with no bosonic excitations at any site and with fermionic modes at the sites alternating between completely empty and completely full. 
Thus the strong-coupling vacuum belongs to the global symmetry sector characterized by
\begin{align}
    \sum_{r=1}^{N} \nui(r) = \sum_{r=1}^{N} \num(r) = \sum_{r=1}^{N} \nuo(r) = \frac{N}{2} ,
    \label{eq: strong coupling vacuum charges LSH}
\end{align}
or equivalently,
\begin{align}
    \mathcal{F} = \frac{3N}{2}, \quad (\Delta\mathcal{P},\Delta\mathcal{Q}) = (0,0).
    \label{eq: strong coupling vacuum charges FPQ}
\end{align}
This is the sector that contains all states in the Hilbert space that can be reached from the strong coupling vacuum via the dynamics governed by the Hamiltonian in Eqs.~\eqref{eq: HM in LSH operators}-\eqref{eq: HI in LSH operators}.
This sector is especially important for periodic boundary conditions where $(\Delta\mathcal{P},\Delta\mathcal{Q})=(0,0)$ necessarily.

\section{Numerical benchmarking of energy eigenstates
\label{subsec: result-numerics}
}
\begin{figure}[t]
    \centering
    \begin{subfigure}{0.4\textwidth}
        \captionsetup{justification=centering}
        \includegraphics[width=\textwidth]{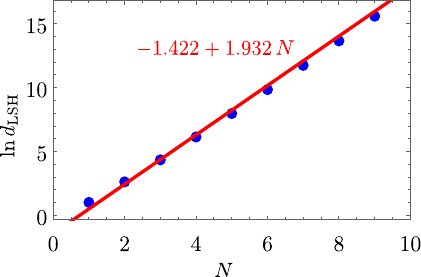}
        \caption{}
        \label{subfig: log dim vs N}
    \end{subfigure}
    \begin{subfigure}{0.584\textwidth}
        \captionsetup{justification=centering}
        \includegraphics[width=\textwidth]{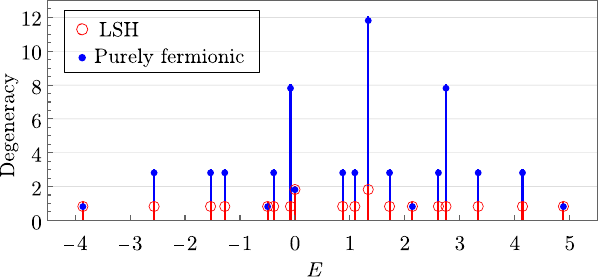}
        \caption{}
        \label{subfig: eigenavlues}
    \end{subfigure}
    \caption{(a) The dependence of the logarithm of the dimension of the physical Hilbert space within the LSH formulation, $d_{\rm LSH}$, is plotted for a range of lattice size values, $N$. For the purposes of this dimensionality comparison, the range of $N$ has been artificially extended to include odd values.   (b) Degeneracy in eigenvalues are plotted against their values for the Hamiltonian with $\mu = x = 1$ that is defined on a lattice of size $N=2$ with open boundary condition and zero incoming flux. Note that the eigenvalues are those of the dimensionless Hamiltonian in Eqs.~\eqref{eq: HM in LSH operators}-\eqref{eq: HI in LSH operators}. The eigenvalues are calculated with the LSH framework (red) and the gauge fixed purely fermionic framework (blue). The eigenvalues in both formulations match exactly with each other up to machine precision, however the degeneracy factor is different as discussed in Sec.~\ref{subsec: result-numerics}.\label{fig: spectrum}}
\end{figure}
The dynamics of the (1+1)-dimensional SU(3) lattice gauge theory with staggered quarks is completely captured by the LSH Hilbert space and the LSH Hamiltonian introduced in the previous subsections.
The derivation of the LSH formulation from the SU(3) Schwinger boson formulation will come in Sec.~\ref{sec: LSH-framework}, but first its validity will be confirmed by comparing its energy eigenvalues with another formulation that was recently considered in Refs.~\cite{Farrell:2022wyt,Atas:2022dqm}.
This formulation, which we refer to as ``purely fermionic,'' is the result of taking the KS formulation of the model with open boundary conditions and completely integrating out the gauge field (leaving only the colored fermionic field degrees of freedom);
a brief derivation is outlined in Appendix~\ref{app: purely fermionic formulation}.
Below, we discuss the results of this numerical benchmarking against the purely fermionic formulation.

The Hilbert space of this model is spanned by the LSH occupation-number basis states given in Eq.~\eqref{eq: LSH global basis ket}, that are subjected to the Abelian Gauss's law constraints given in Eqs.~\eqref{eq: AGL} on all links. Note that Eqs.~\eqref{eq: P Q 1bar side in LSH variables} and~\eqref{eq: P Q 1 side in LSH variables} imply that the matter content at each site can increase or decrease the $P$-type or $Q$-type gauge fluxes by at most 1 unit when traversing a site.
As a result, the physical Hilbert space of an $N$-site lattice with open boundary condition and given gauge fluxes fixed at either one of the boundaries (e.g., the first lattice site) is finite-dimensional.
That said, the dimension of the physical Hilbert space still grows exponentially with the lattice size.
This is illustrated empirically in the plot in Fig.~\ref{subfig: log dim vs N} for the case of $N\leq10$ lattices with zero background flux.
Here, the LSH physical Hilbert space dimension, $d_{\rm LSH}$, is observed to scale with $N$ as 
\begin{equation}
    d_{\rm LSH}\approx \exp (-1.422+1.932N).
    \label{eq: LSH Hilbert space dim fit}
\end{equation}

The Hilbert space in the purely fermionic formulation for the same background flux is spanned by a basis that is characterized by the fermionic occupation number for each color index at each lattice site (see Eq.~\ref{eq: Fermionic basis state} in Appendix~\ref{app: purely fermionic formulation}).
The dimension of this Hilbert space is then $8^N$ that is given by the number of all possible colored fermionic occupation states, which implies that the dimension of the Hilbert space in purely fermionic formulation is always greater than the dimension of the LSH Hilbert space.
To compare, the ratio of the Hilbert space dimension for the fermionic formulation to that of the LSH formulation for lattices in Fig.~\ref{subfig: log dim vs N} is given by
\begin{equation*}
    \ln{d_F/d_{LSH}} = N(\ln 8 -1.932) +1.422 \Rightarrow d_F/d_{LSH}\propto \exp{0.148N}.
\end{equation*}

Despite their different dimensions, the energy eigenvalues in both formulations are the same.
This is demonstrated in the plot in Fig.~\ref{subfig: eigenavlues} that compares the energy spectra of the Hamiltonian with $\mu=x=1$ and zero flux to the left of site $r=1$ for an $N=2$ lattice, calculated in both the LSH and purely fermionic formulations.
The energy eigenvalues match up to machine precision, providing a numerical benchmark for the LSH Hamiltonian. 

The excess of states in the Hilbert space of the purely fermionic formulation, seen in Fig.~\ref{subfig: eigenavlues}, is attributed to the fact that the basis states belong to irreps of the global SU(3) symmetry as discussed in Appendix~\ref{app: purely fermionic formulation}.
For the simplified case of zero boundary flux at the leftmost site, the purely fermionic states in each $\mathcal{F}$ sector are multiplets identified by the remaining two global charges $(\Delta\mathcal{P},\Delta\mathcal{Q})$ which in this case are non-negative integers and characterize the global SU(3) irrep.
In the more general case with a non-zero left-boundary flux in the $(\mathcal{P}_0,\mathcal{Q}_0)$ irrep, the multiplets are now identified by two non-negative integers $(\mathcal{P}_f,\mathcal{Q}_f)$  such that $(\mathcal{P}_f,\mathcal{Q}_f) =(\mathcal{P}_0+\Delta\mathcal{P}, \mathcal{Q}_0+\Delta\mathcal{Q})$.
Each state in the set of degenerate eigenstates with $(\mathcal{P}_f,\mathcal{Q}_f)$ charges is then further distinguished by its isospin and hypercharge quantum numbers, leading to a multiplicity $d({\mathcal{P}_f,\mathcal{Q}_f})$ given by the dimension of the irrep $(\mathcal{P}_f,\mathcal{Q}_f)$:
\begin{equation}
    d(\mathcal{P}_f, \mathcal{Q}_f) = \frac{1}{2}(\mathcal{P}_f+1)(\mathcal{Q}_f+1)(\mathcal{P}_f+\mathcal{Q}_f+2) .
    \label{eq: dim of irrep}
\end{equation}
On the other hand, the same global symmetry sectors in LSH formulation are now described by U(1) charges $\mathcal{F}$ and $(\Delta\mathcal{P},\Delta\mathcal{Q})$ as mentioned in Sec.~\ref{subsec: super-selection}. 
This formulation has no degeneracy associated with SU(3) gauge transformations because the states are all fundamentally constructed in terms of SU(3)-singlet operators.
As a result, for every LSH eigenstate with a $(\mathcal{P}_f,\mathcal{Q}_f)$ global charge, there are $d(\mathcal{P}_f,\mathcal{Q}_f)$ degenerate purely fermionic states. This is further demonstrated for the case of eigenvalues in Fig.~\ref{subfig: eigenavlues} in Table~\ref{tab: Eigenvalues} in Appendix~\ref{app: purely fermionic formulation}.
\section{Summary
\label{sec: LSH conclusions}
}
\noindent
The LSH framework has been shown to have several advantages relative to other frameworks present in the literature for Hamiltonian simulations~\cite{Davoudi:2020yln}.
One of the essential benefits offered by the LSH formalism is its concise solution to the problem of building the gauge invariant Hilbert space and keeping the dynamics inherently confined within that space~\cite{Mathew:2022nep}.
Especially notable is the fact that, despite the 
considerably more involved intermediate steps necessary to derive the SU(3) formulation as compared with SU(2),
the end result still turns out to be a Hamiltonian and Hilbert space that is remarkably similar to that which is obtained in the simpler SU(2) theory:
(i) instead of one bosonic variable and two fermionic variables per site, one has two bosonic variables and three fermionic variables per site, and (ii) the gauge-matter interaction in Eq.~\eqref{eq: HI in LSH operators} naturally separates into three decoupled fermionic hopping terms, instead of two as in the case of SU(2).
Apart from these minimal extensions, the SU(3) LSH framework otherwise contains each and every important feature of the SU(2) framework that make it useful for practical applications.

The primary ingredients of the SU(3) LSH framework presented in this chapter can be summarized as:
(i) strictly SU(3)-invariant, site-local degrees of freedom consisting of two bosonic fluxes and three quark modes,
(ii) the Hamiltonian of (1+1)-dimensional Yang-Mills theory coupled to staggered quarks, written explicitly in terms of the LSH degrees of freedom and furthermore expressed with respect to an LSH occupation-number basis,
(iii) two Abelian constraints per link of the lattice,
and (iv) the super-selection rules for LSH dynamics.
This framework was also supplemented with a pictorial representation of the states and operators in terms of directional gauge fluxes that may be sourced or sinked by the quarks.
Finally, the SU(3) LSH framework is numerically validated by a benchmark comparison with another formulation of the same theory that is currently being considered in the context of quantum simulation for QCD.

The structural similarity of the SU(3) framework to its SU(2) counterpart opens up the possibility of transferring many of the techniques developed for SU(2) over to the gauge group of QCD.
\begin{itemize}
    \item For example, past work that explained how to digitize physical states of SU(2) lattice gauge theory~\cite{Raychowdhury:2018osk} is immediately applicable to the SU(3) framework now developed.
    \item In a recent work concerning digital quantum algorithms for SU(2)~\cite{Davoudi:2022xmb}, it has been discovered that the Trotterization of a hopping term can be reduced from $2^3$ constituent steps (in the KS formulation) to only two;
    the division of the SU(3) LSH hopping terms into three fermionic couplings suggests the possibility of an analogous reduction from $3^3$ costly subterms to only three.
    \item In the context of analog quantum simulation for the $1+1$-dimensional LSH Hamiltonian, it is indeed possible to simulate exact dynamics of the theory using a purely fermionic analog simulation platform, with nearest neighbour hopping terms, provided the simulating Hamiltonian can be set up to mimic the dynamics of fermionic LSH excitations.
    Such a proposal has already been made for the SU(2) theory in Ref. \cite{Dasgupta:2020itb} and may be extendable to simulating SU(3) LSH dynamics. 
    \item 
    An important feature of the SU(2) LSH framework was its strictly Abelian symmetries pertaining to conservation of a single type of flux along links;
    the same feature recurs for SU(3) with the only difference being that there are two types (directions) of Abelian flux now. Whether the calculations are quantum simulations or tensor network calculations, the continuity of these fluxes must be protected in the dynamics.
    A recently proposed symmetry protection scheme for the SU(2) LSH formulation \cite{Mathew:2022nep} should be generalizable to the SU(3) framework, which would be useful in simulating LSH dynamics using classical (such as tensor network) or quantum (analog or digital) architectures. 

    \item The purely fermionic formulation that was used as a benchmark in Sec.~\ref{subsec: result-numerics} and described in Appendix~\ref{app: purely fermionic formulation}, despite being extremely useful in 1+1D, has a non-local Hamiltonian, is difficult to apply to periodic boundary conditions, and simply has no higher-dimensional counterpart given the actual presence of transverse waves. The LSH formulation, by contrast, can accommodate both open and periodic boundary conditions, is local, and is expected to be generalizeable to higher dimensions like its SU(2) counterpart.
\end{itemize}

The driving motivation for the work presented was the need for a convenient Hamiltonian framework for performing quantum simulations of lattice QCD.
With the framework introduced here, there are immediate research opportunities for investigating the advantages that should be offered by the SU(3) LSH formulation.
Any such advantages should have immediate relevance toward the actual goal of quantumly simulating lattice QCD. 
In the near future, the development of quantum simulation algorithms or even tensor network algorithms for SU(3) gauge theory using the 1+1D LSH framework would be interesting applications that can be pursued using just the present work as a basis.
The construction of the LSH formulation of an SU(3) LGT presented here marks a concrete step forward towards the goal of making quantum simulations of QCD a reality, with some of the most important generalizations to be explored in coming years being those of higher spatial dimensions and the extension to multiple fermion flavors.

\renewcommand{\thechapter}{4}

\chapter{Conclusion and Outlook
\label{ch: Conclusion and outlook}
}
\noindent
This thesis furthered theoretical developments in lattice gauge theories (LGTs) on two fronts.
First, by extending the finite volume (FV) formalism in lattice quantum chromodynamics (LQCD) to provide prescriptions for constraining the rare nuclear process of double-beta decay in the two-nucleon sector, both in its observed two-neutrino mode ($2\nu\beta\beta$) and the hypothetical neutrinoless mode ($0\nu\beta\beta$). 
Second, by providing a loop-string-hadron (LSH) formulation of an SU(3) LGT with continuous time and one spatial dimension, aiming to identify a computationally efficient Hamiltonian LGT approach for QCD observables.

The importance of theoretical predictions of decay rates of the $2\nu\beta\beta$ decay process and its counterpart $0\nu\beta\beta$ decay was highlighted, and the role of their accuracy in understanding neutrinos as Majorana fermions and probing physics beyond the standard model (BSM) was emphasized.
The dominant source of uncertainty in predictions of their rates comes from the evaluation of their corresponding nuclear matrix elements (MEs), which are calculated from \textit{ab initio} nuclear many-body methods using the associated two-nucleon decay amplitude.
The framework presented here utilizes LQCD and low-energy effective field theories (EFTs) to provide a way of improving constraints on few-body decay amplitudes that could result in improved accuracy on nuclear ME for these processes.

Prescriptions provided here match the nuclear MEs and energy eigenvalues accessible via numerical LQCD calculations to two-nucleon double-beta decay amplitudes evaluated using the non-relativistic pionless EFT.
It involves a matching relation that maps time separated nuclear MEs of two electroweak currents with Minkowski time in a FV to the corresponding physical decay amplitudes parameterized by the low energy constants (LECs) of the EFT.
The $2\nu\beta\beta$ decay amplitude was evaluated within the standard model (SM), while the light neutrino exchange scenario was considered for the $0\nu\beta\beta$ amplitude.
The $L_{1,A}$ LEC concerned in the former case appears at the next-to-leading order in pionless EFT, and it is associated with a two-body isovector operator.
While for the latter, an unknown LEC $g^{NN}_\nu$, that was discovered in recent years~\cite{Cirigliano:2017tvr, Cirigliano:2018hja}, appears at the LO.
$L_{1,A}$ and $g^{NN}_\nu$ are the dominant sources of uncertainty in evaluating the respective two-nucleon nuclear MEs.
LQCD implementations of matching relations provided in this thesis can mitigate this uncertainty, especially for the $0\nu\beta\beta$ case where $g^{NN}_\nu$ leaves the two-nucleon amplitude undetermined, but the first principles LQCD calculations in conjunction with the matching prescription provided here can pin down its value.

The FV formalism was utilized in deriving these matching relations.
Complexities in the FV formalism originating from light neutrino exchanges were carefully treated via regulating the infrared physics by removing the zero mode in neutrino propagator.
Furthermore, issues with naive analytical continuation of the Minkowski-time separated nuclear MEs of two electroweak currents in the matching relations to their Euclidean-time counterparts accessible from LQCD were discussed.
The problematic divergent contributions from intermediate states that are allowed to go on-shell were identified and removed to achieve a divergence-free analytical continuation.
Steps were laid out to systematically include those missing contributions by manually adding them via analytical construction.

Utilizing the prescriptions provided in this thesis for constraining the corresponding LECs requires LQCD calculations of low-energy two-nucleon (NN) spectra and the relevant NN MEs.
Performing these calculations is challenging, however, efforts for understanding and mitigating these challenges are ongoing~\cite{Davoudi:2022bnl,Cirigliano:2022rmf,Cirigliano:2022oqy}. 
The impact of uncertainties in future LQCD calculations on the accuracy of constraining  $g^{NN}_\nu$ and $L_{1,A}$ was analyzed in this thesis.
This analysis is performed using the above-mentioned $0\nu\beta\beta$ decay matching relation for the LEC $g^{NN}_\nu$.
On the other hand, the matching relation for the hadronic amplitude of single beta decay in the two-nucleon sector provided in Ref.~\cite{Briceno:2015tza} was used for performing a similar analysis on LEC $L_{1,A}$. 
It was found that the LQCD constraints on $L_{1,A}$ will likely be worse than the current experimental constraints for the range of volumes and plausible input uncertainties considered here, and may require (sub)percent-level precision on the finite-volume energies and MEs.
However, for precision levels on the LQCD energies and the ME below 10\%, the constraint on $g^{NN}_\nu$ is likely to improve the existing indirect phenomenological constraint in Refs.~\cite{Cirigliano:2020dmx,Cirigliano:2021qko} and will therefore provide a direct precise determination of it.

The next generation of $0\nu\beta\beta$ decay experiments are expected to probe the region of the parameter space that will significantly impact our understanding of lepton number violating BSM physics~\cite{Dolinski:2019nrj}.
However, it is essential to have accurate calculations of relevant nuclear MEs to draw reliable conclusions from these experiments.
Towards that goal, the work in this thesis can determine the direction for the LQCD community by providing target precisions on the required LQCD calculations.
Furthermore, the extension of the FV formalism provided here can be applied, with appropriate modifications, to other hadronic observables involving two local currents that are displaced in time and two hadrons in the initial and final states, e.g., electromagnetic corrections to two-pion scattering amplitude.

Finally, the complexity and limitations of obtaining physical observables from LQCD were pointed out in this thesis, and the motivation for exploring the method of Hamiltonian simulation of QCD to circumvent them was discussed.
As a step towards that goal, this thesis explored a loop-string-hadron (LSH) formulation of an SU(3) LGT that is expected to be a computational-resource efficient approach for its simulation.
The construction of gauge-invariant states that leads to simpler Gauss's laws is the key feature of LSH that reduces redundancies in its Hilbert space.
The methodology employed in constructing the LSH framework can offer valuable insights into the underlying mechanisms of LGTs, which may be applicable to further investigations within the field of LGTs.
The formulation provided here is restricted to continuous time and one spatial dimension, which is a significant advancement in the direction of making quantum simulations of QCD a reality.
Extensions of this formulation to higher spatial dimensions and incorporating multiple fermion flavors are key to achieving that goal, and such extensions will be explored in coming years.
\titleformat{\chapter}
{\normalfont\large}{Appendix \thechapter:}{1em}{}
\appendix
\renewcommand{\thechapter}{A}
\renewcommand{\chaptername}{Appendix}

\chapter{Two-loop sum-integral difference form for numerical implementation
\label{app: Two-loop sum-integral difference}
}
The steps involved in performing the two-loop sum-integral difference defined in Eq.~(\ref{eq:deltaJV}) for an expedited convergence will be outlined in this section. To simplify the notation, the conventions $n \equiv |\bm{n}|$ and $n^2 \equiv |\bm{n}|^2$ are used for any three-vector $\bm{n}$.

The two-loop integral involving the neutrino propagator is given by
\begin{equation}
\label{eq: definition J(epsilon) in inf V}
J^{\infty}(p_1^2,p_2^2) =
M^2 \int \frac{d^3k_1}{(2\pi)^3} \, \frac{d^3k_2}{(2\pi)^3}
\frac{1}{p_1^2-k_1^2+i\epsilon} \, \frac{1}{p_2^2-k_2^2+i\epsilon} \, \frac{1}{|\bm{k}_1 - \bm{k}_2|^2}.
\end{equation}
This integral is divergent in the UV region of integrating variables and must be regulated. While for the discussion of the physical amplitude in the main text, dimensional regularization is a natural choice as presented in Eq.~(\ref{eq:Jinfty})~\cite{Cirigliano:2019vdj}, for the evaluation of the sum-integral difference, a cutoff regulator  $\Lambda$ proves most useful. As the physical amplitude, as well as the matching condition are UV convergent, both choices can be used in the matching framework. In particular, with the cutoff regularization, $J^{\infty}$ evaluates to
\begin{align}
J^\infty(p_1^2,p_2^2)  &=
\frac{M^2}{8\pi^4}
\int_{0}^{\Lambda} dk_1 \, 
\int_{0}^{\infty}  dk_2 \,
\frac{k_1^2}{p_1^2-k_1^2} \, \frac{k_2^2}{p_2^2-k_2^2}
\int_{-1}^{1} dx \, \frac{1}{k_1^2+k_2^2-2\,k_1\,k_2\,x}\nonumber\\
&\hspace{2 cm}+ \frac{iM^2}{32\pi} - \frac{M^2}{32 \pi^2} \ln\Bigg(\frac{p_1+p_2}{|p_1-p_2|}\Bigg)\nonumber\\
&=M^2 \bigg[
\frac{\ln{\Lambda}}{16\pi^2} + \frac{i}{32\pi} - \frac{\ln{(p_1+p_2)}}{16 \pi^2}\bigg].
\label{eq: cutoff regulated inf V}
\end{align}

In a finite volume with cubic geometry and spatial extent $L$ along each Cartesian coordinate and with periodic boundary conditions, the analog of Eq.~\eqref{eq: definition J(epsilon) in inf V} is given by replacing integrals with sums over quantized three-momenta $\bm{k}=2\pi\bm{n}/L$ with $\bm{n} \in \mathbb{Z}^3$:
\begin{align}
\label{eq: definition J in fin V}
J^V(p_1^2,p_2^2)&=\frac{M^2}{L^6}
\sum_{\substack{
\bm{k}_1\\ 
k_1^2\neq p_1^2}}
\sum_{\substack{
\bm{k}_2\neq\bm{k}_1\\ 
k_2^2\neq p_2^2}}
\frac{1}{p_1^2-k_1^2} \, \frac{1}{p_2^2-k_2^2} \, \frac{1}{|\bm{k}_1 - \bm{k}_2|^2} .
\end{align}
Here, the $i\epsilon$ terms are dropped from the denominators since discrete sums are defined over non-singular values of $\bm{k}_1$ and $\bm{k}_2$.

Equation~\eqref{eq: definition J in fin V} differs from Eq.~\eqref{eq: definition J(epsilon) in inf V} by power-law correction in $1/L$ which can be isolated from the difference
\begin{equation}
\label{eq: S FV as S(epsilon) and delta S}
 \delta J^V(p_1^2,p_2^2) \equiv J^V(p_1^2,p_2^2) - J^\infty(p_1^2,p_2^2) \; .
\end{equation}
To evaluate $\delta J^V$, let us first convert the summation variable in Eq.~\eqref{eq: definition J in fin V} from $\bm{k}_{1(2)}$ to $\bm{n}_{1(2)}$ and rescale $p_{1(2)}$ as $\tilde{p}_{1(2)}=p_{1(2)}\,L/2\pi$. Next, one can observe that the UV divergence in Eq.~\eqref{eq: definition J in fin V} is the same as that occurred in the sum when $p_1^2=p_2^2=0$. Using a cutoff regulator $\Lambda$, this sum reads
\begin{equation}
\label{eq: cutoff regulated in FV}
J^V(0,0)=\frac{M^2}{(2\pi)^6}
\sum_{\bm{n}_1\neq 0}^{\tilde{\Lambda}} \;
\sum_{\bm{n}_2\neq 0,\bm{n}_1}
\frac{1}{n_1^2} \, \frac{1}{n_2^2} \, \frac{1}{|\bm{n}_1 - \bm{n}_2|^2}.
\end{equation}
The upper bound on the sum over $\bm{n}_1$ indicates that only integer triplets that satisfy $n_1 \leq \tilde{\Lambda}\,(=\Lambda L/2\pi)$ must be included. The sum over $\bm{n}_2$ is left unbounded. Now adding and subtracting $J^V(0,0)$ and upon using Eq.~\eqref{eq: cutoff regulated inf V}, Eq.~\eqref{eq: S FV as S(epsilon) and delta S} becomes
\begin{equation}
\label{eq: definition delta S in R X3 X6}
\delta J^V(p_1^2,p_2^2) = \frac{M^2}{(2\pi)^6}\bigg[ 
\mathcal{R}
-\mathcal{X}_1(\tilde{p}_1^2,\tilde{p}_2^2)
-\frac{\mathcal{X}_3(\tilde{p}_2^2)}{\tilde{p}_1}
-\frac{\mathcal{X}_3(\tilde{p}_1^2)}{\tilde{p}_2}
 + \mathcal{X}_6(\tilde{p}_1^2,\tilde{p}_2^2)
\bigg]
+ \frac{M^2}{16\pi^2}\ln{(\tilde{p}_1+\tilde{p}_2)}
-\frac{iM^2}{32\pi} \; ,
\end{equation}
where
\begin{align}
\label{eq: definition R}
&\mathcal{R} \equiv \lim\limits_{\tilde{\Lambda}\to\infty} \Bigg[ \;
\sum_{\bm{n}_1\neq 0}^{\tilde{\Lambda}} \;
\sum_{\bm{n}_2\neq 0,\bm{n}_1}
\frac{1}{n_1^2} \, \frac{1}{n_2^2} \, \frac{1}{|\bm{n}_1 - \bm{n}_2|^2}
- 4\pi^4\ln{\tilde{\Lambda}}
\Bigg] ,
\\
& \mathcal{X}_1(\tilde{p}_1^2,\tilde{p}_2^2)=
\frac{1}{\tilde{p}_1^2\,\tilde{p}_2^2}
\sum_{\substack{
\bm{n}_1\\
n_1^2=\tilde{p}_1^2
}
} \;
\sum_{\substack{
\bm{n}_2\neq \bm{n}_1\\
n_2^2=\tilde{p}_2^2
}
}\frac{1}{|\bm{n}_1- \bm{n}_2|^2},
\\
\label{eq: definition X3}
&\mathcal{X}_3 (\tilde{p}) \equiv
\sum_{\substack{
\bm{n}\neq 0\\ 
n^2\neq \tilde{p}^2}} \,
\frac{1}{n^2(n^2-\tilde{p}^2)},
\\
\label{eq: definition X6}
& \mathcal{X}_6 (\tilde{p}_1^2,\tilde{p}_2^2) \equiv 
\sum_{\substack{
\bm{n}_1\neq 0\\ 
n_1^2\neq \tilde{p}_1^2}}
\sum_{\substack{
\bm{n}_2\neq 0,\bm{n}_1\\ 
n_2^2\neq \tilde{p}_2^2}}
\bigg[\frac{1}{\tilde{p}_1^2-n_1^2} \, \frac{1}{\tilde{p}_2^2-n_2^2} 
-\frac{1}{n_1^2} \, \frac{1}{n_2^2}\bigg]\, \frac{1}{|\bm{n}_1
- \bm{n}_2|^2}.
\end{align}

To evaluate the lattice sums in Eqs.~\eqref{eq: definition R}-\eqref{eq: definition X6}, one can use the method of tail-singularity separation (TSS) described in Ref.~\cite{Tan:2007bg}.
In this method, the sum is split into two pieces: one containing the singular contributions and the other containing a power-law tail which is sufficiently smooth such that it can be approximated by its integral counterpart.
As an example, let us sketch out the details for evaluating the lattice sum in Eq.~\eqref{eq: definition X3}. The TSS scheme can be achieved by introducing exponential factors containing a small positive number, $\alpha$, and rewriting $\mathcal{X}_3$ as
\begin{equation}
\mathcal{X}_3(\tilde{p}^2) \, = \, 
\sum_{\substack{
\bm{n}\neq 0\\ 
n^2\neq \tilde{p}^2}} \,
\frac{\big[e^{-\alpha n^2}+1-e^{- \alpha n^2}\big]}{n^2} \frac{\big[e^{-\alpha(n^2-\tilde{p}^2)}+1-e^{-\alpha(n^2-\tilde{p}^2)}\big]}{n^2-\tilde{p}^2} .
\end{equation}
The smooth function containing a power-law tail is obtain by gathering $\big[1-e^{-\alpha n^2}\big]$ and $\big[1-e^{-\alpha (n^2-\tilde{p}^2)}\big]$ factors, which is then approximated, up to $\mathcal{O}(e^{-\pi^2/\alpha})$ corrections, by $\sum_{\bm{n}}\to\int d^3n$ with values at the poles removed. This gives
\begin{align}
\nonumber
\mathcal{X}_3(\tilde{p}^2) \, &= \, 
\sum_{\substack{
\bm{n}\neq 0\\ 
n^2\neq \tilde{p}^2}}
\frac{\big[e^{-\alpha(n^2-\tilde{p}^2)}+ e^{-\alpha n^2}-e^{-\alpha(2n^2-\tilde{p}^2)}\big]}{n^2(n^2-\tilde{p}^2)}\\
&+ \int d^3n \; \frac{\big[1-e^{-\alpha n^2}\big]\big[1-e^{-\alpha(n^2-\tilde{p}^2)}\big]}{n^2(n^2-\tilde{p}^2)}
-\frac{2\alpha}{\tilde{p}^2}\sinh{(\alpha \tilde{p}^2)} + \mathcal{O}(e^{-\pi^2/\alpha}).
\label{eq: TSS on X3}
\end{align}
The convergence is obtained as $\alpha\to0^+$ and the converged value is independent of $\alpha$ up to exponential corrections in $1/\alpha$.
For $\tilde{p}=1$ and $\alpha$ as large as 0.1, Eq.~\eqref{eq: TSS on X3} converges to $\mathcal{X}_3(1)=14.7$. $\alpha=0.01$ gives $\mathcal{X}_3(1)=14.702$, which is in agreement with Eq.~(A1) of Ref.~\cite{Beane:2014qha} up to five significant figures.

This method can be extended to double sums as demonstrated in Ref.~\cite{Tan:2007bg}. $\mathcal{R}$ defined in Eq.~\eqref{eq: definition R} has been evaluated using TSS in Eq.~(30) of Ref.~\cite{Beane:2014qha}:
\begin{equation}\label{eq: value of R}
\mathcal{R}=-178.42 .
\end{equation}

For the special case of $\tilde{p}_1=\tilde{p}_2=\tilde{p}$, $\mathcal{X}_1$ in Eq.~\eqref{eq: definition X6} straightforwardly evaluates to $27/2$.
On the other hand, $\mathcal{X}_6$ can be rewritten as
\begin{equation}
\label{eq: X6 expanded}
\mathcal{X}_6(\tilde{p}^2,\tilde{p}^2) =  
\, \sum_{\substack{
\bm{n}_1\neq 0\\ 
n_1^2\neq \tilde{p}^2}}
\,\frac{2 \, \tilde{p}^2\,\mathcal{X}_6^{(1)}(\bm{n}_1,\tilde{p}^2)-\tilde{p}^4\,\mathcal{X}_6^{(2)}(\bm{n}_1,\tilde{p}^2)}
{n_1^2\,(n_1^2-\tilde{p}^2)},
\end{equation}
with
\begin{align}
\label{eq: definition X6-1 and X6-2}
\mathcal{X}_6^{(1)}(\bm{n}_1,\tilde{p}^2) = 
\sum_{\substack{
\bm{n}_2\neq 0,\bm{n}_1\\ 
n_2^2\neq \tilde{p}^2}}
\frac{1}{(n_2^2-\tilde{p}^2)|\bm{n}_1-\bm{n}_2|^2} ,
\qquad
\mathcal{X}_6^{(2)}(\bm{n}_1,\tilde{p}^2) =
\sum_{\substack{
\bm{n}_2\neq 0,\bm{n}_1\\ 
n_2^2\neq \tilde{p}^2}}
\frac{1}{n_2^2\,(n_2^2-\tilde{p}^2)|\bm{n}_1-\bm{n}_2|^2}.
\end{align}
TSS can now be used on $\mathcal{X}_6^{(1)}(\bm{n}_1,\tilde{p}^2)$ and $\mathcal{X}_6^{(2)}(\bm{n}_1,\tilde{p}^2)$ in Eq.~\eqref{eq: definition X6-1 and X6-2} just like in Eq.~\eqref{eq: TSS on X3}:
\begin{align}
\nonumber
\mathcal{X}_6^{(1)}(\bm{n}_1,\tilde{p}^2) &=
\sum_{\substack{
\bm{n}_2\neq 0,\bm{n}_1\\ 
n_2^2\neq p^2}}
\frac{1}{(n_2^2-\tilde{p}^2)|\bm{n}_1-\bm{n}_2|^2}\big[e^{-\alpha |\bm{n}_1-\bm{n}_2|^2}+e^{-\alpha (n_2^2-\tilde{p}^2)}-e^{-\alpha (|\bm{n}_1-\bm{n}_2|^2+n_2^2-\tilde{p}^2)}\big]
\\
\label{eq: TSS for X6-1}
&+\mathcal{P}\,\int\,d^3n_2 \, 
\frac{\big[1-e^{-\alpha |\bm{n}_1-\bm{n}_2|^2}\big]}{|\bm{n}_1-\bm{n}_2|^2}
\frac{\big[1-e^{-\alpha (n_2^2-\tilde{p}^2)}\big]}{n_2^2-\tilde{p}^2}
-\alpha\frac{\big[1-e^{-\alpha (n_1^2-\tilde{p}^2)}\big]}{n_1^2-\tilde{p}^2}
+\mathcal{O}(e^{-\pi^2/\alpha}),\\
\nonumber
\mathcal{X}_6^{(2)}(\bm{n}_1,\tilde{p}^2) &=
\sum_{\substack{
\bm{n}_2\neq 0,\bm{n}_1\\ 
n_2^2\neq \tilde{p}^2}}
\frac{1}{n_2^2(n_2^2-\tilde{p}^2)|\bm{n}_1-\bm{n}_2|^2}
\bigg[ e^{-\alpha |\bm{n}_1-\bm{n}_2|^2}
+ e^{-\alpha (n_2^2-\tilde{p}^2)}+e^{-\alpha n_2^2}
\\
\nonumber
& + e^{-\alpha (|\bm{n}_1-\bm{n}_2|^2+2n_2^2-\tilde{p}^2)}-e^{-\alpha( |\bm{n}_1-\bm{n}_2|^2+n_2^2)}
-e^{-\alpha (|\bm{n}_1-\bm{n}_2|^2+n_2^2-\tilde{p}^2)}
-e^{-\alpha (2n_2^2-\tilde{p}^2)}
\bigg]
\\
\nonumber
&+\mathcal{P}\,\int\,d^3n_2 \, 
\frac{\big[1-e^{-\alpha |\bm{n}_1-\bm{n}_2|^2}\big]}{|\bm{n}_1-\bm{n}_2|^2}
\frac{\big[1-e^{-\alpha n_2^2}\big]}{n_2^2}
\frac{\big[1-e^{-\alpha (n_2^2-\tilde{p}^2)}\big]}{n_2^2-\tilde{p}^2}\\
\label{eq: TSS for X6-2}
&+\alpha\frac{\big[1-e^{\alpha \tilde{p}^2}\big]}{\tilde{p}^2}\frac{\big[1-e^{-\alpha n_1^2}\big]}{n_1^2}
-\alpha\frac{\big[1-e^{-\alpha (n_1^2-\tilde{p}^2)}\big]}{n_1^2-\tilde{p}^2}\frac{\big[1-e^{-\alpha n_1^2}\big]}{n_1^2}
+\mathcal{O}(e^{-\pi^2/\alpha}).
\end{align}
Here, $\mathcal{P}$ denotes the Cauchy principal value of the radial integration in $\bm{n}_2$ for the pole $n_2^2=\tilde{p}^2$.
Using Eqs.~\eqref{eq: TSS for X6-1} and \eqref{eq: TSS for X6-2} in Eq.~\eqref{eq: X6 expanded}, the outer sum over $\bm{n}_1$ is then split into two parts: terms containing exponentially suppressed terms in $\alpha$ and terms independent of $\alpha$.
The former can be summed directly while the latter, which stems from evaluating the $\alpha$-independent integrals in Eq.~\eqref{eq: TSS for X6-1} and \eqref{eq: TSS for X6-2}, can be calculated using TSS just like Eq.~\eqref{eq: TSS on X3}. As an example, with this procedure, Eq.~\eqref{eq: X6 expanded} for $\tilde{p}^2=1$ evaluates to
\begin{equation}
\mathcal{X}_6(1,1) = 264,
\end{equation}
which agrees with Eq.~(A10) of Ref.~\cite{Beane:2014qha} up to three significant figures.
Arbitrary accuracy can be achieved by decreasing the value of $\alpha$ and increasing the number of integer-triplets used in the sums with increasing magnitude.

\renewcommand{\thechapter}{B}
\renewcommand{\chaptername}{Appendix}

\chapter{Numerical values associated with the figures in Sec.~\ref{sec: DBD sensitivity analysis}
\label{app: numerical tables for sensitivity analysis}
}

Tables~\ref{tab: E and R table}-\ref{tab: gvNN} below contain many representative numerical values associated with the plots throughout the main text.
\begin{table}[h!]
    \renewcommand{\arraystretch}{1.5}
    \centering
    \begin{tabular}{C{1cm}|C{1.25cm}|C{1.75cm}|C{1.25cm}|C{1.75cm}|C{1.25cm}|C{1.75cm}|C{1.25cm}|C{1.75cm}}
        \hline
        \hline
        \rule{0pt}{4ex}
         $L$ & $E_0$ & $\left| \mathcal{R}(E_0) \right|$ & $E_1$ & $\left| \mathcal{R}(E_1) \right|$ &  $\widetilde{E}_0$ & $\left| \widetilde{\mathcal{R}}(\widetilde{E}_0) \right|$ & $\widetilde{E}_1$ & $\left| \widetilde{\mathcal{R}}(\widetilde{E}_1) \right|$ \\
         $\left[\text{fm}\right]$ &  $\left[\text{MeV}\right]$ &  $\left[\text{MeV}^3\right]$ &   $\left[\text{MeV}\right]$ & $\left[\text{MeV}^3\right]$ & $\left[\text{MeV}\right]$ & $\left[\text{MeV}^3\right]$ &  $\left[\text{MeV}\right]$ & $\left[\text{MeV}^3\right]$ \\
         \hline
         8 & -2.728 & $5.04\times 10^3$ & 19.043 & $9.73\times 10^4$ & -5.579 & $1.88\times 10^3$ & 13.688 & $1.01\times 10^5$ \\
         10 & -1.618 & $2.62\times 10^3$ & 11.606 & $4.87\times 10^4$ & -4.004 & $6.90\times 10^2$ & 7.364 & $4.94\times 10^4$ \\
         12 & -1.067 & $1.55\times 10^3$ & 7.772 & $2.75\times 10^4$ & -3.218 & $2.65\times 10^2$ & 4.299 & $2.70\times 10^4$ \\
         14 & -0.752 & $9.93\times 10^2$ & 5.560 & $1.69\times 10^4$ & -2.788 & $1.02\times 10^2$ & 2.655 & $1.59\times 10^4$ \\
         16 & -0.556 & $6.78\times 10^2$ & 4.176 & $1.11\times 10^4$ & -2.544 & $3.77\times 10^1$ & 1.712 & $9.85\times 10^3$ \\
         \hline
         \hline
    \end{tabular}
    \caption{Numerical values of the FV ground- and first excited-state energies in the $^1S_0$ and $^3S_1$ channels for a range of $L$ values along with the absolute values of the corresponding LL residue functions evaluated at those energies.}
    \label{tab: E and R table}
\end{table}
\begin{table}[hbt!]
    \renewcommand{\arraystretch}{1.75}
    \centering
    \begin{tabular}{C{1.5cm}|C{1.5cm}|C{2.25cm}|C{2.25cm}|C{2.25cm}|C{2.25cm}|C{2.25cm}}
        \hline
        \hline
        &  \multirow{2}{*}{$\Delta_{E(\widetilde{E})}$} & \multicolumn{5}{C{11.25cm}}{$L\;\left[\text{fm}\right]$}\\
        \cline{3-7}
        & & 8 & 10 & 12 & 14 & 16\\
        \hline
         \multirow{3}{*}{\rotatebox[origin=c]{90}{$a^{-1}\;\left[\text{fm}^{-1}\right]$}} & $1\%$ & $-0.043_{-0.004}^{+0.004}$ &	$-0.043_{-0.003}^{+0.004}$ &	$-0.043_{-0.003}^{+0.003}$ &	$-0.043_{-0.003}^{+0.003}$ &	$-0.043_{-0.002}^{+0.002}$ \\
        & $5\%$ & $-0.044_{-0.023}^{+0.022}$ &	$-0.044_{-0.017}^{+0.018}$ &	$-0.043_{-0.015}^{+0.014}$ &	$-0.043_{-0.014}^{+0.012}$ &	$-0.044_{-0.011}^{+0.011}$ \\
        & $10\%$ & $-0.049_{-0.052}^{+0.044}$ &	$-0.047_{-0.039}^{+0.035}$ &	$-0.046_{-0.032}^{+0.029}$ &	$-0.045_{-0.030}^{+0.024}$ &	$-0.045_{-0.024}^{+0.021}$ \\
        \hline
        \multirow{3}{*}{\rotatebox[origin=c]{90}{$r_0\;\left[\text{fm}\right]$}} & $1\%$ & $2.751_{-0.097}^{+0.097}$ &	$2.751_{-0.103}^{+0.105}$ &	$2.745_{-0.112}^{+0.119}$ &	$2.753_{-0.130}^{+0.125}$ &	$2.743_{-0.136}^{+0.136}$ \\
        & $5\%$ &
        $2.753_{-0.455}^{+0.533}$ &	$2.751_{-0.488}^{+0.562}$ &	$2.721_{-0.545}^{+0.620}$ &	$2.762_{-0.626}^{+0.666}$ &	$2.715_{-0.668}^{+0.696}$ \\
        & $10\%$&
        $2.751_{-0.839}^{+1.228}$ &	$2.744_{-0.934}^{+1.222}$ &	$2.689_{-1.057}^{+1.362}$ &	$2.771_{-1.254}^{+1.407}$ &	$2.670_{-1.346}^{+1.477}$ \\
         \hline
         \multirow{3}{*}{\rotatebox[origin=c]{90}{$\widetilde{a}^{-1}\;\left[\text{fm}^{-1}\right]$}} & $1\%$ & $0.184_{-0.002}^{+0.002}$ &	$0.184_{-0.002}^{+0.002}$ &	$0.184_{-0.001}^{+0.001}$ &	$0.184_{-0.001}^{+0.001}$ &	$0.184_{-0.001}^{+0.001}$ \\
         & $5\%$ & 
        $0.184_{-0.011}^{+0.012}$ &	$0.184_{-0.009}^{+0.008}$ &	$0.184_{-0.007}^{+0.007}$ &	$0.184_{-0.006}^{+0.006}$ &	$0.184_{-0.005}^{+0.005}$ \\
        & $10\%$ & 
        $0.182_{-0.024}^{+0.023}$ &	$0.183_{-0.018}^{+0.017}$ &	$0.184_{-0.014}^{+0.014}$ &	$0.183_{-0.011}^{+0.012}$ &	$0.184_{-0.010}^{+0.012}$ \\
        \hline
        \multirow{3}{*}{\rotatebox[origin=c]{90}{$\widetilde{r}_0\;\left[\text{fm}\right]$}} & $1\%$ & 
        $1.750_{-0.030}^{+0.028}$ &	$1.749_{-0.030}^{+0.031}$ &	$1.748_{-0.033}^{+0.033}$ &	$1.751_{-0.037}^{+0.034}$ &	$1.747_{-0.037}^{+0.038}$ \\
        & $5\%$ &
        $1.751_{-0.150}^{+0.145}$ &	$1.746_{-0.154}^{+0.150}$ &	$1.740_{-0.171}^{+0.160}$ &	$1.752_{-0.189}^{+0.161}$ &	$1.732_{-0.192}^{+0.187}$ \\
        & $10\%$ &
        $1.747_{-0.303}^{+0.296}$ &	$1.739_{-0.318}^{+0.299}$ &	$1.722_{-0.364}^{+0.318}$ &	$1.745_{-0.403}^{+0.313}$ &	$1.710_{-0.418}^{+0.355}$ \\
         \hline
         \hline
    \end{tabular}
    \caption{Numerical values associated with Fig. \ref{fig: band ERE}.}
    \label{tab: ERE}
\end{table}
\begin{table}[hbt!]
    \renewcommand{\arraystretch}{1.75}
    \centering
    \begin{tabular}{C{0.75cm}|C{1cm}|C{1.5cm}|C{2.25cm}|C{2.25cm}|C{2.25cm}|C{2.25cm}|C{2.25cm}}
        \hline
        \hline
        &  \multirow{2}{*}{$\Delta_\beta$} & \multirow{2}{*}{$\Delta_{E(\widetilde{E})}$} & \multicolumn{5}{C{11.25cm}}{$L\;\left[\text{fm}\right]$}\\
        \cline{4-8}
        & & & 8 & 10 & 12 & 14 & 16\\
        \hline
        \multirow{12}{*}{\rotatebox[origin=c]{90}{$\widetilde{L}_{1,A}\;\left[\text{MeV}\right]$}} &  \multirow{2}{*}{$1\%$}  & \multirow{2}{*}{$1\%$}
        & $-449.4_{-47.0}^{+47.9}$	& $-449.3_{-59.5}^{+70.1}$	& $-447.7_{-82.8}^{+90.5}$	& $-452.4_{-112.6}^{+122.3}$	& $-446.6_{-153.7}^{+134.0}$\\
         & & & $-432.3_{-274.0}^{+208.0}$	& $-451.8_{-136.0}^{+133.8}$	& $-445.1_{-107.1}^{+88.2}$	& $-446.4_{-84.2}^{+71.6}$	& $-450.6_{-68.9}^{+69.3}$\\
         \cline{2-8}
         &  \multirow{2}{*}{$1\%$}  & \multirow{2}{*}{$5\%$}
         & $-451.3_{-135.0}^{+143.0}$	& $-448.0_{-175.6}^{+178.1}$	& $-433.9_{-207.3}^{+205.8}$	& $-441.7_{-257.3}^{+257.3}$	& $-434.5_{-263.4}^{+242.6}$\\
         & & & $-455.4_{-413.1}^{+366.4}$	& $-445.4_{-273.1}^{+223.7}$	& $-444.2_{-206.7}^{+195.2}$	& $-444.5_{-182.8}^{+173.9}$	& $-440.9_{-170.8}^{+153.7}$\\
         \cline{2-8}
         &  \multirow{2}{*}{$1\%$}  & \multirow{2}{*}{$10\%$}
        & $-454.9_{-262.6}^{+273.4}$	& $-456.9_{-324.7}^{+354.6}$	& $-429.9_{-385.6}^{+404.6}$	& $-435.5_{-457.6}^{+465.0}$	& $-415.1_{-470.1}^{+437.5}$ \\
        & & & $-444.0_{-802.8}^{+579.0}$	& $-455.1_{-462.3}^{+407.5}$	& $-447.2_{-376.1}^{+350.8}$	& $-438.4_{-335.8}^{+323.5}$	& $-428.8_{-323.0}^{+286.1}$\\
         \cline{2-8}
         &  \multirow{2}{*}{$5\%$}  & \multirow{2}{*}{$1\%$}& $-439.4_{-195.4}^{+196.8}$	& $-441.5_{-274.8}^{+300.0}$	& $-458.7_{-359.3}^{+447.7}$	& $-466.8_{-541.2}^{+562.6}$	& $-474.5_{-720.3}^{+703.8}$ \\
         & & & $-403.0_{-1952.7}^{+764.7}$	& $-473.3_{-966.3}^{+532.3}$	& $-434.4_{-630.7}^{+358.1}$	& $-440.1_{-527.2}^{+292.6}$	& $-457.3_{-389.6}^{+259.7}$\\
         \cline{2-8}
         &  \multirow{2}{*}{$5\%$}  & \multirow{2}{*}{$5\%$}
        & $-449.2_{-231.1}^{+241.0}$	& $-456.3_{-283.3}^{+363.6}$	& $-450.3_{-397.9}^{+477.0}$	& $-473.6_{-538.8}^{+634.4}$	& $-444.0_{-750.4}^{+681.4}$ \\
        & & & $-387.0_{-1963.6}^{+774.5}$	& $-470.1_{-1039.2}^{+561.2}$	& $-430.1_{-666.8}^{+390.7}$	& $-435.1_{-517.4}^{+316.5}$	& $-467.0_{-397.2}^{+320.6}$\\
         \cline{2-8}
         &  \multirow{2}{*}{$10\%$}  & \multirow{2}{*}{$10\%$}
        & $-466.5_{-433.3}^{+505.0}$	& $-471.0_{-538.4}^{+773.6}$	& $-469.7_{-770.4}^{+991.4}$	& $-522.2_{-1043.3}^{+1300.8}$	& $-474.3_{-1423.8}^{+1407.7}$\\
        & & & $-345.5_{-2191.0}^{+1207.2}$	& $-477.2_{-1981.7}^{+916.3}$	& $-417.7_{-2036.4}^{+690.5}$	& $-423.6_{-1705.4}^{+568.6}$	& $-487.1_{-1167.7}^{+573.3}$\\
        \hline
        \hline
    \end{tabular}
    \caption{Numerical values associated with Fig. \ref{fig: band L1A}. The top (bottom) value in each cell corresponds to the ground-state to ground-state (first excited-state to first excited-state) transition.}
    \label{tab: L1A}
\end{table}
\begin{table}[hbt!]
    \renewcommand{\arraystretch}{1.5}
    \centering
    \begin{tabular}{C{1cm}|C{2.5cm}|C{2cm}|C{2.5cm}|C{2cm}|C{2.5cm}}
        \hline
        \hline
         $L$ & $J^{\infty}(E_0,E_0;m_\pi)$ & $\delta J^V(E_0,E_0)$ & $\big| \mathcal{M}^{\rm (Int.)}_{nn\to pp}\big|$ & $\big| \mathcal{M}^{0\nu,V}_{nn\to pp}\big|$& $\big|\mathcal{T}_L^{(M)}\big|$\\
         $\left[\text{fm}\right]$ & $\left[\text{MeV}^2\right]$ & $\left[\text{MeV}^2\right]$ & $\left[\text{MeV}^{-1}\right]$ & $\left[\text{MeV}^{-1}\right]$ & $\left[\text{MeV}^5\right]$ \\
         \hline
        8 & $9.84\times 10^3$ & $-1.2\times 10^3$ & $2.2\times 10^{-3}$ & $1.86\times 10^{-3}$ & $1.4\times 10^5$ \\
         10 & $1.13\times 10^4$ & $-1.\times 10^3$ & $4.1\times 10^{-3}$ & $3.64\times 10^{-3}$ & $7.3\times 10^4$ \\
         12 & $1.25\times 10^4$ & $-9.\times 10^2$ & $6.5\times 10^{-3}$ & $5.93\times 10^{-3}$ & $4.1\times 10^4$ \\
         14 & $1.34\times 10^4$ & $-7.8\times 10^2$ & $9.4\times 10^{-3}$ & $8.71\times 10^{-3}$ & $2.4\times 10^4$ \\
         16 & $1.43\times 10^4$ & $-6.8\times 10^2$ & $1.3\times 10^{-2}$ & $1.19\times 10^{-2}$ & $1.5\times 10^4$ \\
         \hline
         \hline
    \end{tabular}
    \caption{Numerical values of the finite- and infinite-volume quantities in the matching relation for the $0\nu\beta\beta$ process in Eq. \eqref{eq: matching relation gvNN}. These quantities are evaluated at the ground-state FV energy eigenvalues in the corresponding volumes.}
    \label{tab: 0vbb Table}
\end{table}
\begin{table}[]
    \renewcommand{\arraystretch}{1.75}
    \centering
    \begin{tabular}{C{1cm}|C{1cm}|C{1.25cm}|C{1.5cm}|C{1.5cm}|C{1.5cm}|C{1.5cm}|C{1.5cm}}
        \hline
        \hline
        &  \multirow{2}{*}{$\Delta_{\beta\beta}$} & \multirow{2}{*}{$\Delta_{E}$} & \multicolumn{5}{C{7.5cm}}{$L\;\left[\text{fm}\right]$}\\
        \cline{4-8}
        & & & 8 & 10 & 12 & 14 & 16\\
        \hline
         \multirow{6}{*}{\rotatebox[origin=c]{90}{$\widetilde{g}_{\nu}^{NN}$}} &  $1\%$  & $1\%$
        & $1.3_{-0.1}^{+0.1}$
        & $1.3_{-0.1}^{+0.1}$
        & $1.3_{-0.1}^{+0.1}$
        & $1.3_{-0.1}^{+0.1}$
        & $1.3_{-0.1}^{+0.1}$\\
         \cline{2-8}
        &  $1\%$  & $5\%$
        & $1.3_{-0.2}^{+0.2}$
        & $1.3_{-0.2}^{+0.2}$
        & $1.3_{-0.2}^{+0.2}$
        & $1.3_{-0.3}^{+0.3}$
        & $1.3_{-0.3}^{+0.3}$\\
        \cline{2-8}
        &  $1\%$  & $10\%$
        & $1.3_{-0.3}^{+0.3}$
        & $1.3_{-0.4}^{+0.4}$
        & $1.3_{-0.5}^{+0.5}$
        & $1.3_{-0.6}^{+0.5}$
        & $1.2_{-0.6}^{+0.6}$\\
         \cline{2-8}
        &  $5\%$  & $1\%$
        & $1.3_{-0.2}^{+0.2}$
        & $1.3_{-0.2}^{+0.2}$
        & $1.3_{-0.2}^{+0.2}$
        & $1.3_{-0.3}^{+0.2}$
        & $1.3_{-0.3}^{+0.3}$\\
         \cline{2-8}
         &  $5\%$  & $5\%$
        & $1.3_{-0.3}^{+0.2}$
        & $1.3_{-0.3}^{+0.3}$
        & $1.3_{-0.3}^{+0.3}$
        & $1.3_{-0.4}^{+0.3}$
        & $1.2_{-0.4}^{+0.4}$\\
         \cline{2-8}
        &  $10\%$  & $10\%$
        & $1.3_{-0.6}^{+0.5}$
        & $1.3_{-0.7}^{+0.6}$
        & $1.2_{-0.7}^{+0.6}$
        & $1.2_{-0.8}^{+0.7}$
        & $1.2_{-0.9}^{+0.8}$\\
        \hline
        \hline
    \end{tabular}
    \caption{Numerical values associated with Fig. \ref{fig: band gvNN}.}
    \label{tab: gvNN}
\end{table}
\renewcommand{\thechapter}{C}
\renewcommand{\chaptername}{Appendix}

\chapter{Purely fermionic formulation of SU(3) gauge theory
\label{app: purely fermionic formulation}
}
%
\begin{table}[t]
    \renewcommand{\arraystretch}{1.5}
    \centering
    \begin{tabular}{C{1cm}  C{2.5cm}  C{2cm} C{2.5cm}}
        \hline
        \hline
        \rule{0pt}{2ex}
        $\mathcal{F}$ & $(\mathcal{P}_f, \mathcal{Q}_f)$ & $d(\mathcal{P}_f, \mathcal{Q}_f)$ & Eigenvalue\\
        \hline
        0 & (0, 0) & 1 & 0.000 \\
        \hline
        \multirow{2}{*}{1} & \multirow{2}{*}{(1, 0)} & \multirow{2}{*}{3} & -0.387 \\
        & & & 1.721\\
        \hline
        \multirow{3}{*}{2} & \multirow{3}{*}{(0, 1)} & \multirow{3}{*}{3} & -1.535 \\
        & & & 0.868 \\
        & & & 3.333 \\
        \hline
        2 & (2, 0) & 6 & 1.333 \\
        \hline
        \multirow{4}{*}{3} & \multirow{4}{*}{(0, 0)} & \multirow{4}{*}{1} & -3.858 \\
        & & & -0.497 \\
        & & & 2.137 \\
        & & & 4.884 \\
        \hline
        \multirow{2}{*}{3} & \multirow{2}{*}{(1, 1)} & \multirow{2}{*}{8} & -0.081 \\
        & & & 2.747\\
        \hline
        \multirow{3}{*}{4} & \multirow{3}{*}{(1, 0)} & \multirow{3}{*}{3} & -2.562 \\
        & & & 1.089 \\
        & & & 4.140 \\
        \hline
        4 & (0, 2) & 6 & 1.333 \\
        \hline
        \multirow{2}{*}{5} & \multirow{2}{*}{(0, 1)} & \multirow{2}{*}{3} & -1.277 \\
        & & & 2.610\\
        \hline
        6 & (0, 0) & 1 & 0.000 \\
        \hline
    \end{tabular}
    \caption{Energy eigenvalues of the Hamiltonian in Eqs.~\eqref{eq: HM in LSH operators}-\eqref{eq: HI in LSH operators} with $N=2$ sites and $\mu= x = 1$ are shown in this table. Without any loss of generality, we  consider zero background flux, i.e. choosing $E^{\rm a}(R,0)=0$, $\forall a$ in Eq.~\eqref{eq: HE-ferm}.
    The eigenvalues are categorized according to according to total fermion number $\mathcal{F}$ and global charges $(\mathcal{P}_f,\mathcal{Q}_f)$ of the corresponding eigenstates.
    The dimension $d(\mathcal{P}_f,\mathcal{Q}_f)$ of the irrep $(\mathcal{P}_f,\mathcal{Q}_f)$ is given by Eq.~\eqref{eq: dim of irrep}, which is also the degeneracy of the eigenvalues in the purely fermionic formulation.
    Thus, each eigenvalue when counted with its corresponding degeneracy factor add up to $8^N=64$, which is the dimension of the Hilbert space of this system in the purely fermionic formulation.
    In the LSH framework, each eigenvalue in each global symmetry sector  has only one eigenstate as explained in Sec.~\ref{subsec: result-numerics}. \label{tab: Eigenvalues}}
\end{table}
The LSH Hamiltonian along with the associated Hilbert space presented in section \ref{sec: LSH-framework} reformulates the KS Hamiltonian in Eq.~\eqref{eq: KS-ham} in terms of Schwinger bosons and staggered fermions.
However, there are other formulations of the same Hamiltonian in terms of different degrees of freedom~\cite{Farrell:2022wyt, Farrell:2022vyh, Atas:2022dqm}.
We present one such formulation in 1+1D with open boundary condition in which the gauge degrees of freedom are removed by working with a particular gauge choice, and the resulting Hamiltonian has only colored fermionic degrees of freedom.
This is known as the gauge fixed formulation or purely fermionic formulation.

Under the Gauss's law constraint in Eq.~\eqref{eq: gauss-law}, the physical Hilbert space of the SU(3) KS Hamiltonian in 1+1D has no dynamical gauge degrees of freedom except for the possible closed flux loops in the case of a periodic boundary condition.
Thus, imposing the open boundary condition, where the closed loops are absent, and fixing the incoming right chromo-electric flux at the first site, denoted by $E^{\rm a}(R,0)$, the chromo-electric field in the entire lattice is completely determined for a given fermionic configuration.
Equation~\eqref{eq: gauss-law} can then be used with a suitable choice of gauge fixing to eliminate all the intermediate gauge-link degrees of freedom as shown below.
As a result, the SU(3) Kogut-Susskind Hamiltonian acting on the physical Hilbert space can be expressed in terms of purely fermionic degrees of freedom and the left boundary ${E}^{\rm a}(R,0)$ flux.
The derivation shown here is extension of the similar formulation for SU(2) theories given in Refs.~\cite{Atas:2021ext,Davoudi:2020yln}. 

Consider the Hamiltonian in Eq.~\eqref{eq: KS-ham} with the following gauge transformation of the fermionic fields:
\begin{equation}
    \psi(r)\to\psi'(r) = \left[\prod_{y<r}U(y)\right] \psi(r),~~\quad~~
    \psi^\dagger(r)\to\psi^{\dagger'}(r)= \psi^\dagger(r)\left[\prod_{y<r}U(y)\right]^\dagger,
    \label{eq: gauge-transform-ferm}
\end{equation}
where $\prod_{y<r}U(y) = U(1)U(2)\cdots U(r-1)$. The corresponding gauge transformation of the gauge links is given by
\begin{equation}
    U(r)\to U'(r) =  \left[\prod_{y<r}U(y)\right]U(r)\left[\prod_{z<r+1}U(z)\right]^\dagger,
    \label{eq: gauge-transform-U}
\end{equation}
such that the gauge-matter interaction term, $H_{I}$, in Eq.~\eqref{eq: KS-ham} remains invariant. Since the gauge links satisfy the unitarity condition $U(r)U^\dagger(r)=\mathds{1}_{3\times3}$, $U'(r)$ in Eq.~\eqref{eq: gauge-transform-U} simplifies to $U'(r)=\mathds{1}_{3\times3}$.
This transforms $H_{I}$ to
\begin{equation}
    H^{\text(F)}_{I} = \sum_{r=1}^{N-1}\psi^{\dagger'}(r)\cdot \psi'(r+1) + \text{H.c.},
    \label{eq: HI-ferm}
\end{equation}
where the superscript $F$ denotes the purely fermionic formulation.
Similarly, the mass term, $H_M$, in Eq.~\eqref{eq: KS-ham} is given by
\begin{equation}
    H^{\text(F)}_M = \mu \sum_{r=1}^{N}(-1)^r\psi^{\dagger'}(r)\cdot \psi'(r).
    \label{eq: HM-ferm}
\end{equation}
Finally, the electric field energy term, $H_E$, in Eq.~\eqref{eq: KS-ham} can be re-expressed in terms of the fermionic field and the left boundary electric flux, $E^{\rm a}(R,0)$, using the Gauss's law stated in Eq.~\eqref{eq: gauss-law}.
For a physical state that satisfies Eq.~\eqref{eq: gauss-law}, the electric field operator to the right of any site $r>1$ can be evaluated by iteratively using the Gauss's law operator defined in Eq.~\eqref{eq: gauss-op}:
\begin{equation}
    \hat{E}^{\rm a}(L,r) = E^{\rm a}(R,0) + \sum_{r'=1}^{r-1} \psi^{\dagger}(r') T^{\rm a} \psi(r').
    \label{eq: Ex-gauge-fixed}
\end{equation}
Thus, $H_E$ in the purely fermionic formulation is given by
\begin{equation}
    H^{\text(F)}_E = \sum_{r=2}^{N-1}\sum_{{\rm a}=1}^8 \left[E^{\rm a}(R,0) + \sum_{r'=1}^{r-1} \psi^{\dagger}(r') T^{\rm a} \psi(r')\right]^2.
     \label{eq: HE-ferm}
\end{equation}
As a result, the final Hamiltonian given by Eqs.\eqref{eq: HI-ferm},~\eqref{eq: HM-ferm}, and~\eqref{eq: HE-ferm} has only the fermionic degrees of freedom and $E^{\rm a}(R,0)$ acts as the background chromo-electric field.
This Hamiltonian is identical to the original Kogut-Susskind Hamiltonian only in the physical Hilbert space.
Thus, any state in the physical Hilbert space in this formulation, $|\Psi\rangle^{(F)}$, can be characterized by its fermionic occupation number at each site:
\begin{equation}
    |\Psi\rangle^{(F)}=\goldieotimes_{r=1}^{N}|f_1,f_2,f_3\rangle_{r}= \prod_{r=1}^{N}(\psi^\dagger_1(r))^{f_1(r)}(\psi^\dagger_2(r))^{f_2(r)}(\psi^\dagger_3(r))^{f_3(r)}\goldieotimes_{r'=1}^{N}|0,0,0\rangle_{r'},
    \label{eq: Fermionic basis state}
\end{equation}
where the subscript $r$ on the state vector denotes the local state at site $r$, $f_{i}(r)=0$ or $1$ refers to the occupation number of the (anti)matter field at site $r$ with color index $i=1,2,3$, and the vacuum state $|0,0,0\rangle_r$ is defined by $\psi^\alpha |0,0,0\rangle_r =0 $ for $\alpha = 1,2,3$.

Equation~\eqref{eq: Fermionic basis state} implies that the dimension of the fermionic Hilbert space of an $N$ site lattice in this formulation is given by $8^N$. 
Furthermore, the fermionic creation operators, $\psi^\dagger_\alpha(r)$, transform in the $(1,0)$ irrep under the local SU(3) gauge group as indicated in Table~\ref{tab: prepotential irreps}, and satisfy anti-commutation relations in Eq.~\eqref{eq: ferm_anticomm}.
Thus, the states $|\Psi\rangle^{(F)}$ with total fermion number $\mathcal{F}$ has $\mathcal{F}$ color indices, and their linear combinations with appropriate Clebsh-Gordon coefficients form the multiplets that transform according to different $(\mathcal{P}_f,\mathcal{Q}_f)$ irreps under the global SU(3) transformations as discussed in Sec.~\ref{subsec: result-numerics}.
As an example, in Table~\ref{tab: Eigenvalues}, we show the degeneracy factors for energy eigenvalues for vanishing left boundary flux as shown in Fig.~\ref{subfig: eigenavlues}. 
The eigenvalues are categorized according to $\mathcal{F}$ and $(\mathcal{P}_f,\mathcal{Q}_f)$ global charges of their eigenstates with degeneracy factor for each eigenvalue given by $d(\mathcal{P}_f,\mathcal{Q}_f)$, which is $d(\Delta\mathcal{P},\Delta\mathcal{Q})$ for $(\mathcal{P}_0,\mathcal{Q}_0)=(0,0)$.
\renewcommand{\thechapter}{D}
\renewcommand{\chaptername}{Appendix}

\chapter{Additional LSH operator factorizations
\label{app: extra factorizations}
}

 In addition to the local SU(3) singlet operator factorizations that are directly relevant to expressing the Hamiltonian (see Eqs.~\eqref{eq: on site operator factorinzation}), one may wish to have a complete suite of on-site SU(3) singlet factorizations for constructing the local Hilbert space as originally given in Eqs.~\eqref{eq: LSH states unnormalized def}-\eqref{eq: Unnormalized LSH-basis-state}.
 The following factorizations supplement Eqs.~\eqref{eq: on site operator factorinzation} to complete that set:
\begin{align}
    \ApBp &\quad\mapsto\quad \hat{\ladder}_P^\dagger \sqrt{\hat{n}_P+1} \sqrt{\hat{n}_P+3-\delta_{\hnum, \hat{\nu}_1}} \sqrt{ 1 + \frac{1}{\hat{n}_P+\hat{n}_Q+3-\delta_{\hnui, \hnum}\delta_{\hnum , \hat{\nu}_1}} } \\
    \BpAp &\quad\mapsto\quad \hat{\ladder}_Q^\dagger \sqrt{\hat{n}_Q+1} \sqrt{\hat{n}_Q+3-\delta_{\hnui, \hnum}} \sqrt{ 1 + \frac{1}{\hat{n}_P+\hat{n}_Q+3-\delta_{\hnui, \hnum}\delta_{\hnum , \hat{\nu}_1}} } \\
    \tfrac{1}{2} \psi^\dagger \cdot \psi^\dagger \wedge A(\obar)^\dagger &\quad\mapsto\quad \hat{\chi}_{\obar}^\dagger \hat{\chi}_0^\dagger ( \hat{\Lambda}_P^+ )^{\hat{\nu}_1} \sqrt{\hat{n}_P+2-\hat{\nu}_1} \sqrt{1+\frac{1-\hat{\nu}_1}{\hat{n}_P+\hat{n}_Q+2}} \\
    \tfrac{1}{2} \psi^\dagger \cdot \psi^\dagger \wedge A(1)^\dagger &\quad\mapsto\quad \hat{\chi}_0^\dagger \hat{\chi}_1^\dagger ( \hat{\Lambda}_Q^+ )^{\hat{\nu}_{\obar}} \sqrt{\hat{n}_Q+2-\hat{\nu}_{\obar}} \sqrt{1+\frac{1-\hat{\nu}_{\obar}}{\hat{n}_P+\hat{n}_Q+2}} \\
    \psi^\dagger \cdot A(\obar)^\dagger \wedge A(1)^\dagger &\quad\mapsto\quad \hat{\chi}_0^\dagger (\hat{\Lambda}_P^+)^{\hat{\nu}_1} (\hat{\Lambda}_Q^+)^{\hat{\nu}_{\obar}} \sqrt{\hat{n}_P+2-\hat{\nu}_1} \sqrt{\hat{n}_Q+2-\hat{\nu}_{\obar}} \nonumber \\
    & \qquad \qquad \times \sqrt{1+\frac{1}{\hat{n}_P+\hat{n}_Q+2+\hat{\nu}_{\obar}+\hat{\nu}_1-\hat{\nu}_{\obar} \hat{\nu}_1}} 
\end{align}

\renewcommand{\baselinestretch}{1}
\small\normalsize

\addcontentsline{toc}{chapter}{Bibliography}
\bibliographystyle{apsrev4-1}
\nocite{apsrev41Control}
\bibliography{QCDandEFT,FiniteVolume,DBD,Hamiltonian,LQCD, revtex-custom} 

\end{document}